\documentclass[12pt,psf,epsf]{article}

\usepackage[centertags]{amsmath}
\usepackage{graphicx}
\usepackage{epsfig}
\usepackage{ulem}
\usepackage[english]{babel}
\usepackage{array}
\usepackage{amsthm}
\usepackage{latexsym}
\usepackage[mathcal]{euscript}
\pdfoutput=1
\usepackage{epsfig}
\usepackage{color}
\usepackage{jheppub}

\usepackage{hyperref}

\newcommand{\nn}{\nonumber}
\newcommand{\be}{\begin{equation}}
\newcommand{\ee}{\end{equation}}
\newcommand{\bea}{\begin{eqnarray}}
\newcommand{\eea}{\end{eqnarray}}

\newcommand{\fG}{\mathfrak{G}}
\newcommand{\fF}{\mathfrak{F}}

\newcommand{\fV}{\mathfrak{V}}

\newcommand{\bi}{\begin{itemize}}
\newcommand{\ei}{\end{itemize}}
\newcommand{\re}{\mathop{\mathrm{Re}}}

\title{Holographic Anisotropic Background with
  Confinement-Deconfinement Phase Transition}

\author{Irina Aref'eva$^a$ and Kristina Rannu$^{a,b}$}

\affiliation{$^a$Steklov Mathematical Institute, Russian Academy of
  Sciences,\\Gubkina str. 8, 119991, Moscow, Russia\\
  $^b$Peoples' Friendship University of Russia, Miklukho-Maklaya
  str.6, 117198, Moscow, Russia}

\emailAdd{arefeva@mi.ras.ru}

\emailAdd{rannukristina@gmail.com}

\abstract{We present new anisotropic black brane solutions in 5D
  Einstein-dilaton-two-Maxwell system. The anisotropic background is
  specified by an arbitrary dynamical exponent $\nu$, a nontrivial
  warp factor, a non-zero dilaton field, a non-zero time component of
  the first Maxwell field and a non-zero longitudinal magnetic
  component of the second Maxwell field. The blackening function
  supports the Van der Waals-like phase transition between small and
  large black holes for a suitable first Maxwell field charge. The
  isotropic case corresponding to $\nu = 1$ and zero magnetic field
  reproduces previously known solutions. We investigate the anisotropy
  influence on the thermodynamic properties of our background, in
  particular, on the small/large black holes phase transition
  diagram.

  We discuss applications of the model to the bottom-up holographic
  QCD. The RG flow interpolates between the UV section with two
  suppressed transversal coordinates and the IR section with the
  suppressed time and longitudinal coordinates due to anisotropic
  character of our solution. We study the temporal Wilson loops,
  extended in longitudinal and transversal directions, by calculating
  the minimal surfaces of the corresponding probing open string
  world-sheet in anisotropic backgrounds with various temperatures and
  chemical potentials. We find that dynamical wall locations depend on
  the orientation of the quark pairs, that gives a crossover
  transition line  between  confinement/deconfinement phases in the
  dual gauge theory. Instability of the background leads to the
  appearance of the critical points $(\mu_{\vartheta,b},
  T_{\vartheta,b})$ depending on the orientation $\vartheta$ of
  quark-antiquark pairs in respect to the heavy ions collision line.

\keywords{AdS/QCD, holography, phase transition}}

\begin{document}
\maketitle

\section{Introduction}  

Study of the phase diagram of QCD, as a function of temperature T and
chemical potential $\mu$, corresponding to baryon density or some
other conserved charge,  is one of the great modern challenge.  The
diagram of QCD displays a rich structure
\cite{Lattice,0402115}. Particularly interesting and important
features of the phase diagram are the nature of the chiral phase
transition, the location of the chiral critical point and its
properties. There are well known obstacles to study this problem by
usual tools, such as perturbative theory or lattice calculations.

The gravity/gauge duality provides an alternative tool for
understanding dynamics of the strong coupling system, where standard
methods are not enough. One such system is the quark-gluon plasma
(QGP) produced in the heavy-ions-collisions (HIC) \cite{Solana, IA,
  DeWolf}. It is believed that the QGP is created in a very short time
after the collision ($\tau_{\rm therm} \approx $ few $0.1 \, {\rm
  fm}/c$) and there are indications that during this time the QGP is
anisotropic \cite{Strickland:2013uga}. In \cite{AG} we have considered
a special anisotropic backgrounds, parametrized by the dynamical
exponent $\nu$, and found out that the shock wave model of HIC for
$\nu = 4.45$ reproduces the experimental energy dependence of the
total multiplicity, ${\cal M} \sim s^{0.155}$ \cite{1512.06104}. Note
that all attempts to reproduce this dependence in isotropic models
failed \cite{Gubser, Gubser:2009sx,AlvarezGaume:2008fx, Lin:2009pn,
  Albacete:2009ji, ABG, ABJ, Kovchegov:2009du, Kovchegov:2010zg, KT,
  APP,Arefeva:2014uqa, Ageev:2014mma}.
  
On the other hand, static holographic models perfectly reproduce
main properties of QCD \cite{Solana, 0812.0792, IA,
  DeWolf}. Therefore, it would be interesting to know how these
properties are changed in anisotropic backgrounds. Some of these
questions have been already addressed in \cite{Mateos:2011tv,Mateos:2011ix, Chernicoff:2012iq,Chernicoff:2012bu,Chernicoff:2012gu,Patino:2012py,Giataganas:2013lga, Jahnke:2013rca,Avila:2016mno,1503.02185,AGG,Aref'eva:2016doe,
  Arefeva:2016rob, Giataganas:2017koz} and refs therein. In
particular, in \cite{Arefeva:2016rob} the confinement-deconfinement 
crossover transition in the temperature-chemical potential plane has
been observed. However, in this paper, analogous to \cite{AZ} and many
others \cite{Karch:2006pv, White:2007tu, Pirner:2009gr, Karch:2010eg,
  1001.4414, 1008.3116}, it was not assumed that the metric is
supported by some Lagrangian. Let us remind, that for the isotropic
case, $\nu = 1$, the background used in \cite{AZ} was constructed in
\cite {1103.5389} as a solution of equations of motion in the
Einstein-dilaton theory. Later, this background has been generalized
for non-zero electro-field \cite{1301.0385, yang2015} in the
Einstein-dilaton-Maxwell theory in order to describe the
confinement-deconfinement phase transition holographically, (see also
\cite{1008.3116, 1001.4414, 1201.0820, 1209.4512, Li:2017tdz} and
early papers \cite{9902170,1012.1864,1108.2029}). The anisotropic
$\nu \neq 1$ black brane background with a trivial warp factor and a
non-zero electro-magnetic field has been constructed in \cite{AGG}. As
it already has been mentioned, the collision of shock waves in this
5-dimensional dual vacuum background gives the total multiplicity
dependence on energy in the form  ${\cal   M} \sim s^{0.155}$, which
reproduces the experimental dependence \cite{1512.06104}.
 
This paper is devoted to the construction of the anisotropic
5-dimensional background specified by an arbitrary dynamical exponent
$\nu$, a nontrivial warp factor, a non-zero time component of the
first Maxwell field and a non-zero longitudinal magnetic component of
the second Maxwell field. To see clearly possible new effects caused
by anisotropy, it is useful to deal with an explicit analytical
solution. For this purpose we take the particular case of the simplest
warp factor $b(z) = e^{\frac{c z^2}{2}}$. We find the dilaton
potential by the potential reconstruction method similar to the
isotropic case \cite{yang2015} (and refs therein). We show that only
$c \leq 0$ guarantees real solutions for the dilaton (compare with
\cite{APP} and refs therein). We construct the blackening function
that supports the Van der Waals-like phase transition between small
and large black holes for a suitable first Maxwell field charge. The
isotropic case corresponding to $\nu = 1$ and zero magnetic field
reproduces previously known solutions \cite{1301.0385, yang2015}. We
investigate the anisotropy influence on the thermodynamic properties
of our background, in particular, on the small/large black holes phase
transition diagram. We find that the anisotropy changes the location
of the domain of instability.

We also discuss applications of the model to the bottom-up holographic
QCD. We note that the RG flow interpolates between the UV section with
two transversal suppressed coordinates and the IR section with the
suppressed time and longitudinal coordinates due to anisotropic
character of our solution. We study the temporal Wilson loops extended
in longitudinal and transversal directions by calculating the minimal
surfaces of the corresponding probing open string world-sheet for
various values of temperature and chemical potential. We find that for
particular sets of the model parameters the dynamical wall
appears. The appearance of dynamical walls also depends on the
orientation of the temporal Wilson loop, that gives 
a  crossover transition line  between  confinement-deconfinement
phases in the dual gauge theory. This effect has been also observed in
the anisotropic model considered in \cite{Arefeva:2016rob}. In the
background, investigated in the present paper, there are two more
anisotropic effects. Namely, the instability of the background
restricts leads to the appearance of critical points
$(\mu_{\vartheta,b},T_{\vartheta,b})$. Each critical point is located
at intersection of the confinement/deconfinement open string phase
transition line and the small/large black holes phase transition line
of our background. The lines of the first type depend on orientation
of the quark-antiquark pairs and the lines of the second type are
fixed for the given anisotropy parameter $\nu$. In other words,
positions of the critical points $(\mu_{\vartheta,b},T_{\vartheta,b})$
depend on the orientation of quark-antiquark pair in respect to the
heavy ions collision line. Averaging on all possible orientations on
the quark-antiquark pairs, one gets a family of the critical
points. In our model it also happens, that the
confinement/deconfinement transition line, oriented along the
transversal direction, is below the small/large black holes phase
transition line. This means that near the small chemical potential the
small/large black hole transition line is hidden by the
confinement/deconfinement transition line for the pair of quarks
oriented in the transverse direction. Recall that a small/large black
holes transition line near the top of the holographic phase diagram is
treated as a problem, since this behavior is not supported neither by
experimental data nor by calculations performed in the framework of
effective theories. Let us also remind, that most
of the effective models suggest the existence of a QCD critical point
$(\mu_{CEP},T_{CEP})$ somewhere in the middle of the phase diagram,
where the crossover line becomes a first order transition line. There
were attempts to relate $(\mu_{CEP},T_{CEP})$ with the small/large
black holes background transition \cite{1012.1864, 1108.2029}, but
here  there is a problem with the first order phase transition in the
top of the QCD phase diagram, that is not present at the QCD phase
diagram \cite{Lattice,0402115}. The isotropic holographic model
improving this has been proposed in \cite{Li:2017tdz} just by removing
the small/large black holes background  transition, see also
\cite{Knaute:2017opk,Brehm:2017dmt}. We note, that the presence of the small/large black
holes background transition endows our anisotropic model by a rich
phase structure.

The paper is organized as follows. In Sect. \ref{Sect:BBAS} we
construct the anisotropic 5-dimensional solution with an arbitrary
dynamical exponent $\nu$, a nontrivial warp factor, a non-zero time
component of the first Maxwell field and a non-zero longitudinal
magnetic component of the second Maxwell field. In
Sect. \ref{Sec:c-nu} we consider exponential warp factors with
quadratic exponent and show that only negative definite quadratic form
guarantees the real solutions for the dilaton. In
Sect. \ref{Sec:Thermo} we discuss the thermodynamics of the
constructed background and find out the small/large black holes
transition line in the $(\mu,T)$-plane. Sections \ref{Sec:RG} and
\ref{Sec:Ph-Tr} are devoted to applications to QCD. In
Sect. \ref{Sec:RG} we shortly discuss the RG flows corresponding to
constructed solutions. In Sect. \ref{Sec:DDW} we find dynamical walls
corresponding to the temporal Wilson loops extended in the
longitudinal and transversal directions. In Sect. \ref{Sec:PTL} we
determine the relative position of the background and the
confinement-deconfinement phase transition lines and discuss the
corresponding critical points. In Appendix \ref{Sec:Appendix} we
derive E.O.M. and in Appendix \ref{Sec:simplest-sol} we present
simplest solutions, the black hole solutions for $c = 0$, with zero
and non-zero chemical potential, and the vacuum solution for $c < 0$
for completeness.

\section{Black brane  anisotropic solutions}\label{Sect:BBAS}

\subsection{The equations of motion and boundary conditions}

We consider a 5-dimensional Einstein-dilaton-two-Maxwell system. In
the Einstein frame the action of the system is specified as
\be
  S = \int \cfrac{d^5x}{16\pi G_5} \, \sqrt{-\det(g_{\mu\nu})} \left[
    R - \cfrac{f_1(\phi)}{4} \ F_{(1)}^2 -  \cfrac{f_2(\phi)}{4} \
    F_{(2)}^2 - \cfrac{1}{2} \ \partial_{\mu} \phi \partial^{\mu} \phi
    - V(\phi) \right], \label{action}
\ee
where $F_{(1)}^2$ and $F_{(2)}^2$ are the squares of the Maxwell
fields $F_{\mu\nu}^{(1)} = \partial_{\mu} A_{\nu} - \partial_{\nu}
A_{\mu}$ and $F_{\mu\nu}^{(2)} = q \ dy^1 \wedge dy^2$, $f_1(\phi)$
and $f_2(\phi)$ are the gauge kinetic functions associated with the
corresponding Maxwell fields, $V(\phi)$ is the potential of the scalar
field $\phi$.

We search the black brane solution in the anisotropic background. For
this purpose we use the metric ansatz in the following form:
\bea
  ds^2 = \frac{L^2 \, b(z)}{z^2} \left[ - \ g(z) dt^2 + dx^2 +
    z^{2-\frac{2}{\nu}} \left( dy_1^2 + dy_2^2 \right) +
    \cfrac{dz^2}{g(z)} \right], \label{eq:2.02} \\
  \phi = \phi(z), \qquad A_{\mu}^{(1)} = A_t (z)
  \delta_\mu^0, \label{eq:2.03} \\
  F_{\mu\nu}^{(2)} = q \ dy^1 \wedge dy^2,
\eea
where $b(z)$ is the warp factor and $g(z)$ is the blackening
function; we set the AdS radius $L = 1$ and all the quantities in
formulas and figures are presented in dimensionless units.

The variation of the action (\ref{action}) over metric components
$g_{\mu\nu}$ gives 4 independent equations, corresponding to 00-, 11-,
22- and 44-components of the Einstein tensor, that are presented in
the Appendix \ref{Sec:Appendix}. These equations can be transformed to
the following ones:
\bea
  g'' + g' \left( \cfrac{3 b'}{2 b} - \cfrac{1}{z} - \cfrac{2}{\nu z}
  \right) - \cfrac{z^2}{b} \, f_1 A_t^{\prime \,2} &=&
  0, \label{1-st-t} \\
  b'' - \frac{3 (b')^2}{2 b} + \frac{2 b'}{z} - \frac{4b}{3 \nu z^2}
  \left( 1 - \frac{1}{\nu} \right) + \frac{b}{3} \, (\phi')^2 &=&
  0, \label{2-nd-t} \\
  2 g' \left( 1 - \cfrac{1}{\nu} \right) + g \left( 1 - \cfrac{1}{\nu}
  \right) \left( \cfrac{3 b'}{b} - \cfrac{4}{z} - \cfrac{4}{\nu z}
  \right) + \cfrac{q^2 z^{- 1 + \frac{4}{\nu}}}{b} \, f_2 &=&
  0, \label{3-rd-t} \\
  - V - \frac{z^4 }{2 b^2}\,A_t^{\prime \, 2} f_1-\frac{3 z^2 b' g'}{2
    b^2} - \frac{3 z^2 g b^{\prime \, 2}}{b^3} + \frac{9 z g b'}{2 \nu
    b^2} + \frac{15 z g b'}{2b^2} + \frac{z g'}{\nu b} + \frac{2 z
    g'}{b} &+& \nn\\
  + \, \frac{z^2 g \phi^{\prime \,2}}{2b} - \frac{8 g}{\nu b} -
  \frac{4 g}{b} &=& 0. \label{4-th-t}
\eea
Here and below $' = d/dz$. The variation of the action \eqref{action}
over the scalar field $\phi$ and the components $A^{(1)}_{\mu}$ of the
first Maxwell field leads to the following EOM:
\bea
  \phi'' + \phi' \left( \cfrac{g'}{g} + \cfrac{3 b'}{2 b} - \cfrac{\nu
      + 2}{\nu z} \right) + \cfrac{z^2 A_t^{\prime \, 2}}{2 b g} \
  \cfrac{\partial f_1}{\partial \phi} - \cfrac{q^2 z^{- 2 +
      \frac{4}{\nu}}}{2 b g} \ \cfrac{\partial f_2}{\partial \phi} -
  \cfrac{b}{z^2 g} \ \cfrac{\partial V}{\partial \phi} &=&
  0, \label{eq:2.25} \\
  A_t'' + A_t' \left( \cfrac{b'}{2 b} + \cfrac{f_1'}{f_1} - \cfrac{2 -
      \nu}{\nu z}\right) &=& 0 \label{A0} 
\eea
The EOM for the second Maxwell field doesn't give any contribution
into system \eqref{1-st-t}--\eqref{A0} as its left-hand side is
identically zero:

\be
  \partial_{\mu} \left(\sqrt{-g} \, f_2(\phi) \, F^{\mu\nu}_{(2)}
  \right) = \partial_{4} \left(\sqrt{-g} \, f_2(\phi) \,
    F^{4\nu}_{(2)} \right) \equiv 0. \nn
\ee
We also impose the boundary conditions in the form:
\bea
  b(0) &=& 1 \label{bbc}\\
  g(0) &=& 1 \quad \mbox{and} \quad g(z_h) = 0, \label{gbc} \\
  A_t(0) &=& \mu \quad \mbox{and} \quad A_t(z_h) = 0, \label{Atbc}
\eea
where $z_h$ is the horizon. As to the scalar field, it is natural to
require that $\phi(z)$ is real for $0 < z \leq z_h$ and that
\be
  \phi(z_h) = 0. \label{phibc}
\ee

\subsection{Solutions with factor $b(z) = \exp{P(z)}$ and spatial
  anisotropy}\label{Sec:P}

One can use the following strategy to find particular solutions of the
system of equations \eqref{1-st-t}--\eqref{A0}.
\bi
\item Choose the form of functions $b(z)$ and $f_1(z)$.
\item Using these $b(z)$ and $f_1(z)$, find the time component of the
  electric field $A_t(z)$ from \eqref{A0}.
\item Using $f_1(z)$, $b(z)$ and $A'_t(z)$, get the blackening
  function $g(z)$ from \eqref{1-st-t}.
\item Using $b(z)$, find the derivative of the scalar field $\phi'(z)$
  from \eqref{2-nd-t}. To have a solution one has to be sure that
  \be
    b'' - \cfrac{3 b^{\prime \,2 }}{2 b} + \cfrac{2 b'}{z} - \cfrac{4
      b(\nu -1)}{3 \nu^2 z^2} \leq 0. \label{positive}
  \ee
\item Using $g(z)$ and $b(z)$, get $f_2(z)$ from \eqref{3-rd-t}.
\item Finally, get $V = V(z)$ from \eqref{4-th-t}.
\ei

Let us express the warp factor $b(z)$ via a polynomial
$P(z)$\footnote{For the isotropic case this form of factor $b$ has
  been considered, in particular, in \cite{1301.0385,yang2015}, other
  form of this factor has been also considered in \cite{Dudal:2017}.}:
\bea
  b(z) &=& e^{P(z)}, \label{bP}
\eea
and take the coupling factor $f_1(z)$:
\bea
  f_1(z) &=& z^{- 2 + \frac{2}{\nu}}. \label{f1}
\eea
In this case the equation \eqref{A0} becomes
\be
  A_t'' + A_t' \left( \cfrac{P'(z)}{2} - \cfrac{1}{z} \right) =
  0 \label{eq:2.202}
\ee
and it's solution has the form
\be
  A_t = C_1 \int_0^z e^{-\frac{P(\xi)}{2}} \xi d \xi +
  C_2. \label{eq:2.204}
\ee
If we take into account the boundary conditions \eqref{Atbc}, the
integration constants equal to
\be
  C_1 = - \, \cfrac{\mu}{\int_0^{z_H} e^{-\frac{P(\xi)}{2}} \xi d
    \xi} = - \, \tilde{\mu}, \quad C_2 = \mu \label{eq:2.205}
\ee
and the solution (\ref{eq:2.204}) becomes
\be
  A_t = \mu \left( 1 - \cfrac{\int_0^z e^{-\frac{P(\xi)}{2}} \xi d
      \xi}{\int_0^{z_H} e^{-\frac{P(\xi)}{2}} \xi d \xi}
  \right) = \tilde{\mu} \int_z^{z_H} e^{-\frac{P(\xi)}{2}} \xi d
  \xi.\label{eq:2.206}
\ee
In a similar way equation \eqref{1-st-t} takes the form
\be
  g'' + g' \left(\cfrac{3}{2} \, P'(z) - \cfrac{1}{z} - \cfrac{2}{\nu
      z} \right) - \tilde{\mu} \, e^{-2 P(z)} z^{2+\frac{2}{\nu}} =
  0, \label{eq:2.207}
\ee
and its solution is:
\be
  g = C_4 + \int_0^z e^{- \frac{3 P(\xi)}{2}} \left( C_3 +
    \tilde{\mu}^2 \int_0^\xi e^{- \frac{P(\chi)}{2}} \chi d \chi
  \right) \xi^{1+\frac{2}{\nu}} d\xi. \label{eq:2.209}
\ee
Using the first boundary condition in \eqref{gbc} we can determine the
integration constant $C_4 = 1$, and taking into account the boundary
condition on the horizon we find $C_3$
\be
  C_3 = - \, \cfrac{1 + \tilde{\mu}^2 \int_0^{z_H} e^{- \frac{3
        P(\xi)}{2}} \left( \int_0^\xi e^{- \frac{P(\chi)}{2}} \chi d
      \chi \right) \xi^{1+\frac{2}{\nu}} d\xi}{\int_0^{z_H} e^{-
      \frac{3 P(\xi)}{2}} \xi^{1+\frac{2}{\nu}} d\xi}, \label{C3}
\ee
therefore the solution (\ref{eq:2.209}) becomes:
\bea
  g = 1 &+& \tilde{\mu}^2 \int_0^{z} e^{- \frac{3 P(\xi)}{2}} \left(
    \int_0^\xi e^{- \frac{P(\chi)}{2}} \chi d \chi \right)
  \xi^{1+\frac{2}{\nu}} d\xi \, \label{eq:2.212} \\
  &-& \cfrac{1 + \tilde{\mu}^2 \int_0^{z_H} e^{- \frac{3 P(\xi)}{2}}
    \left( \int_0^\xi e^{- \frac{P(\chi)}{2}} \chi d \chi \right)
    \xi^{1+\frac{2}{\nu}} d\xi}{\int_0^{z_H} e^{- \frac{3
        P(\xi)}{2}} \xi^{1+\frac{2}{\nu}} d\xi} \int_0^{z} e^{-
    \frac{3 P(\xi)}{2}} \xi^{1+\frac{2}{\nu}} d\xi. \nn
\eea

Equation \eqref{3-rd-t} gives the following expression
for the coupling function $f_2(z)$:
\bea
  &&f_2(z) = - \, \cfrac{\nu - 1}{q^2 \nu} \ e^{P(z)}
  z^{1-\frac{4}{\nu}} \left[ 2 e^{-\frac{3 P(z)}{2}}
    z^{1+\frac{2}{\nu}} \left( C_3 + \tilde{\mu}^2 \int_0^z e^{-
        \frac{P(\xi)}{2}} \xi d \xi \right) \right. \label{eq:2.215}
  \\
  &&+ \left. \cfrac{3 z \nu P'(z) - 4 (\nu + 1)}{z \nu} \left( 1 + C_3
      \int_0^{z} e^{- \frac{3 P(\xi)}{2}} \xi^{1+\frac{2}{\nu}} d\xi +
      \tilde{\mu}^2 \int_0^{z} e^{- \frac{3 P(\xi)}{2}} \left(
        \int_0^\xi e^{- \frac{P(\chi)}{2}} \chi d \chi \right)
      \xi^{1+\frac{2}{\nu}} d\xi \right) \right]. \nn
\eea

Equation \eqref{2-nd-t} allows to find the scalar field $\phi(z)$:
\be
  \phi(z) = C_5 + \int_0^z \sqrt{- \, 3 P''(\xi) + \cfrac{3}{2} \,
    P'^2(\xi) - \cfrac{6}{\xi} \, P'(\xi) + 4 \, \cfrac{\nu - 1}{\xi^2
      \nu^2}} \ d \xi. \label{eq:2.217}
\ee
Using the boundary condition \eqref{phibc}, we get
\be
  C_5 = - \int_0^{z_h} \sqrt{- \, 3 P''(\xi) + \cfrac{3}{2} \, P'^2(\xi)
    - \cfrac{6}{\xi} \, P'(\xi) + 4 \, \cfrac{\nu - 1}{\xi^2 \nu^2}} \
  d \xi \label{eq:2.217b}
\ee
and therefore
\be
  \phi(z) = \int_{z_h}^z \sqrt{- \, 3 P''(\xi) + \cfrac{3}{2} \,
    P'^2(\xi) - \cfrac{6}{\xi} \, P'(\xi) + 4 \, \cfrac{\nu - 1}{\xi^2
      \nu^2}} \ d \xi. \label{eq:2.217c}
\ee

Finally, equation \eqref{4-th-t} gives the scalar potential
$V(\phi(z))$:
\bea
  V(\phi(z)) &=& \cfrac{9}{2} \, P'(z) \, e^{-P(z)} z \left[ 1 +
    \cfrac{1}{\nu} \, - C_3 \left( \cfrac{e^{-\frac{3 P(z)}{2}}}{2} \
      z^{2+\frac{2}{\nu}} - \left(1 + \cfrac{1}{\nu} \right) \int_0^z
      e^{-\frac{3 P(\xi)}{2}} \xi^{1+\frac{2}{\nu}} d \xi \right)
  \right. \nn \\
  &+& \left. \tilde{\mu}^2 \left( \cfrac{e^{-\frac{3 P(z)}{2}}}{2} \
      z^{2+\frac{2}{\nu}} \int_0^{z} e^{-\frac{P(\xi)}{2}} \xi d\xi -
      \left(1 + \cfrac{1}{\nu} \right) \int_0^{z} e^{-\frac{3
          P(\xi)}{2}} \left( \int_0^\xi e^{- \frac{P(\chi)}{2}} \chi d
        \chi \right) \xi^{1+\frac{2}{\nu}} d\xi \right) \right] \nn \\
  &+&\cfrac{3}{2} \ e^{-P(z)} z^2 \left( P''(z) + \cfrac{3}{2} \
    P'^2(z) + 4 \, \cfrac{1 + 3 \nu + 2 \nu^2}{2 z^2 \nu^2}
  \right) \label{eq:2.218} \\
  &\times& \left[ 1 + C_3 \int_0^{z} e^{-\frac{3 P(\xi)}{2}} \xi^{1 +
      \frac{2}{\nu}} d\xi + \tilde{\mu}^2 \int_0^{z} e^{-\frac{3
        P(\xi)}{2}} \left( \int_0^\xi e^{- \frac{P(\chi)}{2}} \chi d
      \chi \right) \xi^{1+\frac{2}{\nu}} d\xi \right] \nn \\
  &+& e^{- \frac{5 P(z)}{2}} z^{2 +\frac{2}{\nu}} \left[ \left(
      \cfrac{3}{4} \, P'(z) \, z + 2 + \cfrac{1}{\nu} \right) \left(
      C_3 + \tilde{\mu}^2 \int_0^{z} e^{-\frac{P(\xi)}{2}} \xi d\xi
    \right) - \tilde{\mu}^2 \, e^{-\frac{P(z)}{2}} \, \cfrac{z^2}{2}
  \right]. \nn
\eea

If we substitute the expressions \eqref{bP}, \eqref{f1},
\eqref{eq:2.206} and \eqref{eq:2.212}--\eqref{eq:2.218} into the
constraint \eqref{eq:2.25} and take into account that $\partial
V/\partial \phi = V'/\phi'$, we can make certain that the left-hand
side of \eqref{eq:2.25} disappears. Therefore the system
\eqref{1-st-t}--\eqref{A0} is self-consistent and satisfied by the
general solution \eqref{bP}, \eqref{f1}, \eqref{eq:2.206} and
\eqref{eq:2.212}--\eqref{eq:2.218}.

\subsection{Solutions with factor $b(z) = \exp(c z^2/2)$ and spatial
  anisotropy}\label{Sec:c-nu}

As we are interested in effects that can be caused by the anisotropy
of the chosen metric ansatz, it is needed to find some particular
solution of the system \eqref{1-st-t}--\eqref{A0} and investigate it's
properties explicitly. For this purpose we preferred to start from the
simplest form of the warp-factor, the same as in \cite{AZ}:
\bea
  b(z) &=& e^{\frac{c z^2}{2}}, \label{bc}
\eea
and take the factor $f_1(z)$:
\bea
  f_1 &=& z^{- 2 + \frac{2}{\nu}}.
\eea
In this case the equation \eqref{1-st-t} becomes
\be
  A_t'' + A_t' \left( \cfrac{c z}{2} - \cfrac{1}{z} \right) =
  0, \label{eqA0}
\ee
and together with the boundary conditions
\be
  A_t(0) = \mu, \qquad A_t(z_h) = 0 \label{bcA0}
\ee
this gives
\be
  A_t(z) = \mu \frac{e^{- \frac{c z^2}{4}} - e^{- \frac{c
        z_h^2}{4}}}{1 - e^{- \frac{c z_h^2}{4}}}. \label{solA0}
\ee

\subsubsection{Blackening function}

Using the solution \eqref{solA0} and $f_1$ given by \eqref{f1}, we
rewrite equation \eqref{1-st-t} for the blackening function:
\be
  g'' + g' \left( \cfrac{3 c z}{2} - \cfrac{1}{z} - \cfrac{2}{\nu z}
  \right) - \cfrac{\mu^2 c^2 \ z^{2+\frac{2}{\nu}} \ e^{- c z^2}}{4
    \left( 1 - e^{- \frac{c z_H^2}{4}} \right)^2} = 0. \label{black}
\ee
Taking into account the boundary conditions \eqref{gbc}, we get
\bea
  g(z) = 1 &-& \frac{\gamma(1 + \frac{1}{\nu} ; \frac34 c
    z^2)}{\gamma(1 + \frac1{\nu} ; \frac{3}{4} c z_h^2)} - \frac{\mu^2
    e^{\frac{1}{2} c z_h^2}}{4 c^{1/\nu}} \ \frac{\gamma(1 +
    \frac{1}{\nu} ; c z^2)}{(1 - e^{\frac{1}{4} c z_h^2})^2} \nn \\
  &+& \frac{\mu^2 e^{\frac{1}{2} c z_h^2}}{4 c^{1/\nu}} \
  \frac{\gamma(1 + \frac{1}{\nu} ; c z_h^2 )}{(1 - e^{\frac{1}{4} c
      z_h^2})^2} \ \frac{\gamma\left(1 + \frac{1}{\nu} ; \frac{3}{4} c
      z^2 \right)}{\gamma(1 + \frac{1}{\nu} ; \frac{3}{4} c z_h^2)}, 
  \label{sol-g}
\eea
where $\gamma(1 + \frac{1}{\nu}, x)$ is the incomplete gamma
function.
    
There is no problem with solution in such a form for $c > 0$, but for
$c < 0$ some ingredients of this presentation seem to fail. Indeed,
$c^{1/\nu}$ is ill-defined for $c < 0$. By this reason we rewrite the
above formula as
\bea
  g(z) = 1 - \frac{z^{2+\frac{2}{\nu}}}{z^{2+\frac{2}{\nu}}_h} \
  \frac{\fG(\frac34 c z^2)}{\fG(\frac34 c z_h^2)} - \cfrac{\mu^2 c
    z^{2+\frac{2}{\nu}} e^{\frac{c z_h^2}{2}}}{{4 \left( 1 -
        e^{\frac{c z_h^2}{4}} \right)^2}} \ \fG( c z^2 ) +
  \cfrac{\mu^2 c z^{2+\frac{2}{\nu}} e^{\frac{c z_h^2}{2}}}{4 \left( 1
      - e^{\frac{c z_h^2}{4}} \right)^2} \ \frac{\fG(\frac34 c
    z^2)}{\fG(\frac34 c z_h^2)} \ \fG(c z_h^2), \nn \\
  \label{bl-sum}
\eea
where
\be
  \fG(x) = x^{-1-\frac{1}{\nu}} \ \gamma \left( 1 + \frac{1}{\nu}, x
  \right) \label{expan1}
\ee
and the function $\fG(x)$ has the following expansion (see \cite{GR},
p. 1377):
\be
  \fG(x) = \sum_{n=0}^{\infty} \cfrac{(-1)^n \, x^n}{n! (1 + n +
    \frac{1}{\nu})}. \label{expan}
\ee
Taking into account the first two terms of the expansion
\eqref{expan}, we get
\be
  g_{appr}(z) = 1 - \cfrac{z^{2+\frac{2}{\nu}}}{z_h^{2+\frac{2}{\nu}}}
    \left( \rho + Q z_h^2 - Q z^2 \right), \label{g-aprox}
\ee
where
\bea
  &\rho = \cfrac{4 (1 + 2 \nu) - 3 c z^2 (1 + \nu)}{4 (1 + 2 \nu) - 3
    c z_h^2 (1 + \nu)}, \\
  &Q = \cfrac{\mu^2 c^2 \nu \, z_h^{2+\frac2{\nu}} e^{\frac{c
        z_h^2}2}}{4 \left( 1 - e^{\frac{c z_h^2}{4}} \right)^2 \Bigl(
      4 (1 + 2 \nu) - 3 c z_h^2 (1 + \nu) \Bigr)}.
\eea

This expression can be recast into a form
\be
  g_{appr}(z) = 1 - z^{2+\frac{2}{\nu}} \left( \left( \frac{\rho}{z_h}
    \right)^{2+\frac{2}{\nu}} + \frac{Q}{z_h^{2/\nu}} \right) + Q \
  \frac{z^{4+\frac{2}{\nu}}}{z_h^{2+\frac{2}{\nu}}}, \label{appr-g}
\ee
and after the redefinition $z_h = \tilde z_h \rho$, $Q = \tilde Q
z_h^{2+\frac{2}{\nu}}$ it becomes
\be
  g_{appr}(z) = 1 - z^{2+\frac{2}{\nu}} \left(\cfrac{1}{\tilde
      z_h^{2+\frac{2}{\nu}}} + \tilde Q \rho^2 \tilde z_h^{2} \right)
  + \tilde Q\,z^{4+\frac{2}{\nu}}. \label{appr-g1}
\ee
\begin{figure}[h!]
  \centering
  \includegraphics[scale=0.8]{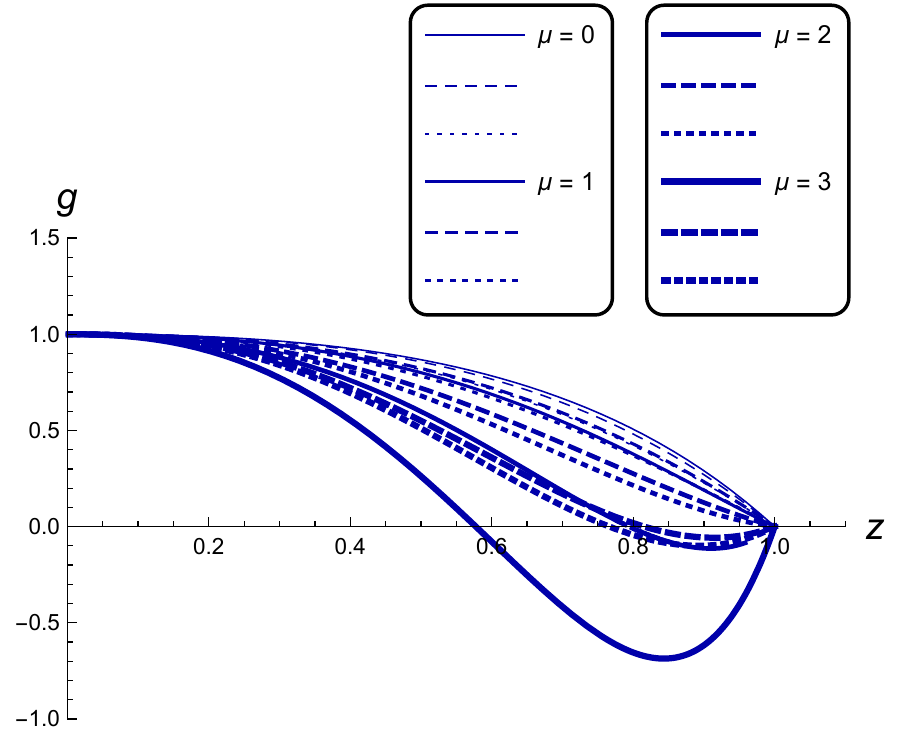} \quad
  \includegraphics[scale=0.8]{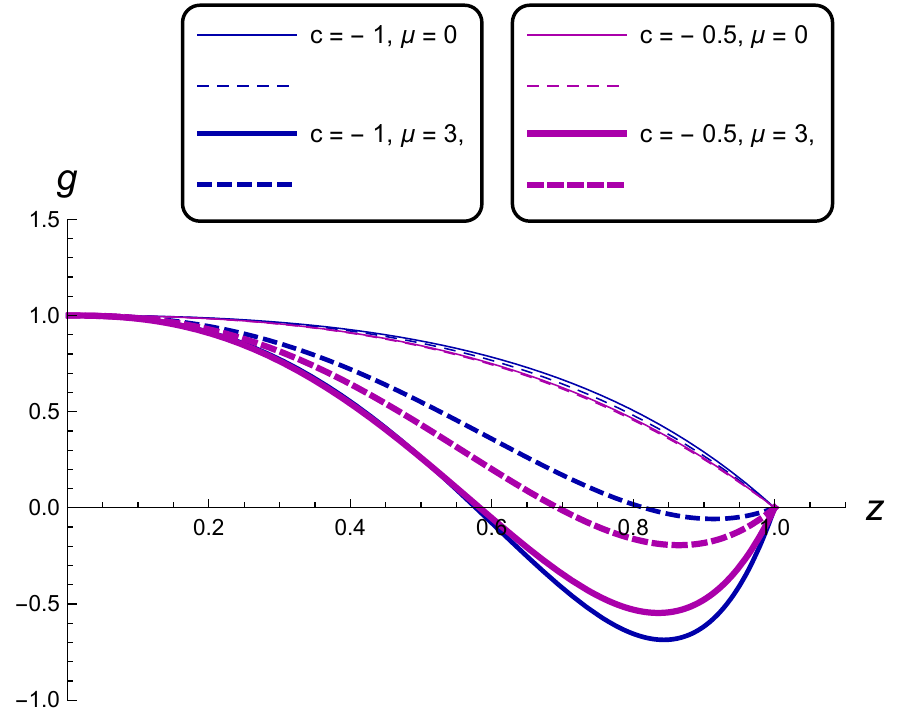}\\
  A\hspace{200pt}B \\ \ \\
  \includegraphics[scale=0.8]{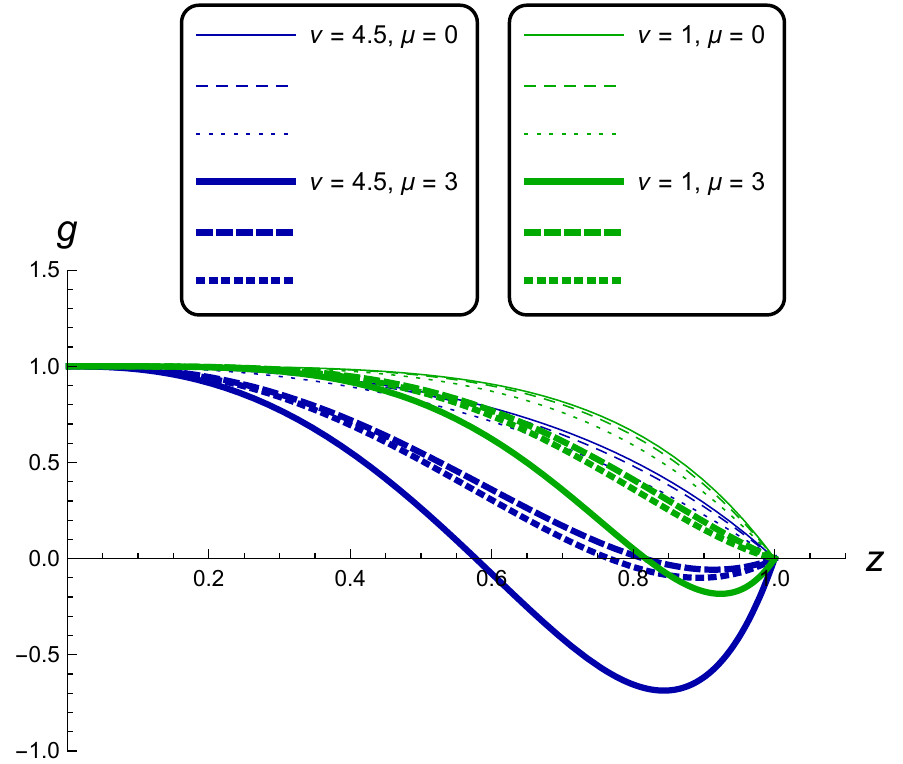} \quad
  \includegraphics[scale=0.8]{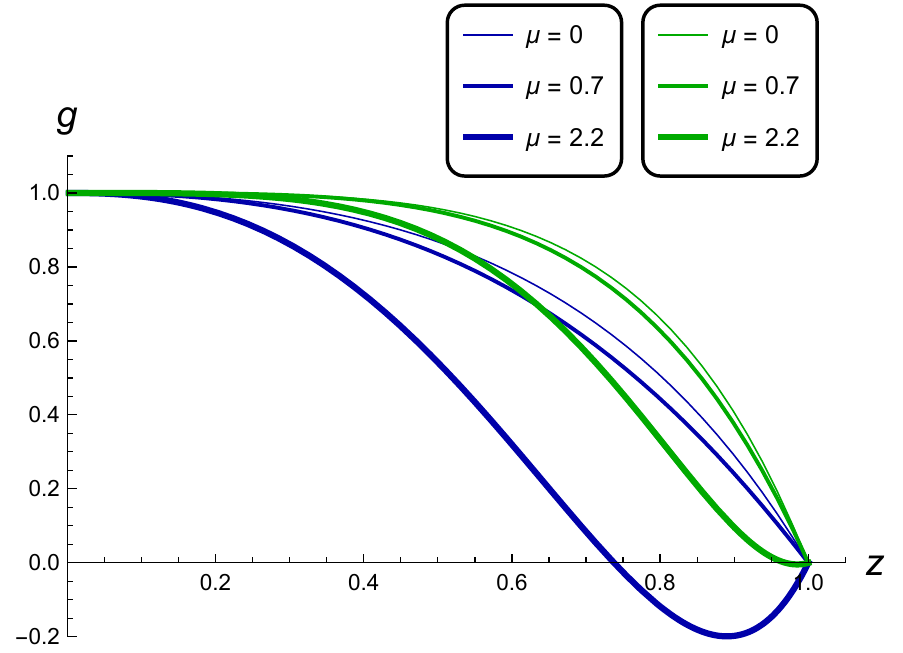}\\
  C\hspace{200pt}D
  \caption{Blackening functions $g(z)$ (solid lines), $g_{approx}(z)$
    (same thickness dashed lines) and $g_{approx,[2]}(z)$ (same
    thickness dotted lines) for $z_h = 1$, $\nu = 4.5$, $c = -1$ (blue
    lines), $\nu = 4.5$, $c = -0.5$ (magenta lines), $\nu = 1$, $c =
    -1$ (green lines) and different $\mu = 0, \, \dots, \, 3$.}
  \label{Fig:black}
\end{figure}
The blackening function used in \cite{IA} 
\be
  g_{\small[2]}(z) = 1 - z^{2+\frac{2}{\nu}} \left( \cfrac{1}{z_h^{2 +
        \frac{2}{\nu}}} + Q z_h^2 \right) z^{2+\frac{2}{\nu}} + Q
  z^{4+\frac{2}{\nu}} \label{gz-aprox}
\ee
is different by the factor $\rho^2$ in the second coefficient in front
of $z^{2+\frac{2}{\nu}}$. Near the horizon this factor is
approximately equal to 1:

\be
  \rho \approx 1 - \cfrac{6 c z_h (1 + \nu)}{4 (1 + 2 \nu) - 3 c z_h^2
    (1 + \nu)} \ (z - z_h) + {\cal O}((z - z_h)^2).
\ee

The behavior of the blackening function from the holographic
coordinate $z$ till horizon is depicted on Fig.\ref{Fig:black}. The
main feature is that the blackening function values decrease faster
for larger chemical potential $\mu$ (Fig.\ref{Fig:black}.A) and for
smaller warp factor coefficient $c$ (Fig.\ref{Fig:black}.B). The
difference between the approximations \eqref{appr-g1} and
\eqref{gz-aprox} and the exact expression \eqref{sol-g} is irregular
and depends on the model parameters (Fig.\ref{Fig:black}.C). In the
isotropic case the blackening function values are larger than in the
anisotropic ones (Fig.\ref{Fig:black}.C and D). For $\mu$ close to
zero it is the desreasing function of $z$ till the horizon, but for
growing $\mu$ the local minimums and the second horizons small than
the original ones appear. Changing the values of $c$ almost does not
influence on the horizon position.

\subsubsection{Coupling function $f_2$}

We can substitute the expression for the blackening function
\eqref{bl-sum} into \eqref{3-rd-t}, take into account
\bea
g'(z) = - \, 2 \, \cfrac{z^{1+\frac{2}{\nu}}}{z_h^{2+\frac{2}{\nu}}}
  \ \cfrac{e^{-\frac{3 c z^2}{4}}}{\fG(\frac34 c z_h^2)} -
  \cfrac{\mu^2 c \, z^{1+\frac{2}{\nu}} \, e^{- c z^2 + \frac{c
        z_h^2}{2}}}{2 \left( 1 - e^{\frac{c z_h^2}{4}}\right)^2}
  \left( 1 - e^{\frac{c z^2}{4}} \cfrac{\fG(c z_h^2)}{\fG(\frac34 c
      z_h^2)} \right)
  \label{dg}
\eea
and get
\bea
  f_2(z) &=& \cfrac{\nu - 1}{q^2 \nu^2} \ z^{-\frac{4}{\nu}}
  e^{\frac{c z^2}{2}} \Biggl[ 4 (1 + \nu) - 3 c \, \nu \, z^2 + 4 \,
  \cfrac{z^{2+\frac{2}{\nu}}}{z_h^{2+\frac{2}{\nu}}} \, \Bigg\{
    \cfrac{\nu \, e^{-\frac{3 c z^2}{4}}}{\fG\left(\frac34 c
        z_h^2\right)} - (1 + \nu) \ \cfrac{\fG\left(\frac34 c
        z^2\right)}{\fG\left(\frac34 c z_h^2\right)} \ \fF \nn \\
  &+& \cfrac{\mu^2 c \, \nu \, z_h^{2+\frac{2}{\nu}} e^{-c z^2
        + \frac{c z_h^2}{2}}}{4 \left( 1 - e^{\frac{c z_h^2}{4}}
      \right)^2} \left( 1 - e^{\frac{c z^2}{4}} \, \cfrac{\fG\left(c
          z_h^2\right)}{\fG\left(\frac34 c z_h^2\right)} \right)
  \Bigg\} + 3 c \, \nu \,
  \cfrac{z^{4+\frac{2}{\nu}}}{z_h^{2+\frac{2}{\nu}}} \,
  \cfrac{\fG\left(\frac34 c z^2\right)}{\fG\left(\frac34 c
      z_h^2\right)} \ \fF \Biggr], \label{eq:2.118}
\eea
where
\be
  \fF = 1 - \cfrac{\mu^2 c \, z_h^{2+\frac{2}{\nu}} e^{\frac{c
        z_h^2}{2}}}{4 \left( 1 - e^{\frac{c z_h^2}{4}} \right)^2}
  \left( \fG\left(c z_h^2\right) - \fG\left(c z^2\right) \,
    \cfrac{\fG\left(\frac34 c z_h^2\right)}{\fG\left(\frac34 c
        z^2\right)} \right). \label{fgoth}
\ee
At the horizon
\be
  f_2(z_h) = 4 \, \cfrac{\nu - 1}{q^2 \nu} \ z^{-\frac{4}{\nu}}
  \Bigg\{ \cfrac{e^{-\frac{c z_h^2}{4}}}{\fG\left(\frac34 c
      z_h^2\right)} + \cfrac{\mu^2 c \, z_h^{2+\frac{2}{\nu}}}{4
    \left( 1 - e^{\frac{c z_h^2}{4}} \right)^2} \left( 1 - e^{\frac{c
        z_h^2}{4}} \, \cfrac{\fG\left(c z_h^2\right)}{\fG\left(\frac34
        c z_h^2\right)} \right) \Bigg\}. \label{f2zH}
\ee

\begin{figure}[h!]
  \centering
  \includegraphics[scale=0.8]{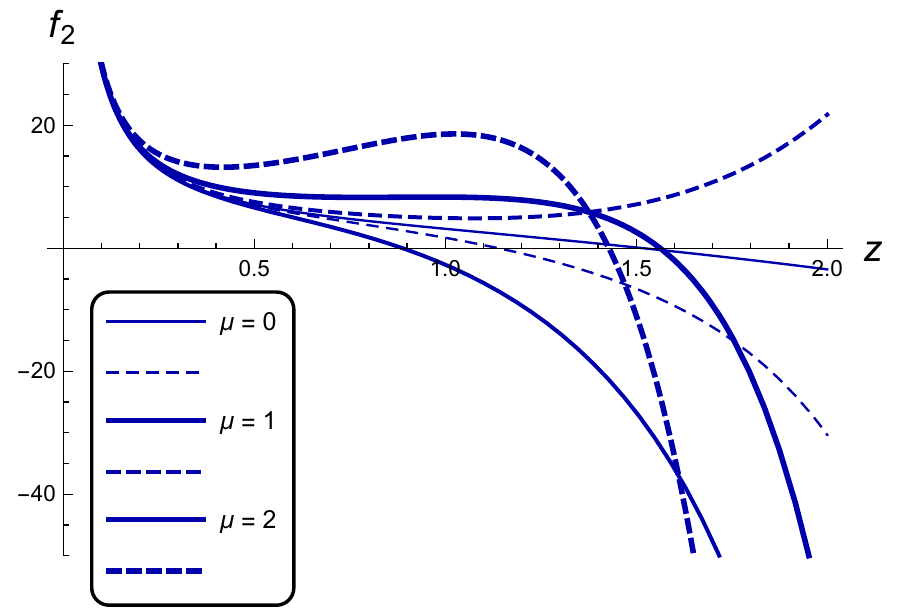} \quad
  \includegraphics[scale=0.8]{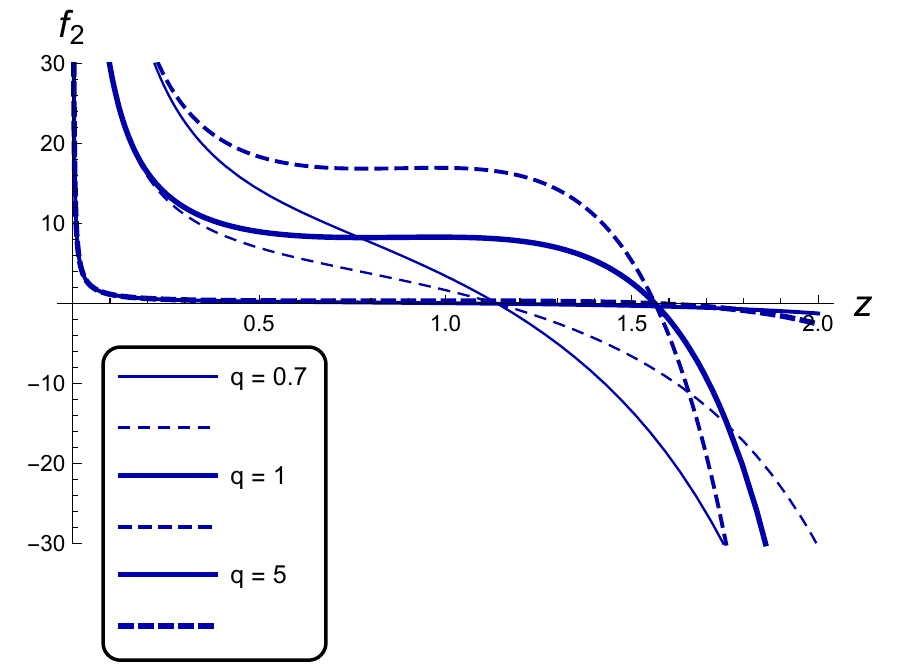}\\ A\hspace{200pt}B \\
  \caption{Coupling function $f_2(z)$ (solid lines) and $f_{2\
      approx}$ (same thickness dashed lines) for $z_h = 1$, $\nu =
    4.5$, $c = - 1$, $q = 1$, $\mu = 0, \ 1, \ 5$ (A) and $\mu = 1$,
    $q = 0.7, \ 1, \ 5$ (B).}
  \label{Fig:f2}
\end{figure}

Using the first two terms of the expansion \eqref{expan} on $z$ and
$z_h$
\bea
  \fG(c z_{(h)}^2) &=& \cfrac{\nu}{\nu + 1} - \cfrac{c \, \nu \,
    z_{(h)}^2}{2 \nu + 1}, \label{fG1} \\
  \fG\left(\frac34 \, c z_{(h)}^2\right) &=& \left( \frac34
  \right)^{1+\frac{1}{\nu}} \left( \cfrac{\nu}{\nu + 1} - \cfrac{3}{4}
    \ \cfrac{c \nu z_{(h)}^2}{2 \nu + 1}\right), \label{fG2}
\eea
we have the following expansion of coupling factor $f_2$:
\bea
  f_{2\ approx}(z) &=& \cfrac{\nu - 1}{q^2 \nu^2} \ z^{-\frac{4}{\nu}}
  e^{\frac{c z^2}{2}} \Biggl[ 4 (1 + \nu) - 3 c \nu z^2 \label{f2appr}
  \\
  &+& \cfrac{z^{2+\frac{2}{\nu}}}{z_h^{2+\frac{2}{\nu}}} \Bigg\{
    \cfrac{16 (1 + \nu) (1 + 2 \nu)}{3 c z_h^2 (1 + \nu) - 4 (1 + 2
      \nu)} \left( 1 -
      \cfrac{4^{1+\frac{1}{\nu}}}{3^{1+\frac{1}{\nu}}} \ e^{-\frac{3 c
          z^2}{4}} \right) \nn \\
  &-& \cfrac{\mu^2 c \nu z_h^{2+\frac{2}{\nu}} e^{\frac{c
        z_h^2}{2}}}{\left( 1 - e^{\frac{c z_h^2}{4}} \right)^2} 
    \left( 1 - e^{-c z^2} - 4 \ \cfrac{c z_h^2 (1 + \nu) - (1 + 2
        \nu)}{3 c z_h^2 (1 + \nu) - 4 (1 + 2 \nu)} \left( 1 -
        \cfrac{4^{1+\frac{1}{\nu}}}{3^{1+\frac{1}{\nu}}} \ e^{-\frac{3
            c z^2}{4}} \right) \right) \Bigg\} \, \nn \\
  &-& \cfrac{z^{4+\frac{2}{\nu}}}{z_h^{2+\frac{2}{\nu}}} \Bigg\{
    \cfrac{12 c (1 + 3 \nu (1 + \nu))}{3 c z_h^2 (1 + \nu) - 4 (1 + 2
      \nu)} + \cfrac{\mu^2 c^2 \nu z_h^{2+\frac{2}{\nu}} e^{\frac{c
          z_h^2}{2}}}{4 \left( 1 - e^{\frac{c z_h^2}{4}} \right)^2}  \
    \cfrac{3 c \nu z_h^2 + 4 (1 + 2 \nu)}{3 c z_h^2 (1 + \nu) - 4 (1 +
      2 \nu)} \Bigg\} \nn \\
  &+& \cfrac{z^{6+\frac{2}{\nu}}}{z_h^{2+\frac{2}{\nu}}} \ \cfrac{9
    c^2 \nu (1 + \nu)}{3 c z_h^2 (1 + \nu) - 4 (1 + 2 \nu)} \Bigg( 1 +
    \cfrac{\mu^2 c \nu z_h^{2+\frac{2}{\nu}} e^{\frac{c z_h^2}{2}}}{12
      \left( 1 - e^{\frac{c z_h^2}{4}} \right)^2 (1 + \nu)} \Bigg)
  \Biggr] \nn
\eea
and at the horizon
\bea
  f_{2\ approx}(z_h) &=& \cfrac{1 - \nu}{q^2 \nu^2} \
  z_h^{-\frac{4}{\nu}} e^{-\frac{c z_h^2}{4}} \Bigg\{
    \cfrac{4^{3+\frac{1}{\nu}}}{3^{1+\frac{1}{\nu}}} \ \cfrac{(1 +
      \nu) (1 + 2 \nu)}{3 c z_h^2 (1 + \nu) - 4 (1 + 2 \nu)}
    \label{f2apprzH} \\
  &-& \cfrac{\mu^2 c \nu z_h^{2+\frac{2}{\nu}} e^{\frac{c
        z_h^2}{4}}}{\left( 1 - e^{\frac{c z_h^2}{4}} \right)^2} \left(
    1 - \cfrac{4^{2+\frac{1}{\nu}}}{3^{1+\frac{1}{\nu}}} \ \cfrac{c
      z_h^2 (1 + \nu) - (1 + 2 \nu)}{3 c z_h^2 (1 + \nu) - 4 (1 + 2
      \nu)} \ e^{\frac{c z^2}{4}} \right) \Bigg\}. \nn
\eea
If we substitute $g_{appr}(z)$ into $f_2$, the result coincides with
\eqref{f2appr}.

\subsubsection{Scalar Field}

Substitution of \eqref{bc} into \eqref{2-nd-t} leads to
\bea
  \phi' = \cfrac{1}{\nu z} \ \sqrt{\cfrac{3}{2} \ \nu^2 c^2 z^4 - 9
    \nu^2 c z^2 + 4 \nu - 4}. \label{eq:2.119}
\eea
Here we should take into account, that the radicand in
\eqref{eq:2.119} shouldn't be negative. Therefore we have different
cases of parameter $c$ value.

\begin{itemize}
\item \noindent{$\bf{c < 0}$}
\end{itemize}

For $c < 0$ this requirement is fullfiled without any restrictions and
for the scalar field we have
\bea
  \phi &=& \frac{1}{2 \sqrt{2} \, \nu} \left\{ \sqrt{3 c^2 \nu^2 z^4
      - 18 c \, \nu^2 z^2 + 8 \, (\nu - 1)} - \sqrt{3 c^2 \nu^2 z_h^4
      - 18 c \, \nu^2 z_h^2 + 8 \, (\nu - 1)}
  \right. \label{phicneg} \\
  &+& \left. 2 \sqrt{2 \, (\nu - 1)} \ln \left( \cfrac{z^2}{z_h^2}
    \right) - 3 \, \sqrt{3} \, \nu \ln\left( \cfrac{\sqrt{3 c^2 \nu^2
          z^4 - 18 c \, \nu^2 z^2 + 8 \, (\nu - 1)} - \sqrt{3} \, \nu
        \, (3 - c z^2)}{\sqrt{3 c^2 \nu^2 z_h^4 - 18 c \, \nu^2 z_h^2
          + 8 \, (\nu - 1)} - \sqrt{3} \, \nu \, (3 - c z_h^2)}
    \right) \right. \nn \\
  &-& \left. 2 \sqrt{2 \, (\nu - 1)} \ln{\left( \cfrac{9 c \, \nu^2
          z^2 - 8 \, (\nu - 1) - \sqrt{2 \, (\nu - 1)} \, \sqrt{3 c^2
            \nu^2 z^4 - 18 c \, \nu^2 z^2 + 8 \, (\nu - 1)}}{9 c \,
          \nu^2 z_h^2 - 8 \, (\nu - 1) - \sqrt{2 \, (\nu - 1)} \,
          \sqrt{3 c^2 \nu^2 z_h^4 - 18 c \, \nu^2 z_h^2 + 8 \, (\nu -
            1)}} \right)} \right\}. \nn
\eea
For small $z$ the scalar field can be approximated as
\be
  \phi \sim - \, k(z_h,\nu,c) + \frac{2 \, \sqrt{\nu -1}}{\nu} \, \log
  \left(\cfrac{z}{z_h}\right)
\ee
and for large $z$
\be
  \phi \sim \frac{|c|}{2}\, \sqrt{\frac{3}{2}}\,  z^2.
\ee

\begin{figure}[h!]
  \centering
  \includegraphics[scale=0.9]{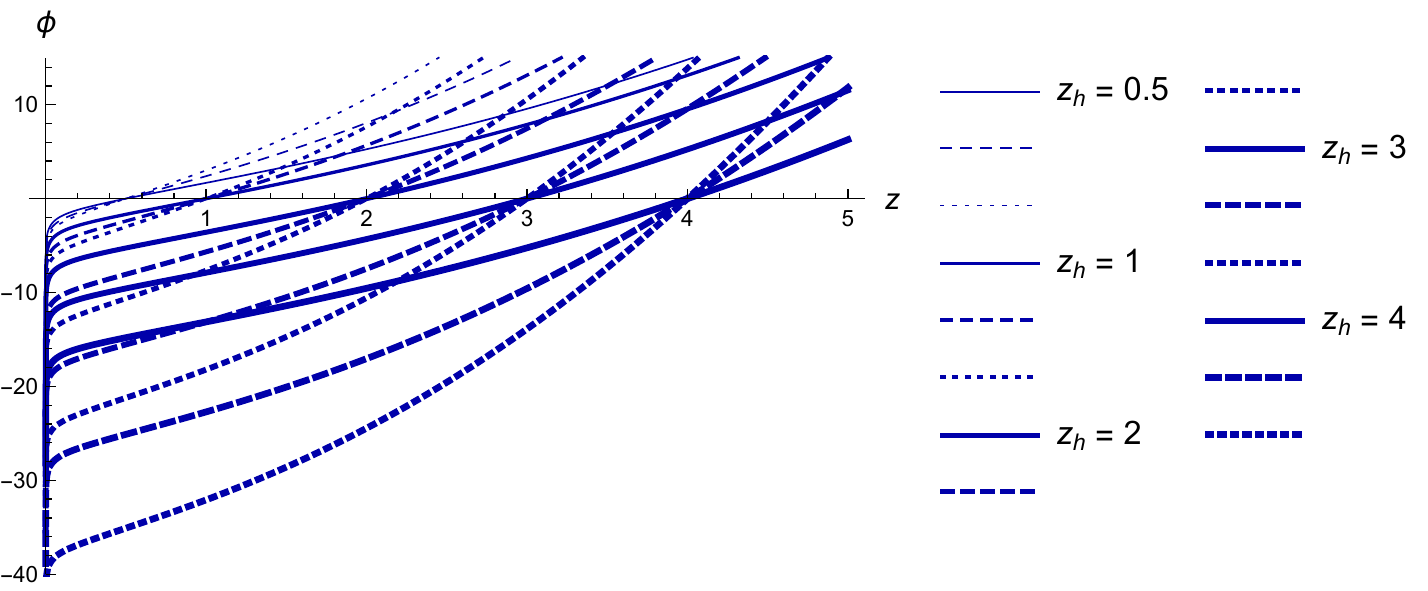}
  \caption{Scalar field $\phi(z)$ for $\nu = 4.5$, $c = -1$ (solid
    lines), $c = -2$ (dashed lines) and $c = -3$ (dotted lines) and
    different $z_h$.}
  \label{Fig:phi-negative}
\end{figure}

\newpage

\begin{itemize}
\item
\noindent{$\bf{c > 0}$}
\end{itemize}

For $c > 0$ expression \eqref{eq:2.119} can be parametrized as
\bea
  \phi' = \cfrac{c}{z} \, \sqrt{\frac{3}{2} \, (\alpha^2 -
    z^2)(\beta^2 - z^2)}, \label{eq:2.120}
\eea
where
\bea
  \alpha^2 \beta^2 = \cfrac{8(\nu - 1)}{3 \nu^2 c^2}, \qquad \alpha^2
  + \beta^2 = \frac{6}{c},
\eea
and 
\be
  \alpha = \sqrt{\cfrac{3}{c} - \cfrac{1}{c} \, \sqrt{9 - \cfrac{8
        (\nu - 1)}{3 \nu^2}}}, \qquad \beta = \sqrt{\cfrac{3}{c} +
    \cfrac{1}{c} \, \sqrt{9 - \cfrac{8 (\nu - 1)}{3 \nu^2}}}.
\ee
Note, that we can get a real solution only for
\bea
  i) \quad \alpha^2 - z^2 > 0, \quad \beta^2 - z^2 > 0, \\
  ii) \quad \alpha^2 - z^2 < 0, \quad \beta^2 - z^2 < 0.
\eea

Integrating \eqref{eq:2.120} we obtain
\bea
  \phi(z) &=& \sqrt{\frac{3}{8}} \, c \, \left\{ \sqrt{(\alpha^2 -
    z^2)(\beta^2 - z^2)} - \sqrt{(\alpha^2 - z_h^2)(\beta^2 - z_h^2)}
  + \alpha \beta \ln \left(\cfrac{z^2}{z_h^2}\right) \right. \nn \\
  &-& \left. \cfrac{\alpha^2 + \beta^2}{2} \, \ln \left(\cfrac{
        \sqrt{(\alpha^2 - z^2)(\beta^2 - z^2)} + z^2 - \cfrac{\alpha^2
          + \beta^2}{2}}{\sqrt{(\alpha^2 - z_h^2)(\beta^2 - z_h^2)} +
        z_h^2 - \cfrac{\alpha^2 + \beta^2}{2}} \right)
  \right. \label{phicp} 
\eea
\bea
  &-& \left. \alpha \beta \ln \left( \cfrac{\alpha^2 \beta^2 -
        \cfrac{\alpha^2 + \beta^2}{2} \, z^2 + \alpha \beta
        \sqrt{(\alpha^2 - z^2)(\beta^2 - z^2)}}{\alpha^2 \beta^2 -
        \cfrac{\alpha^2 + \beta^2}{2} \, z_h^2 + \alpha \beta
        \sqrt{(\alpha^2 - z_h^2)(\beta^2 - z_h^2)}} \right) \right\}
  \nn
\eea
We see that the solution \eqref{phi} becomes complex for $\alpha < z <
\beta$. It leads to an instability region for the scalar field.
\begin{figure}[b!]
  \centering
  \includegraphics[scale=0.8]{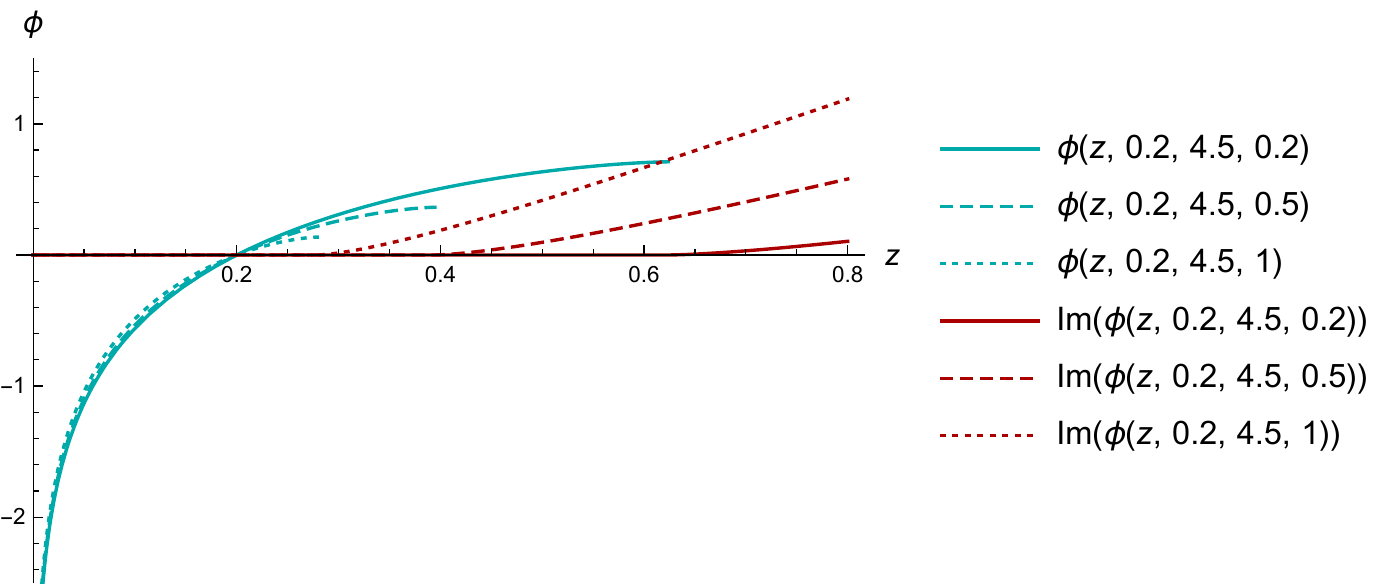}\\A\\
  \includegraphics[scale=0.8]{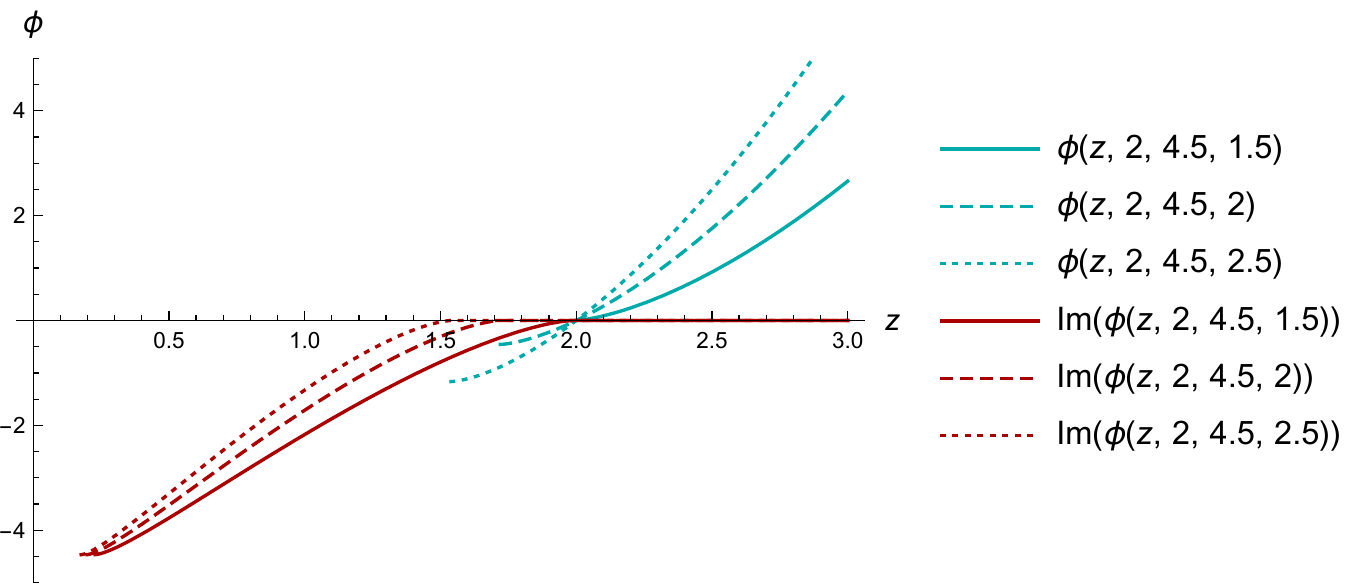}\\B\\
 \caption{Scalar field $\phi(z, \, z_h, \, \nu, \, c)$ and its
   imaginary part for $\nu = 4.5$, different positive $c$, $z_h = 0.2$
   (A) and $z_h = 2$ (B).}
  \label{Fig:phi-positive}
\end{figure}

\begin{itemize}
\item \noindent{$\bf{c = 0}$}
\end{itemize}

For $c = 0$ we get
\bea
  &\phi' = \cfrac{2 \, \sqrt{\nu - 1}}{\nu z}, \label{phic0} \\
  &\phi = \cfrac{2 \, \sqrt{\nu - 1}}{\nu} \, \ln \left(
    \cfrac{z}{z_h} \right).
\eea
\begin{figure}[h!]
  \centering
   \includegraphics[scale=0.85]{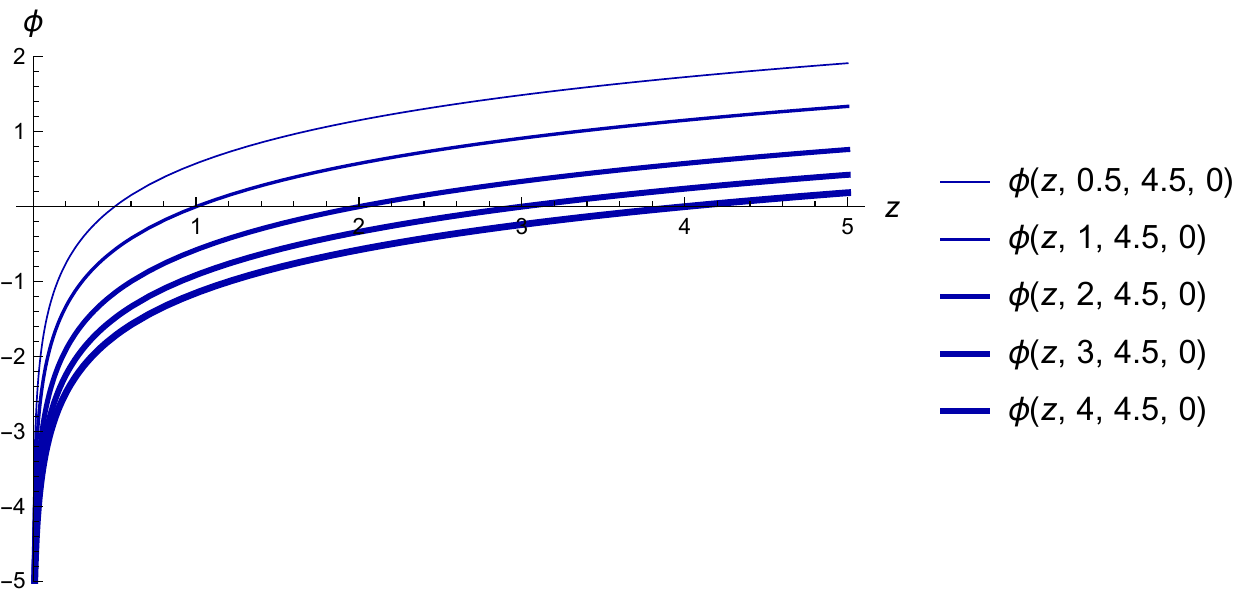}
  \caption{Scalar field $\phi(z, \, z_h, \, \nu, \, c)$ for $\nu =
    4.5$, $c = 0$ and different $z_h$.}
  \label{Fig:phi-negative}
\end{figure}

\subsubsection{Scalar potential }    

From equation \eqref{4-th-t} we get the expression for the scalar
potential $V$ as a function of $z$:
\bea
  V(z) &=& \cfrac{e^{-\frac{c z^2}{2}}}{8 \nu^2} \left\{ - 16 \left( 2
      \nu^2 + 3 \nu + 1 \right) + 12 c \, \nu \, z^2 \left( 2 \nu + 3
    \right) + 16 \left( 2 \nu^2 + 3 \nu + 1 \right)
    z^{2+\frac{2}{\nu}} \, \fV \right. \nn \\
  &-& \left. 16 \nu \left( 2 \nu + 1 \right)
  \cfrac{z^{2+\frac{2}{\nu}}}{z_h^{2+\frac{2}{\nu}}} \ e^{-c z^2}
  \Bigg[ \cfrac{e^{\frac{c z^2}{4}}}{\fG(\frac34 c z_h^2)} +
    \cfrac{\mu^2 c \, z_h^{2+\frac{2}{\nu}} e^{\frac{c z_h^2}{2}}}{4
        \left( 1 - e^{\frac{c z_h^2}{4}} \right)^2} \left( 1 -
        e^{\frac{c z^2}{4}} \frac{\fG(c z_h^2)}{\fG(\frac34 c z_h^2)}
      \right) \Bigg] \right. \nn \\
  &-& \left. 18 c^2 \nu^2 z^4 - 12 c \nu \left( 2 \nu + 3 \right)
    z^{4+\frac{2}{\nu}} \, \fV \right. \nn \\
  &-& \left. \nu \, \cfrac{z^{4+\frac{2}{\nu}}}{z_h^{2+\frac{2}{\nu}}}
    \ e^{-c z^2} \Bigg[ \cfrac{- 24 c \, \nu \, e^{\frac{c
            z^2}{4}}}{\fG(\frac34 c z_h^2)} - \cfrac{5 \mu^2 c^2 \nu
        \, z_h^{2+\frac{2}{\nu}} e^{\frac{c z_h^2}{2}}}{\left( 1 -
          e^{\frac{c z_h^2}{4}} \right)^2} \left( 1 - \cfrac65 \,
        e^{\frac{c z^2}{4}} \frac{\fG(c z_h^2)}{\fG(\frac34 c z_h^2)}
      \right) \Bigg] \right. \nn \\
  &+& \left. 18 c^2 \nu^2 z^{6+\frac{2}{\nu}} \, \fV
  \right\}, \label{Vseries}
\eea
\begin{figure}[h!]
  \centering
  \includegraphics[scale=0.5]{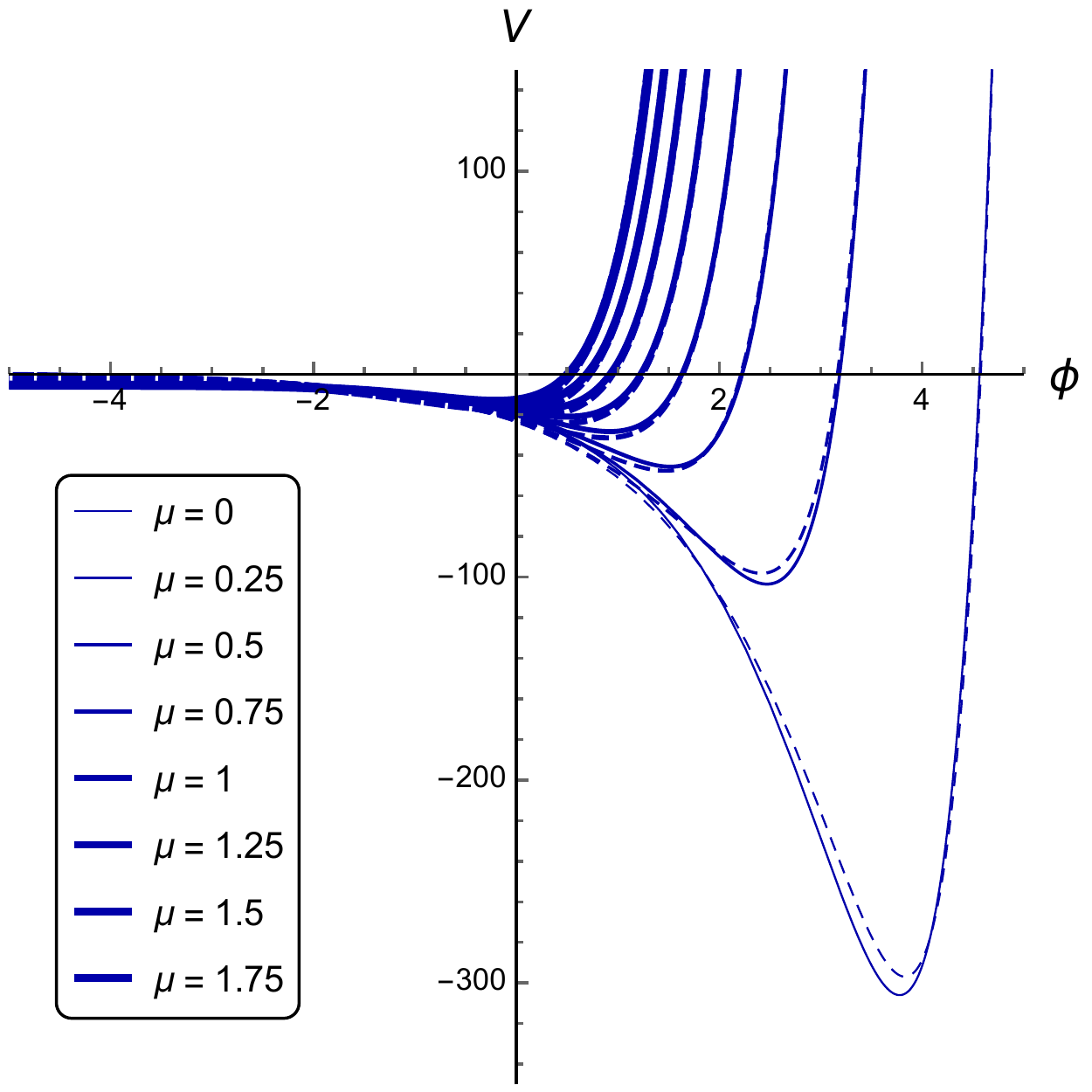} \quad
  \includegraphics[scale=0.5]{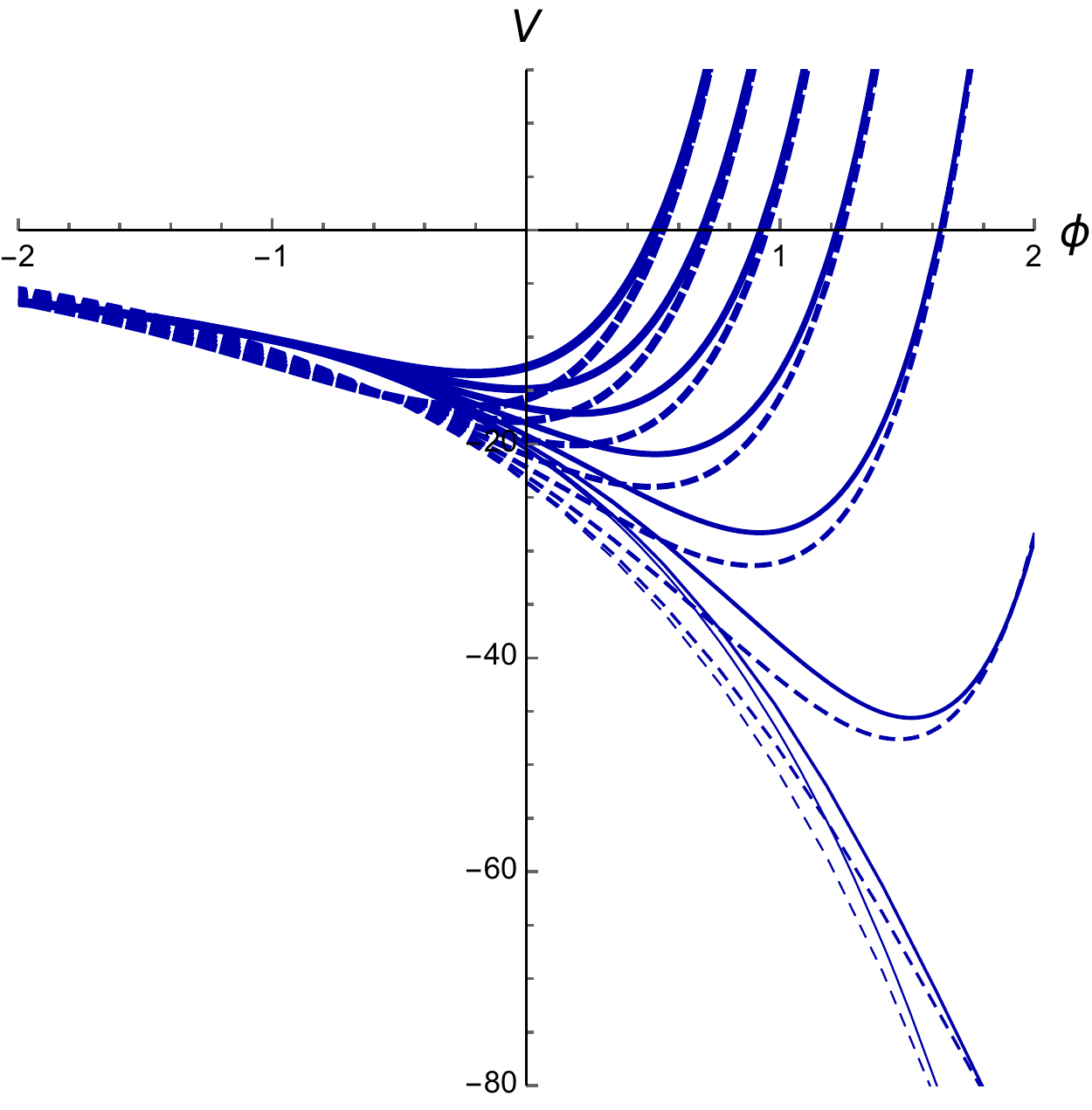}\\
  A\hspace{200pt}B \\ \ \\
  \includegraphics[scale=0.5]{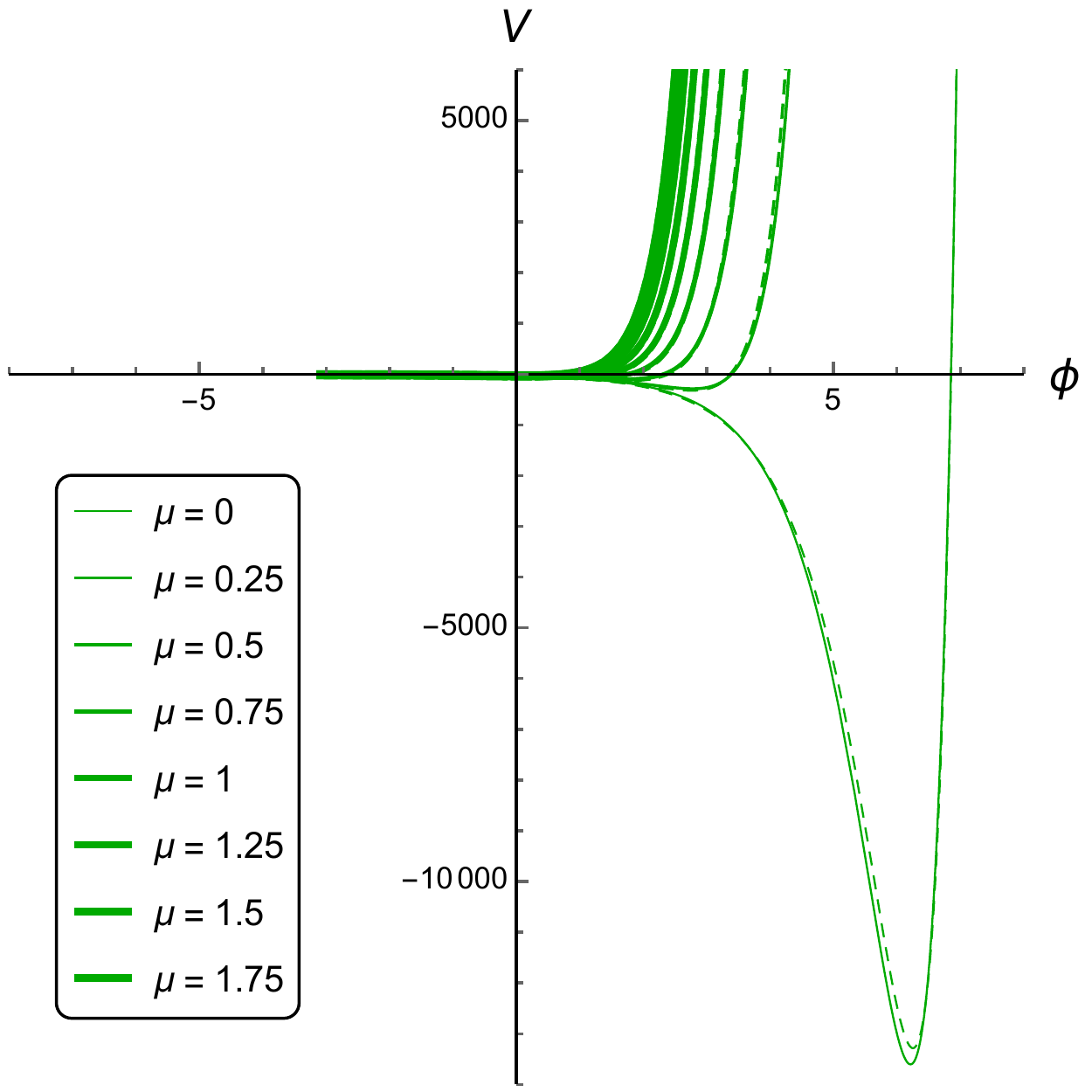} \quad
  \includegraphics[scale=0.5]{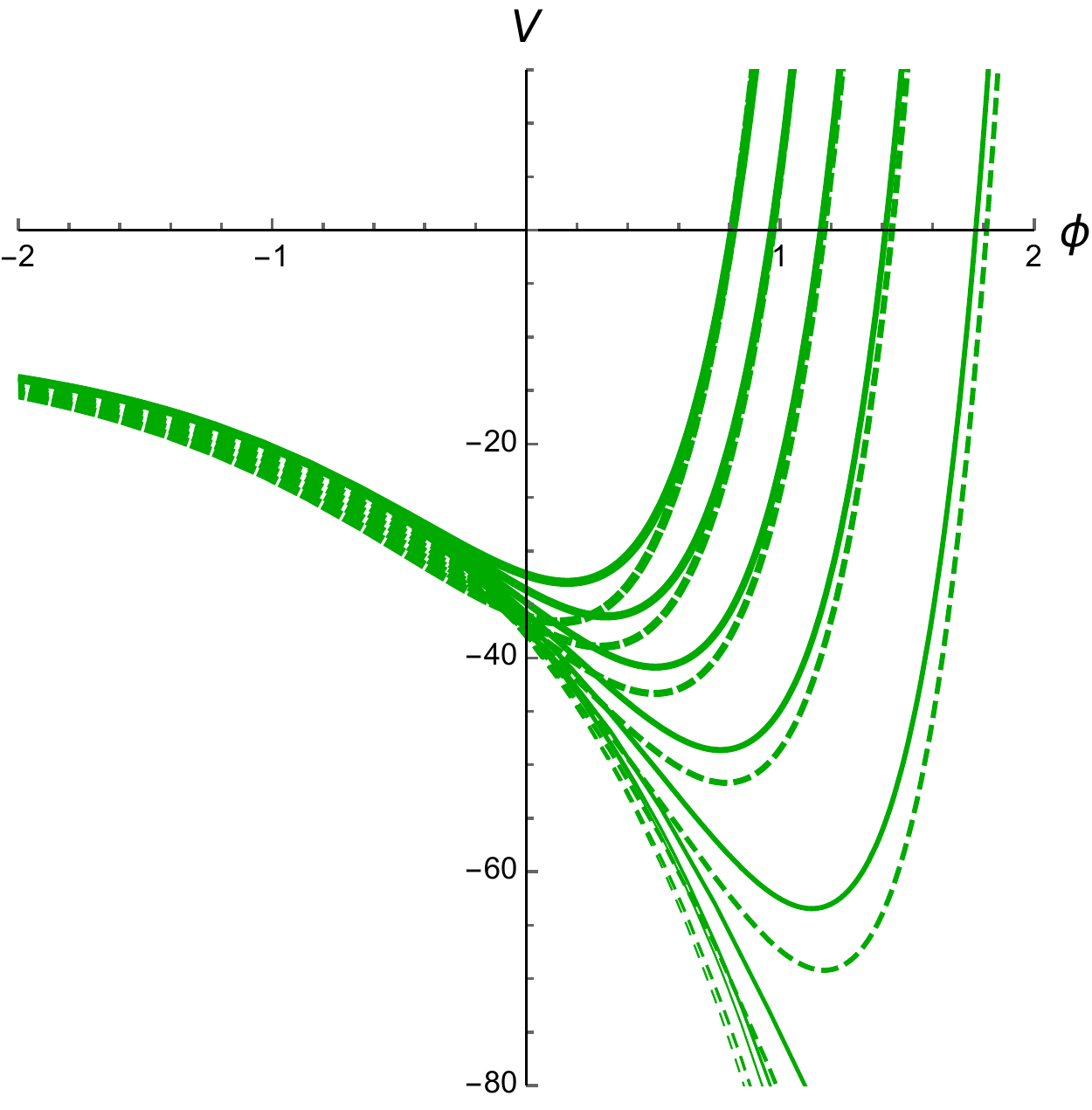}\\
  C\hspace{200pt}D
  \caption{Scalar field potential $V(\phi)$ (solid lines) and its
    approximation (dashed lines) as a sum of two exponents and a
    constant for $z_h = 1$, $c = -1$ and different $\mu$ in
    anisotropic, $\nu = 4.5$, (A) and isotropic (C) cases. The right
    plots (B) and (D) are zoom of the corresponding left ones.}
  \label{Fig:V-approx}
\end{figure}
where
\be
  \fV = \cfrac{1}{z_h^{2+\frac{2}{\nu}}} \, \frac{\fG(\frac34 c
    z^2)}{\fG(\frac34 c z_h^2)} + \cfrac{\mu^2 c \, e^{\frac{c
        z_h^2}{2}}}{4 \left( 1 - e^{\frac{c z_h^2}{4}} \right)^2}
  \left( \fG(c z^2) - \fG(c z_h^2) \, \frac{\fG(\frac34 c
      z^2)}{\fG(\frac34 c z_h^2)} \right).
\ee

The dependence $V(\phi)$ can't be expressed explicitly due to
nontrivial behavior of $\phi(z)$ \eqref{phicneg}, but it can be
displayed graphically (Fig.\ref{Fig:V-approx}). For $c = -1$ it can be
approximated by a sum of two exponents and a negative constant:
\be
  V_{2EA}(\phi,\mu,\nu) = V_0(\nu) - C_7(\mu,\nu) e^{K_1(\nu) \phi} +
  C_8(\mu,\nu) e^{K_2(\nu) \phi}.
\ee
The best fit is given by 
\bea
  V_0(4.5) = - \, 0.5778, \quad K_1(4.5) = 0.7897, \quad K_2(4.5) =
  2.0995
\eea
with the coefficients depending on the chemical potential $\mu$
(Fig.\ref{Fig:V-approx}.A and B):
\bea
  C_7 (\mu,4.5) = 23.0779 + 2.4236 \mu^2, \quad C_8 (\mu,4.5) = 0.0575
  + 4.9919 \mu^2.
\eea
In isotropic case (Fig.\ref{Fig:V-approx}.C and D) the approximation
constants are:
\bea
  &&V_0(1) = - \, 10.8689, \quad K_1(1) = 1.0852, \quad K_2(1) =
  2.4103, \\
  &&C_7 (\mu,1) = 27.2825 + 4.3749 \mu^2, \quad C_8 (\mu,1) = 0.0031
  + 5.03093 \mu^2.
\eea

Note, that in \cite{AGP} an explicit isotropic solution for the
dilaton potential as a sum of two exponents and zero chemical
potential has been constructed. It would be interesting to generalize
this construction to the anisotropic and non-zero chemical potential
cases.

\begin{figure}[h!]
  \centering
  \includegraphics[scale=1]{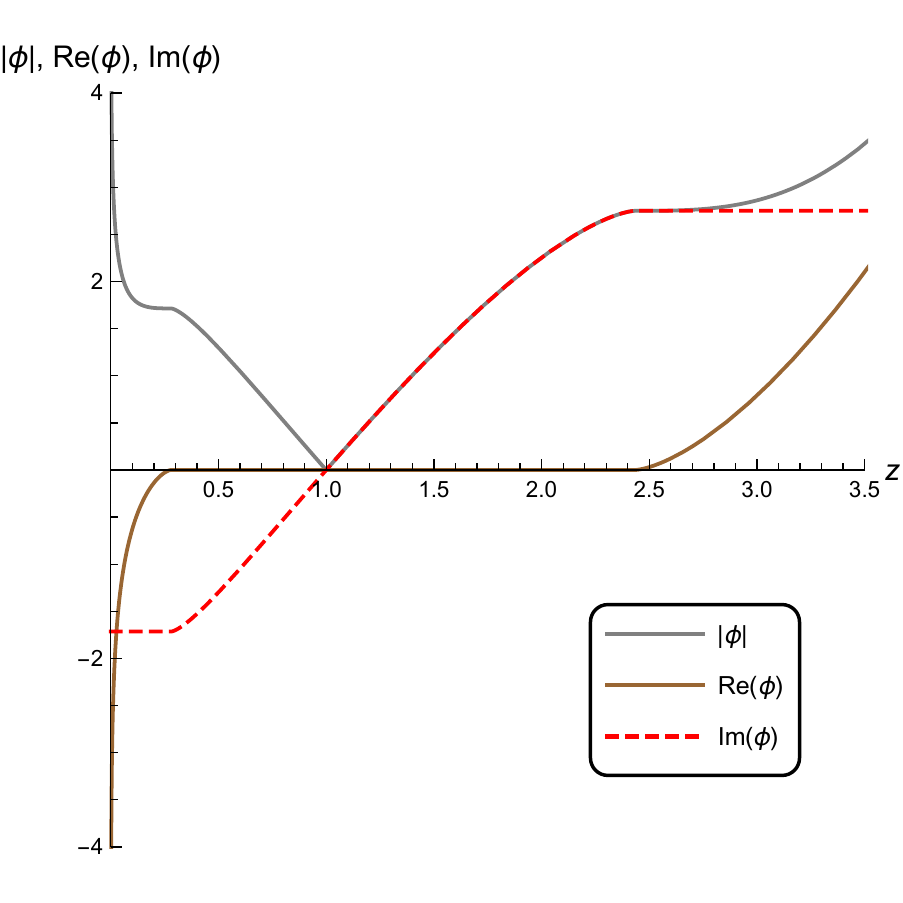}
  \caption{Module, real and imaginary parts of the scalar field $\phi$
    for $\nu = 4.5$, $z_h = 1$, $c = 1$.}
  \label{Fig:phi-positive-c}
\end{figure}

\begin{figure}[h!]
  \centering
  \includegraphics[scale=0.71]{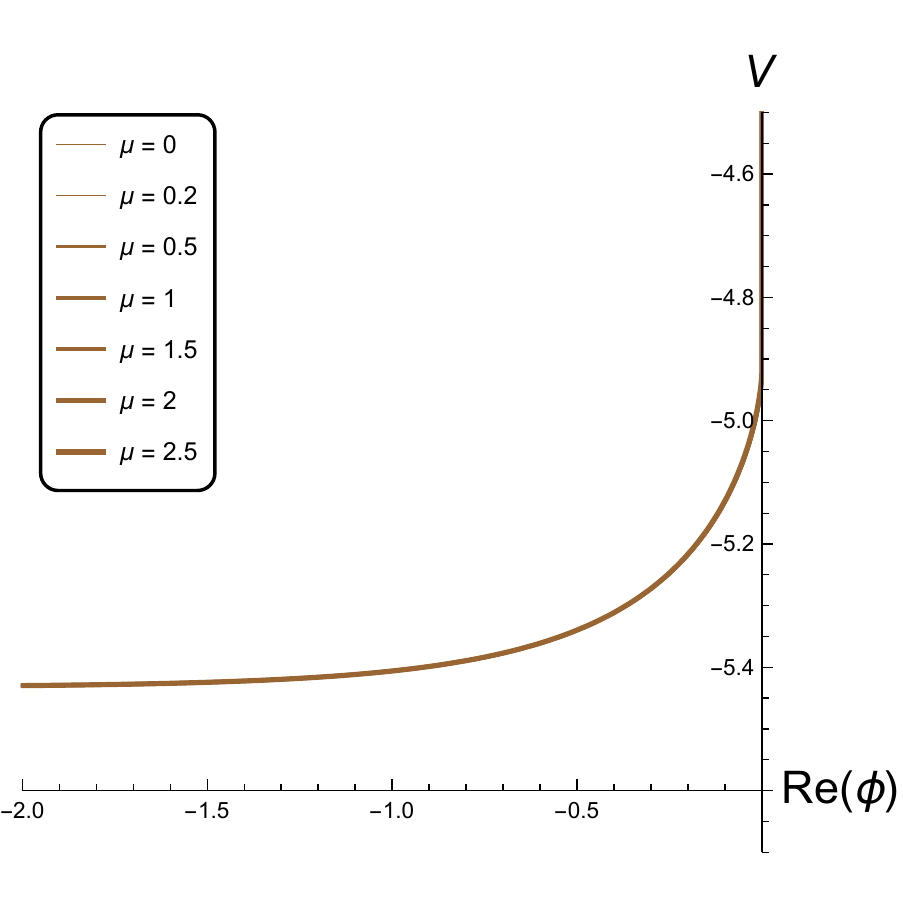} \quad
  \includegraphics[scale=0.68]{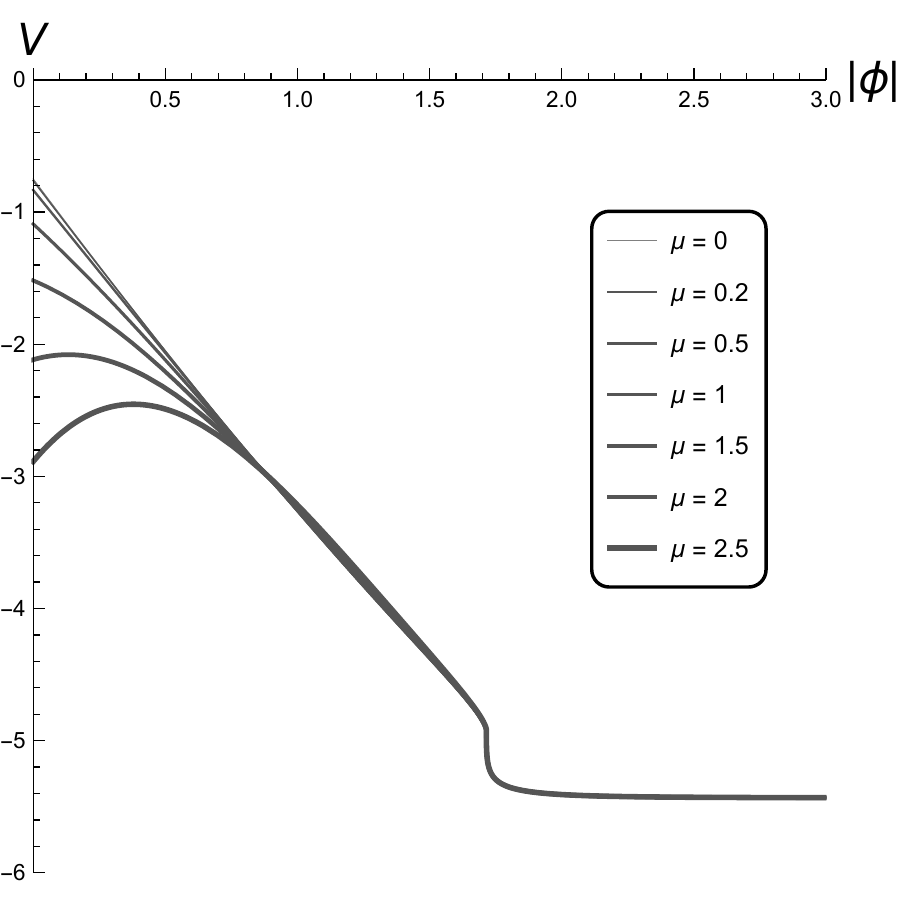}\\
  A\hspace{210pt}B
  \caption{Scalar potential $V\bigl(\re(\phi)\bigr)$ (A) and
    $V(|\phi|)$ (B) for $\nu = 4.5$, $z_h = 1$, $c = 1$ and different
    $\mu$ under horizon, i.e. $0 < z < 1$.}
  \label{Fig:V-phi-positive-c}
\end{figure}

The behavior of $V(\phi)$ for positive warp factor coefficient is
quite different. Let us recall that for $c > 0$ the scalar field
$\phi$ becomes complex under horizon
(Fig.\ref{Fig:phi-positive-c}). The function $V(\re(\phi))$ doesn't
display visible dependence on chemical potential
(Fig.\ref{Fig:V-phi-positive-c}.A) and the function $V(|\phi|)$ stops
to depend on $\mu$ rather soon (Fig.\ref{Fig:V-phi-positive-c}.B).

\newpage
\subsection{Scalar invariants}

\begin{figure}[b!]
  \centering
  \includegraphics[scale=0.6]{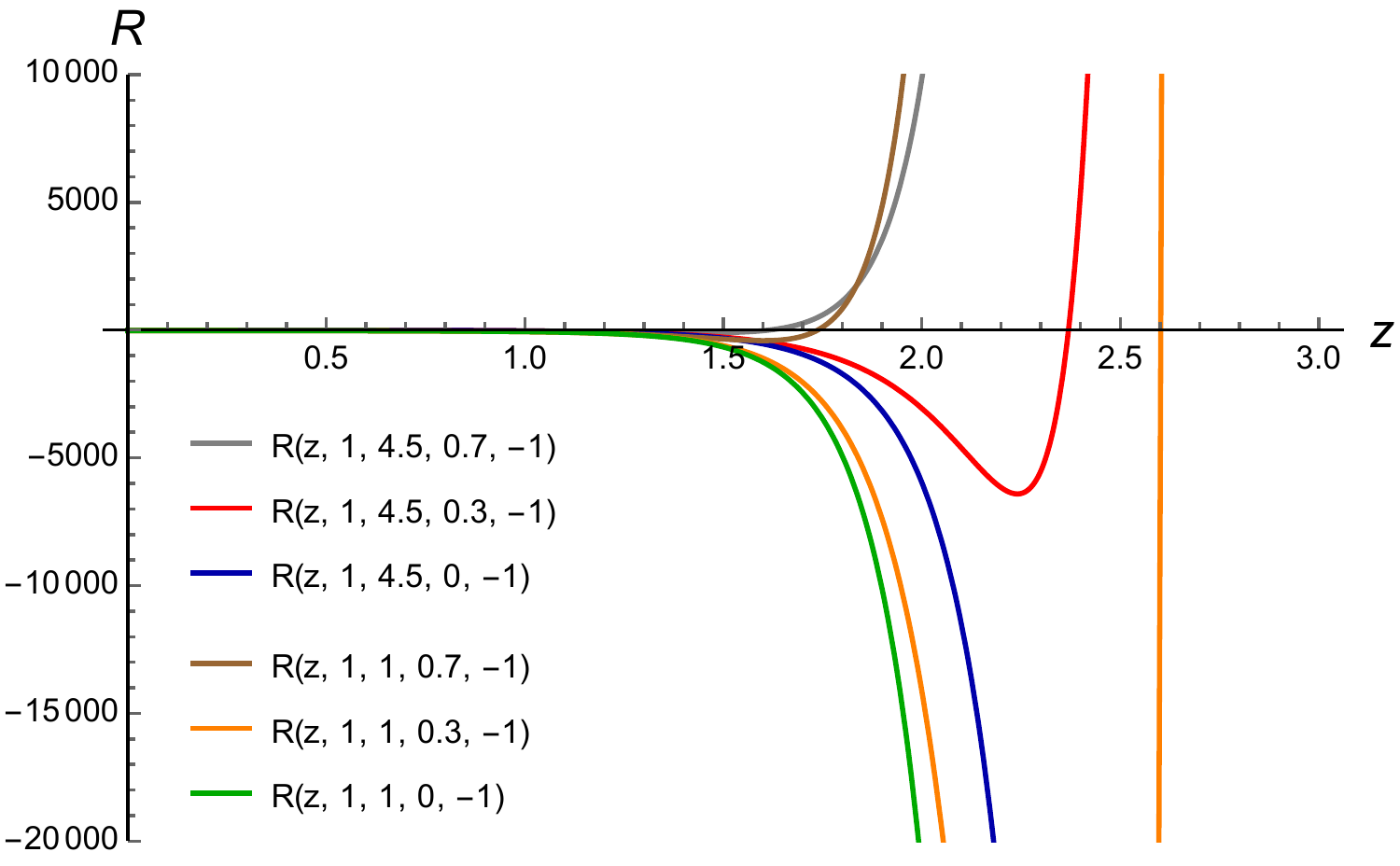}
  \caption{Ricci scalar $R(z)$ for $z_h = 1$, $c =  -1$ in isotropic,
    $\nu = 1$, and anisotropic, $\nu = 4.5$, cases for different
    $\mu$.}
  \label{Fig:sc-curv-negative-c}
\end{figure}

\begin{figure}[h!]
  \centering
  \includegraphics[scale=0.61]{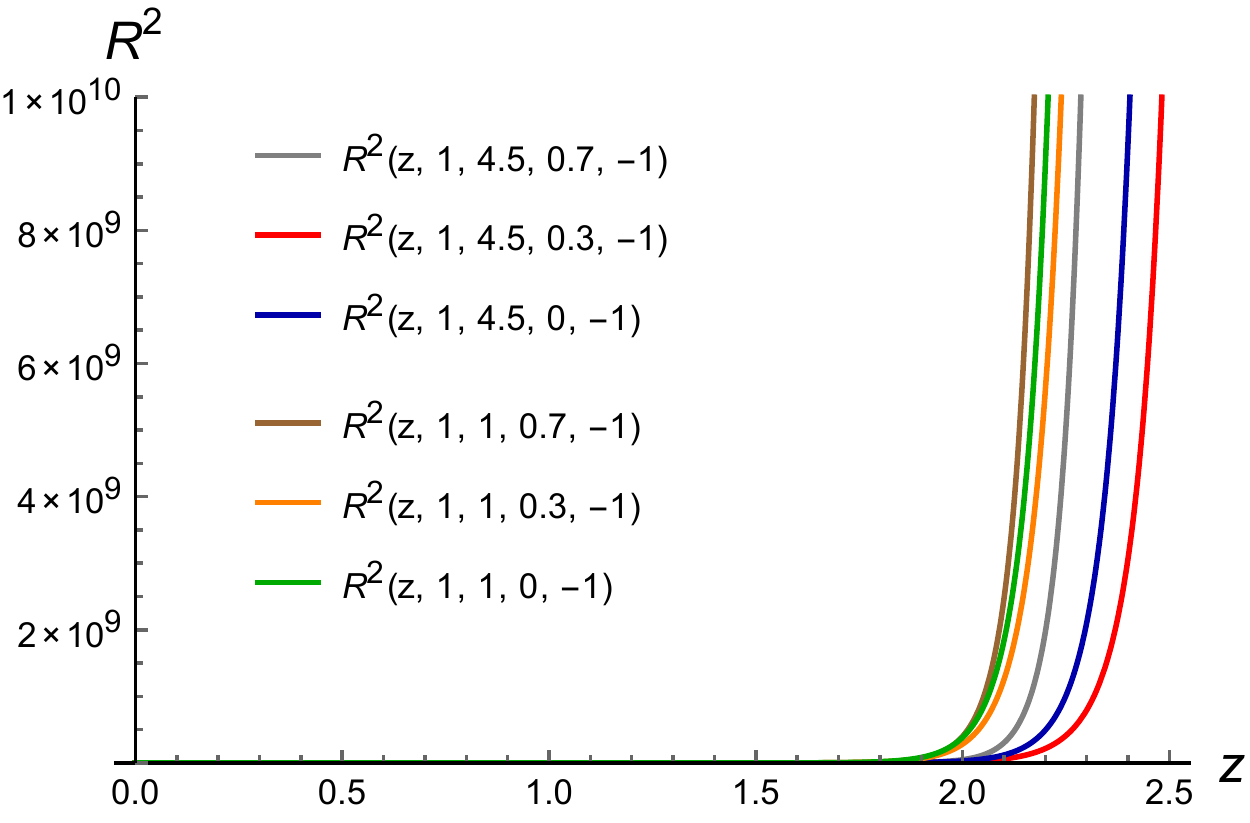} \quad
  \includegraphics[scale=0.56]{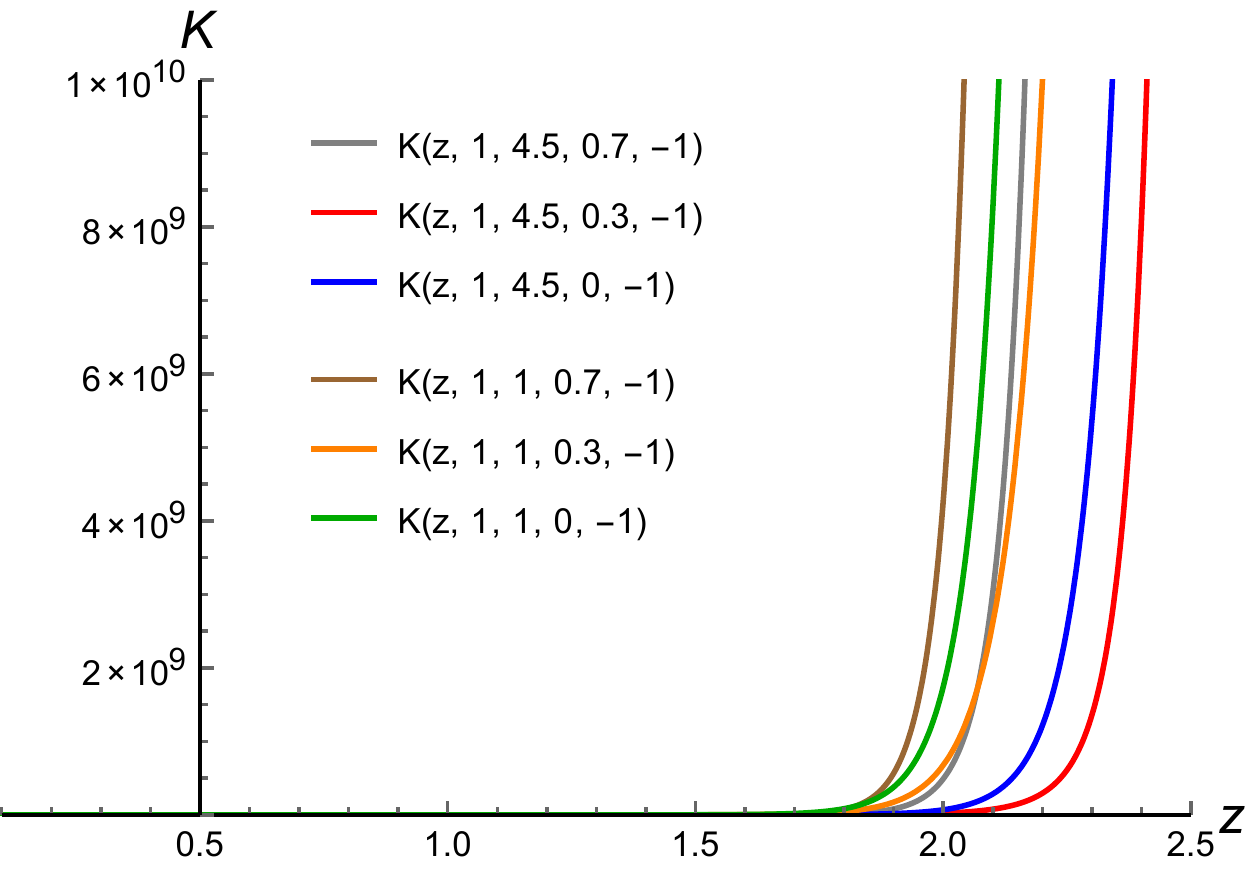} \\
  A\hspace{220pt}B
  \caption{Ricci $R^2(z)$ and Riemann $K(z)$ invariants for $z_h = 1$,
    $c =  -1$ in isotropic, $\nu = 1$, and anisotropic, $\nu = 4.5$,
    cases for different $\mu$.}
  \label{Fig:curv-negative-c}
\end{figure}

For completeness we present here the dependence of the scalar
invariants $R$, $R^2 = R_{\mu\nu}R^{\mu\nu}$ and $K =
R_{\mu\nu\rho\sigma}R^{\mu\nu\rho\sigma}$ on the parameter $\mu$ for
the unit horizon and negative warp factor coefficient $c$. All the
invariants are smooth inside the black hole and start to diverge for
$z > z_h$ (Fig.\ref{Fig:sc-curv-negative-c},
\ref{Fig:curv-negative-c}). In isotropic case it happens earlier,
means for smaller $z$, than is anisotropic one. Thus the horizons of
the blackening function, depicted on Fig.\ref{Fig:black}, are
regular.

\newpage

\subsection{RG flow} \label{Sec:RG}

\begin{figure}[h!]
  \centering
   \includegraphics[scale=0.55]{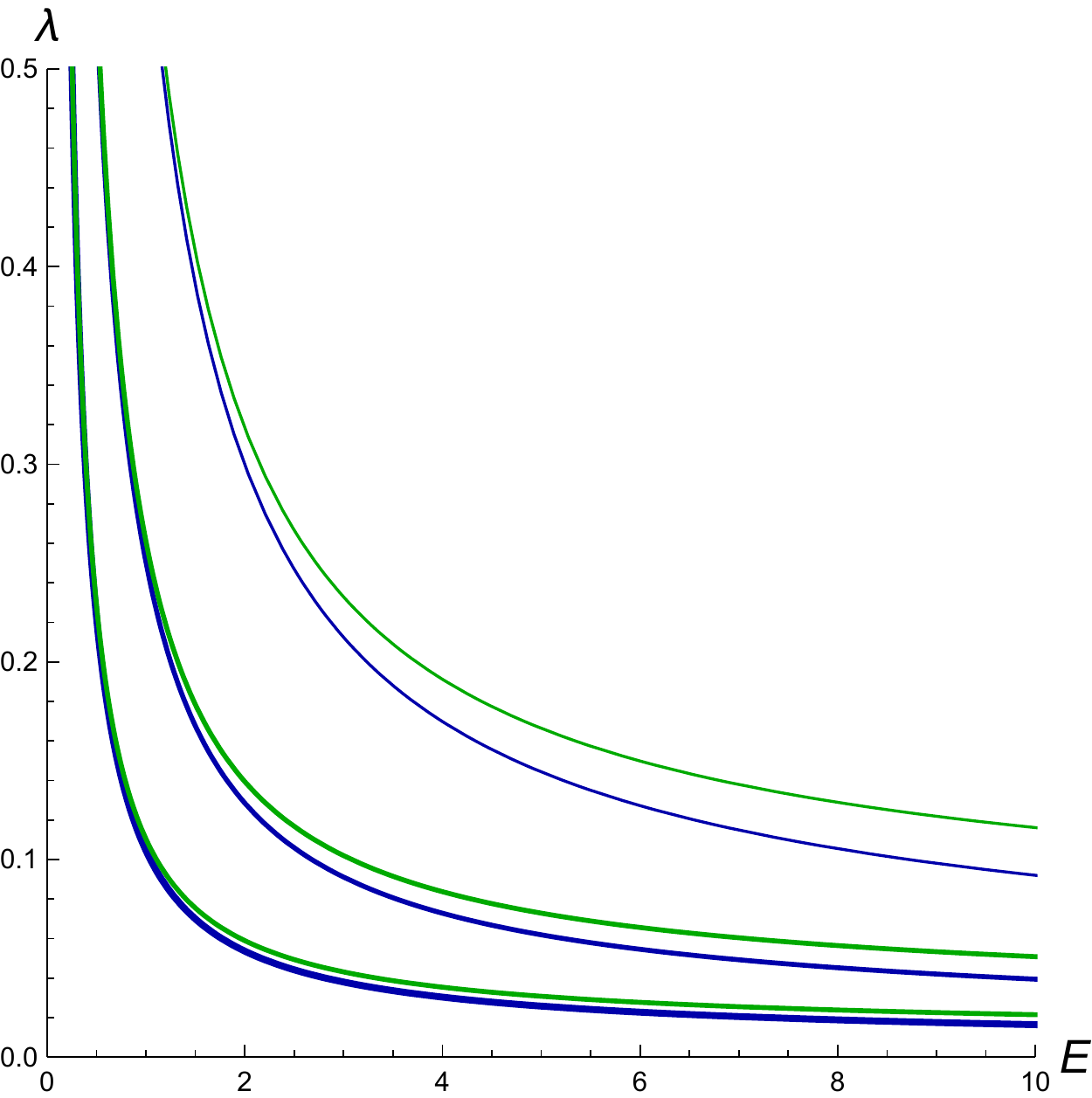} \quad
   \includegraphics[scale=0.75]{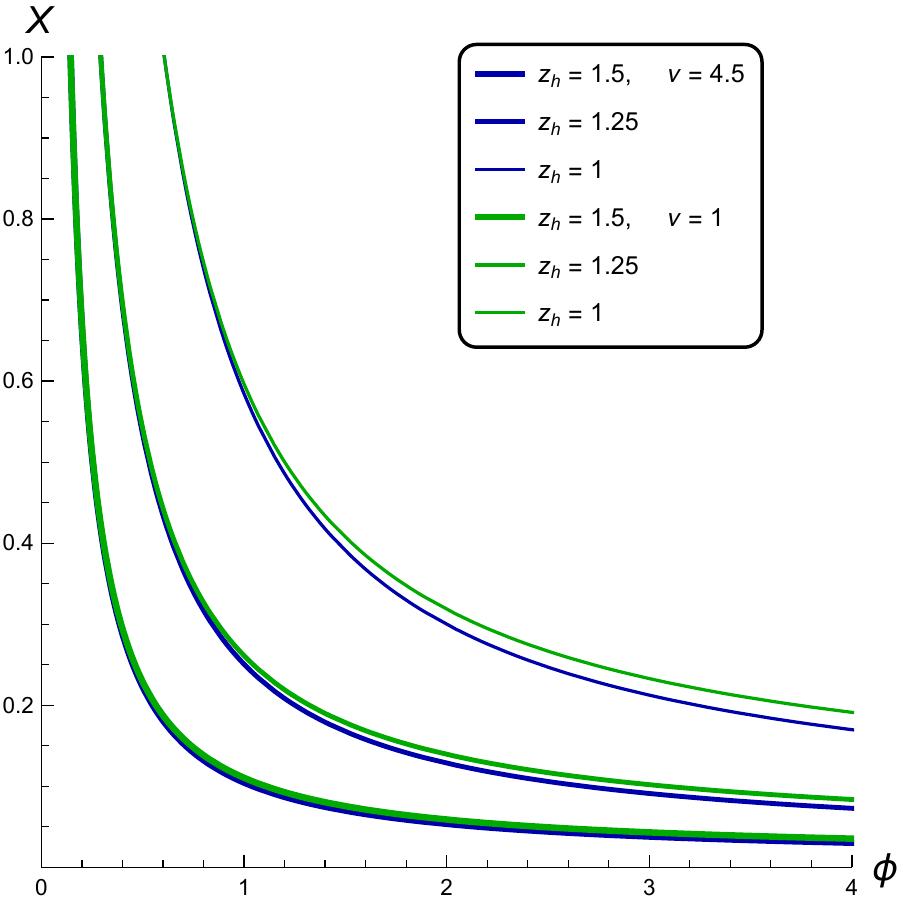}\\
   A\hspace{200pt}B
  \caption{The running coupling as function of the energy scale (A)
    and the RG flow (B) in isotropic and anisotropic $\nu = 4.5$ case;
    the plot legends are the same for both panels.}
  \label{Fig:RG}
\end{figure}

Our background is an anisotropic analog of the background used in the
improved holographic QCD model \cite{0812.0792}. The holographic
coordinate $z$ corresponds to the 4D RG scale. According to
holographic dictionary one identifies the 4D energy scale $E$ with the
metric scalar factor, i.e. $E = E_0 \, L \, \sqrt{b(z)}/z \equiv E_0 L
B(z)$, in what follows we put $E_0 L = 1$. The running 't Hooft
coupling $\lambda_t$ is identified with the string coupling $\lambda =
e^\phi$ up to a factor, $\lambda = \kappa \lambda_t$. In
Fig.\ref{Fig:RG}.A we show the dependence of coupling constant on the
energy parameter for isotropic and anisotpopic cases. We see that the
running coupling constant decreases from the IR region to the UV
region. This behavior reproduces our expectations of the running
coupling view in a nonperturbative QCD. Note, that the anisotropic
case does not differ much from the isotropic one. The difference
becomes more essential for small $z_h$.

The $\beta$-function in terms of the background
is defined as \cite{DeWolf, Kiritsis:2014kua}
\be
  \beta(\lambda) = \cfrac{d \lambda}{d \log E} = \lambda \, \cfrac{d
    \phi}{d \log B}, \qquad B = \cfrac{\sqrt{b(z)}}{z}.
\ee
Introducing the function $X$, related with the $\beta$-function as
\be
  X(\phi) = \cfrac{\beta(\lambda)}{3\lambda}, \label{X}
\ee
the function $Y$, related with the blackening function as
\bea
  Y(\phi) &=& \cfrac14 \, \cfrac{g'}{g} \, \cfrac{B}{B'}, \label{Y}
\eea
and the function $H$, related with the vector field provided by
non-zero chemical potential as
\bea
  H(\phi) &=&\frac{A_t'}{B}, \label{H}
\eea
one can check that in the isotropic case due to E.O.M. these
quantities satisfy the first order differential equations 

\bea
  \cfrac{dX}{d\phi} &=& - \ \cfrac{4}{3} \left( 1 - \cfrac{3}{8} \ X^2
    + Y \right) \left( 1 + \cfrac{1}{X} \ \cfrac{2 \partial_{\phi} V -
      H^2 \ \partial_{\phi} f_1}{2 V + H^2 f_1} \right), \\
  \cfrac{dY}{d\phi} &=& - \ \cfrac{4 Y}{3 X} \left( 1 - \cfrac{3}{8} \
    X^2 + Y \right) \left( 1 + \cfrac{3}{2 Y} \ \cfrac{H^2 f_1}{2 V
      + H^2 f_1} \right), \\
  \cfrac{dH}{d\phi} &=& - \left( \cfrac{1}{X} +
    \cfrac{\partial_{\phi} f_1}{f_1} \right) H,
\eea
where $\partial_\phi = \partial/\partial \phi$. The anisotropic case is
more subtle and will be the subject of a forthcoming paper.

\begin{figure}[h!]
  \centering
  \includegraphics[scale=0.75]{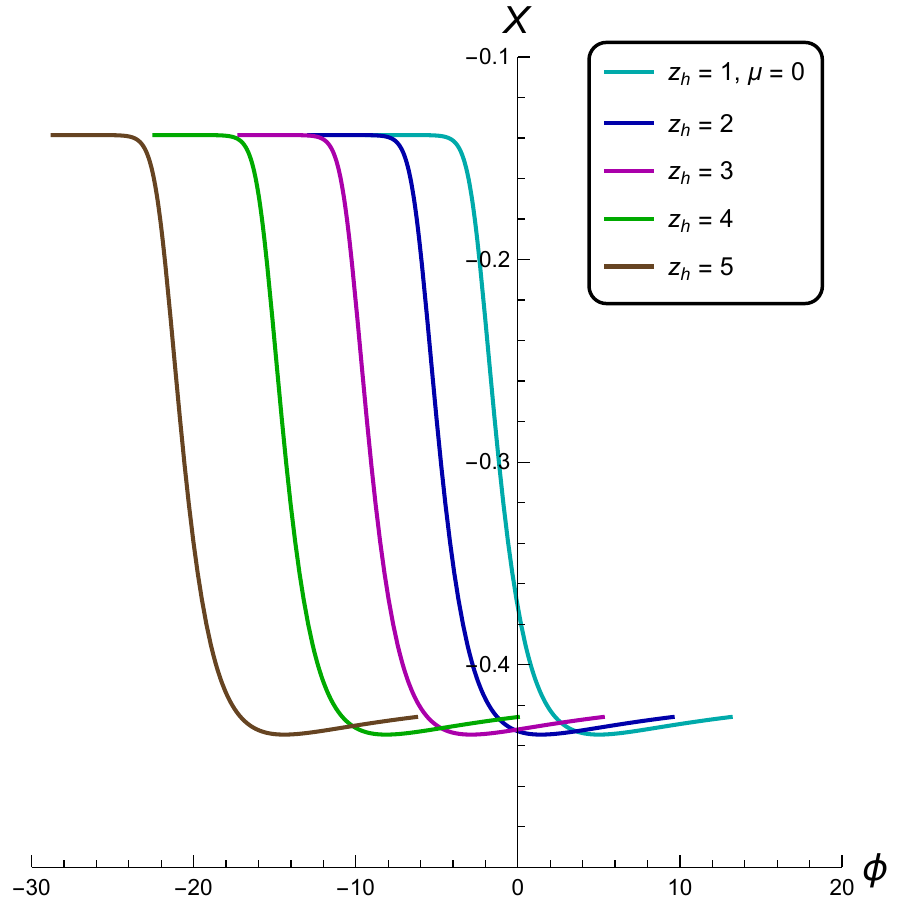}\\
  A\\
  \includegraphics[scale=0.75]{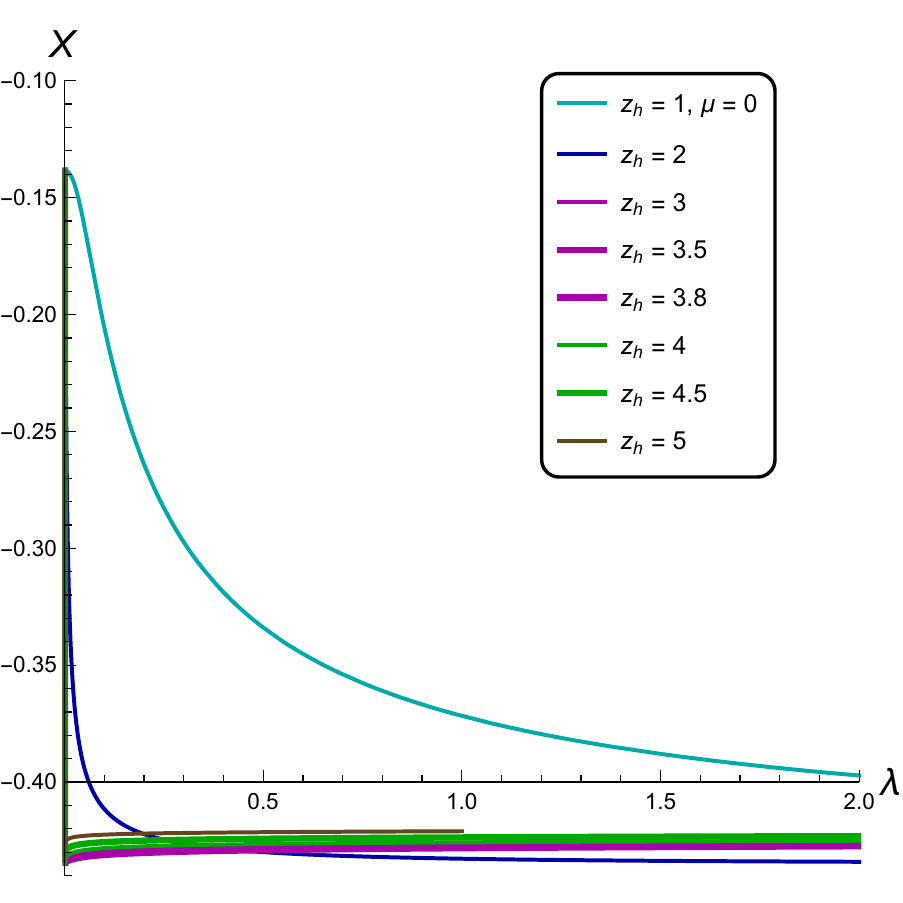} \quad
  \includegraphics[scale=0.75]{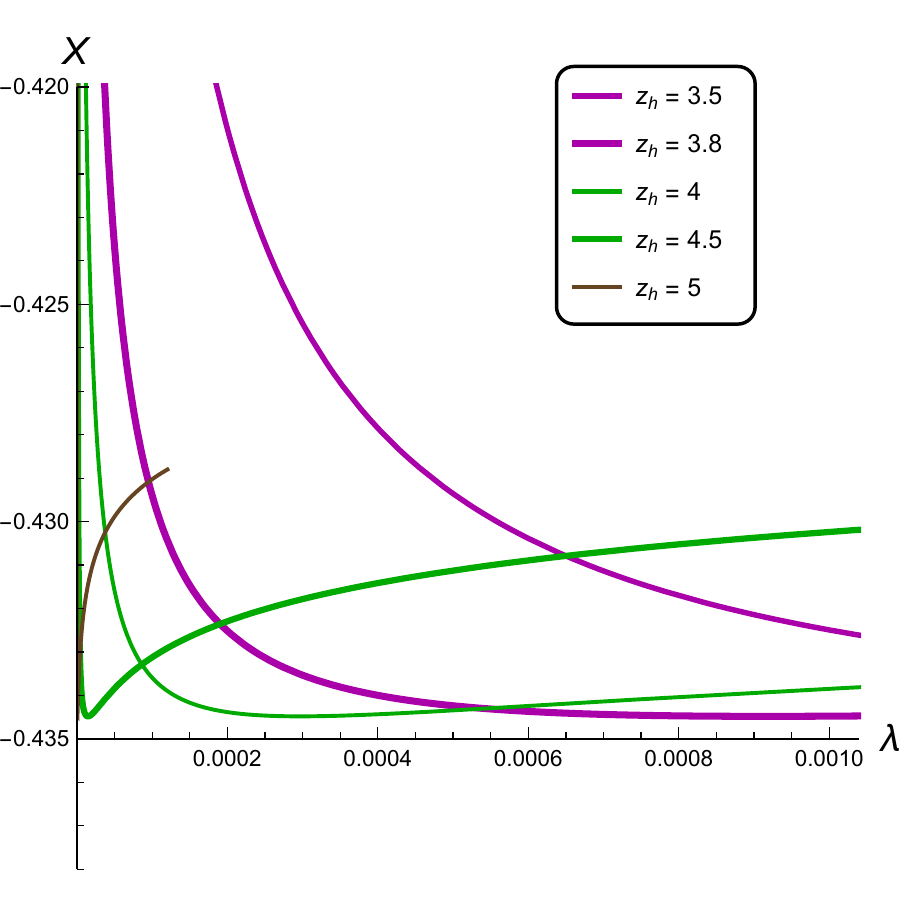}\\
  B \hspace{200pt}C\\
  \caption{The RG flows for anisotropic case $\nu = 4.5$  and
    different $z_h$ in the $(\phi,X)$-plane (A), in the
    $(\lambda,X)$-plane (B) and in a small part of (B) near the origin
    (C).}
  \label{Fig:RGX}
\end{figure}
\begin{figure}[h!]
  \centering
  \includegraphics[scale=0.55]{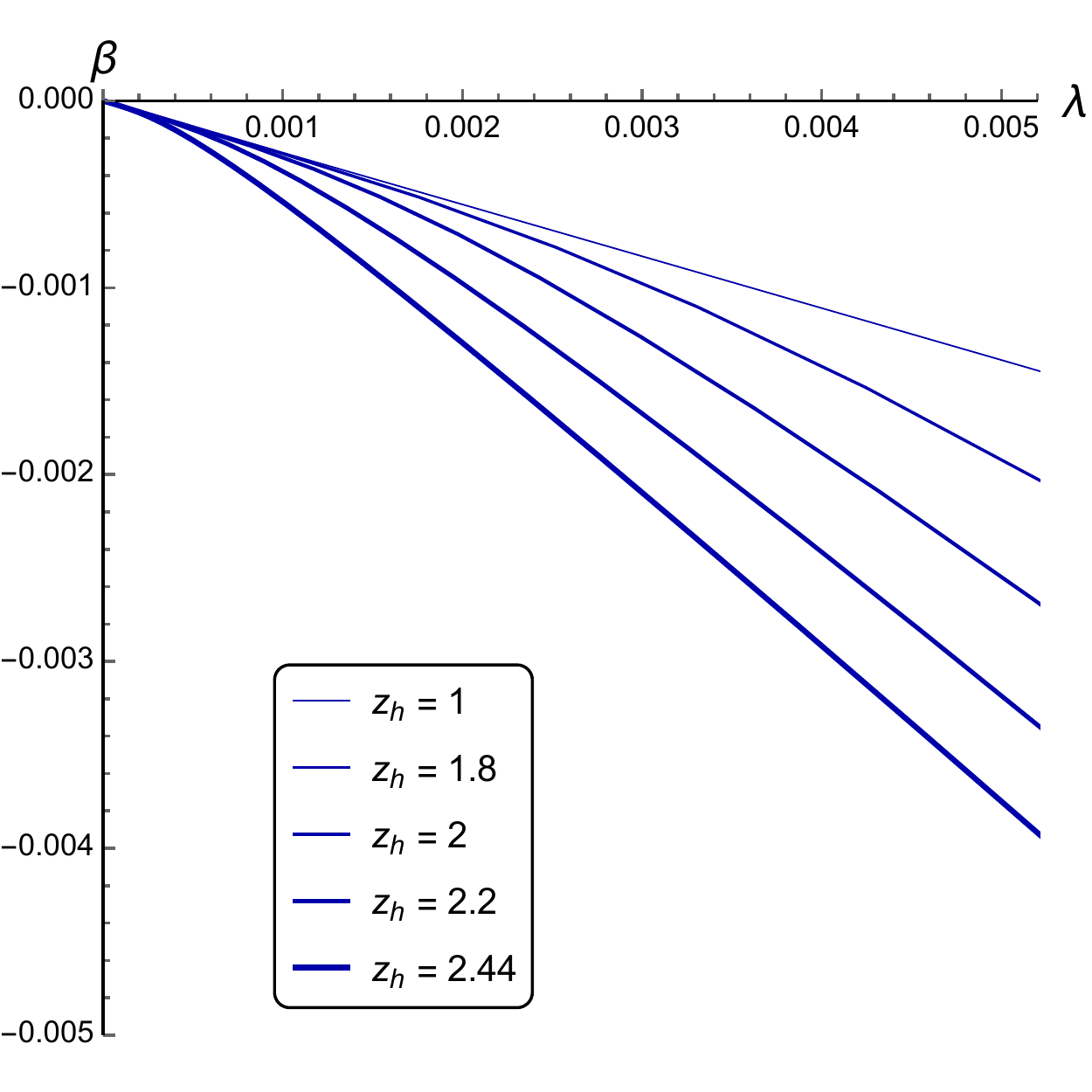} \quad
  \includegraphics[scale=0.55]{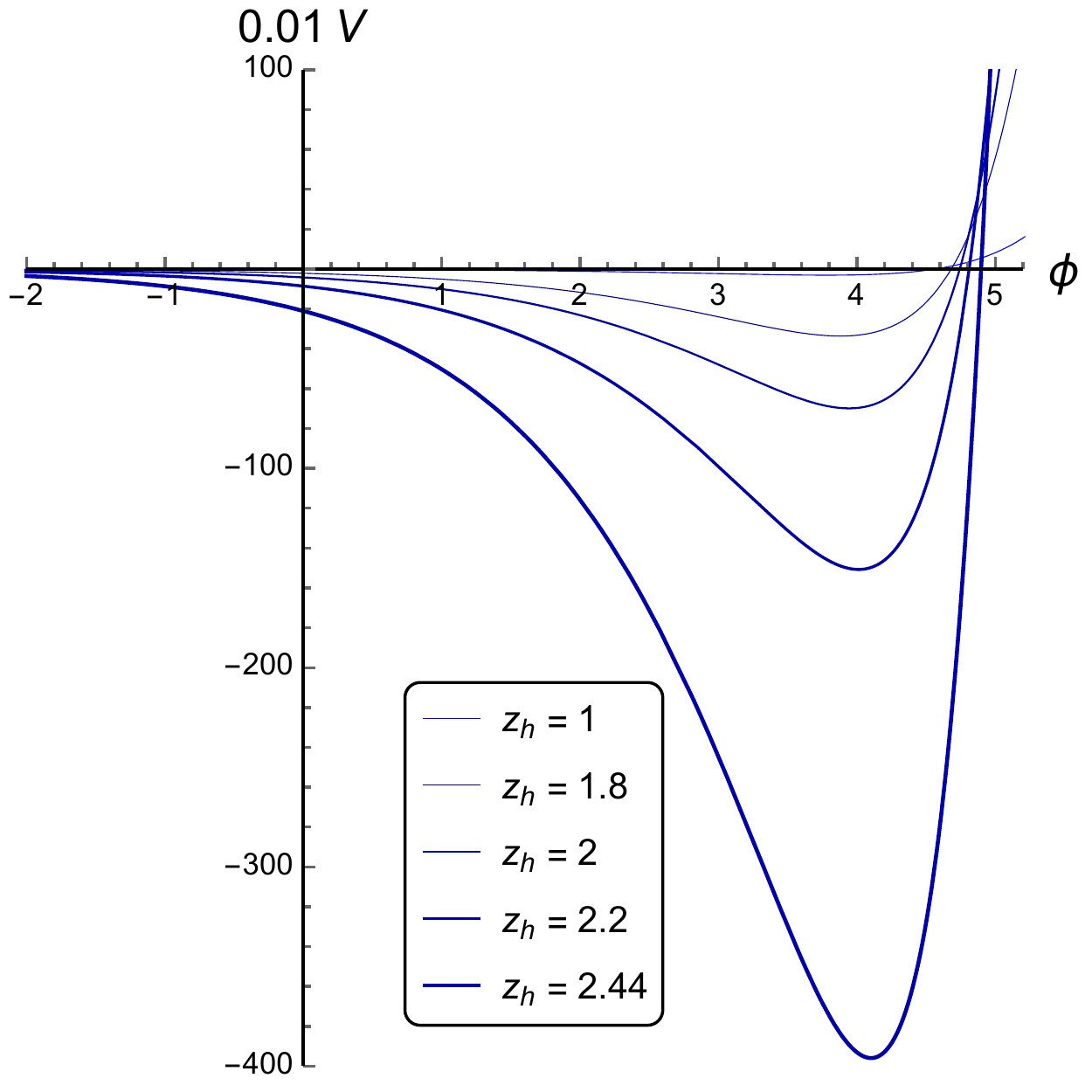}\\
  A \hspace{200pt} B \\
  \includegraphics[scale=0.55]{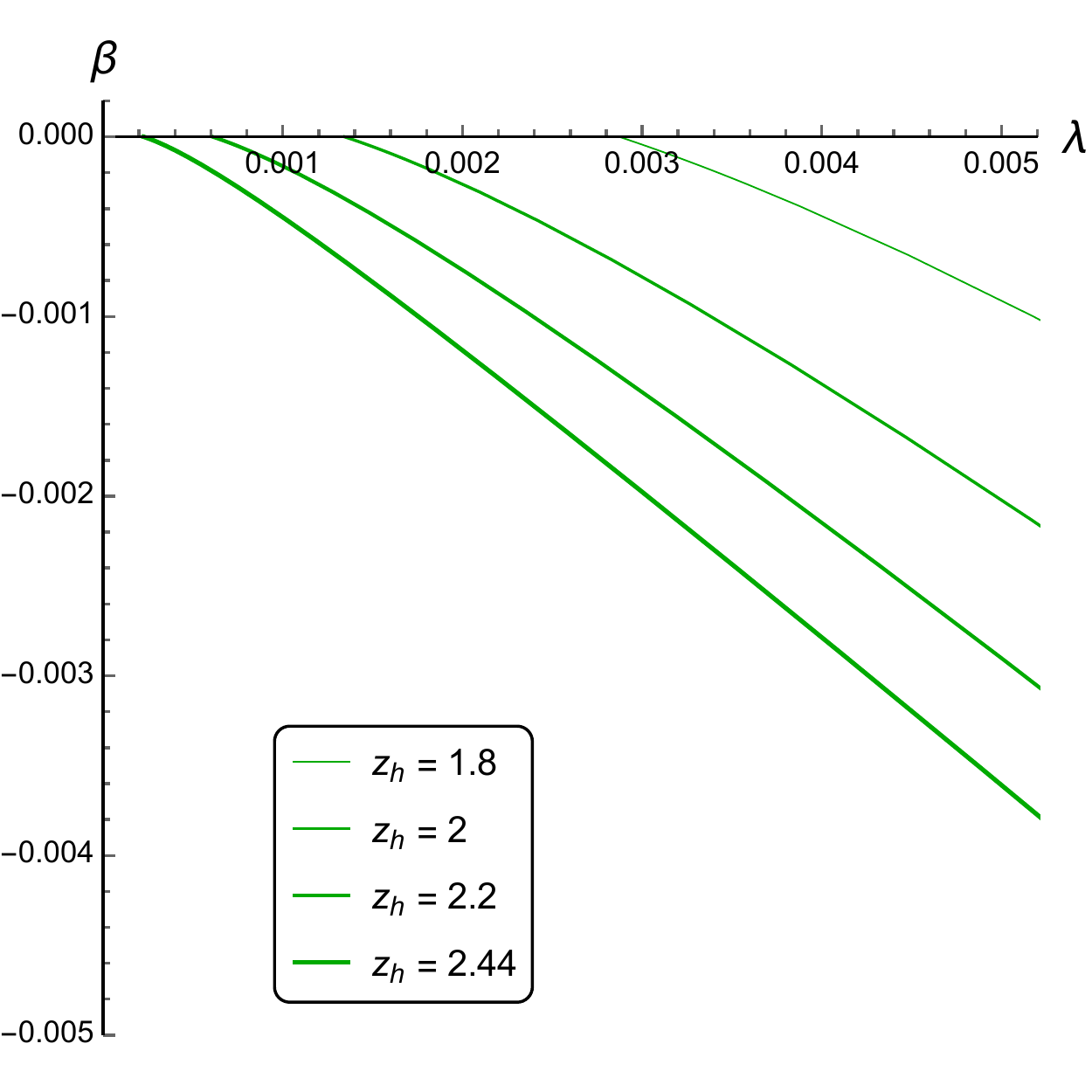} \quad
  \includegraphics[scale=0.55]{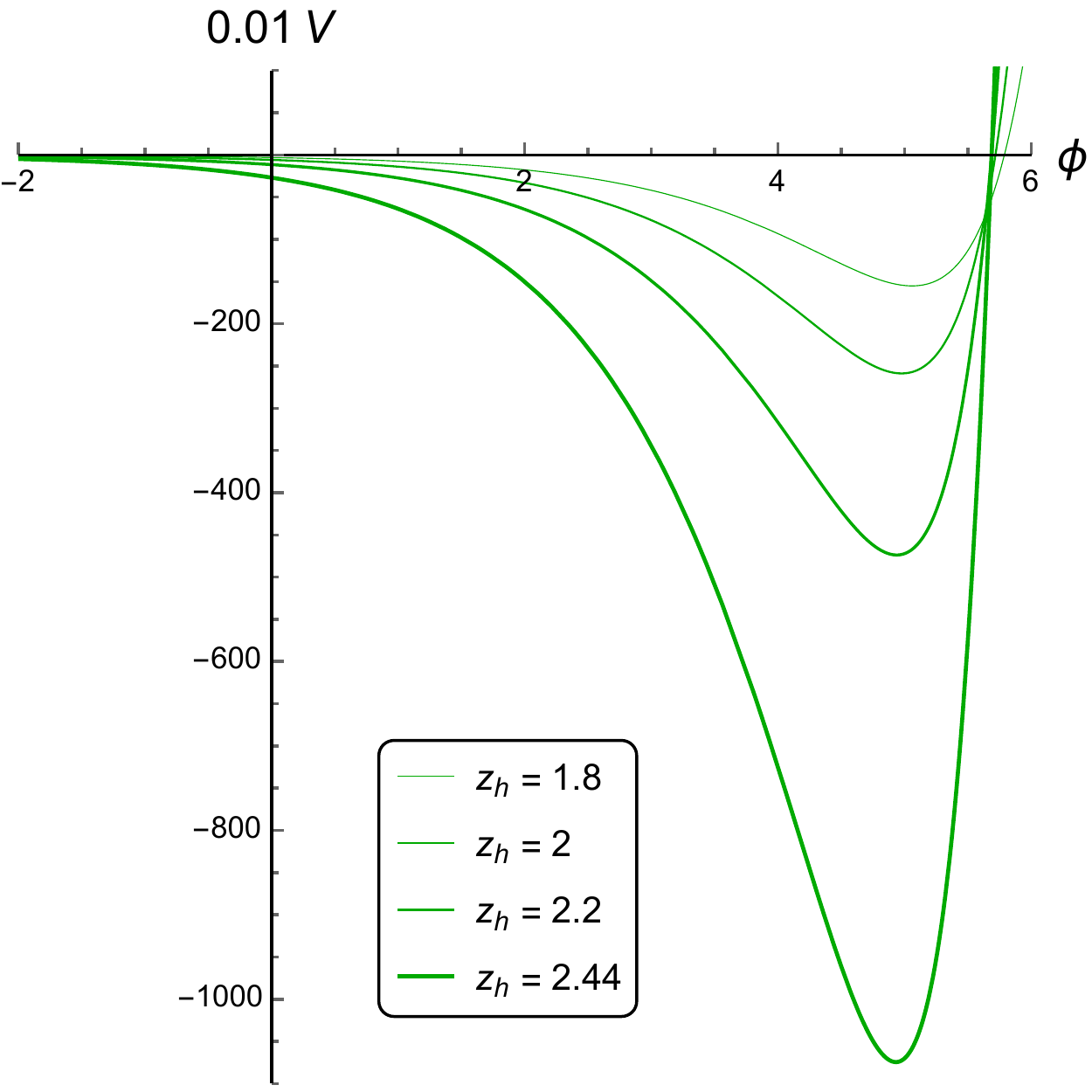}\\
  C \hspace{200pt} D
  \caption{The function $\beta(\lambda)$ in anisotropic, $\nu = 4.5$,
    (A) and isotropic, $\nu = 1$, (C) cases near small $\lambda$;
    scalar potentials $V(\phi)$ for $\mu = 0$ in anisotropic, $\nu =
    4.5$, (B) and isotropic, $\nu = 1$, (D) cases.}
  \label{Fig:beta}
\end{figure}

Fig.\ref{Fig:RGX} shows the $X$-flow in the anisotropic case for $\nu
= 4.5$ and different $z_h$ in the $(\phi,X)$- and in the
$(\lambda,X)$-planes. The function $X(\phi)$ decreases starting from a
constant value up to a local minimum as the argument grows and is
shifted to the right for larger horizon. The function $X(\lambda)$ for
$z_h < 4$ smoothly decreases with increasing $\lambda$ and for $z_h >
4$ the dependence is more complicated.
 
In Fig.\ref{Fig:beta} the behavior of functions $\beta(\lambda)$ in
anisotropic, $\nu = 4.5$, (Fig.\ref{Fig:beta}.A) and isotropic
(Fig.\ref{Fig:beta}.C) cases and the corresponding potentials
$V(\phi)$ (Fig.\ref{Fig:beta}.B and D) for the same horizon values are
shown. We see, that for both cases $\beta(\lambda) < 0$ in an
agreement with the asymptotical freedom. For larger $z_h$ both
functions display more non-linearity and decrease faster with the
argument grow. This tendency is peculiar either for the isotropic or
the anisotropic case. The only special difference is the non-zero
value of $\beta(0)$ for $\nu = 1$. We expect that changing the form of
$P(z)$ we can, as in the isotropic case \cite{DeWolf,
  Kiritsis:2014kua, Pirner:2009gr}, to recover the first orders
expression of the perturbative $\beta$-function.

\begin{figure}[h!]
  \centering
  \includegraphics[scale=0.7]{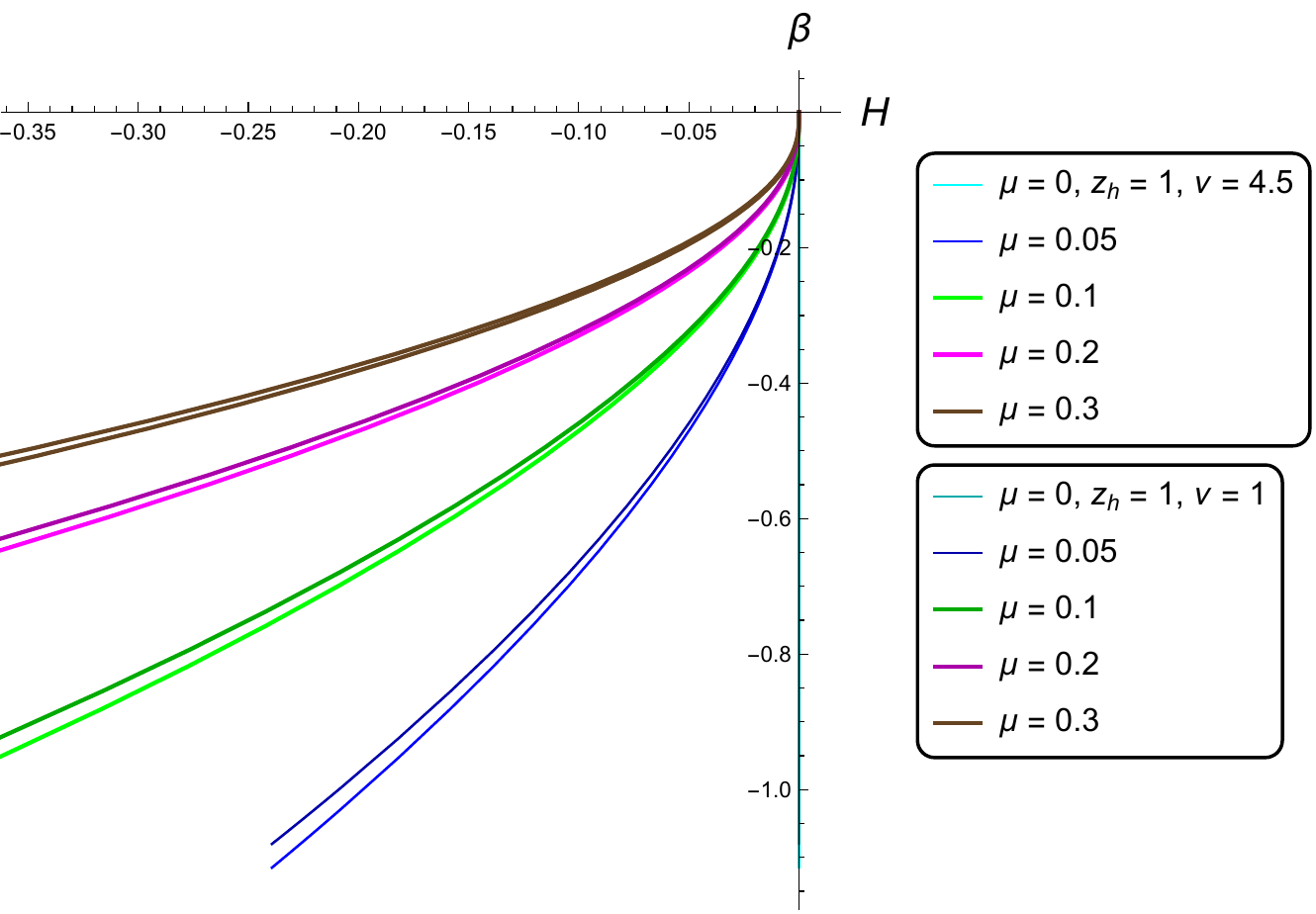}
  \caption{The function $\beta(H)$ in isotropic and anisotropic $\nu =
    4.5$ cases.}
  \label{Fig:betaH}
\end{figure}

In Fig.\ref{Fig:betaH} we show the dependence of the $\beta$-function
on $H$ for the isotropic and anisotropic ($\nu = 4.5$) cases. We see
that $\beta$ is the increasing function of $H$, approximately linear
for large negative argument values and displaying its non-linearity
near zero. The function values are visibly larger for larger chemical
potential, while the anisotropy does not change this picture much.

\begin{figure}[h!]
  \centering
 \includegraphics[scale=0.7]{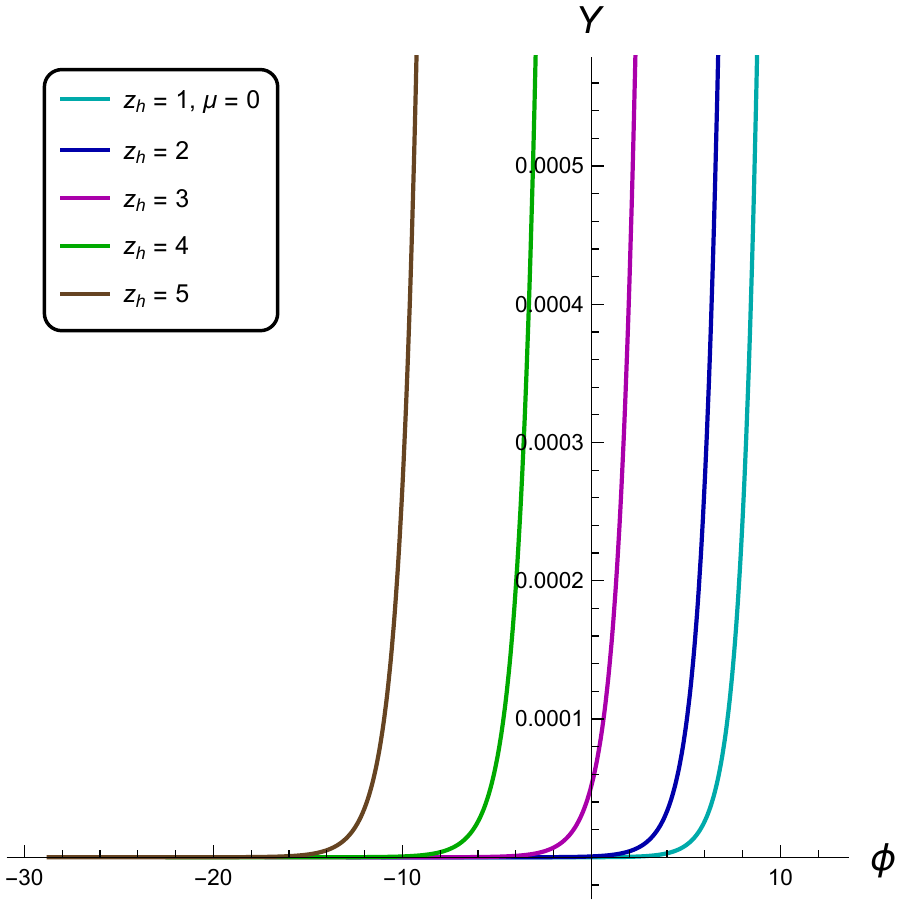} \quad
 \includegraphics[scale=0.7]{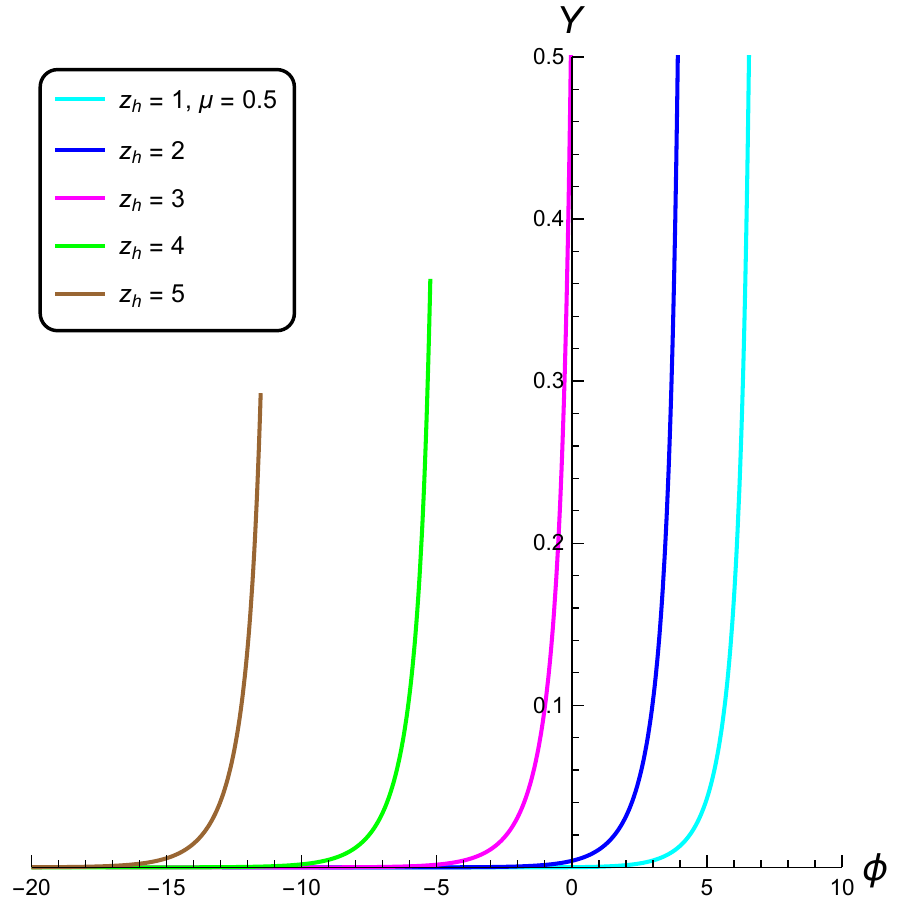}\\
 A \hspace{160pt} B
 \caption{$Y$ flows in anisotropic case $\nu = 4.5$ for different
   $z_h$, $\mu = 0$ (A) and $\mu = 0.5$ (B).}
  \label{Fig:RGY}
\end{figure}

In Fig.\ref{Fig:RGY} the $Y$-flow is shown for anisotropic case with
zero and non-zero chemical potential. The function grows rapidly and
this growth does not essentially depend on the size of the horizon.

\begin{figure}[h!]
  \centering
  \includegraphics[scale=0.54]{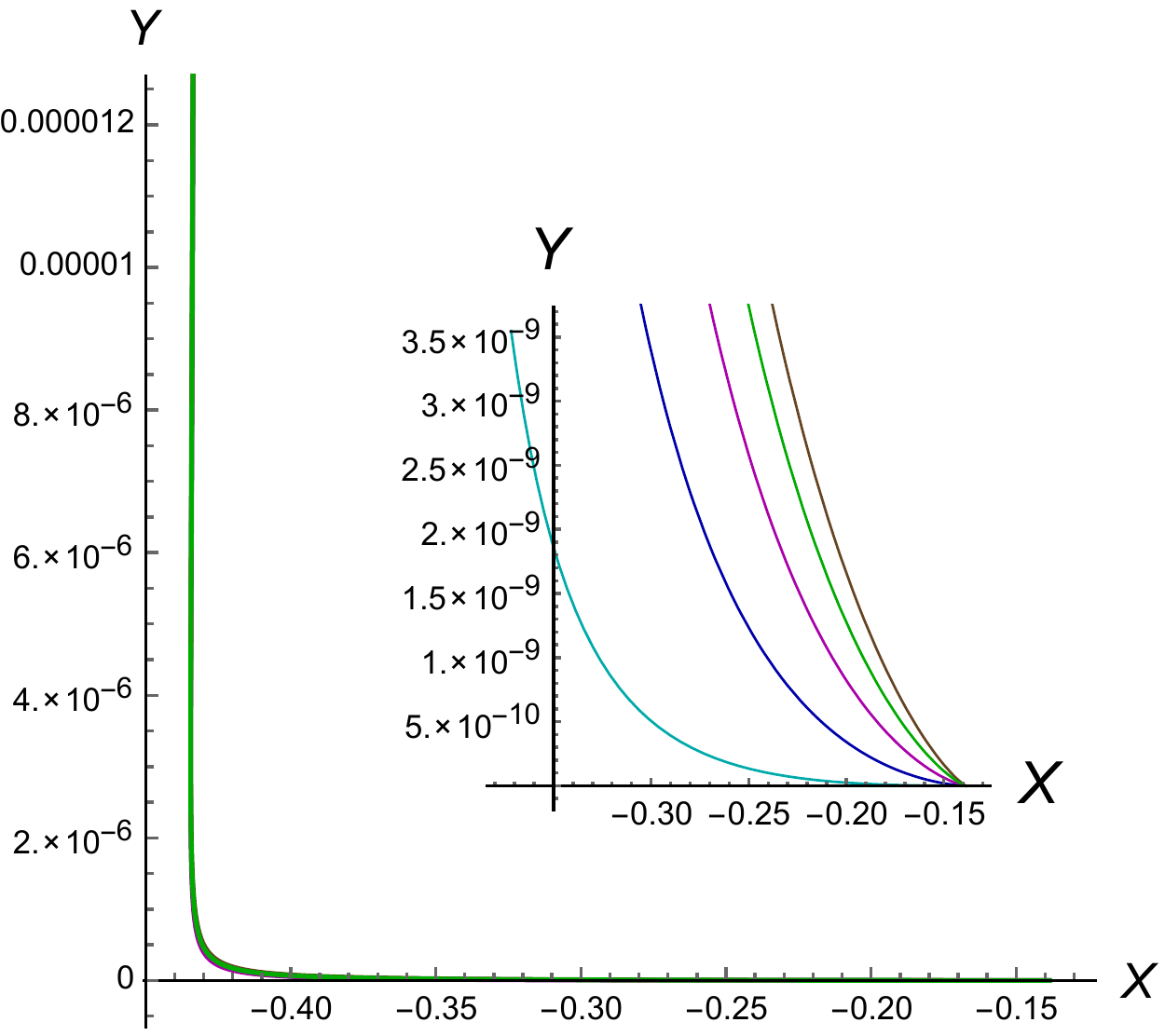} \quad
  \includegraphics[scale=0.5]{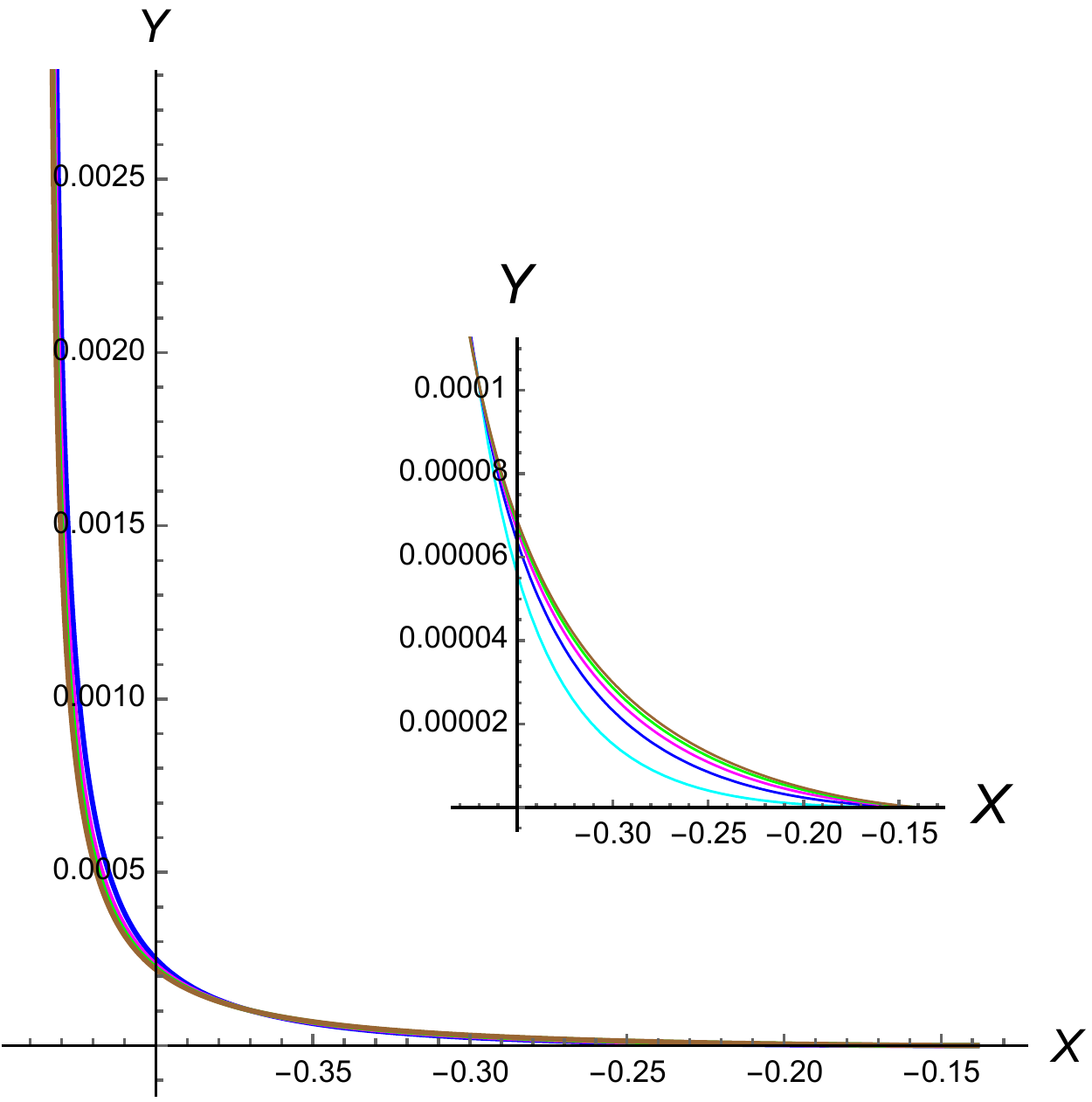}\\
  A \hspace{200pt} B
  \caption{RG flows in the $(X,Y)$-plane in anisotropic case $\nu =
    4.5$ for $\mu = 0$ (A) and $\mu = 0.5$ (B); the plot legends are
    the same as in Fig.\ref{Fig:RGY}.}
  \label{Fig:RGmu}
\end{figure}

In Fig.\ref{Fig:RGmu} we display the RG
flows in the $(X,Y)$-plane for anisotropic case $\nu = 4.5$ and zero
and non-zero chemical potential. The $X$ and $Y$ have the inverse
ratio dependence and does not change much for different $z_h$.

\begin{figure}[h!]
  \centering
  \includegraphics[scale=0.9]{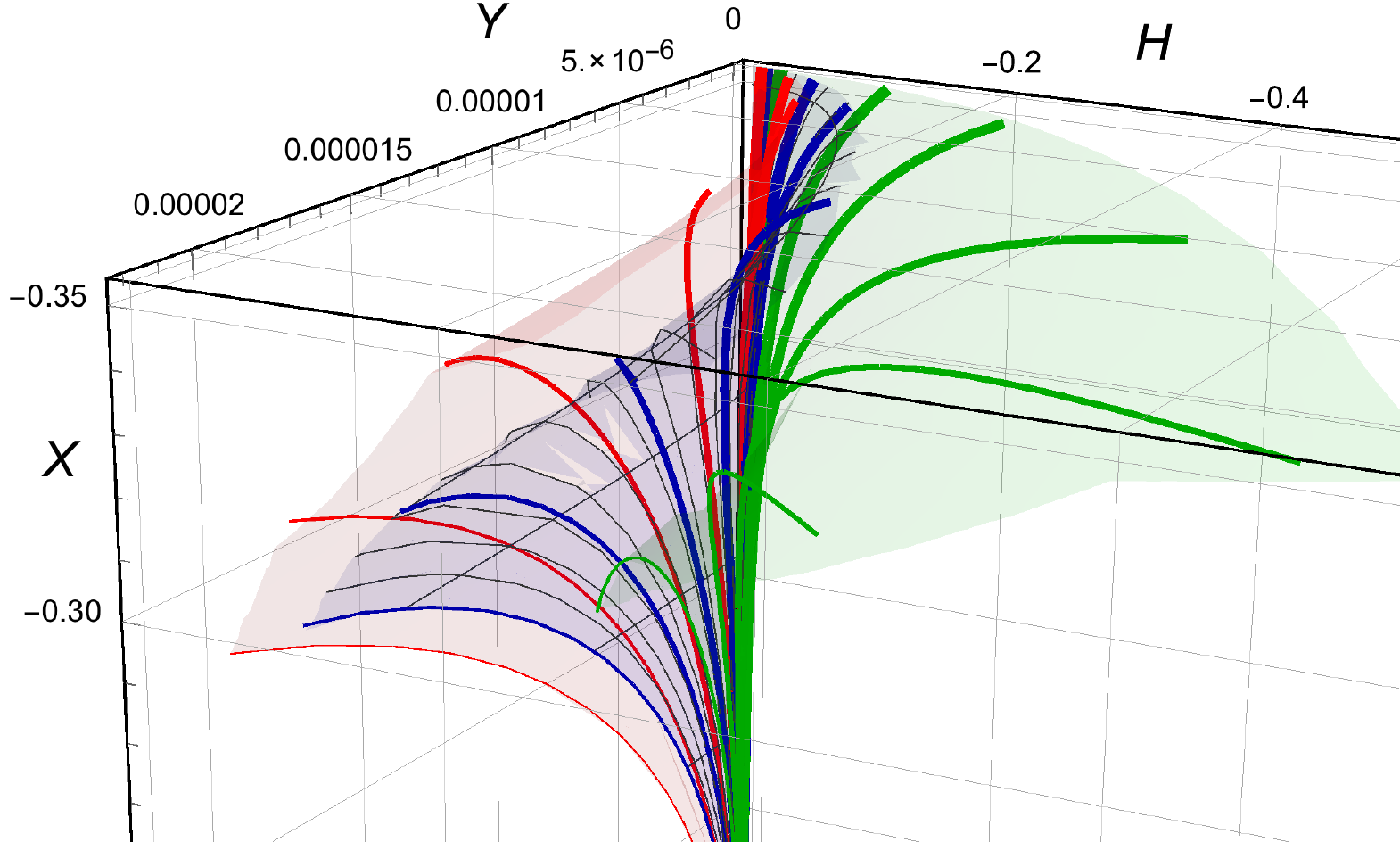} \\ A \\
  \includegraphics[scale=1]{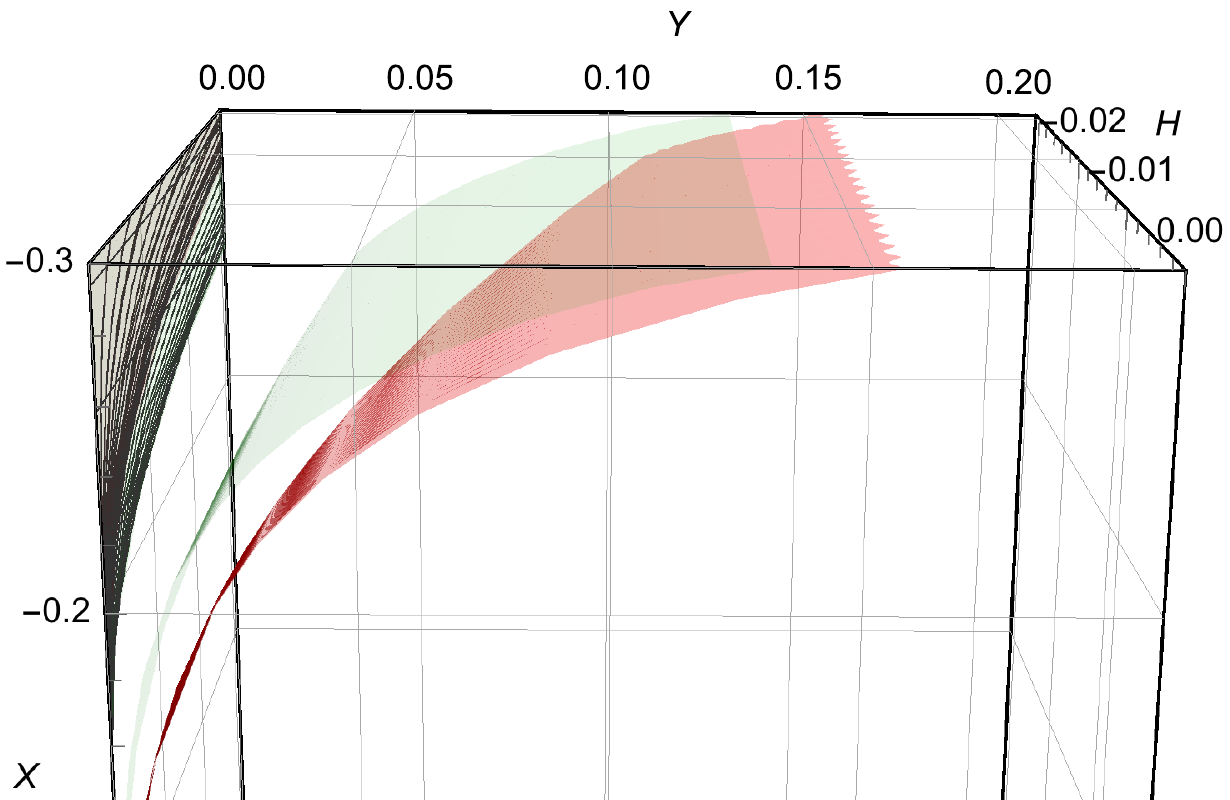} \\ B \\
  \caption{RG flows in the $(X,Y,H)$-space in anisotropic case $\nu =
    4.5$ for $z_h = 2$ (light red), $z_h = 1.5$ (light blue) and $z_h
    = 1$ (lightgreen) (A); RG flows in the $(X,Y,H)$-space in
    anisotropic, $\nu = 4.5$, (meshed lines) and isotropic cases (not
    meshed lines) for $z_h = 1$ (lightgreen) and $z_h = 2$ (lightred)
    (B).}
  \label{Fig:RGXYH}
\end{figure}

Fig.\ref{Fig:RGXYH} shows the RG flows in the $(X,Y,H)$-space. We see
that our anisotropy essentially changes the character of the flow.


\newpage
$$\,$$
\newpage

\section{Thermodynamics of the background}\label{Sec:Thermo}
\subsection{Temperature}

Calculating the derivative of the blackening function \eqref{dg} at
the horizon we get the temperature 
\bea
  T(z_h,\mu,c,\nu) = \cfrac{g'(z_h)}{4\pi} = \frac{e^{-\frac{3 c
        z_h^2}{4}}}{2 \pi z_h} \left| \cfrac{1}{\fG(\frac{3}{4}
      cz_h^2)} + \cfrac{\mu^2 c z_h^{2+\frac{2}{\nu}} e^{\frac{c
          z_h^2}{4}} }{4 \left( 1 - e^{\frac{c z_h^2}{4}}\right)^2}
    \left( 1 - e^{\frac{c z_h^2}{4}} \, \cfrac{\fG(c
        z_h^2)}{\fG(\frac34 c z_h^2)} \right) \right|. \label{temp}
\eea
Here the dependence on $\nu$ is caused by the
  function $\fG$ \eqref{expan1} in the right-hand side of \eqref{temp}.
In particular, for the zero chemical potential, $\mu = 0$,
\bea
  T(z_h,c,\nu) \equiv T(z_h,0,c,\nu) &=& \frac{e^{-\frac{3 c
        z_h^2}{4}}}{2 \pi z_h \fG(\frac{3}{4} c z_h^2)}.
\eea
For $c = 0$ it reproduces the result from \cite{AGG}:
\be
  T(z_h,0,\nu) = \lim_{c \to 0} \cfrac{e^{-\frac{3 c z_h^2}{4}}}{2 \pi
    z_h \fG(\frac{3}{4} c z_h^2)} = \cfrac{1}{2 \pi  z_h} \,
  \cfrac{\nu + 1}{\nu}. \label{fG-0}
\ee

\begin{figure}[h!]
  \centering
  \includegraphics[scale=0.75]{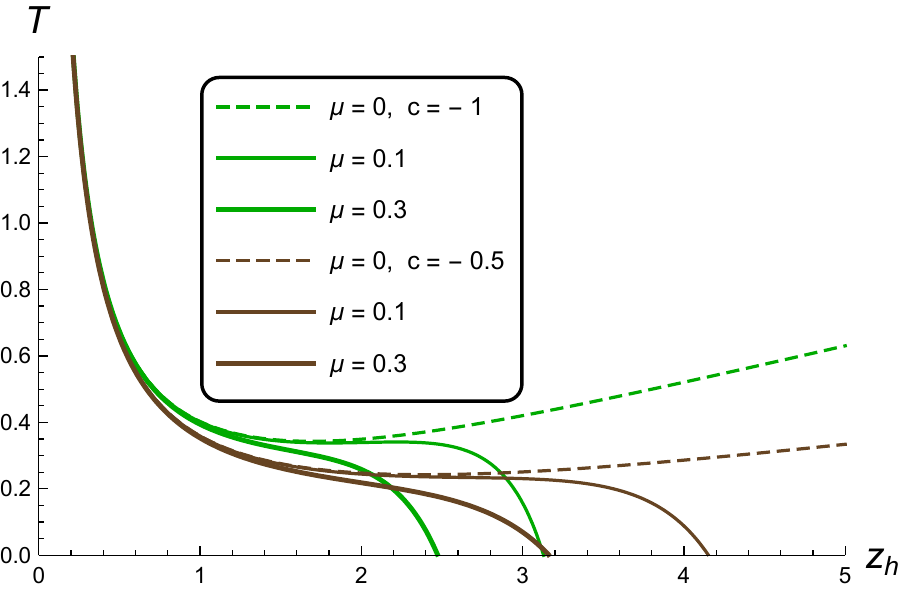} \qquad
  \includegraphics[scale=0.75]{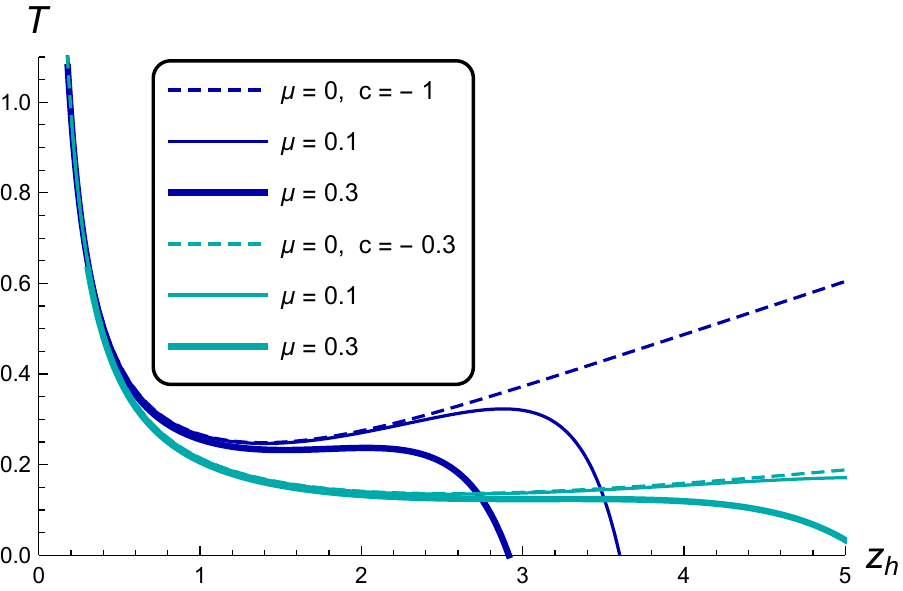} \\
  A \hspace{200pt} B
  \caption{File: {\bf
      Termo-mu-c-zh-original.nb+}. The dependence of
      temperature $T(z_h,\mu,c,\nu)$ on $z_h$ for different $c$ and
    $\mu$ in isotropic (A) and anisotropic, $\nu = 4.5$, (B) cases.}
  \label{Fig:TvszhC}
\end{figure}

\begin{figure}[h!]
  \centering
  \includegraphics[scale=0.57]{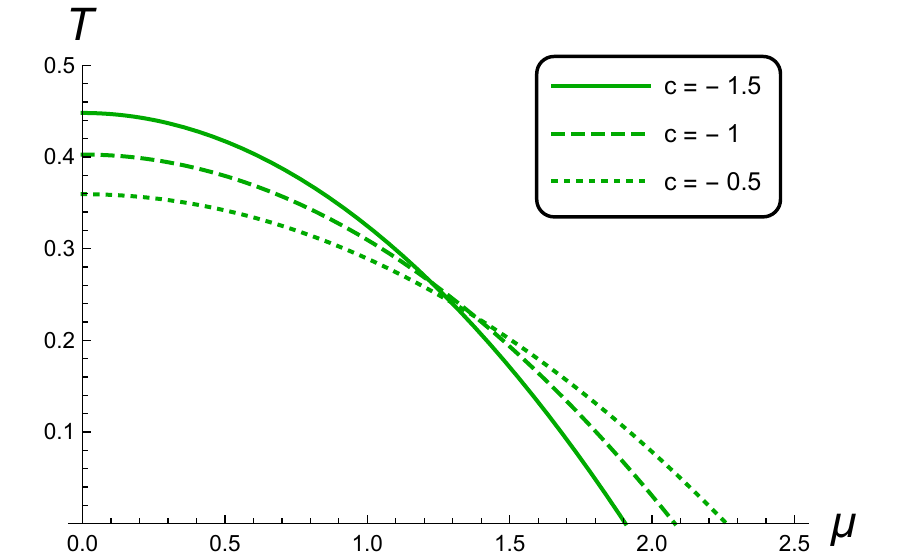}
  \includegraphics[scale=0.57]{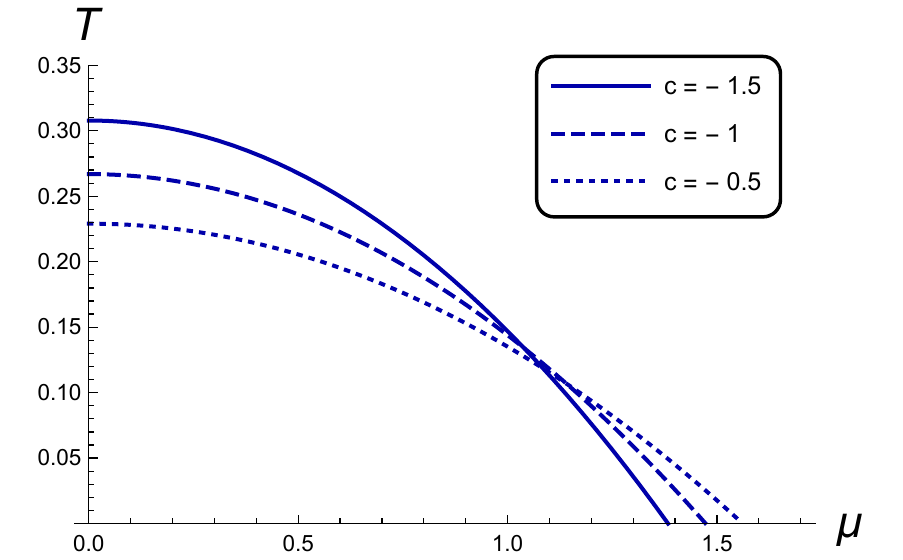}
  \includegraphics[scale=0.37]{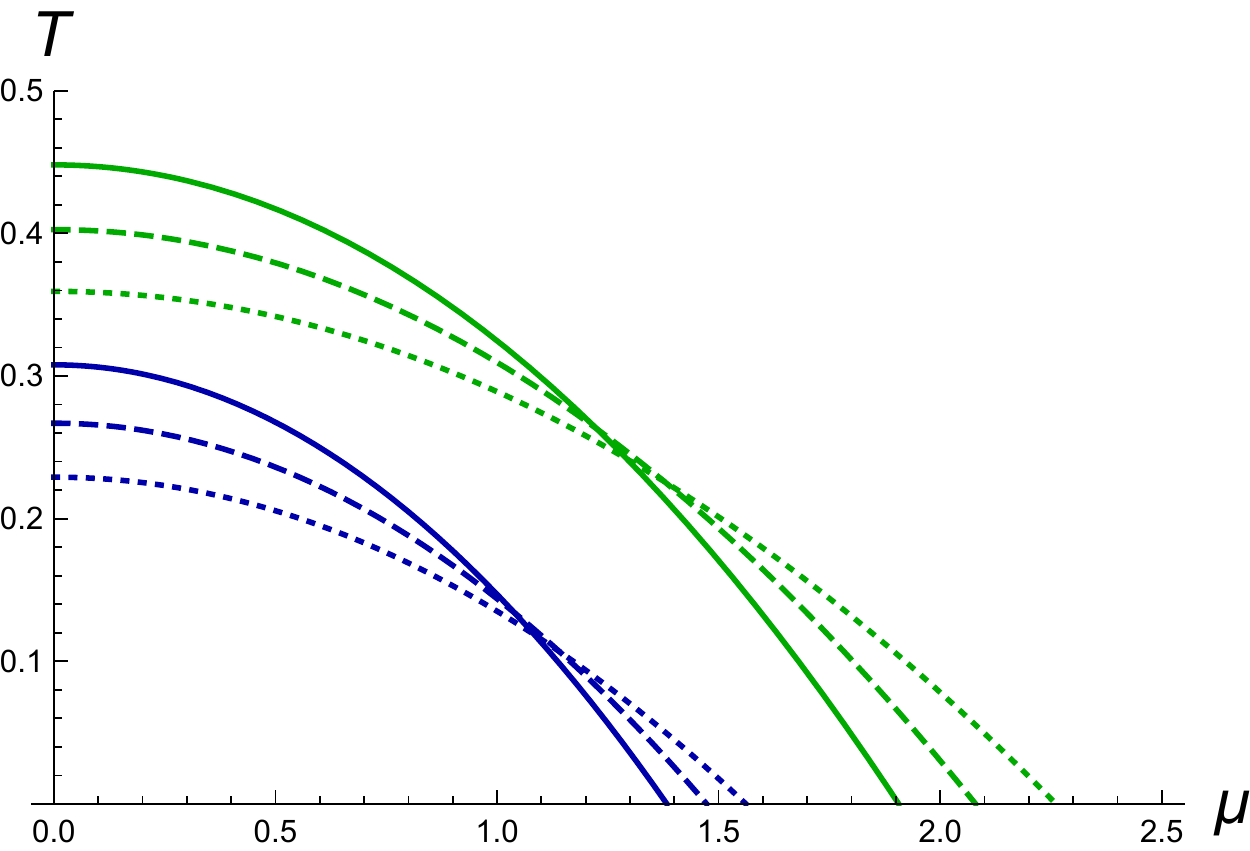}\\
  A \hspace{120pt} B\hspace{120pt} C\\ \ \\
  \includegraphics[scale=0.7]{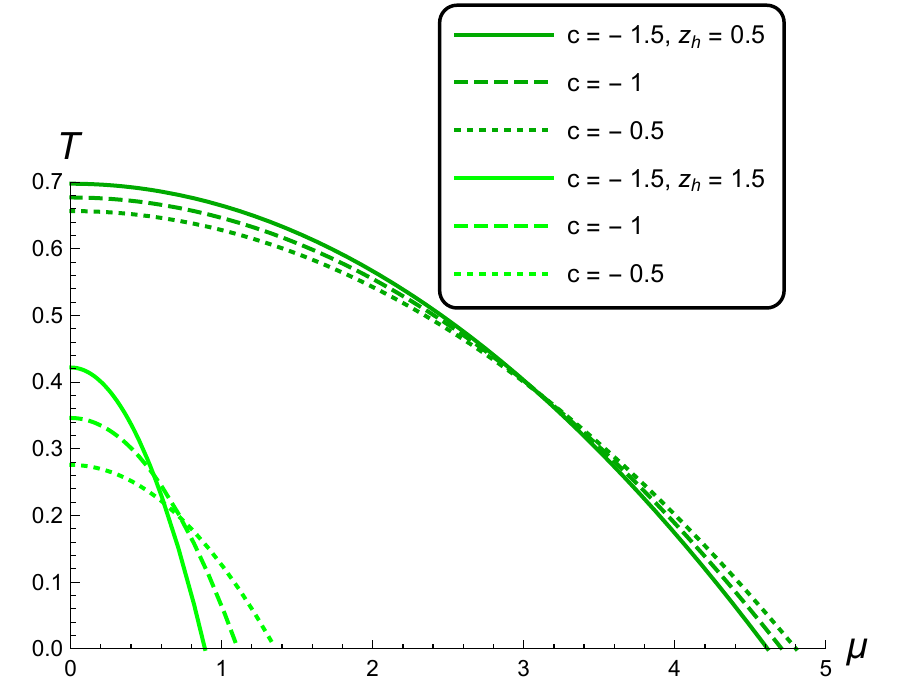}
  \includegraphics[scale=0.7]{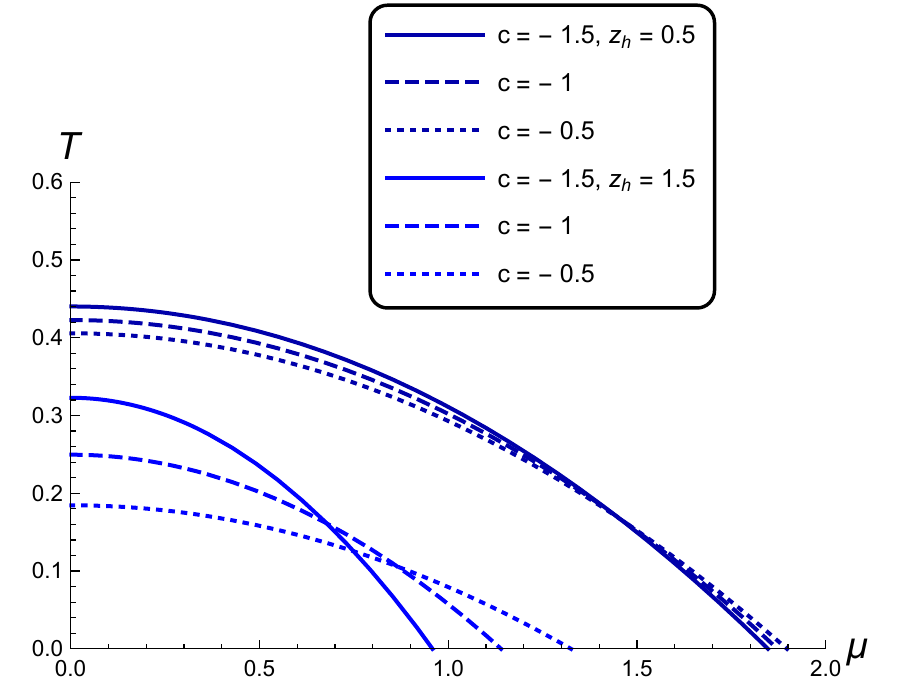}\\
  D \hspace{150pt} E
  \caption{The dependence of temperature $T(z_h,\mu,c,\nu)$ on $\mu$
    for $z_h = 1$ and different $c$ in isotropic (A) and anisotropic,
    $\nu = 4.5$, (B) cases, their the comparison with the same line
    labels (C); for small black holes with $z_h = 1.5$ and large black
    holes with $z_h = 0.5$ and different $c$ in the isotropic (D) and
    anisotropic, $\nu = 4.5$, (E) cases.}
  \label{Fig:Tvsmu}
\end{figure}

\begin{figure}[h!]
  \centering
  \includegraphics[scale=0.65]{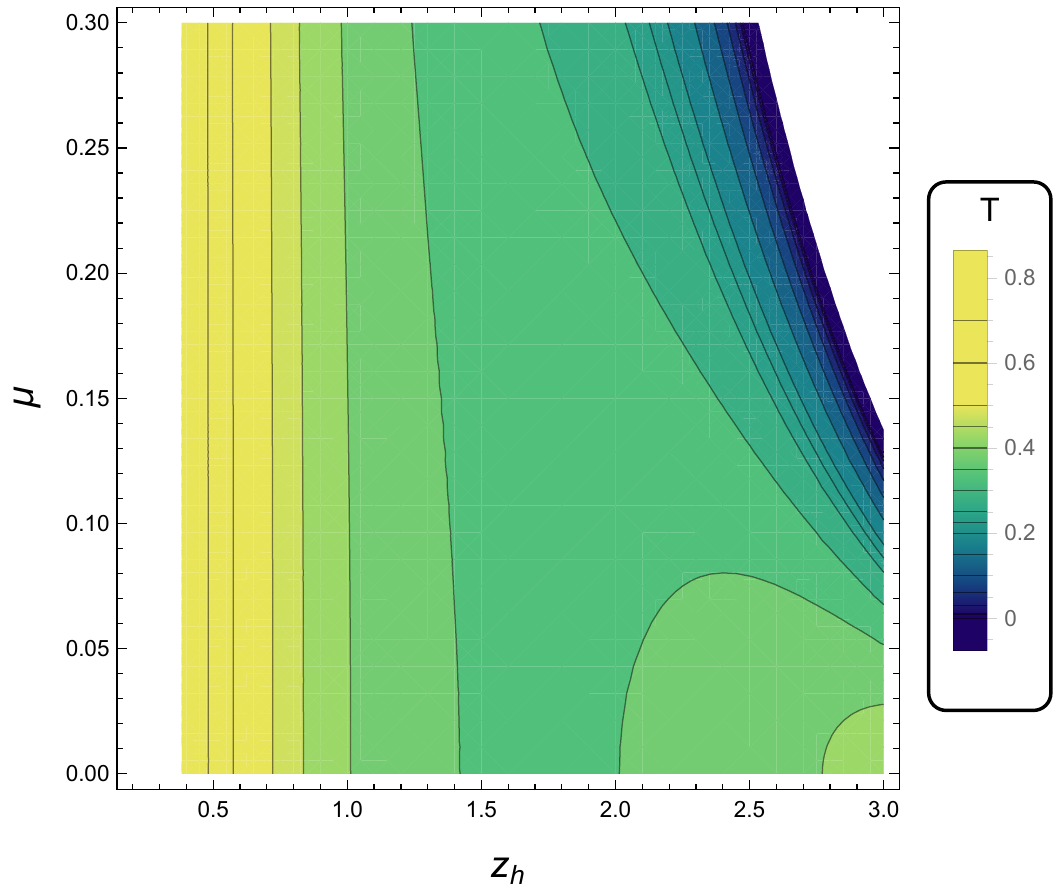} \quad
  \includegraphics[scale=0.65]{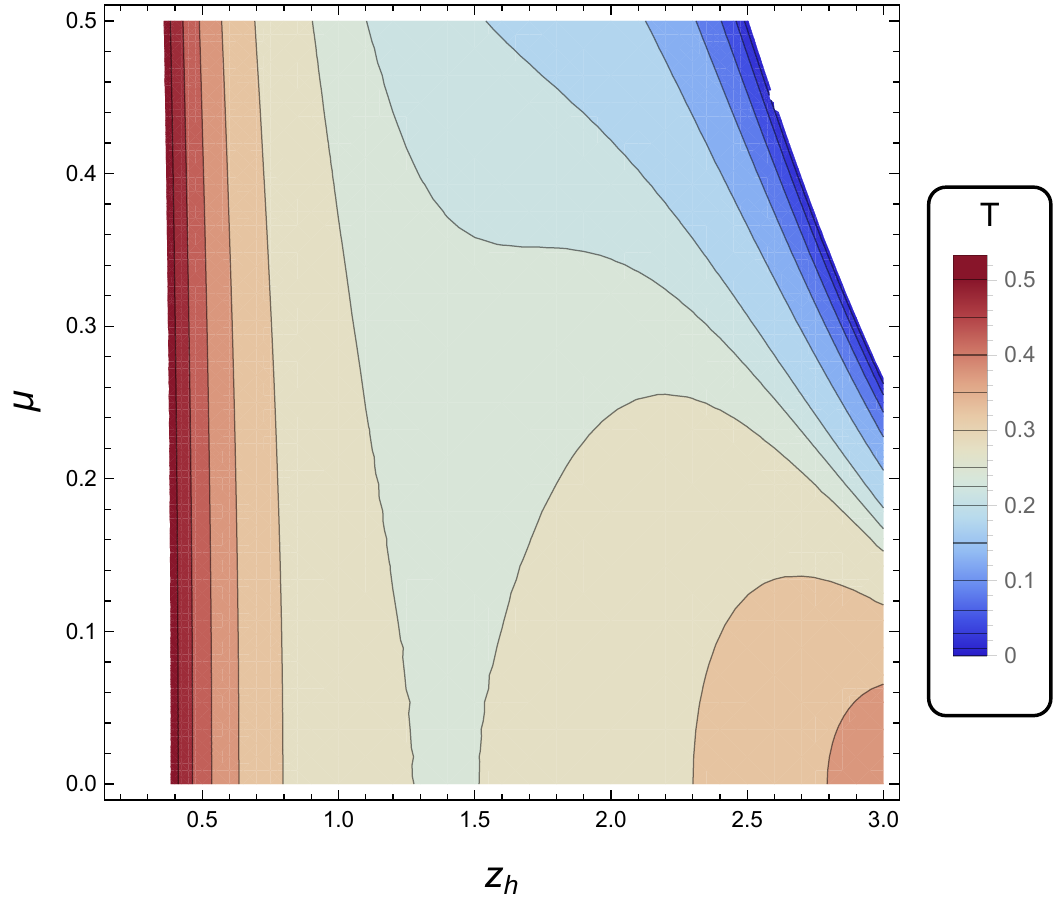}\\
  A \hspace{200pt} B
  \caption{Contour lines of the temperature $T = T(z_h,\mu,c,\nu)$ for
    $c = -1$ in the isotropic (A) and anisotropic $\nu = 4.5$ (B)
    cases.}
  \label{Fig:Tvsmuzh}
\end{figure}

From \eqref{temp} we get the dependence of temperature on $z_h$,
$\mu$, $c$ and $\nu$. In Fig.\ref{Fig:Tvszh}.A and
Fig.\ref{Fig:Tvszh}.B we present the dependence of $T$ on $z_h$ for
different $\mu$ and fixed $c = -1$ for isotropic
  (Fig.\ref{Fig:Tvszh}.A) and anisotropic (Fig.\ref{Fig:Tvszh}.B)
cases, respectively. In Fig.\ref{Fig:Tvszh}.C we compare the plots,
presented in Fig.\ref{Fig:Tvszh}.A and Fig.\ref{Fig:Tvszh}.B. 
Plot in Fig.\ref{Fig:Tvszh}.D is a zoom of Fig.\ref{Fig:Tvszh}.B. In
Fig.\ref{Fig:TvszhC} we present the dependence of temperature on $z_h$
for different $\mu = 0$ and $c < 0$, and in Fig.\ref{Fig:Tvsmu} the
dependence of temperature on $\mu$ keeping $z_h = 1$ for different $c
< 0$ is shown. Fig.\ref{Fig:Tvsmuzh} displays contour plots for the
temperature dependence on the horizon position and chemical potential
for the isotropic (A) and anisotropic (B) cases at fixed $c = -1$.

These plots show the following behavior of the temperature: 
\begin{itemize}
\item
For $\mu=0$ (dashed lines) there is one extremal point (minimum) for
the temperature as a function of the horizon position; we denote the
corresponding horizon $z_{h}$ as $z_{h,min}(0) = z_{h,min}(0,c,\nu)$,
and we get the following picture:
\begin{itemize}
\item for $0 < z_h < z_{h,min}(0)$ (large black holes) the temperature
  drops as $z_h$ grows and for $z_{h,min}(0) < z_h$ (small black holes)
  the temperature increases with the growth of $z_h$;
\item the minimal isotropic horizon $z_{min}^{(iso)}(c) \equiv
  z_{h,min}(0,c,1)$ is larger than the  minimal anisotropic horizon
  $z_{min}^{(\nu)}(c) \equiv z_{min}(0,c,\nu)$,
  i.e. $z_{min}^{(iso)}(c) > z_{min}^{(\nu)}(c)$ and the corresponding
  critical temperature $T_{min}(0)$ is higher for the isotropic case,
  i.e. $T(z_{min}^{(iso)}(c),c,1) \equiv T_{min}^{(iso)} >
  T_{min}^{(\nu)} \equiv T(z_{min}^{(\nu)}(c),c,\nu)$; one can read
  these inequalities from the plot in Fig.\ref{Fig:Tvszh}.C and D;
\item for negative $c$ with decreasing $|c|$ the temperature
  $T_{min}^{(iso)}$ (the brown dashed line) is below the green one and
  its minimum is shifted to the right from the minimum of the green
  one (Fig.\ref{Fig:TvszhC}.A); $T_{min}^{(\nu)}$ (the cyan dashed
  line) is below the blue one and its minimum is shifted to the right
  from that of the blue one Fig.\ref{Fig:Tvszh}.B);
\item for $c = -1$  the  values of the minimal horizons $z_{min}(\nu)$
  are $z_{min}^{(iso)} = 1.547$ and $z_{min}^{(4.5)} = 1.168$ for the
  isotropic and anisotropic cases; the corresponding temperatures are
  $T_{min}^{(iso)} = 0.355$ and $T_{min}^{(4.5)} = 0.245$
  (Fig.\ref{Fig:TvszhC}.A and B).
\end{itemize}
\item For $0 < \mu < \mu_{cr}$ there are two extremal points
  $z_{h,min}(\mu) = z_{h,min}(\mu,c,\nu)$ and $z_{h,max} =
  z_{h,max}(\mu,c,\nu)$; corresponding $T_{min}(\mu)$ and
  $T_{max}(\mu)$ are shown in Fig.\ref{Fig:Tvszh}:
  \begin{itemize}
  \item for $0 < z < z_{h,min}$ the temperature drops with the growth
    of $z_h$;
  \item for $z_{h,min} < z < z_{h,max}$ the temperature increases with
    the growth of $z_h$;
  \item for $z_{h,max} < z < z_{h_{0}}$ the temperature decreases
    again with the growth of $z_h$; here $z_{h_{0}}$ is the position
    of the new horizon, $T(z_{h_0}) = 0$, $z_{h_0} =
    z_{h_0}(\mu,c,\nu)$;
  \item $\mu_{cr}$, $z_{h,min}$ and $z_{h,max}$ depend on the warp
    factor coefficient $c$ and the anisotropic parameter $\nu$;
  \item the anisotropy increases the size of the new horizon,
    $z_{h_0}(\mu, c, 1) < z_{h_0}(\mu, c, \nu)$.
  \end{itemize}
\begin{figure}[h!]
  \centering
  \includegraphics[scale=0.51]{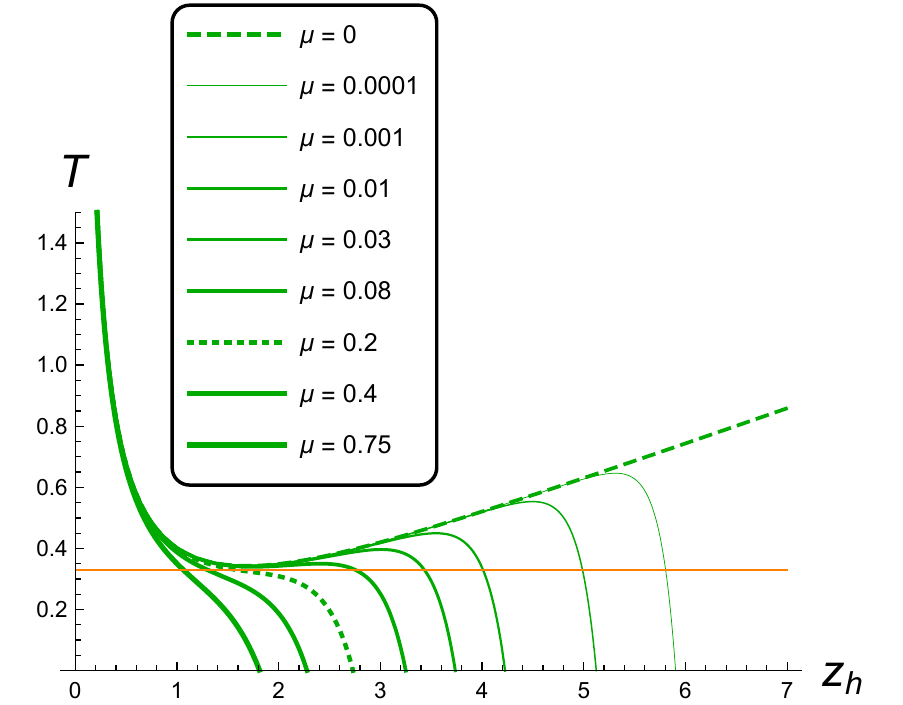}
  \includegraphics[scale=0.51]{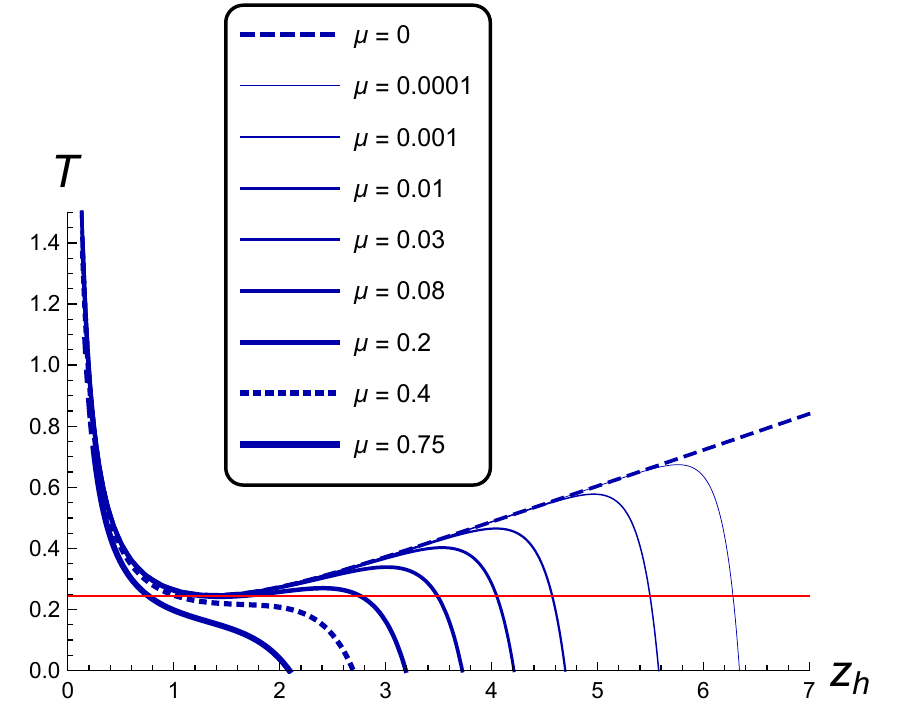}
  \includegraphics[scale=0.41]{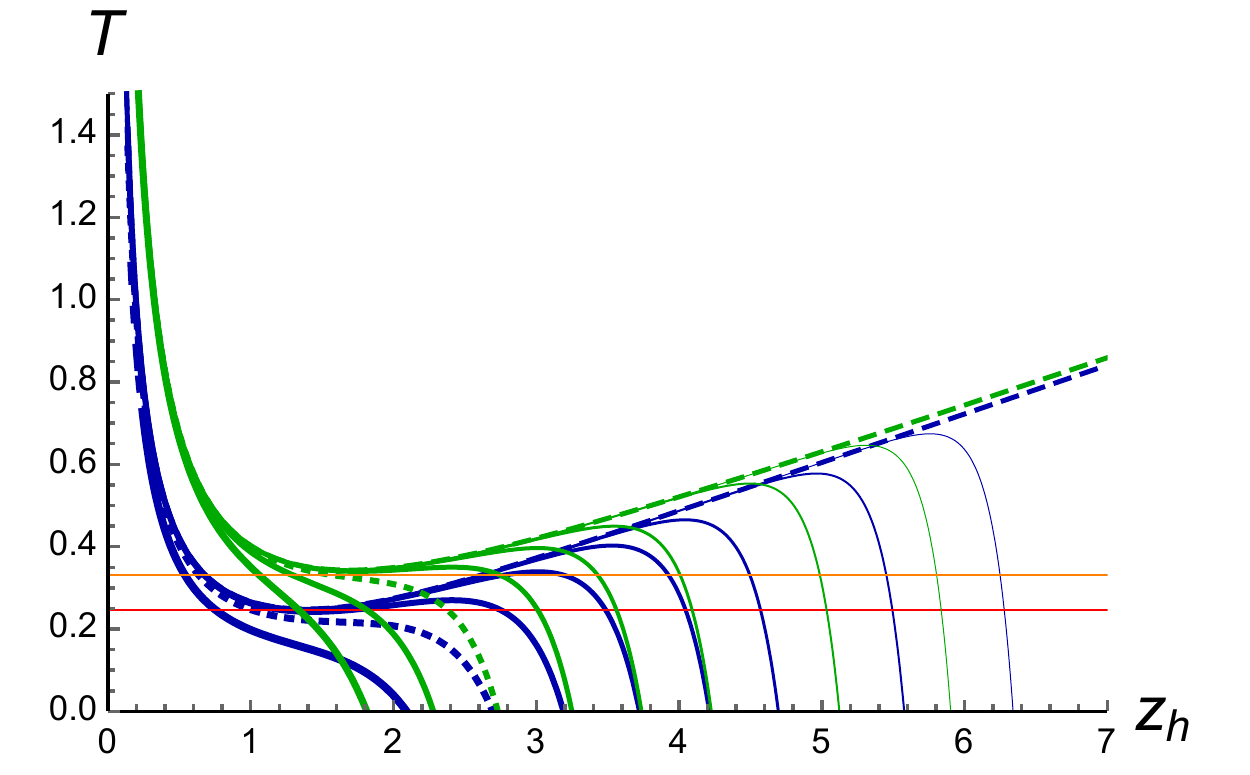}\\
  A \hspace{120pt} B \hspace{120pt} C \\
  \includegraphics[scale=0.7]{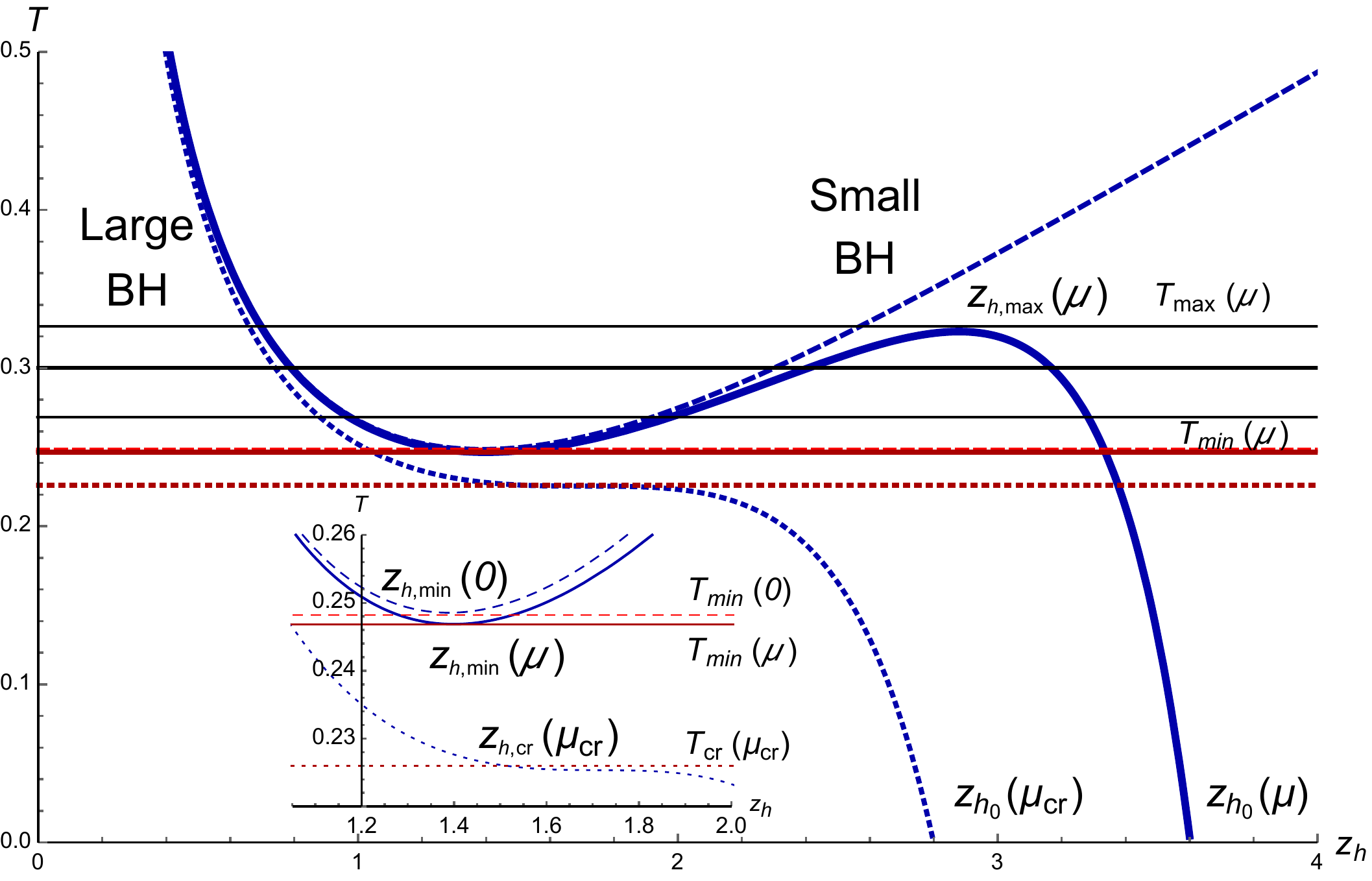} \\ D
  \caption{The dependence of temperature $T(z_h,\mu,c,\nu)$ on $z_h$
    for $c = - 1$ and different $\mu$ for isotropic (A) and
    anisotropic, $\nu = 4.5$, (B) cases, their comparison with the
    same line labels (C); anisotropic case for $c = -1$ and different
    $\mu$ in details (D); the horizontal orange and red lines show
    locations of the global minima in isotropic and anisotropic cases,
    respectively.}
  \label{Fig:Tvszh}
\end{figure}
\item For $\mu = \mu_{cr}$ (dotted lines in Fig.\ref{Fig:Tvszh})
  there is no extremal point, but there is an inflection point
  $(z_{h,cr}, T_{cr})$, $z_{h,cr} = z_{h,cr}(\mu_{cr})$, $T_{cr} =
  T_{cr}(\mu_{cr})$, therefore
  \begin{itemize}
  \item for all values of $z_h$ with its growth the
    temperature decreases;
  \item the temperature becomes equal to zero at a new horizon.
  \end{itemize}
\item For $\mu > \mu_{cr}$ increasing $z_h$ we decrease the
  temperature and there is a point $z_{h_0} = z_{h_0}(\mu,c,\nu)$
  where $T(z_{h_0}) = 0$, i.e. a new horizon appears;
  \begin{itemize}
  \item for negative $c$ with the growth of $|c|$ the values of
    $z_{h_0}(\mu,\nu)(c)$  decrease,
    $z_{h_0}(\mu,\nu)(c_1) < z_{h_0}(\mu,\nu)(c_2)$ for $|c_1| > |c_2|$
    and all $\nu \geq 1$ (Fig.\ref{Fig:TvszhC}).
  \end{itemize}
\end{itemize}

\newpage
\
\newpage

We can also investigate the behavior of $T(\mu)$ using expression
\eqref{temp} and taking some fixed values of the horizon. In
Fig.\ref{Fig:Tvsmu} we plot the curves for different negative values
of the warp factor coefficient $c$ in isotropic (A) and anisotropic
(B) cases for $z_h = 1$. In Fig.\ref{Fig:Tvsmu}.C we compare these
cases plotting them together. Fig.\ref{Fig:Tvsmu}.D and E display
$T(\mu)$ for large black holes with $z_h = 0.5$ and small black holes
with $z_h = 1.5$. The function $T(\mu)$ decreases faster for smaller
$c$. The isotropic case curves lie higher than the anisotropic ones
and reach zero temperature at larger chemical potential values
(Fig.\ref{Fig:Tvsmu}.C). For smaller horizons we have the same picture
(Fig.\ref{Fig:Tvsmu}.D and E).

To summarize, note that the Van der Waals type of the
temperature-horizon dependence $T(z_h)$ observed in \cite{1301.0385, 
  yang2015} for isotropic case, also takes place in the anisotropic
one (see Fig.\ref{Fig:Tvsmuzh}). In both cases this behavior becomes
more pronounced with decreasing of negative $c$, see
Fig.\ref{Fig:TvszhC}. The approximate solution considered in
\cite{Arefeva:2016rob} does not inherit this property.

\ \\

$$\,$$

\newpage

\subsection{Entropy}

\begin{figure}[b]
  \centering
  \includegraphics[scale=0.55]{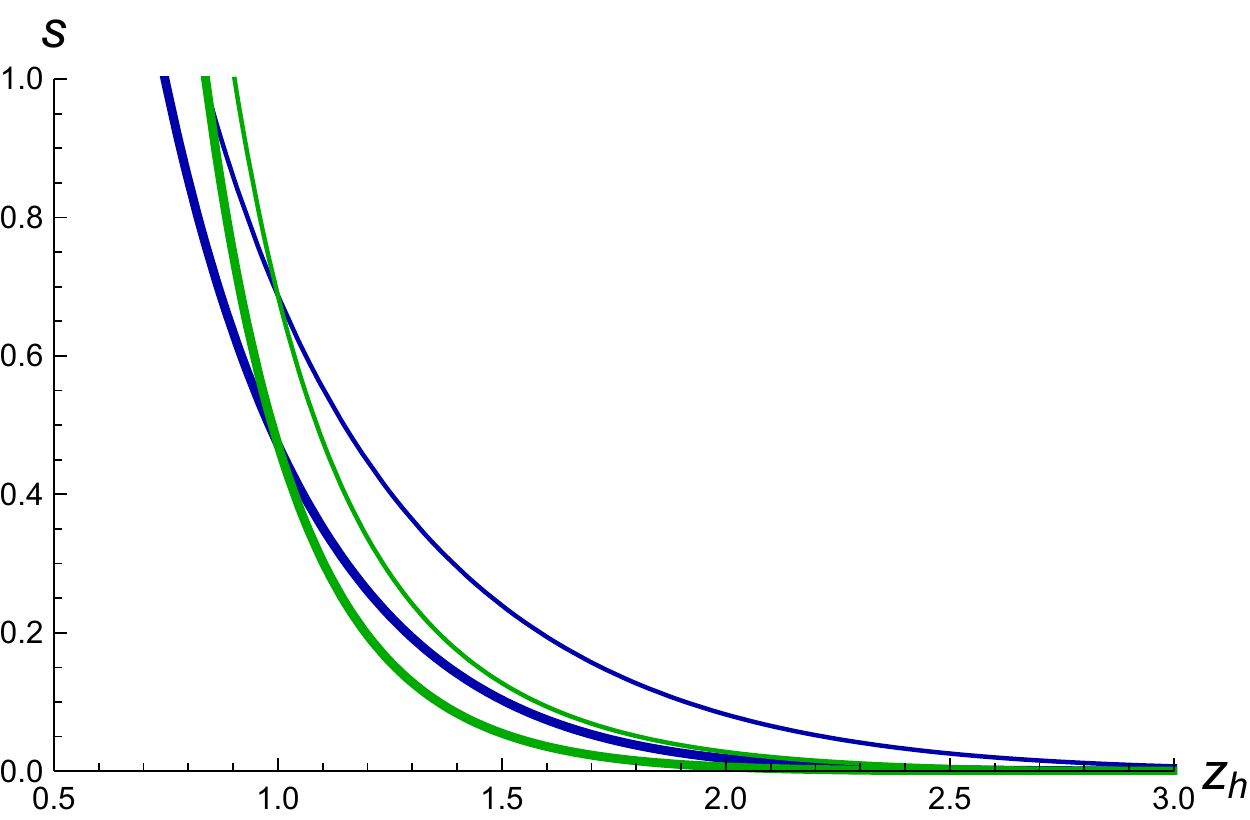} \qquad
  \includegraphics[scale=0.8]{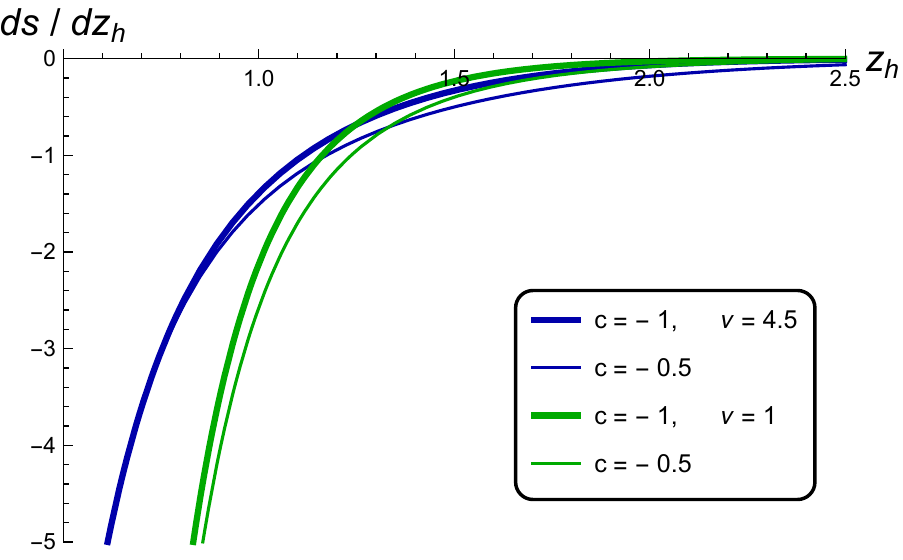}\\
  A\hspace{200pt}B
  \caption{The black hole entropy $s(z_h)$ for $\mu = 0$ and different
    $c$ in isotropic (green lines) and anisotropic (blue lines) (A)
    cases; the velocity of the entropy decreasing $\partial
    s(z_h)/\partial z_h$ (B); the plot legends are the same for both
    panels.}
  \label{Fig:ezh}
\end{figure}

The entropy is given by formula
\be
  s(z_h,c,\nu) = \frac{e^{\frac{3}{4} c z_h^2}}{4} \, z_h^{-\frac{(\nu
      + 2)}{\nu }}
\ee
and is plotted in Fig.\ref{Fig:ezh}. 

Fig.\ref{Fig:ezh}.A shows that the entropy is a monotonously
decreasing function of the horizon $z_h$ both for the isotropic and
the anisotropic cases, in other words the entropy values are bigger for
larger black holes, whose horizons are smaller. As we see from
Fig.\ref{Fig:ezh}.B, the velocity of the entropy decreasing depends on
parameters $c$ and $\nu$. It is interesting to note that absolute
value of this velocity for the same $c$ is bigger in the isotropic
case for large black holes and is smaller for small black holes. More
precisely, for $c = -1$
\bea
  s^\prime_{z_h}(z_h,-1,1) < s^\prime_{z_h}(z_h,-1,4.5) < 0 \quad
  &\mbox{for}& \quad z_{h} < 1.248, \\
  s^\prime_{z_h}(z_h,-1,4.5) < s^\prime_{z_h}(z_h,-1,1) < 0 \quad
  &\mbox{for}& \quad z_{h} > 1.248
\eea
and for $c = -0.5$
\bea
  s^\prime_{z_h}(z_h,-0.5,1) < s^\prime_{z_h}(z_h,-0.5,4.5) < 0
  \quad &\mbox{for}& \quad z_{h} < 1.331, \\
  s^\prime_{z_h}(z_h,-0.5,4.5) < s^\prime_{z_h}(z_h,-0.5,1) < 0
  \quad &\mbox{for}& \quad z_{h} > 1.331.
\eea
Here $s^\prime_{z_h}(z_h,c,\nu) = \partial
  s(z_h,c,\nu)/\partial z_h$.
\begin{figure}[h!]
  \centering
  \includegraphics[scale=0.55]{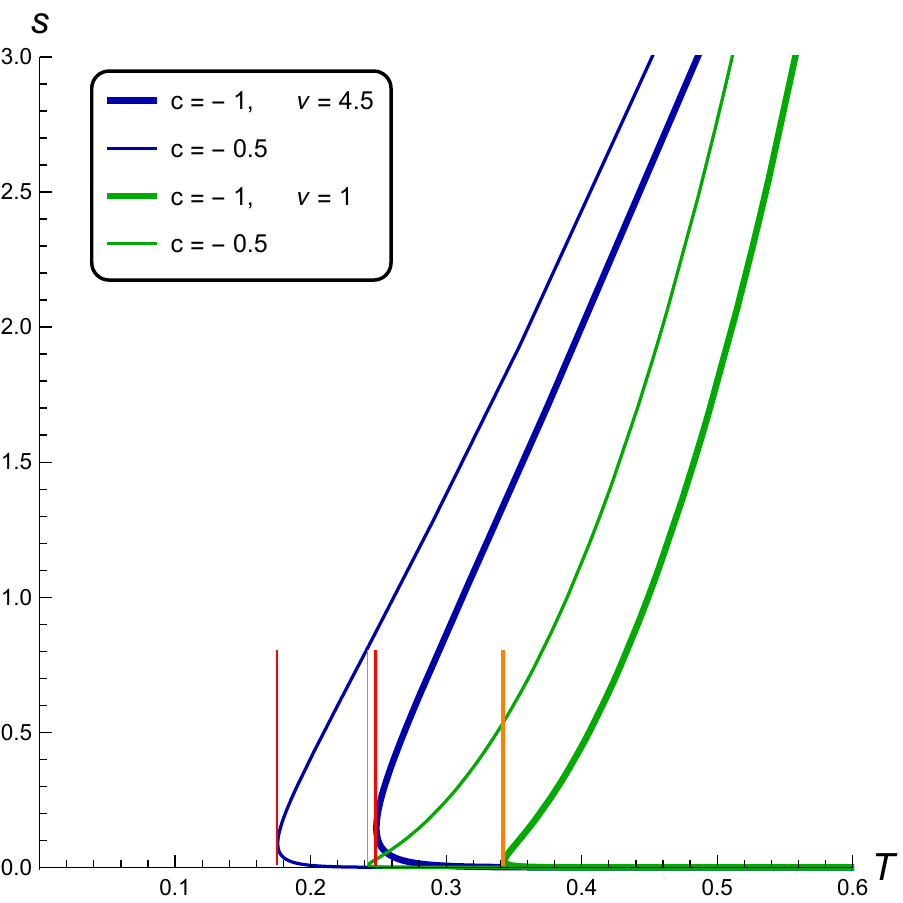}\qquad
  \includegraphics[scale=0.39]{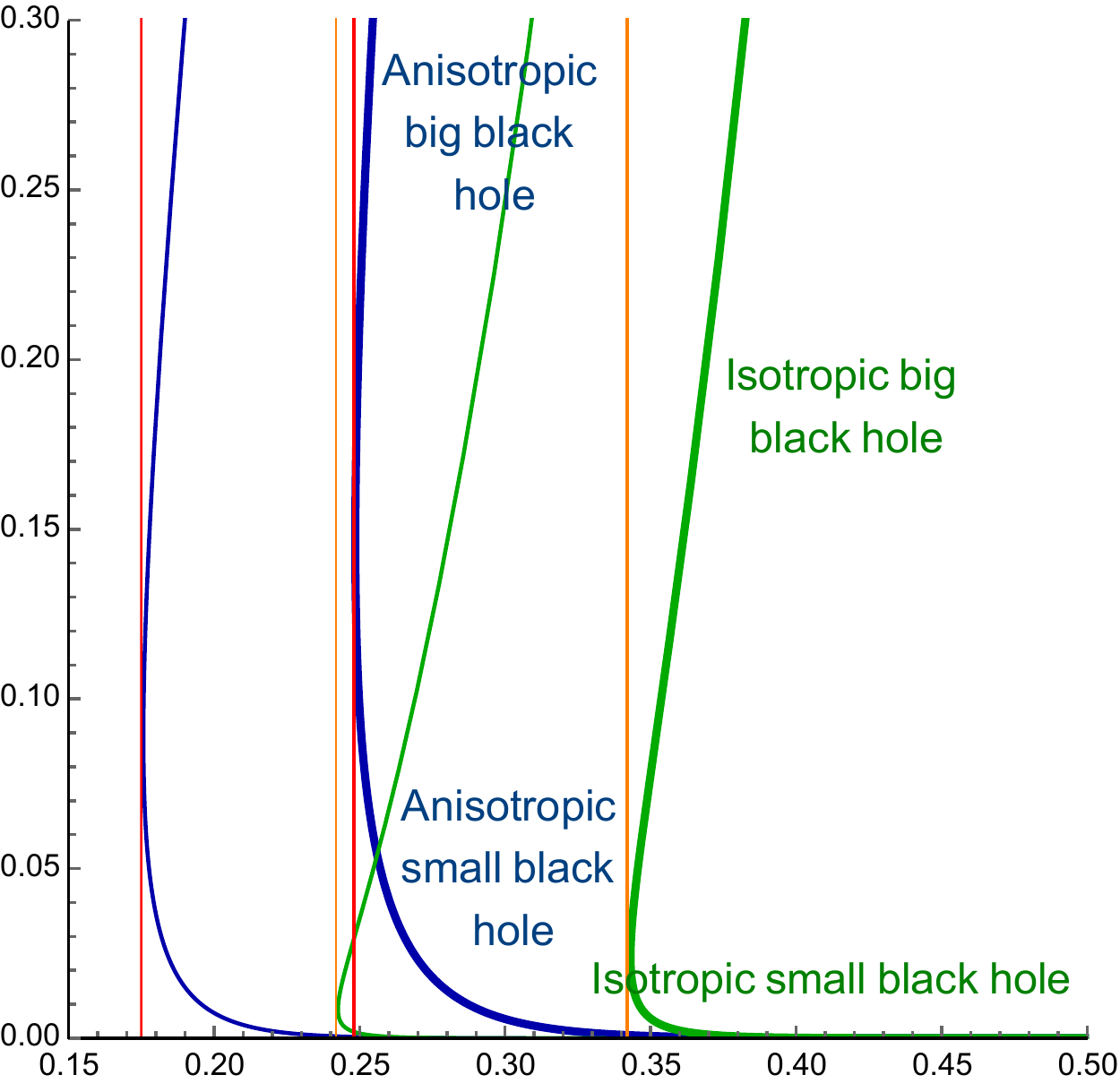}
  \includegraphics[scale=0.35]{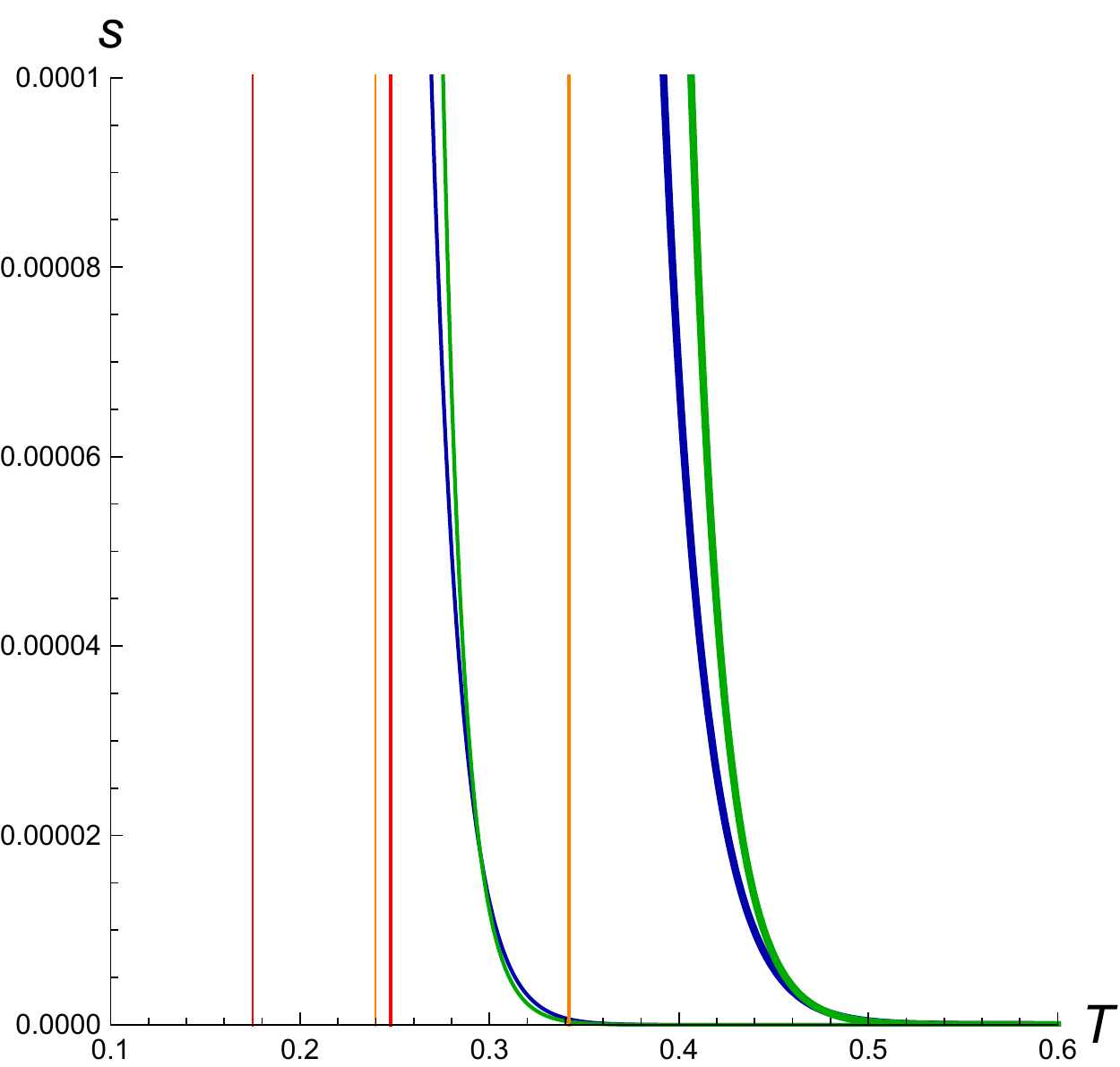}\\
  A\hspace{200pt}B
  \caption{The black hole entropy $s(T)$ for $\mu = 0$ and different
    $c$ in isotropic (green lines) and anisotropic (blue lines) cases
    (A); regions located near the minimum temperatures
    $T_{min}(c,\nu)$ with an increased scale on the right (B); the red
    vertical lines indicate $T_{min}(c,4.5)$ and the orange ones
    indicate $T_{min}(c, 1)$ for $c = -0.5$ and $c = -1$.}
  \label{Fig:eTmu0}
\end{figure}
\begin{figure}[h!]
  \centering
  \includegraphics[scale=0.7]{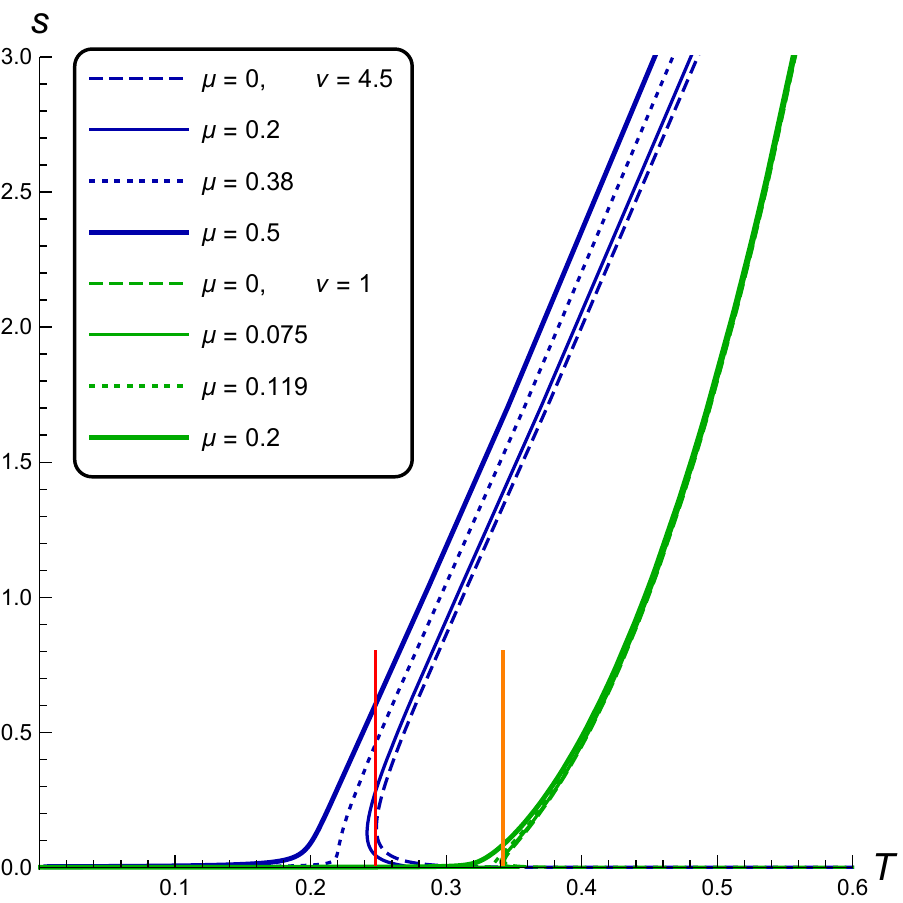} \qquad
  \includegraphics[scale=0.53]{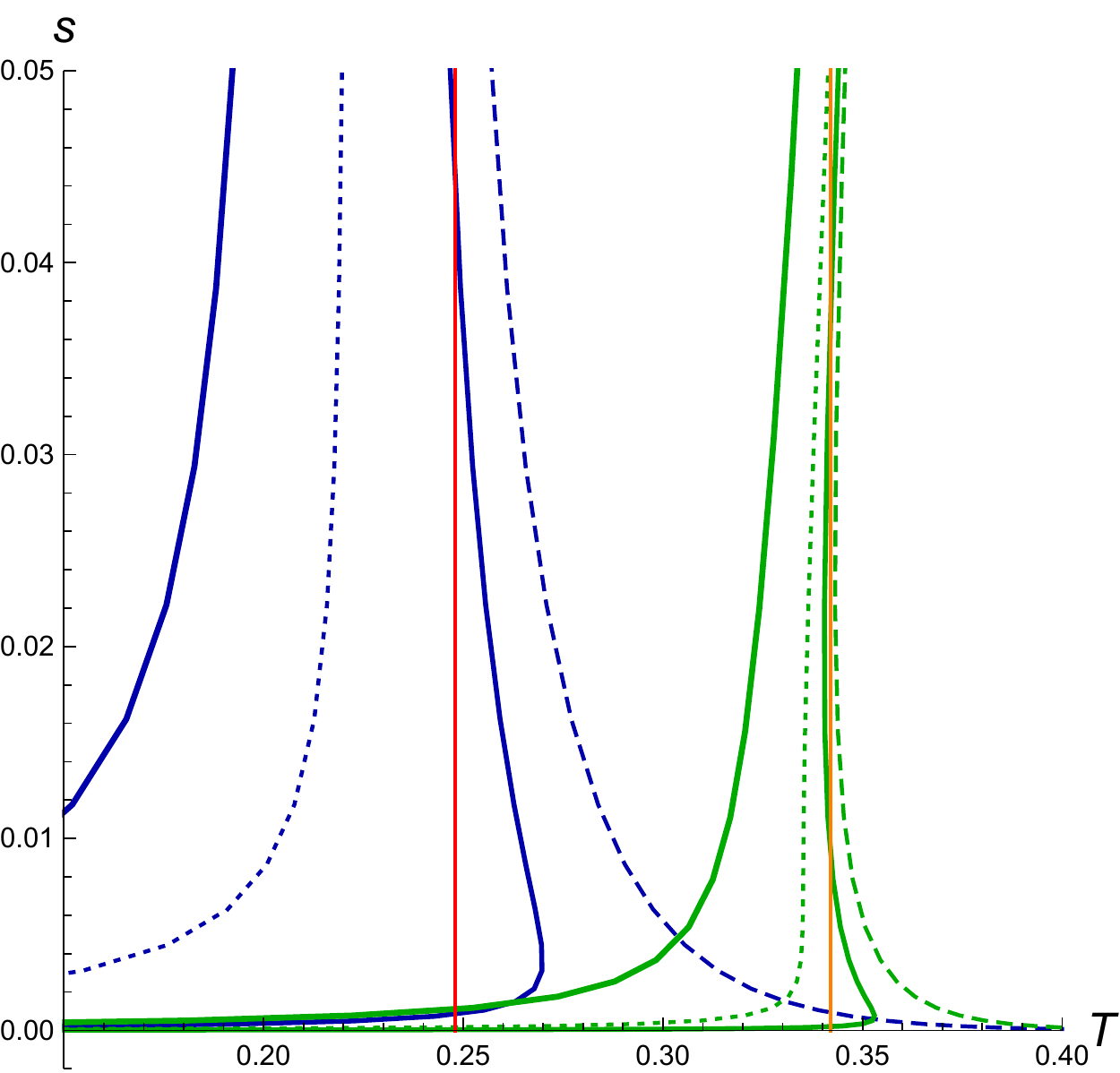}\\
  A\hspace{200pt}B\\
  \includegraphics[scale=0.5]{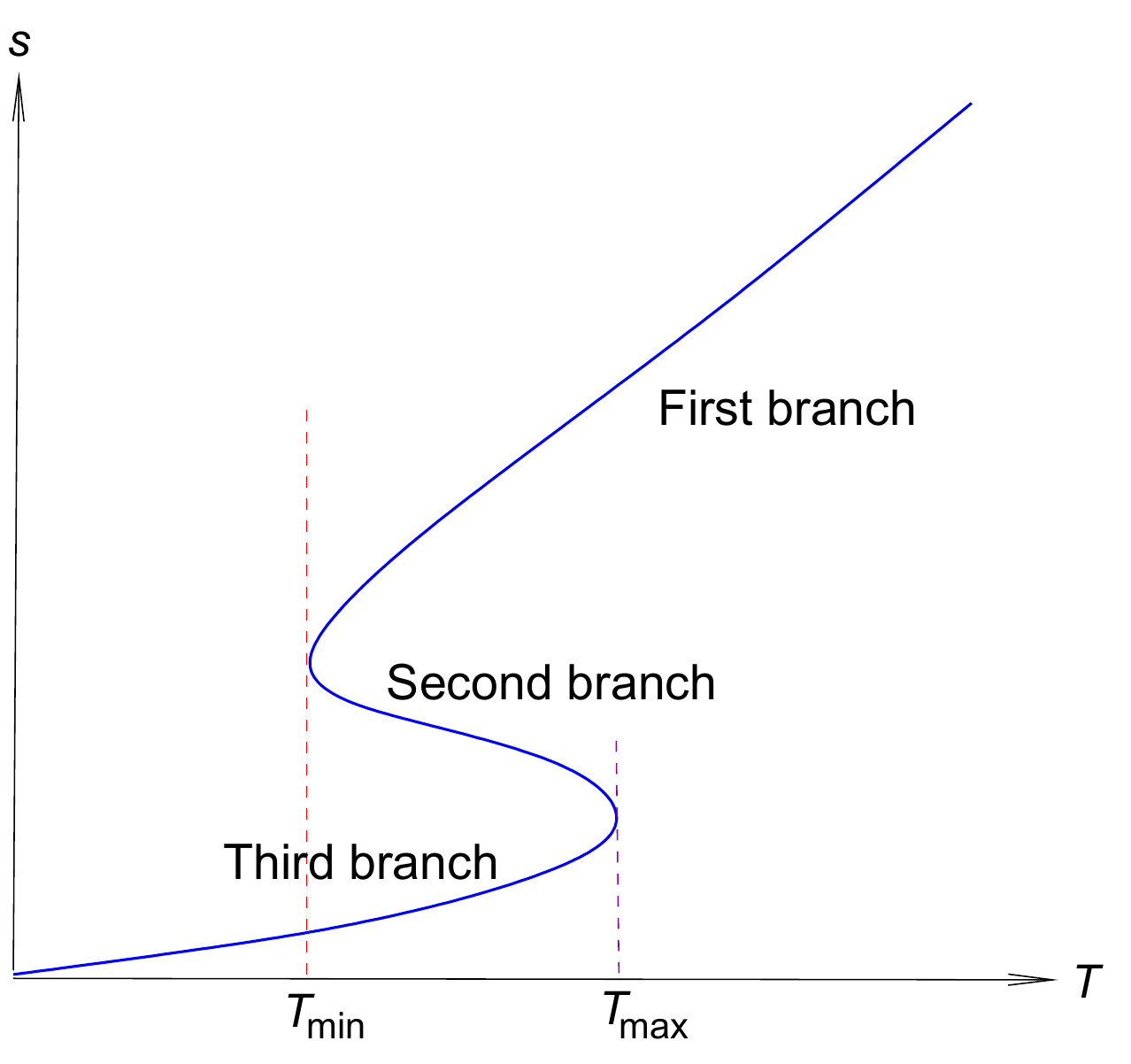}\\
  C
  \caption{The black hole entropy $s(T)$ for different $\mu$ and $c =
    -1$ in isotropic (green lines) and anisotropic case (blue lines)
    (A); a region of (A), located near the minimum temperatures
    $T_{min}(c,\nu)$, in increased scale (B), where the red vertical
    line indicates $T_{min}(-1, 4.5) = 0.245 $ and the orange one
    indicates $T_{min}(-1, 1) = 0.355$ for $\mu = 0$; the schematic
    view of $s(T)$ three-branches behavior (C).}
  \label{Fig:eT}
\end{figure}

In Fig.\ref{Fig:eTmu0} we present the entropy dependence on
temperature $T$ for $\mu = 0$ and different $c$ for isotropic (green
lines) and anisotropic (blue lines) cases. The plots in
Fig.\ref{Fig:eTmu0} show that in both cases there are minimal
temperatures for which the black holes exist. The minimal temperature
in the isotropic case is higher then in the anisotropic one for the
same value of $c < 0$, $T_{min}(c,4.5) < T_{min}(c,1)$, that agrees
with the plots in Fig.\ref{Fig:Tvszh}. In both cases the entropy is a
double-valued function and has a large black holes branches and a
small black holes one. For the small black holes the entropy increases
with decreasing $T$ thus leading to the negativity of the specific
heat $c_v = T ds/dT$. Therefore small black holes are
thermodynamically unstable, whereas entropy of the large black holes
grows while temperature increases and therefore large black holes are
thermodynamically stable.

\begin{figure}[h!]
  \centering
  \includegraphics[scale=0.7]{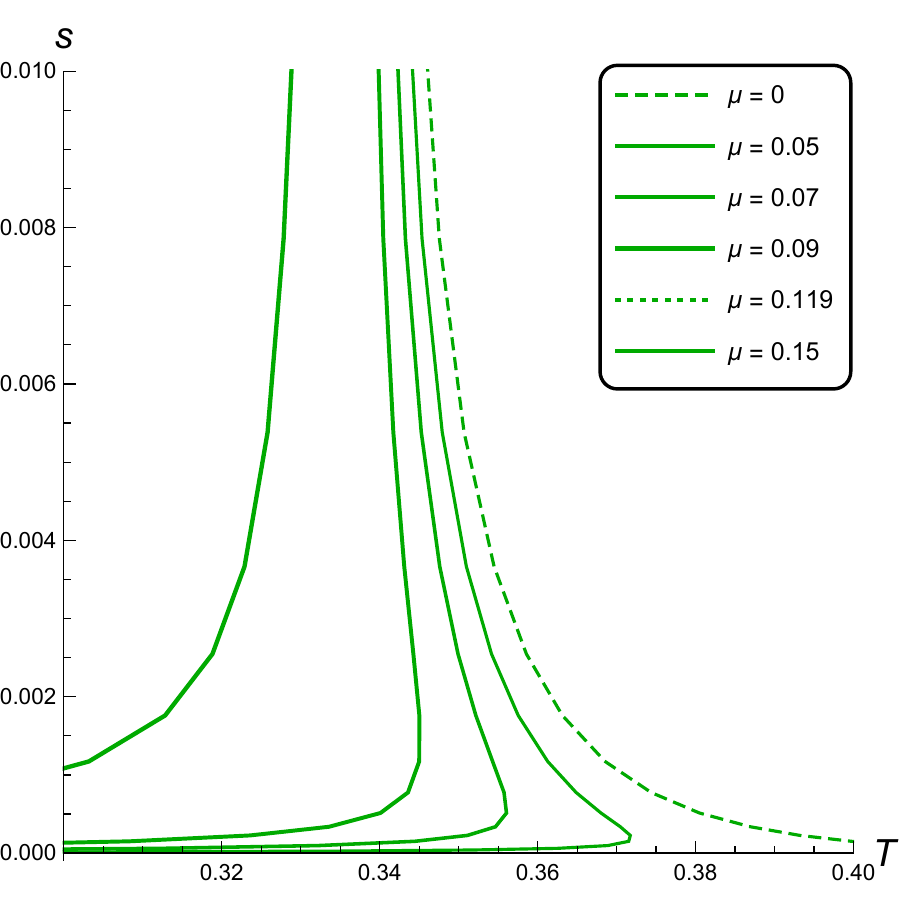} \qquad
  \includegraphics[scale=0.7]{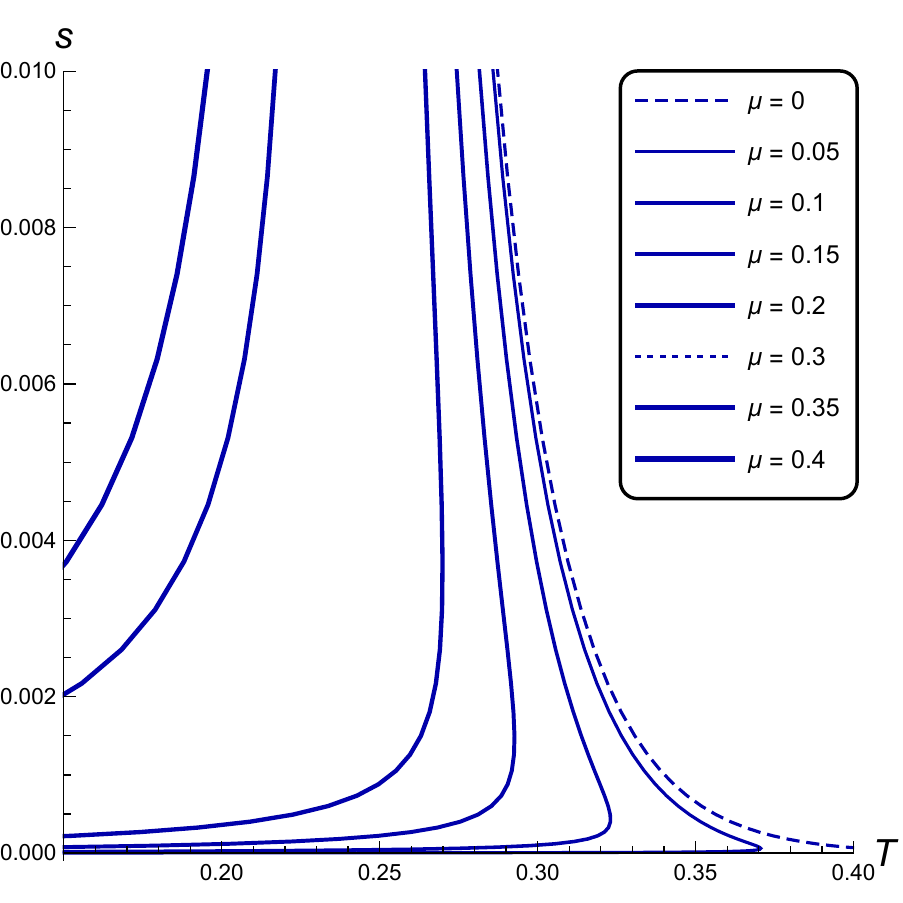}\\
  A\hspace{200pt}B
  \caption{Unstable second branches of $s(T)$ in isotropic (A) and
    anisotropic case (B) cases for $c = -1$ and different $\mu$.}
  \label{Fig:eT2}
\end{figure}

In Fig.\ref{Fig:eT} we present the entropy dependence on temperature
$T$ for $c = -1$ and different $\mu \geq 0$ for isotropic (green
lines) and anisotropic (blue lines) cases. The plots in
Fig.\ref{Fig:eT} show that in both cases for fixed $\mu$, $0 \leq \mu
\leq \mu_{cr}(c,\nu)$, there are minimal $T_{min}(\mu,\nu)$ and
maximal $T_{max}(\mu,\nu)$ temperatures, between which the entropy is
a multivalued function of $T$ with three branches. We see well only two
braches in Fig.\ref{Fig:eT}.A, to see the third one has to draw the
picture Fig.\ref{Fig:eT}.B for small values of $s(T)$. The schematic
picture of three branches is presented in Fig.\ref{Fig:eT}.C. When we
decrease the temperature, the entropy decreases along the first branch
($T_{min}(\mu,\nu) < T < \infty$). Then the entropy decreases along the
second branch with an increase of temperature from $T_{min}(\mu,\nu)$
to $T_{max}(\mu,\nu)$, i.e. here the black holes are unstable. Finally
the entropy increases along the third branch with an increase of
temperature for $0 \leq T < T_{max}(\mu,\nu)$, see also
Fig.\ref{Fig:Tvszh}.D. In plots Fig.\ref{Fig:eT2}.A and 
Fig.\ref{Fig:eT2}.B we show the unstable second branches as well as the
transition of the three-branch solution to the unified one-branch
solutions at $\mu = \mu_{cr}(-1,\nu)$ (dotted green and blue lines for
the isotropic and anisotropic cases). The entropy dependence on the
temperature at unified branches corresponding to $\mu \geq
\mu_{cr}(c,\nu)$ is presented in plots Fig.\ref{Fig:eT2} by lines for
$\mu \ge 0.119$ in isotropic case and $\mu \ge 0.3$ in anisotropic
case.

\subsection{Free energy} 

\begin{figure}[h!]
  \centering
  \includegraphics[scale=0.7]{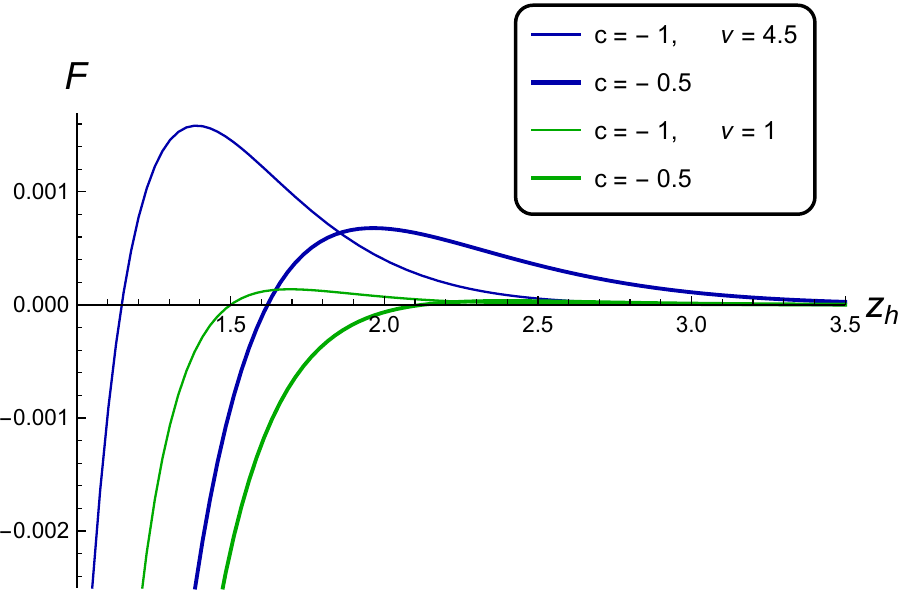}\qquad
  \includegraphics[scale=0.7]{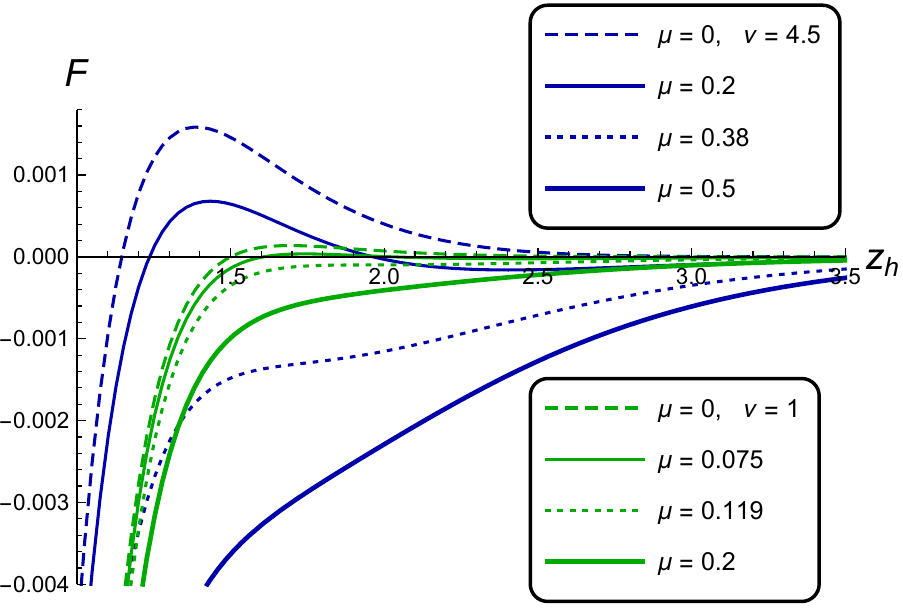}\\
  A\hspace{200pt}B
  \caption{The black hole free energy $F(z_h)$ in isotropic (green
    lines) and anisotropic (blue lines) cases for different $c$, $\mu
    = 0$ (A) and different $\mu \geq 0$, $c = -1$ (B). The
    intersections with the horizontal axis give the values of the
    Hawking-Page horizons $z_{h,HP}(\mu,c,\nu)$.}
  \label{Fig:Fzh}
\end{figure}

\begin{figure}[t]
  \centering
  \includegraphics[scale=0.7]{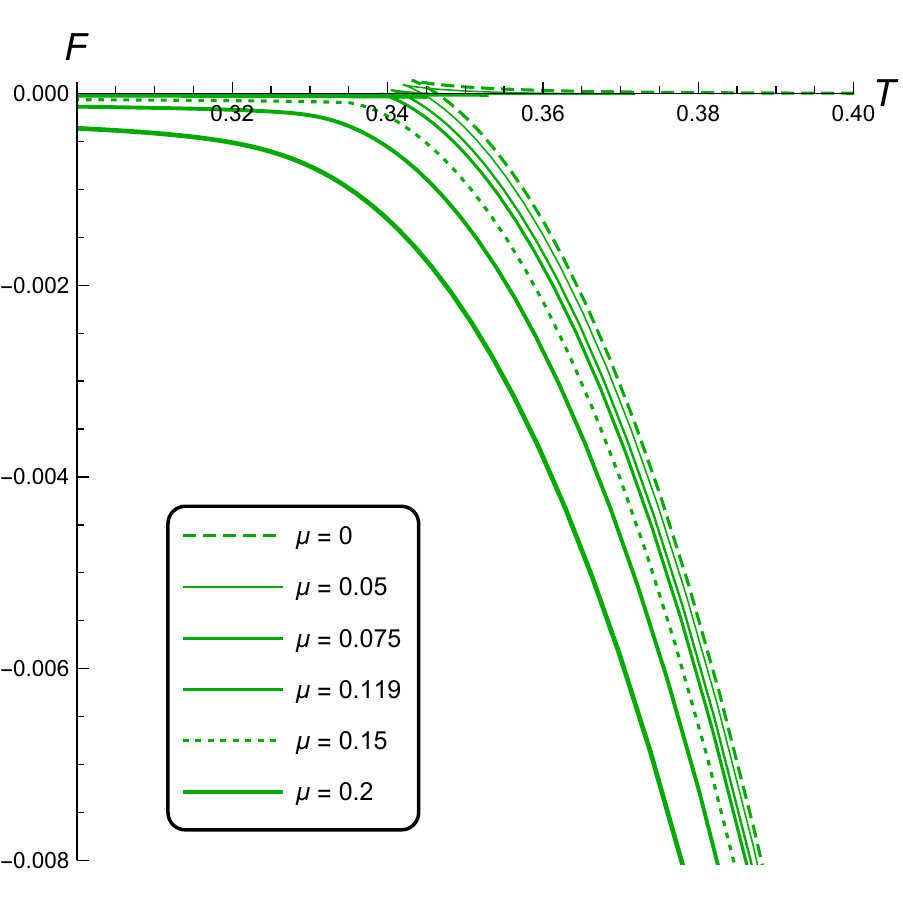} \qquad
  \includegraphics[scale=0.7]{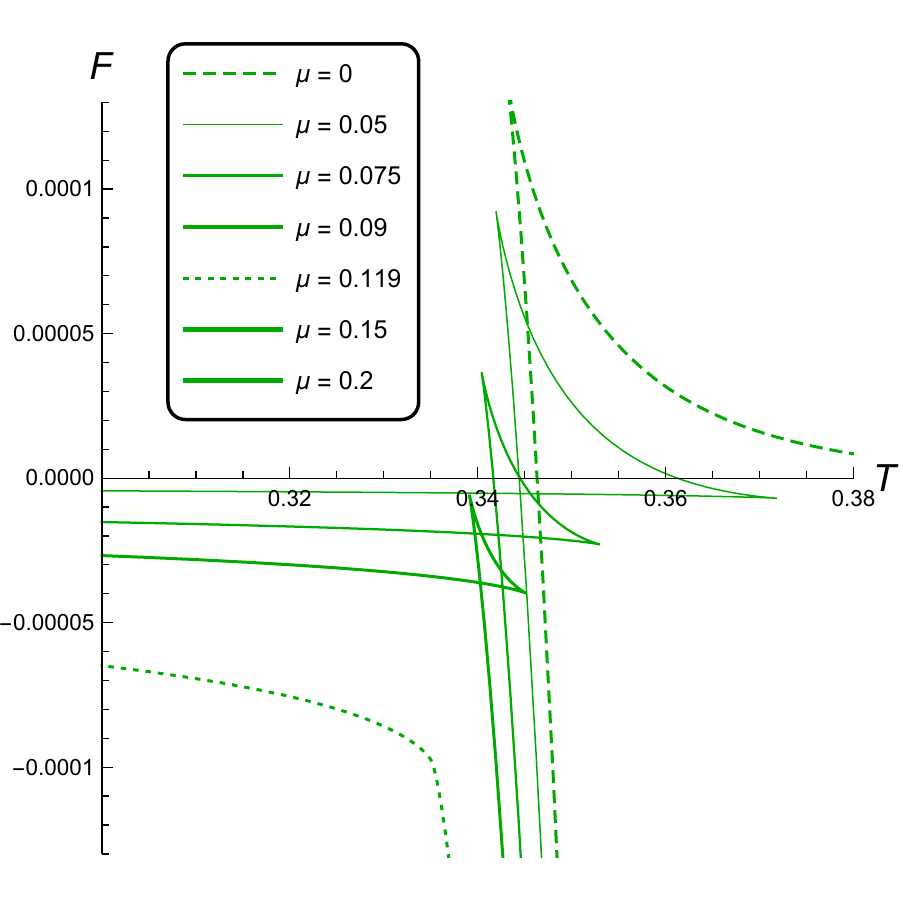}\\
  A\hspace{200pt}B
  \caption{The black hole free energy $F(T)$ for $c = -1$ and
    different $\mu$ in isotropic case (A) and its zoom near $F = 0$
    (B).}
  \label{Fig:FTmuiso}
\end{figure}

\begin{figure}[h!]
  \centering
  \includegraphics[scale=0.65]{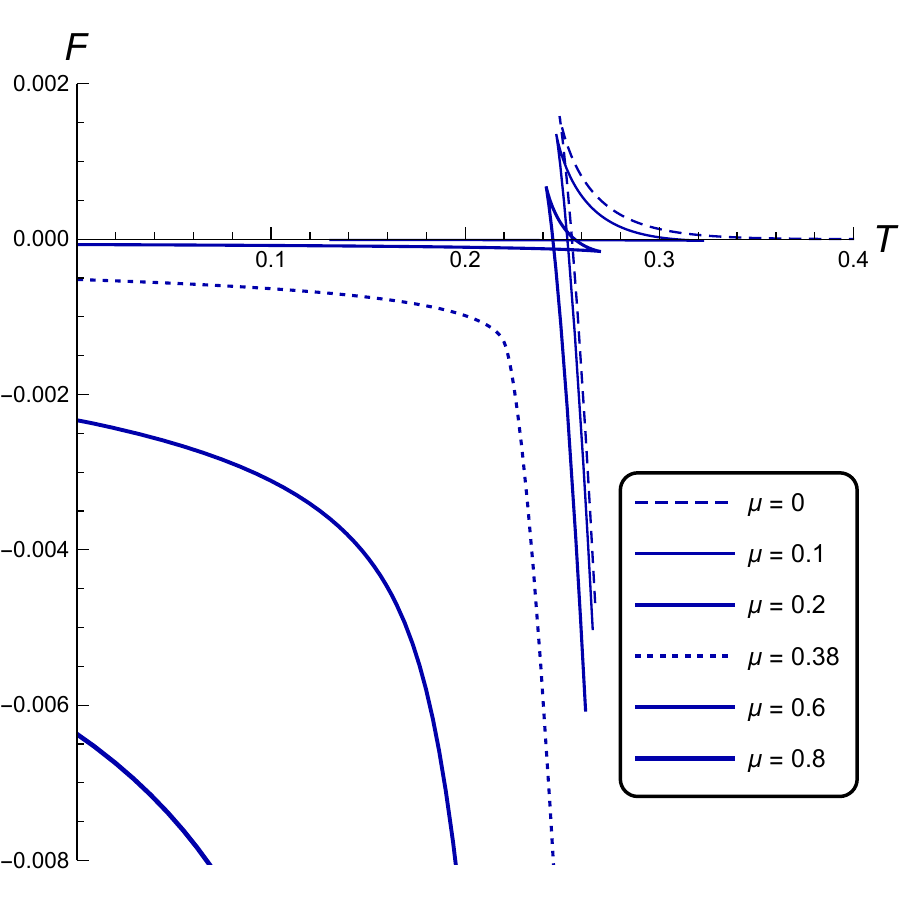}
  \includegraphics[scale=0.6]{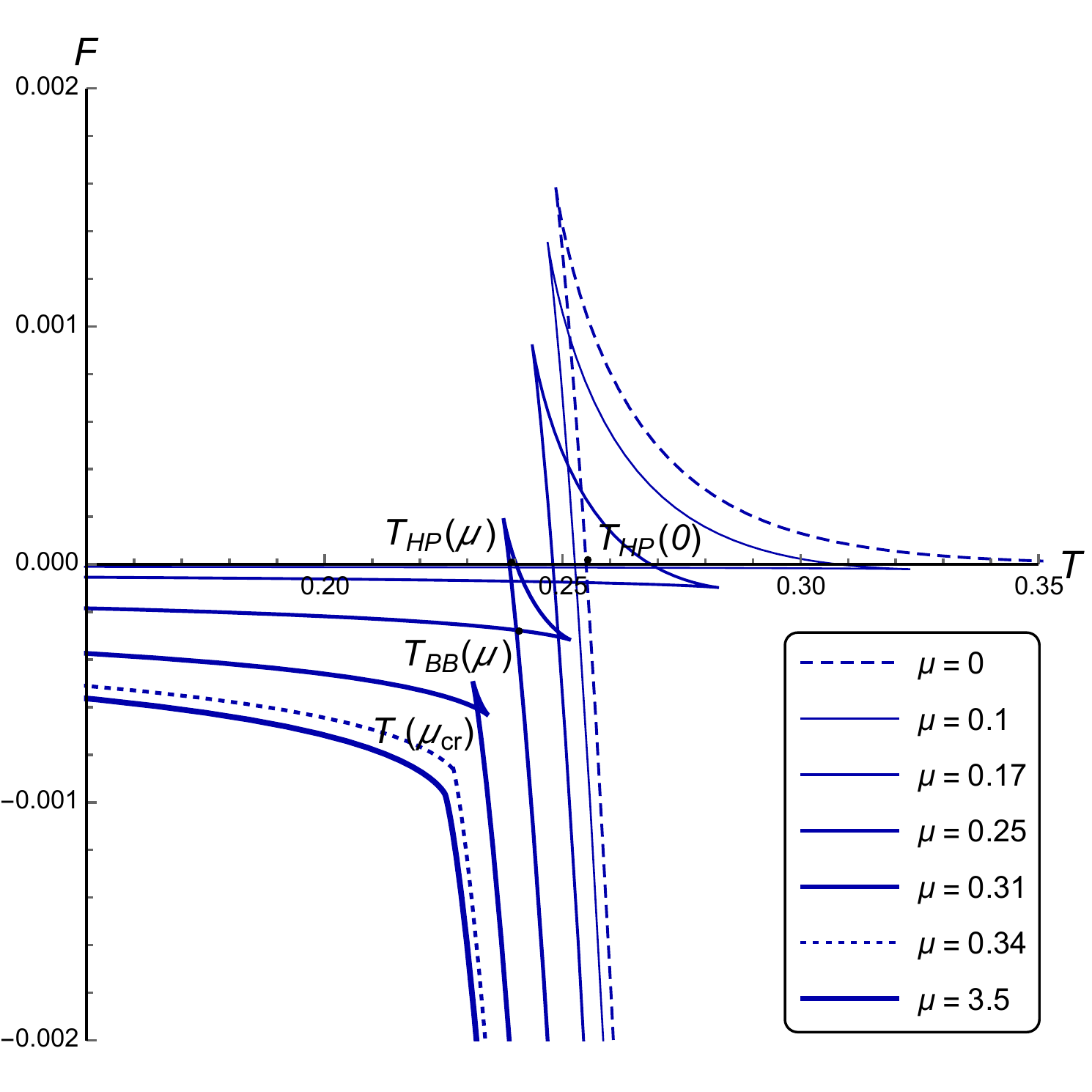}\\
  A\hspace{200pt}B 
  \caption{The black hole free energy $F(T)$ for $c = -1$ and
    different $\mu$ in anisotropic case, $\nu = 4.5$, (A) and its zoom
    near $F = 0$ (B).}
  \label{Fig:FTmu45}
\end{figure}

To study transitions between different branches in more detail it is
reasonable to consider the free energy behavior of the corresponding
solutions. The free energy for a given chemical potential and fixed
volume is related to the entropy as
\be
  dF = - s \, dT \label{dF}
\ee
and can be found by integration of \eqref{dF} that gives
\be
  F(z_h,c,\nu) = \int \, s \, dT = \int_{z_h}^{\infty}
  s(z_h,c,\nu) \, T^\prime (z_h,c,\nu) \, dz_h.
\ee
The dependence of the free energy on the horizon position $z_h$ is
presented in Fig.\ref{Fig:Fzh}, the dependence on $T$ is presented in
Fig.\ref{Fig:FTmuiso} and Fig.\ref{Fig:FTmu45}.
In Fig.\ref{Fig:Fzh}.A, which corresponds to $\mu = 0$, we can see
that the free energy as the function of $z_h$ is equal to zero at
$z_h = z_{h,HP}(0,c,\nu)$, and at this point the
  Hawking-Page phase transition takes place. The value
  $z_{h,HP}(0,c,\nu)$ depends on $c$ and $\nu$, and
  $z_{h,HP}(0,c_1,\nu) < z_{h,HP}(0,c_2,\nu)$ for $c_1 < c_2 < 0$. For
  the anisotropic background the Hawking-Page horizon is less than for
  the isotropic one with the same $c < 0$. In particular,
\be
  1.138 = z_{h,HP}(0,-1,4.5) < z_{h,HP}(0,-1,1) = 1.505.
\ee

Note that the position of horizon, where the temperature gets its
local minimum, exceed the position of the Hawking-Page horizon
$z_{h,HP}(0,-1,\nu) < z_{h,min}(0,-1,\nu)$.

As we can see from the plots in Fig.\ref{Fig:Fzh}.B, for $0 < \mu <
\mu_{cr,HP}(c,\nu)$ the free energy as the function of the horizon
position keeps the same behavior as for $\mu = 0$. At $\mu =
\mu_{cr,HP}(c,\nu)$ the free energy becomes non-positive and for $\mu
> \mu_{cr,HP}(c,\nu)$ the Hawking-Page horizon disappears. But for the
chemical potential values in the interval $\mu_{cr,HP}(c,\nu) < \mu <
\mu_{cr}(c,\nu)$, the free energy still is double-valued what causes
the black hole to black hole phase transition (see below).

In Fig.\ref{Fig:FTmuiso} and Fig.\ref{Fig:FTmu45} we show the
behavior of the free energy as function of the temperature. At $\mu =
0$, as we can see in Fig.\ref{Fig:FTmuiso}.B and
Fig.\ref{Fig:FTmu45}.D, the free energy plots intersect the horizontal
axis at $T_{HP}(c, \nu)$, where the Hawking-Page phase transitions
take place, $T_{HP}(0,-1, 1) = 0.347$ and $T_{HP}(0,-1, 4.5) =
0.256$. At $T_{HP}$ black holes dissolve to thermal gas states with
$F_{gas}=0$. We note that $T_{HP}(0,-1, 4.5) < T_{HP}(0,-1, 1)$, and
comparing to $T_{min}(0,-1, 4.5) = 0.255 $ and $T_{min}(0,-1, 1) =
0.345$ we conclude that in both isotropic and anisotropic cases,
\bea
  0.347 = T_{HP}(0,-1, 1) &>& T_{min}(0,-1, 1) = 0.345\\
  0.256 = T_{HP}(0,-1, 4.5) &>& T_{min}(0,-1, 4.5) = 0.255.
\eea

From Fig.\ref{Fig:Fzh} we see that for zero chemical potential the
free energy increases with $z_h$ growth for large black holes,
i.e. for $z_h < z_{h_{cr}}(c,\nu)$, and decreases for small black
holes.


For $0 < \mu < \mu_{cr}(c,\nu)$ the dependence of the free energy from
the temperature looks like the swallow-tailed shape both in isotropic
and anisotropic cases. When we decrease the temperature from very
large values up to $T_{min}(\mu)$ ($T_{min}(\mu) =
T_{min}(\mu, c,\nu)$, see Fig.\ref{Fig:Tvszh}), the free energy riches
its maximum value, then goes down to its local minimum at
$T_{max}(c,\nu)$ and turns back to increase. It intersects itself at
$T = T_{BB}\left(\mu\right)$, where a large black hole transits to a
small one. Since both free energy values at $T = T_{BB}$ are equal and
negative, meanwhile the free energy of the thermal gas is zero, the
system undergoes the phase transition not to a thermal gaz, but to
small black hole background. When we increase the chemical potential
$\mu$ from zero to $\mu_{cr}$, the loop of the swallow-tailed shape
shrinks to disappear at $\mu = \mu_{cr}(c,\nu)$. For $\mu >
\mu_{cr}(c,\nu)$, the curve of the free energy increases smoothly from
higher to lower values of temperature.

\begin{figure}[b]
$$\,$$
  \centering
  \includegraphics[scale=0.8]{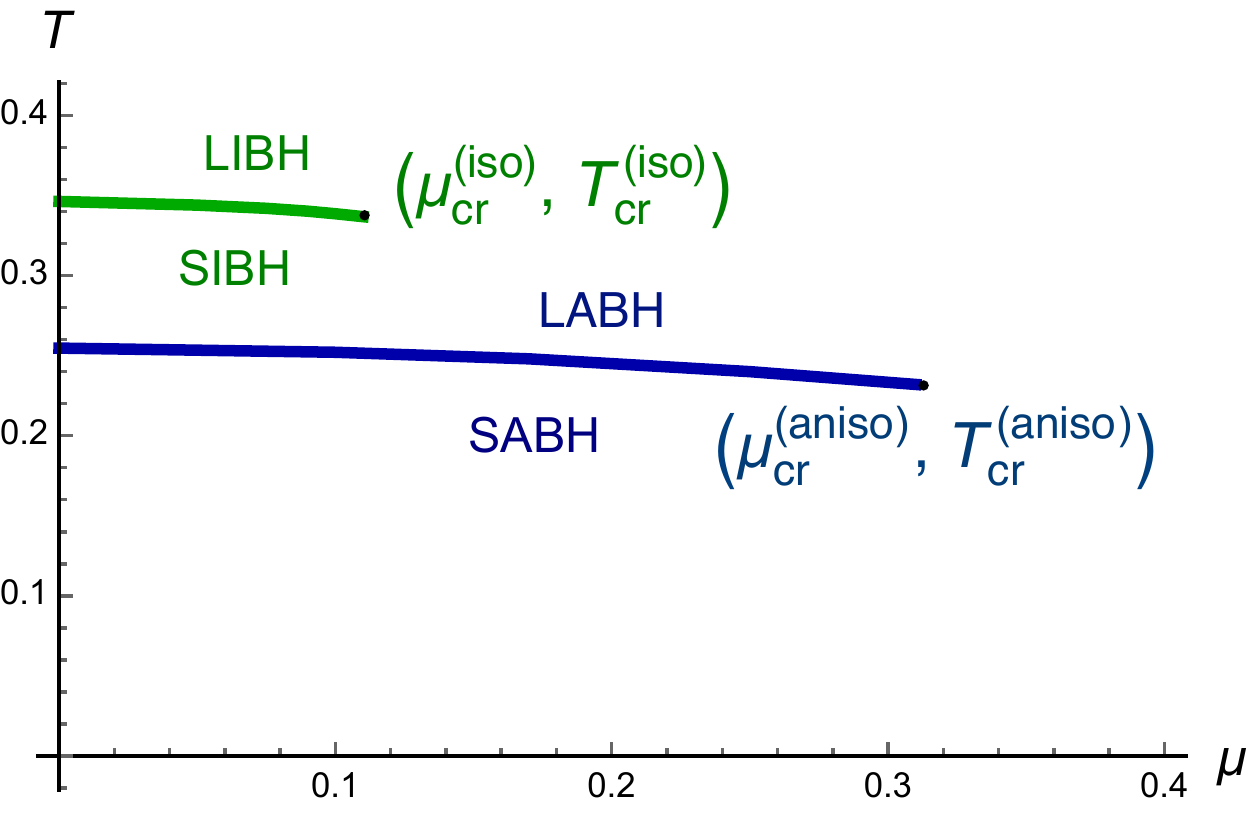} 
  \caption{The phase diagrams in the $(T,\mu)$-plane for the isotropic
    background (green line) and for anisotropic $\nu = 4.5$ background
    (blue line). LIBH and  SIBH (LABH and  SABH) indicate the regions
    of small and large isotropic (anisotropic) black holes.}
   \label{Fig:BCKFT}
\end{figure}

It is interesting to compare the phase diagrams corresponding to
isotropic and anisotropic backgrounds, see Fig.\ref{Fig:BCKFT}. We see 
that the first order phase transitions start at $(0,T_{HP}(0,1))$ and
$(0,T_{HP}(0,4.5))$, so that $T_{HP}(0,4.5) < T_{HP}(0,1)$ and the
transition lines describing transitions from large black holes to
small ones stop at points $(\mu_{cr}^{(iso)},T_{cr}^{(iso)})$ and
$(\mu_{cr}^{(aniso)},T_{cr}^{(aniso)})$, herewith $\mu_{cr}^{(iso)} <
\mu_{cr}^{(aniso)}$ and $T_{cr}^{(iso)} > T_{cr}^{(aniso)}$.

\begin{figure}[h!]
  \centering
  \includegraphics[scale=0.45]{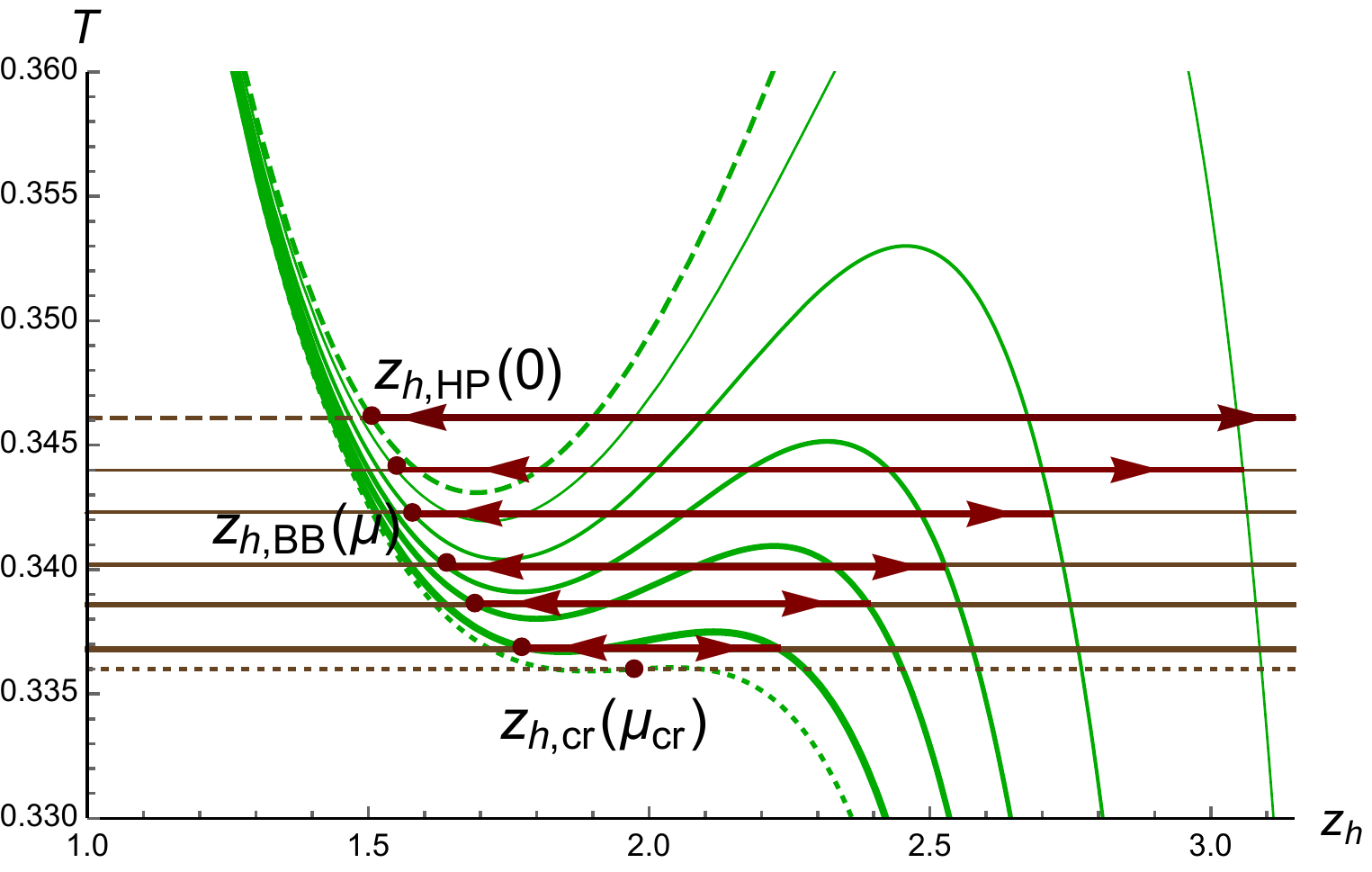} \qquad
  \includegraphics[scale=0.55]{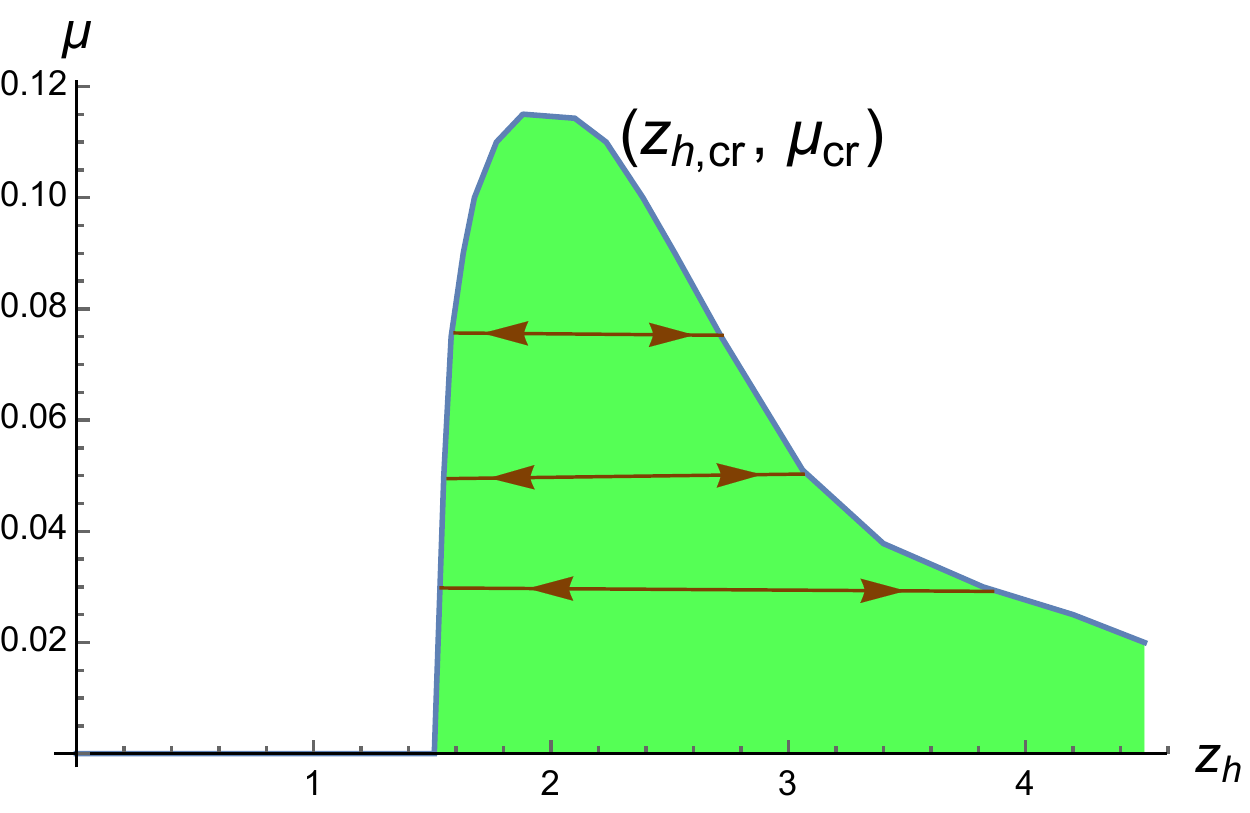}\\
  A \hspace{200pt} B\\
  \includegraphics[scale=0.45]{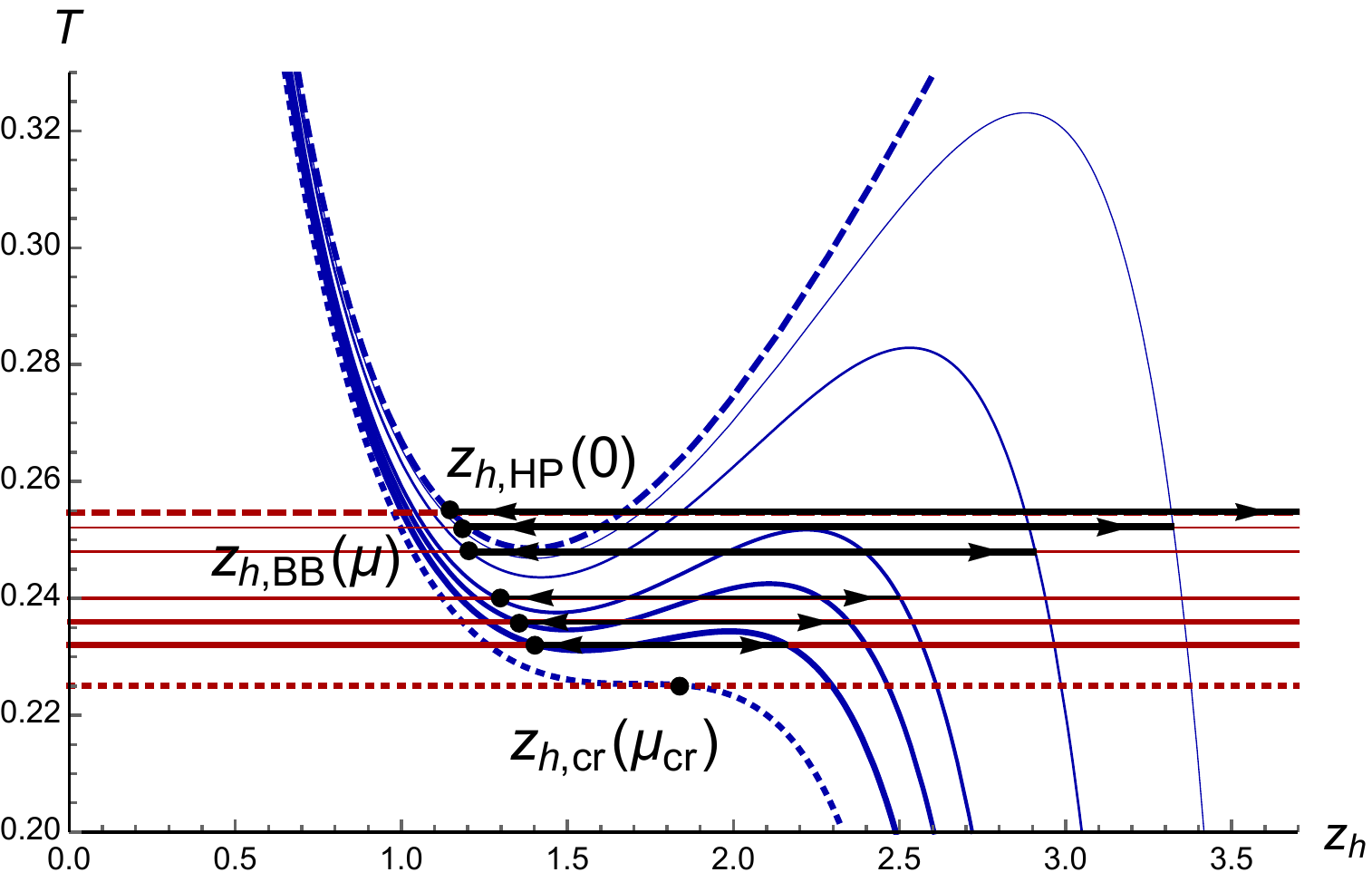} \qquad
  \includegraphics[scale=0.55]{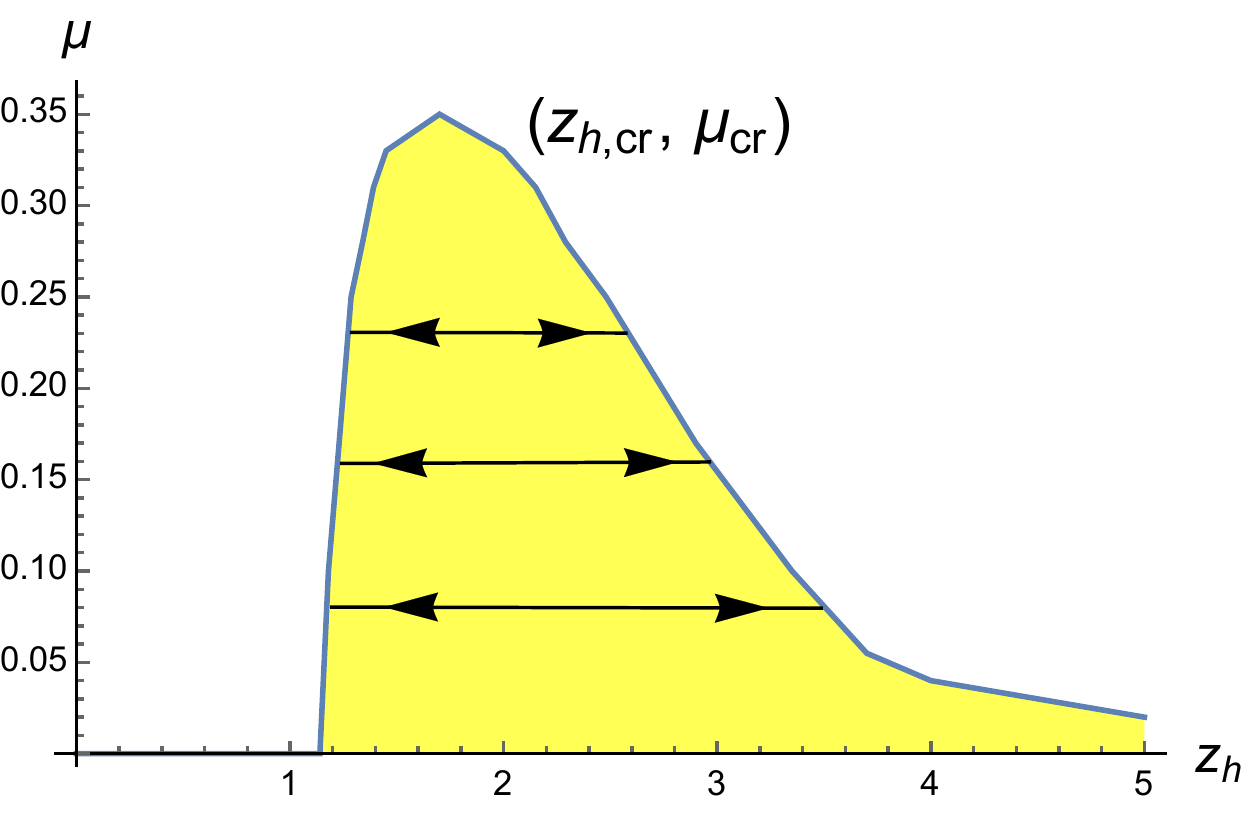}\\
  C \hspace{200pt} D\\
  \caption{The arrows show transitions  large/small black holes for
    the isotropic (A) and anisotropic $\nu = 4.5$ (C) cases in the
    $(z_h,T)$-plane; the large/small holes transitions in the
    $(z_h,\mu)$-plane are shown by arrows for isotropic (B) and
    anisotropic (D) cases.}
  \label{Fig:BB-trans}
\end{figure}

It is also important for us to know the  position of the large black
holes to small black holes transition points  at $(z_h,T)$-plane,
see Fig.\ref{Fig:BB-trans}. In these plots the horizontal arrows show
transitions from the large black holes to small black holes for the
anisotropic $\nu = 4.5$ and isotropic cases. The shaded  by these
arrow areas define the instability zones.
\begin{figure}[h!]
  \centering
  \includegraphics[scale=0.59]{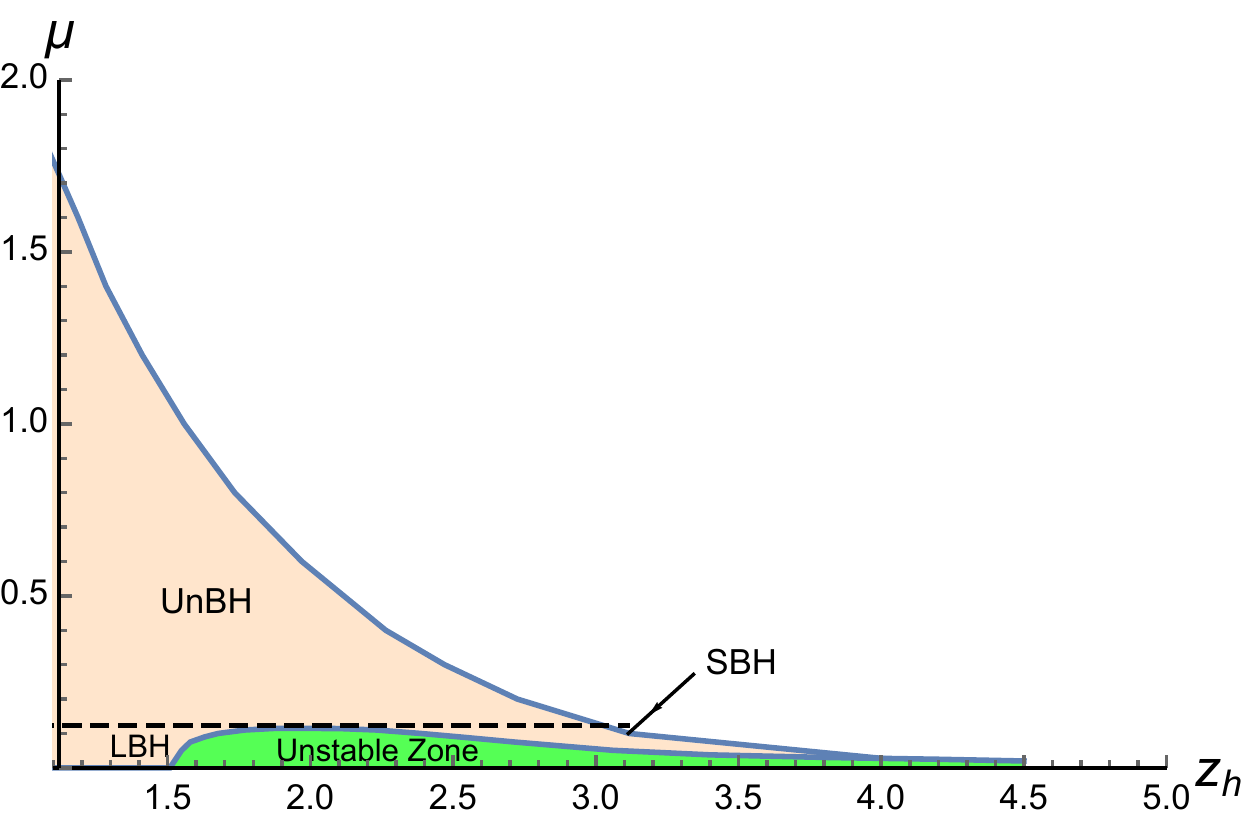}\quad 
  \includegraphics[scale=0.59]{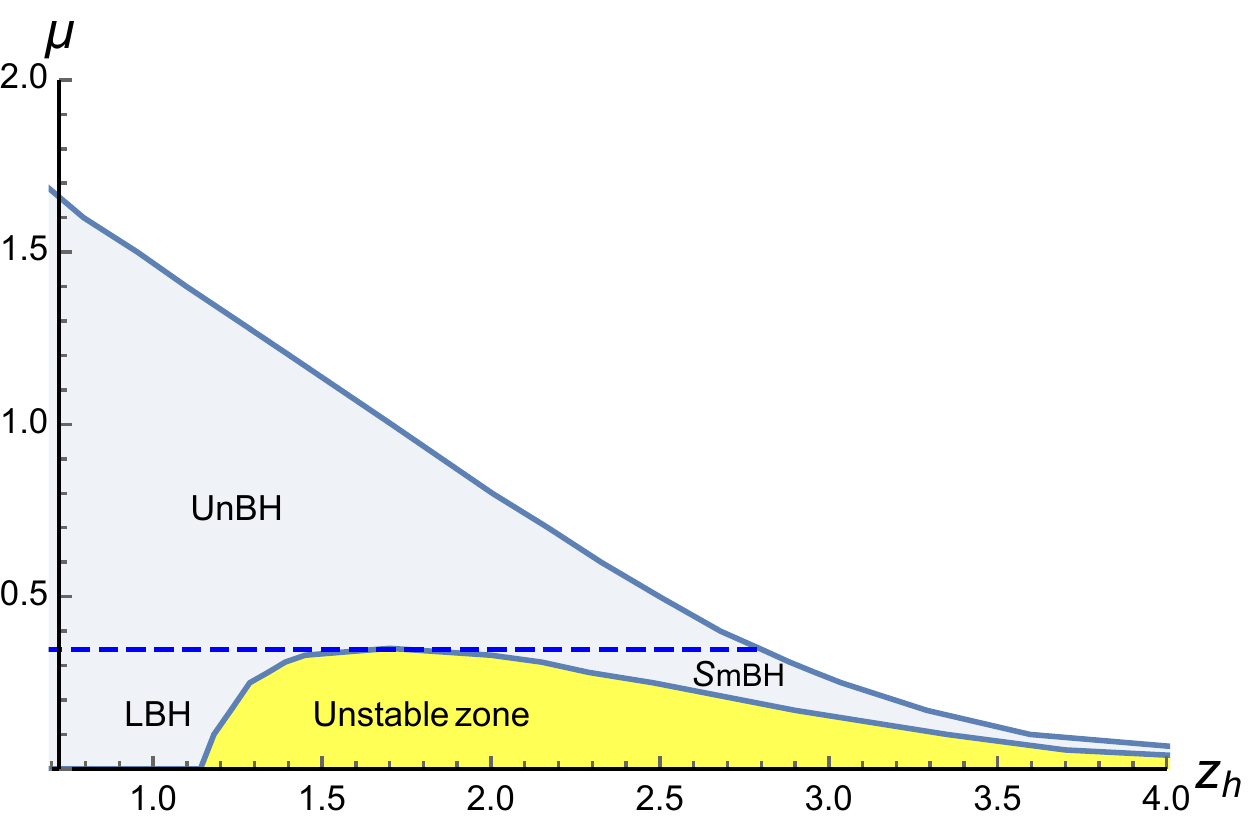}\\
  A \hspace{200pt} B
  \caption{Stability zones at the $(z_h,\mu)$-plane of the blackhole
    isotropic (A) and anisotropic (B) backgrounds. Here SBH indicates
    the regions of small black holes, LBH indicates the regions of
    large black holes, and UnBH  indicates the regions of parameters
    where is no separation on the small and large black holes.}
  \label{Fig:StabZone}
\end{figure}

Fig.\ref{Fig:StabZone} summarizes our discussion of the phase
transitions of our black hole isotropic and anisotropic
backgrounds. In the next section we put  probe strings in these
backgrounds to find out information about the
confinement/deconfinement phase transition.

$$\,$$
\newpage

\section{Confinement-deconfinement phase transition}\label{Sec:Ph-Tr}

\subsection{Equation for the dynamical wall}\label{Sec:DDW}

To guarantee the confinement-deconfinement phase transition one has to
check the existence of the dynamical wall (DW). The dynamical wall
position $z_{DW}$ is defined by the {\it minimal} extremal point of the
effective potential, that depends on the orientation
\cite{Arefeva:2016rob} and is related to the warp factor power $P(z)$
and the scalar field
$\phi$:

\bea
  {\cal V}_x &=& \frac{e^{P(z) + \sqrt{\frac23} \phi(z)}}{z^2} \,
  \sqrt{g(z)}, \\
  {\cal V}_y &=& \frac{e^{P(z) + \sqrt{\frac23} \phi(z)}}{z^{1/\nu +
      1}} \, \sqrt{g(z)}.
\eea
Here the subscribte indexes show the orientation of
  the Wilson loop. Therefore the dynamical wall position is given by
equations:
\bea
  {\cal V}'_x(z_{DWx}) &=& 0 \Rightarrow P'(z) + \sqrt{\frac23} \,
  \phi'(z) - \cfrac{2}{z} + \frac{g'(z)}{2 g(z)} \, \Big|_{z =
    z_{DWx}} \hspace{-15pt} = 0, \label{DWx} \\
  {\cal V}'_y(z_{DWy}) &=& 0 \Rightarrow P'(z) + \sqrt{\frac23} \,
  \phi'(z) - \cfrac{\nu + 1}{\nu z} + \frac{g'(z)}{2 g(z)} \, \Big|_{z
    = z_{DWy}} \hspace{-15pt} = 0. \label{DWy}
\eea
For zero temperature we have
\bea
  P'(z) + \sqrt{\frac23} \, \phi'(z) - \cfrac{2}{z} \, \Big|_{z =
    z_{DWx}} \hspace{-15pt} = 0, \\
  P'(z) + \sqrt{\frac23} \, \phi'(z) - \cfrac{\nu + 1}{\nu z} \,
  \Big|_{z = z_{DWy}} \hspace{-15pt} = 0.
\eea
Substituting $\phi'(z)$ from \eqref{2-nd-t} we get the following
equations for the positions of the dynamical wall corresponding to x-
and y-directions of the quark orientations:
\bea
  &\left. \left[P'(z) + \sqrt{\cfrac23} \, \sqrt{- \, 3 P''(z) +
        \cfrac{3}{2} \, P'^2(z) - \cfrac{6}{z} \, P'(z) +
        \cfrac{4}{z^2} \, \cfrac{\nu - 1}{\nu^2}} - \cfrac{2}{z}
    \right] \right|_{z = z_{DWx}} \hspace{-15pt} = 0, \nn \\
  &\left. \left[P'(z) + \sqrt{\cfrac23} \, \sqrt{- \, 3 P''(z) +
        \cfrac{3}{2} \, P'^2(\xi) - \cfrac{6}{z} \, P'(\xi) +
        \cfrac{4}{z^2} \, \cfrac{\nu - 1}{\nu^2}} - \cfrac{\nu +
        1}{\nu z} \right] \right|_{z = z_{DWy}} \hspace{-15pt} =
  0. \nn
\eea

\subsection{$P(z) = cz^2/2$}

We take the simplest case \eqref{bc} again. Therefore we choose the
expression with positive sign in \eqref{eq:2.119}, and equations
\eqref{DWx} and \eqref{DWy} become:

\bea
  {\cal DW}_x &\equiv& cz + \cfrac{1}{\nu z} \, \sqrt{\cfrac23} \,
  \sqrt{3 c \, \nu^2 z^2 \left( \cfrac{c z^2}{2} - 3 \right) + 4 \nu -
    4} - \cfrac{2}{z} + \cfrac{g'}{2 g} \Big|_{z = z_{DWx}}
  \hspace{-15pt} = 0, \label{spexTcK} \\
  {\cal DW}_y &\equiv& cz + \cfrac{1}{\nu z} \, \sqrt{\cfrac23} \,
  \sqrt{3 c \, \nu^2 z^2 \left( \cfrac{c z^2}{2} - 3 \right) + 4 \nu -
    4} - \cfrac{\nu + 1}{\nu z} + \cfrac{g'}{2 g} \Big|_{z =
    z_{DWy}} \hspace{-15pt} = 0.
  \label{speyTc}
\eea
Note that in both equations there is also a dependence of the
corresponding blackening functions on $\nu$. The phase transition from
confinement to deconfinement occurs when the corresponding equations
loss solutions.

To see the dependence of the dynamical wall position on the parameters
of the metric, it is useful to study the details of dependence of the
different terms defined ${\cal DW}_x$, ${\cal DW}_y$ and ${\cal
  DW}_{iso}$ on these parameters. Here ${\cal DW}_{iso}$ denotes the
left-hand side of the equation, similar to \eqref{speyTc}, in the
isotropic case.

\subsubsection{Zero temperature}

Let us first consider the case of the zero temperature, i.e. $g =
1$. In this case we deal with equations
\bea
  z = z_{DWx}: \quad \sigma(z,\nu,c) &=& \frac{2}{z} \quad or \quad
  \sigma_x(z,\nu,c) = 0, \label{spexTc0K} \\
  z = z_{DWy}: \quad \sigma(z,\nu,c) &=& \frac{\nu + 1}{\nu z} \quad
  or \quad \sigma_y(z,\nu,c) = 0, \label{speyTc0K}
\eea
where
\bea\label{LHS}
  \sigma(z,c,\nu) &\equiv& cz + \cfrac{1}{\nu z} \, \sqrt{\frac23} \,
  \sqrt{3 c \, \nu^2 z^2 \left( \cfrac{c z^2}{2} - 3 \right) + 4 \nu -
    4}.
\eea
We also use notations:
\bea
  \sigma_x(z,c,\nu) &\equiv& cz + \cfrac{1}{\nu z} \, \sqrt{\frac23}
  \, \sqrt{3 c \, \nu^2 z^2 \left( \cfrac{c z^2}{2} - 3 \right) + 4
    \nu - 4} - \frac{2}{z}, \label{LHSx} \\
  \sigma_y(z,c,\nu) &\equiv& cz + \cfrac{1}{\nu z} \, \sqrt{\frac23}
  \, \sqrt{3 c \, \nu^2 z^2 \left( \cfrac{c z^2}{2} - 3 \right) + 4
    \nu - 4} - \cfrac{\nu + 1}{\nu z}. \label{LHSy}
\eea

\begin{figure}[h!]
  \centering
  \includegraphics[scale=0.55]{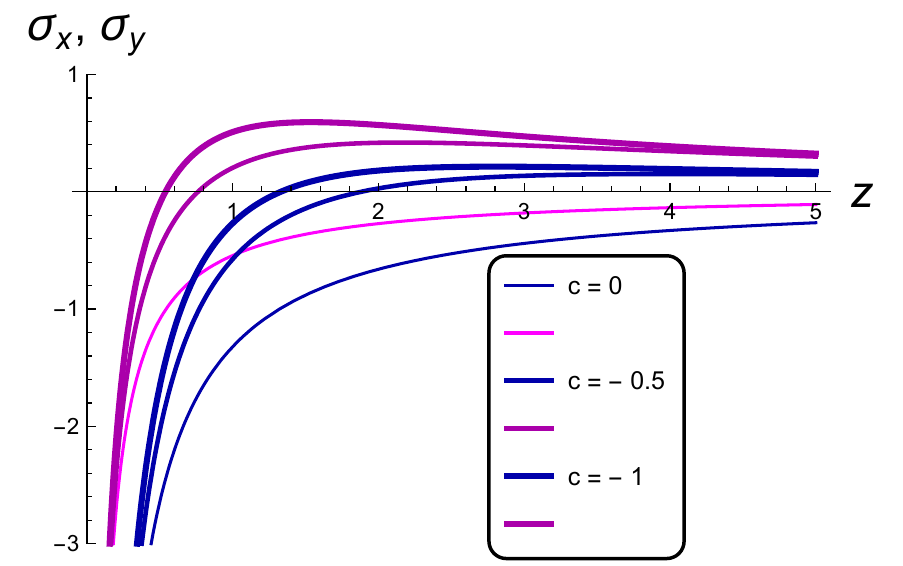}
  \includegraphics[scale=0.5]{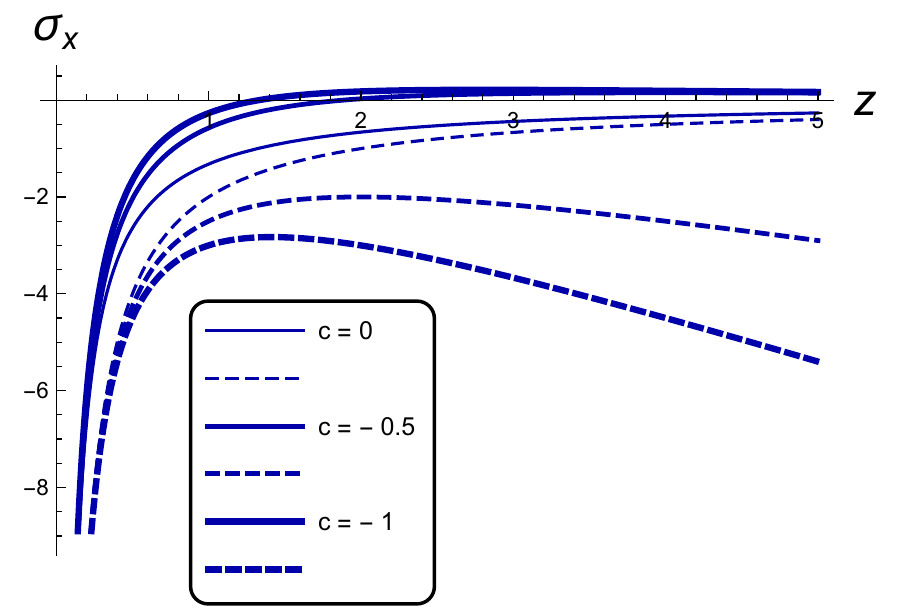}
  \includegraphics[scale=0.5]{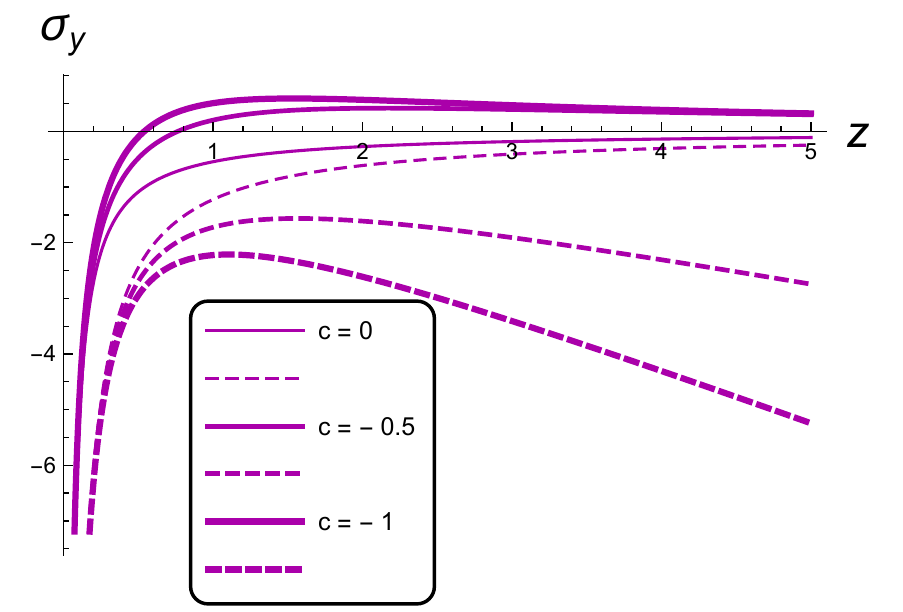}\\
  A\hspace{130pt}B\hspace{130pt}C\\ \ \\
  \includegraphics[scale=0.9]{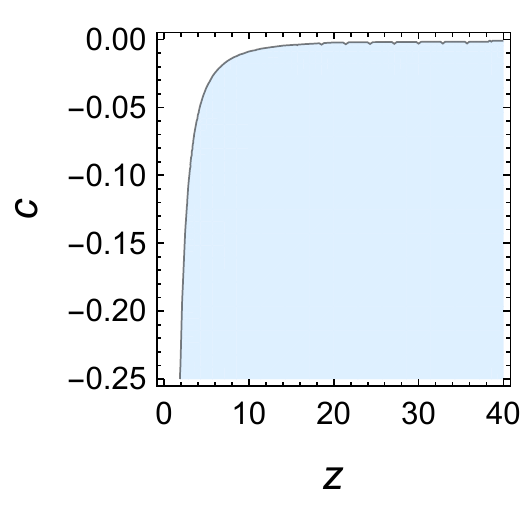}
  \includegraphics[scale=0.9]{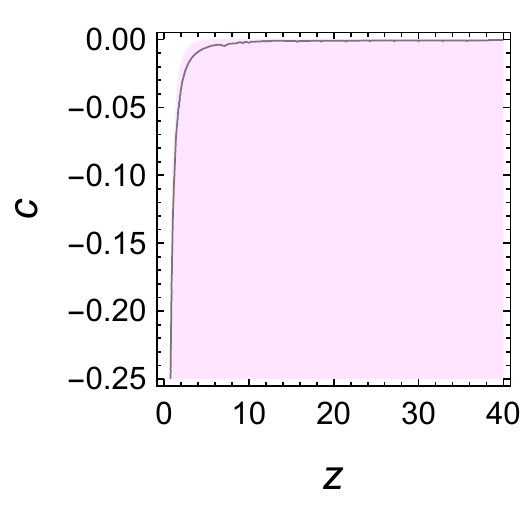}
  \includegraphics[scale=0.9]{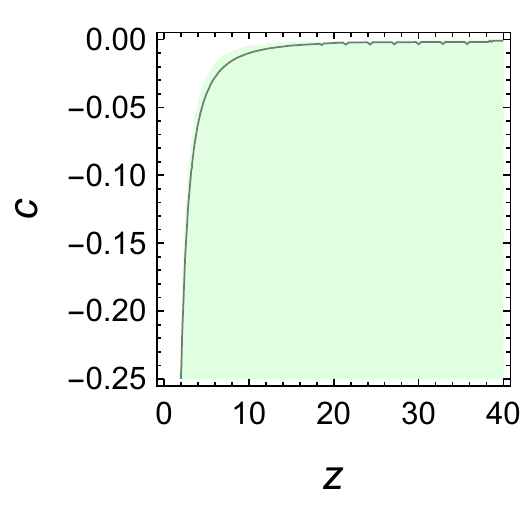}\\D
  \caption{Functions $\sigma_x(z,c,\nu)$ (blue lines) and
    $\sigma_y(z,c,\nu)$ (magenta lines) for different $c$ with dilaton
    contributions (solid lines) and without them (dashed lines)
    together (A) and separately (B, C); solutions to \eqref{spexTc0K}
    (boundary of blue and white areas), to \eqref{speyTc0K} (boundary
    of magenta and white areas) and for the isotropic case (boundary
    of green and white areas) (D).} 
  \label{Fig:DW-T0K}
\end{figure}

Behavior of $\sigma_x(z,c,\nu) $ and $\sigma_y(z,c,\nu) $ as functions
of $z$ are shown in Fig.\ref{Fig:DW-T0K}. The positions of the
dynamical walls are defined by the intersections of
the solid blue and magenta lines representing $\sigma_x(z,c,\nu)$ and
$\sigma_y(z,c,\nu)$ with the horizontal line $\sigma = 0$.

To show that the presence of the dilaton field supports the appearance
of the DW, we display parts of the expressions $\sigma_x(z,c,\nu)$ and
$\sigma_y(z,c,\nu)$ without the square roots, that are originated from
the dilaton fields, by the dashed lines in Fig.\ref{Fig:DW-T0K}.B and
C. We see that these dashed lines never intersect horizontal axis,
therefore in these cases there are no DW solutions.

Solutions to equations \eqref{spexTc0K} and \eqref{speyTc0K} can also
be represented as the boundary between positive and negative values of
functions $\sigma_x(z,c,\nu)$ and $\sigma_y(z,c,\nu)$. Taking $\nu =
1$ provides us with the isotropic case result
(Fig.\ref{Fig:DW-T0K}.D). We see that the critical $c = c_{cr}$, above
which there are no solutions in all cases, is $c_{cr} = 0$. For $c >
0$ our consideration is not valued, since the scalar field becomes
complex.

\subsubsection{Non-zero temperature, zero chemical potential}

Let us take the nontrivial blackening function. The blackening
function modifies the DW equations. It is convenient to present these
equations in the form
\bea
  z = z_{DWx}: \quad \Sigma(z,z_h,\nu,c) &=&
  \cfrac{2}{z}, \label{spexTcKmu0} \\
  z = z_{DWy}: \quad \Sigma(z,z_h,\nu,c) &=& \cfrac{\nu +
    1}{\nu z}, \label{speyTcKmu0}
\eea
where
\bea
  \Sigma(z,z_h,c,\nu) &\equiv& \sigma(z,c,\nu) +
  G(z,z_h,c,\nu). \label{LHST}
\eea
Function $\sigma(z,\nu,c)$ is given by \eqref{LHS} and 
\bea
  G(z,z_h,c,\nu) &\equiv& \cfrac{g'}{2 g}. \label{gg}
\eea

\begin{figure}[h!]
  \centering
   \includegraphics[scale=0.55]{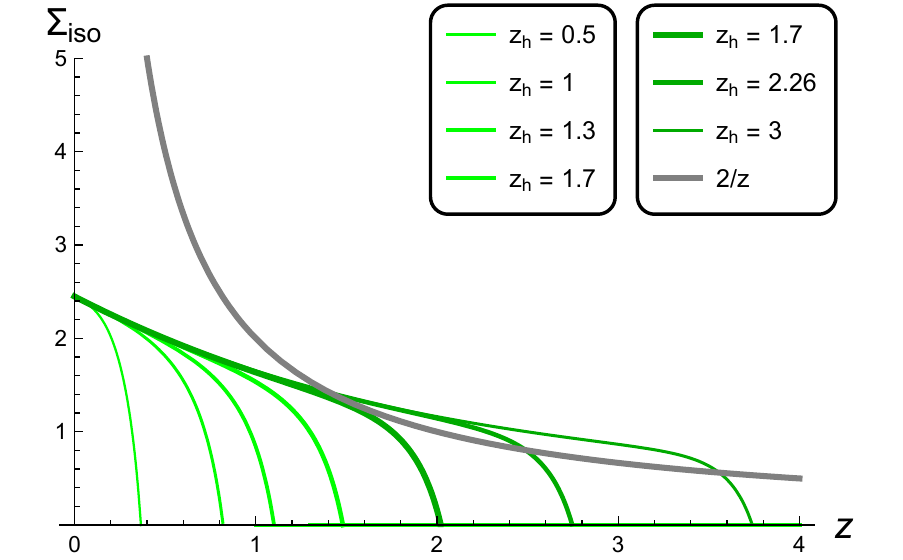}
   \includegraphics[scale=0.55]{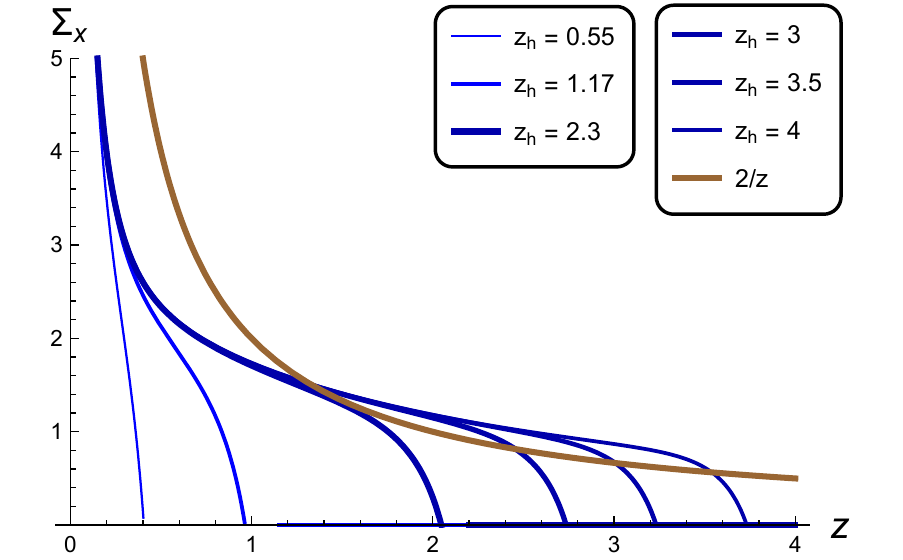} 
   \includegraphics[scale=0.55]{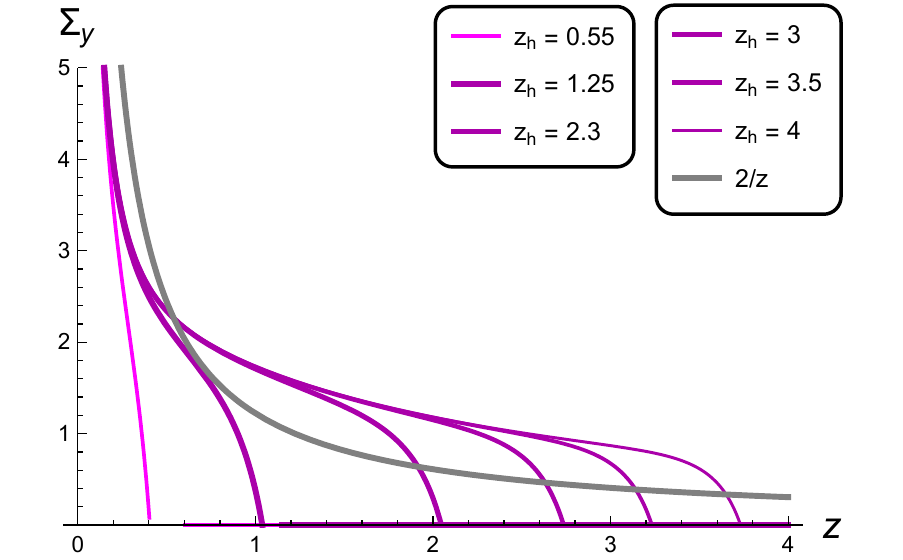}\\
   A\hspace{140pt}B\hspace{140pt}C
   \caption{A) Dynamical walls' positions in isotropic case are given
     by the intersections of green lines representing
     $\Sigma_{iso}(z)$ and the grey line representing $2/z$; B)
     dynamical walls' positions, corresponding to the Wilson loop
     $W_x$ in the anisotropic case $\nu = 4.5$, are given by
     intersections of blue lines representing $\Sigma(z)$ and the
     brown line representing $2/z$; C) dynamical walls' positions,
     corresponding to the Wilson loop $W_y$ in the anisotropic case
     $\nu = 4.5$ are given by the intersections of magenta lines
     representing $\Sigma(z)$ and the grey line representing $(\nu +
     1)/(\nu z)$. Here we vary $z_h$ and $c$. In all cases to get the
     DW position we take the minimal intersection point.}
  \label{Fig:DDW}
\end{figure} 

\begin{figure}[h!]
  \centering
   \includegraphics[scale=0.75]{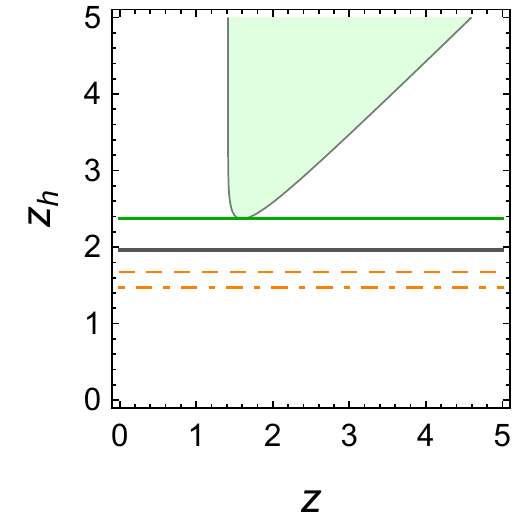}\qquad
   \includegraphics[scale=0.75]{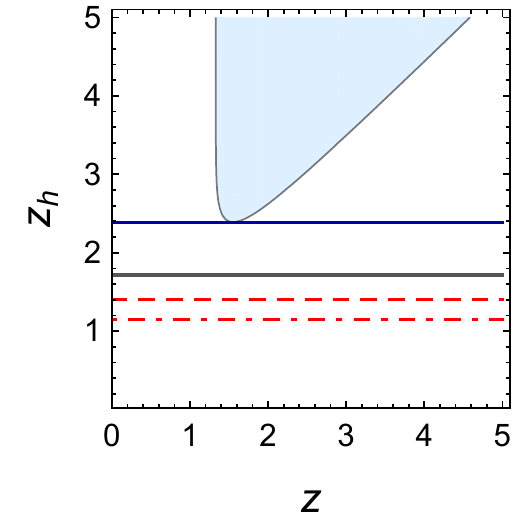}\qquad
   \includegraphics[scale=0.75]{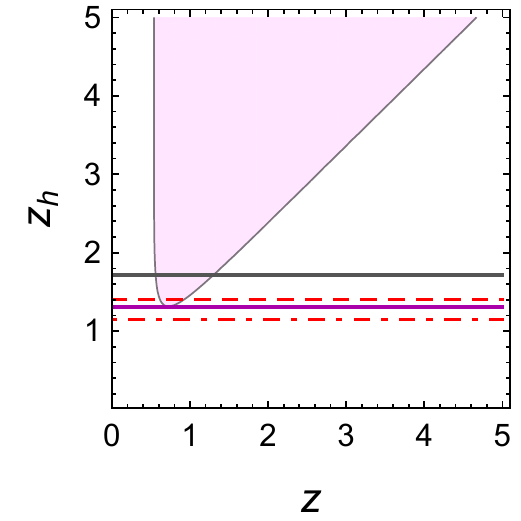}\\
   A\hspace{120pt}B\hspace{120pt}C
   \caption{A) Dynamical walls' positions in isotropic case are
     located on the left part of the boundary of the green and white
     areas. We see that for given $z_h > 2.304$ there are two
     solutions and for $z_h = 2.304$ there is only one $z_{DWiso} =
     1.455$. B) Dynamical walls' positions corresponding to the Wilson
     loop $W_x$ are located on the left part of the boundary of the
     blue and white ares. We see that for given $z_h > 2.376$ there
     are two solutions and for $z_h = 2.376$ only one $z_{DWx} =
     1.503$. C) Dynamical walls' positions corresponding to the Wilson
     loop $W_y$ are located on the left part of the boundary of the
     magenta and white ares. We see that for given $z_h > 1.326$ there
     are two solutions and for $z_h = 1.326$ only one $z_{DWiso} =
     0.6871$. Here $\nu = 4.5$ and $c = -1$. Dashed orange and red
     lines show positions of the critical $z_{h,min}(0)$ in the
     corresponding isotropic and anisotropic backgrounds, and
     point-dashed lines show positions of the horizons $z_{h,HP}(0)$,
     where the HP transitions take place at zero chemical potential in
     the isotropic and anisotropic cases. The dark grey lines show
     $z_{h,cr} = z_{h, cr}(\mu_{cr})$ the value of the horison at the
     inflection points. These plots show that all dynamical walls
     appear below the corresponding horizons, i.e. $z_{DW}(z_{h},\nu)
     < z_{h}$ and these horizons $z_{h}$ are larger than
     $z_{h,min}(0)$ and $z_{h, cr}$ for the isotropic case and for
     $W_x$ case, but they can be smaller than $z_{h,min}(0)$ and
     $z_{h, cr}$ in the anisotropic $W_y$ case. The positions of
     minimal $z_h$ admitted the DW, $z_{h,DW}$ are indicated by dark
     green, dark blue and dark magenta lines.}
  \label{Fig:DDW2}
\end{figure} 

We find the solutions of equations \eqref{spexTcKmu0} and
\eqref{speyTcKmu0} numerically. To visualize the location of
these solutions we plot $\Sigma(z,z_h,c,\nu)$ as a function of $z$ for
different values of parameters $z_h$, $\nu$, $c$ and find its
intersection with $2/z$ for the Wilson loop $W_x$ and with $(\nu +
1)/(\nu z)$ for the Wilson loop $W_y$ in
Fig.\ref{Fig:DDW}. If there are two intersection
  points, we take the minimal one and we call it the minimal
  intersection point. In all cases to get the corresponding DW
  position we take the minimal intersection point. From
Fig.\ref{Fig:DDW}.A we see that for $z_h > z_{h,cr}$ the dynamical
wall always appears, as there are intersections of the grey and
dark green lines. At the critical horizon (the thick dark green
line) there is a touch of these two lines and for $z < z_{h,cr}$
(lighter green lines) there is no intersection at all, therefore
confinement disappears.

The light blue curves in Fig.\ref{Fig:DDW}.B do not cross the grey
line and for these cases there are no dynamical dynamical walls. The
dark blue lines cross the brown one and for the corresponding
temperature there is the quark confinement, meanwhile the thick dark
blue line just touches the grey line and at this temperature the phase
transition occurs. The similar picture can be seen at
Fig.\ref{Fig:DDW}.C, corresponding to different orientation of quark
pairs.

Therefore the dynamical wall always appears for $z_h >
z_{h,cr}$. The particular values of $z_{h,cr}$ are different for
isotropic (A) and anisotropic (B, C) cases and depend on the quark
orientations. This appearance/disappearance of dynamical walls
corresponds to confinement and deconfinement phases. The phase
transition between these two regimes occurs at $z = z_{h,cr}$.

In Fig.\ref{Fig:DDW2} solutions to equations
  \eqref{spexTcKmu0}, \eqref{speyTcKmu0} are located on the boundary
  of the colored and white areas. Since to find the dynamical walls'
  positions we have to take the minimal solutions, the dynamical
  walls' positions are located on the left parts of boundaries between
  the colored and white areas.

\subsubsection{Non-zero chemical potential}

\begin{figure}[h!]
  \centering
  \includegraphics[scale=0.75]{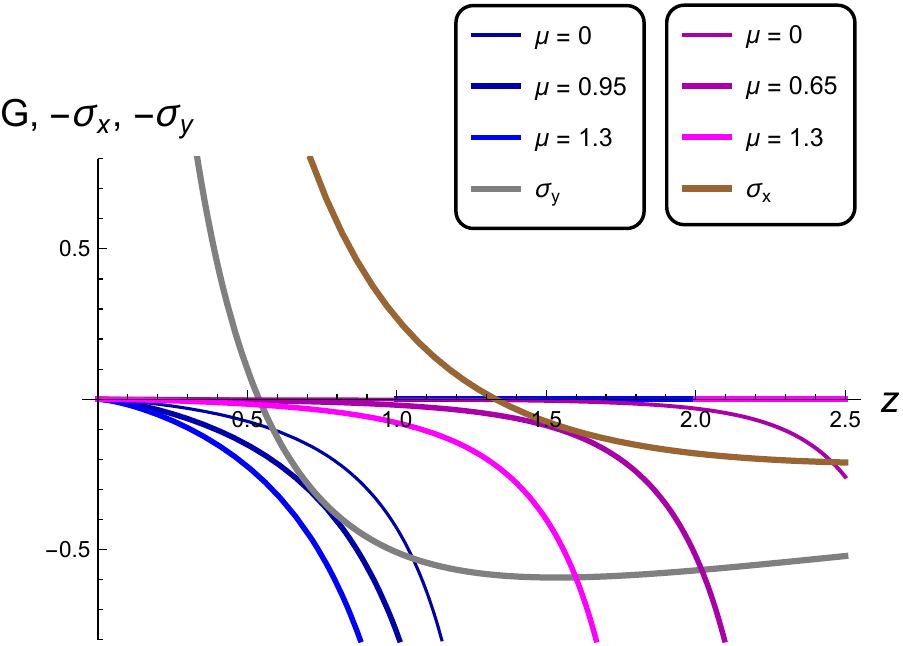}\quad
  \includegraphics[scale=0.75]{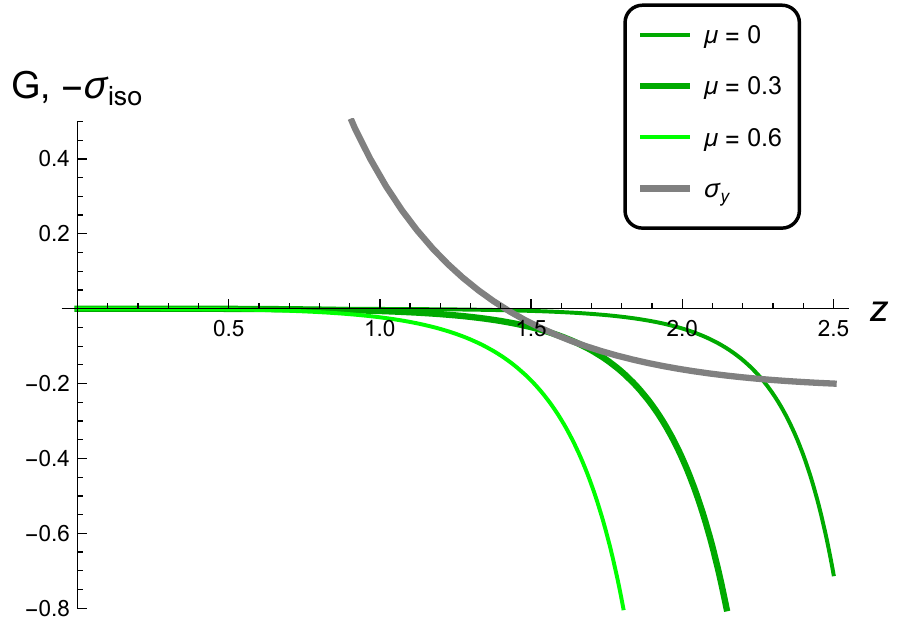}\\
  A\hspace{200pt}B
  \caption{The positions of the dynamical dynamical walls $G = -
    \sigma_x$ for $z_h = 1.5$ and $G = - \sigma_y$ for $z_h = 3$ in
    anisotropic $\nu = 4.5$ (A) and isotropic $G = - \sigma_{iso}(z)$
    (B) cases, $c = -1$.}
 \label{Fig:DDWTemp-mu}
\end{figure}

We can also study how these plots look for non-zero chemical potential.
The positions of the dynamical walls for non-zero temperature and
non-zero chemical potential in isotropic and anisotropic cases are
presented in Fig.\ref{Fig:DDWTemp-mu}.A and Fig.\ref{Fig:DDWTemp-mu}.B
correspondingly.

It is convenient to write equations \eqref{spexTc0K} and
\eqref{speyTc0K} in the form
\bea
  z = z_{DWx}: \qquad \Sigma(z,z_h,\mu,c,\nu) &=&
  \cfrac{2}{z}, \label{spexTcK} \\
  z = z_{DWy}: \qquad \Sigma(z,z_h,\mu,c,\nu) &=& \cfrac{\nu +
    1}{\nu z}, \label{speyTcK}
\eea
where
\bea
  &&\Sigma(z,z_h,\mu,c,\nu) \equiv \sigma(z,c,\nu) +
  G(z,z_h,\mu,c,\nu), \label{LHST} \\
  &&G(z,z_h,\mu,c,\nu) \equiv \cfrac{g'}{2 g} \label{gg}
\eea
and $\sigma(z,c,\nu)$ is defined by \eqref{LHS}.

The light blue and dark blue curves in Fig.\ref{Fig:DDWTemp-mu}.A
represent the function $G(z,z_h,\mu,c,\nu)$ for $\nu = 4.5$, $c =
-1$, $z_h = 1.5$ and different $\mu$. The light magenta and dark
magenta lines correspond to $G(z,z_h,\mu,c,\nu)$ for the same set of
parameters and $z_h = 3$. The thick lines, that touch $\sigma_x$ (grey
line) and $\sigma_y$ (brown line), depict the critical values of
chemical potential $\mu$. Thus the presence and the particular
position of the horizon modifies the position of dynamical walls as
compare with the zero temperature case ($g = 1$) presented in
Fig.\ref{Fig:DW-T0K}. Anisotropy also influences on the dynamical
walls' position. This can be seen from comparing of
Fig.\ref{Fig:DDWTemp-mu}.A and Fig.\ref{Fig:DDWTemp-mu}.B, where the
isotropic case is pictured.

\begin{figure}[h!]
  \centering
  \includegraphics[scale=0.6]{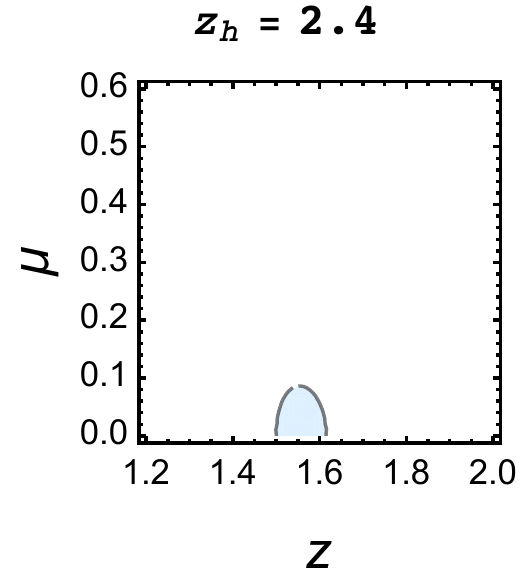}
  \includegraphics[scale=0.6]{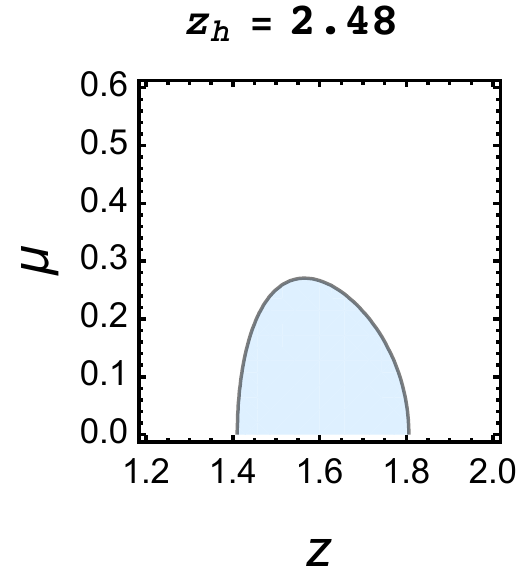}
  \includegraphics[scale=0.6]{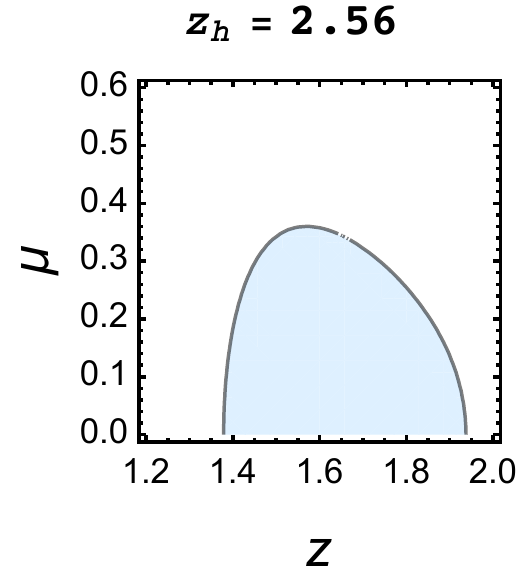}
  \includegraphics[scale=0.6]{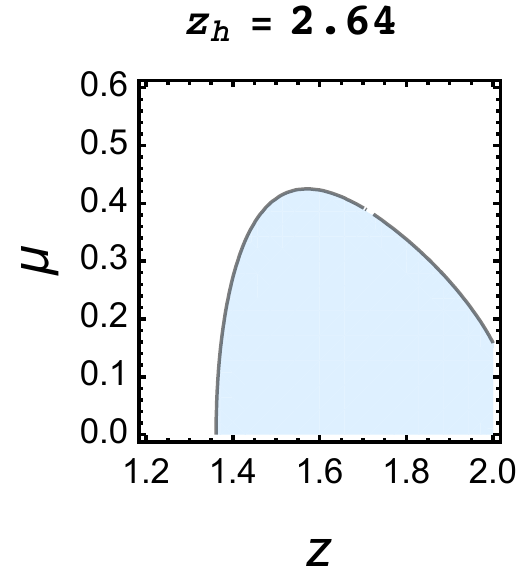}
  \caption{Contour plots for ${\cal DW}_x$ near zero for different
    $z_h$ (labels above).}
  \label{Fig:CP1}
\end{figure}

\begin{figure}[h!]
  \centering
  \includegraphics[scale=0.6]{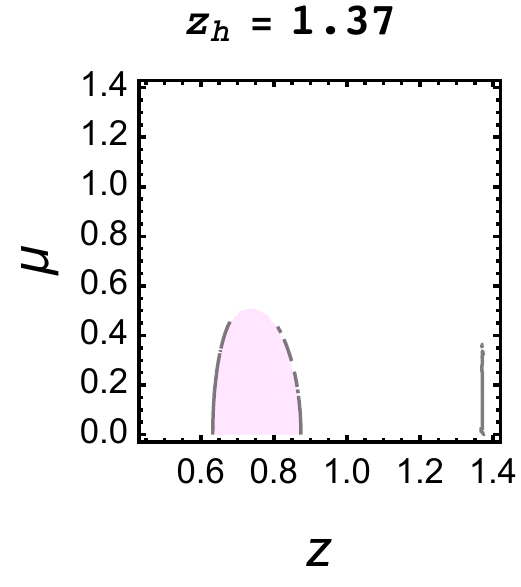}
  \includegraphics[scale=0.6]{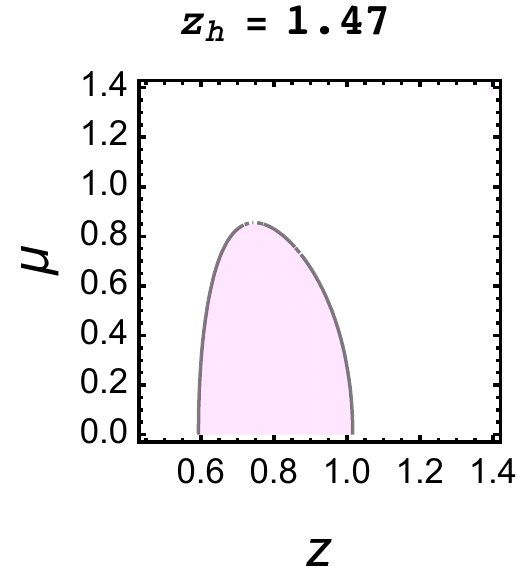}
  \includegraphics[scale=0.6]{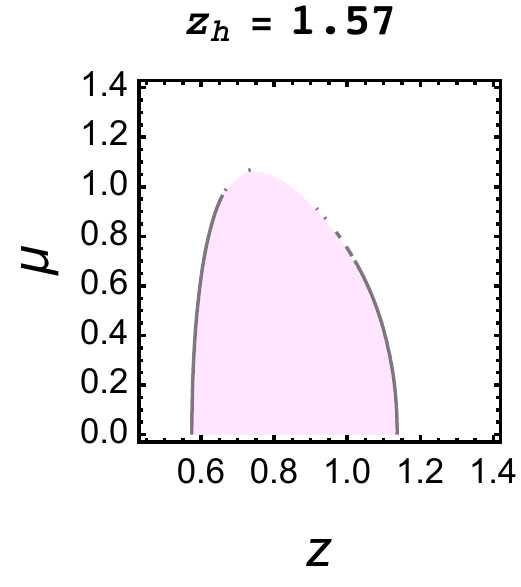}
  \includegraphics[scale=0.6]{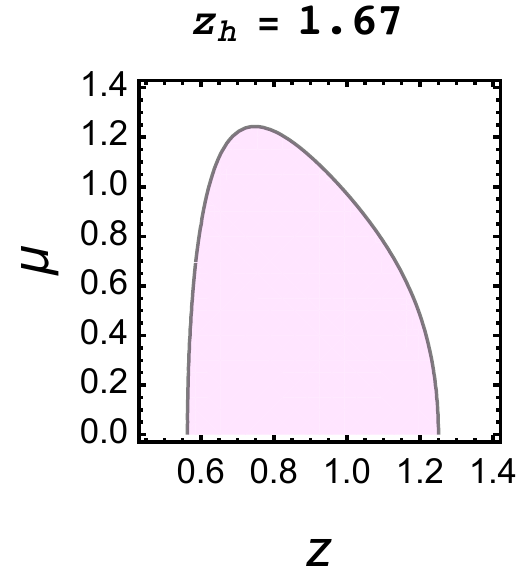}
  \caption{Contour plots for ${\cal DW}_y$  near zero for different
    $z_h$ (labels above).}
  \label{Fig:CP2}
\end{figure}

\begin{figure}[h!]
  \centering
  \includegraphics[scale=0.6]{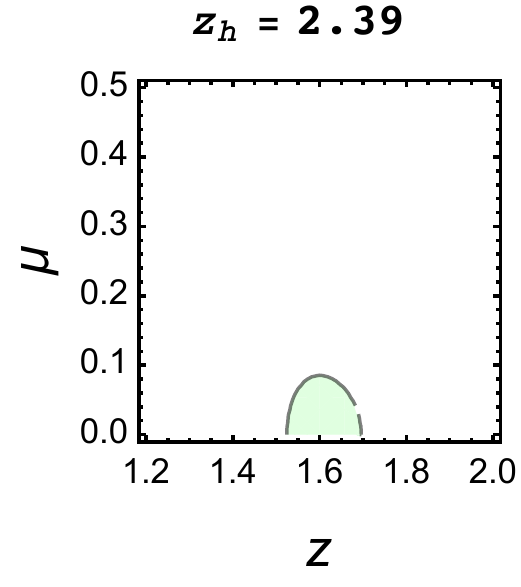}
  \includegraphics[scale=0.6]{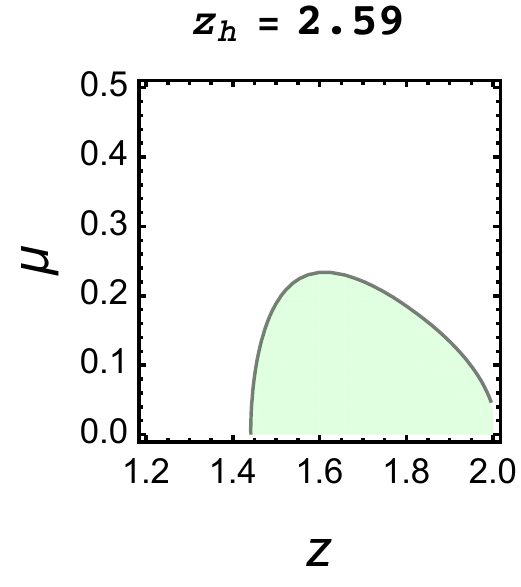}
  \includegraphics[scale=0.6]{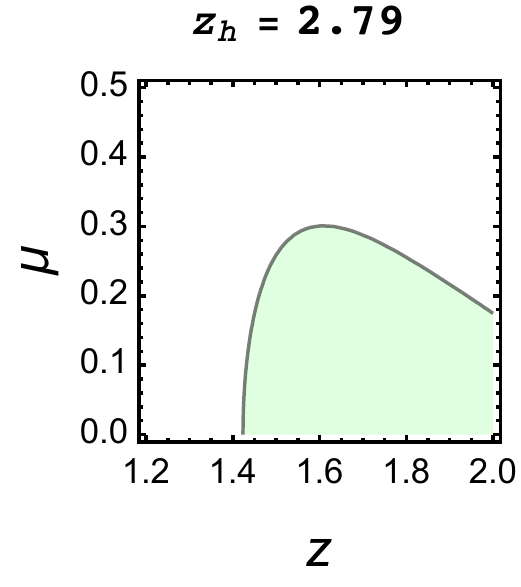}
  \includegraphics[scale=0.6]{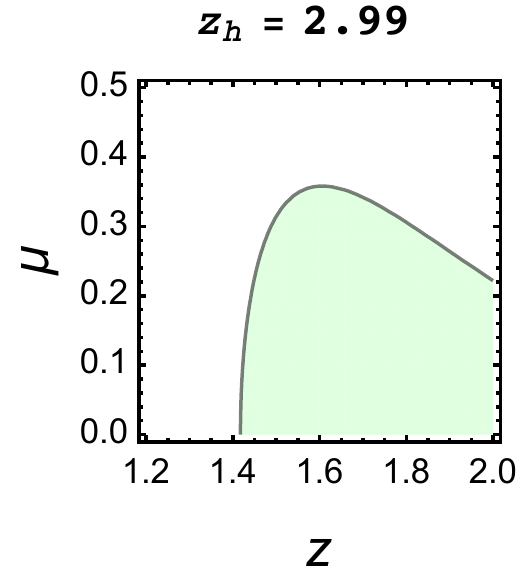}
  \caption{Contour plots for ${\cal DW}_{iso}$ near zero value.}
  \label{Fig:CP-iso}
\end{figure}

To find the phase transition line we have to determine
$\mu(z_h,c,\nu,W_i)$, here $W_i$ indicates the orientation of the 
Wilson line, $W_x$ or $W_y$, for given $z_h$ such that for any $\mu >
\mu(z_h,c,\nu,W_i)$ there is no real solution of equations
\eqref{spexTcK} and \eqref{speyTcK}. To find these points, it is 
convenient to draw the contour plots for functions ${\cal DW}_x$ and
${\cal DW}_y$ near zero. They are presented in Fig.\ref{Fig:CP1} and
Fig.\ref{Fig:CP2} correspondingly. For comparison we present in
Fig.\ref{Fig:CP-iso} the contour plots for ${\cal DW}_{iso}$ near
$0$.


\subsubsection{The dynamical  wall  position}

\begin{figure}[t]
  \centering
  \includegraphics[scale=0.45]{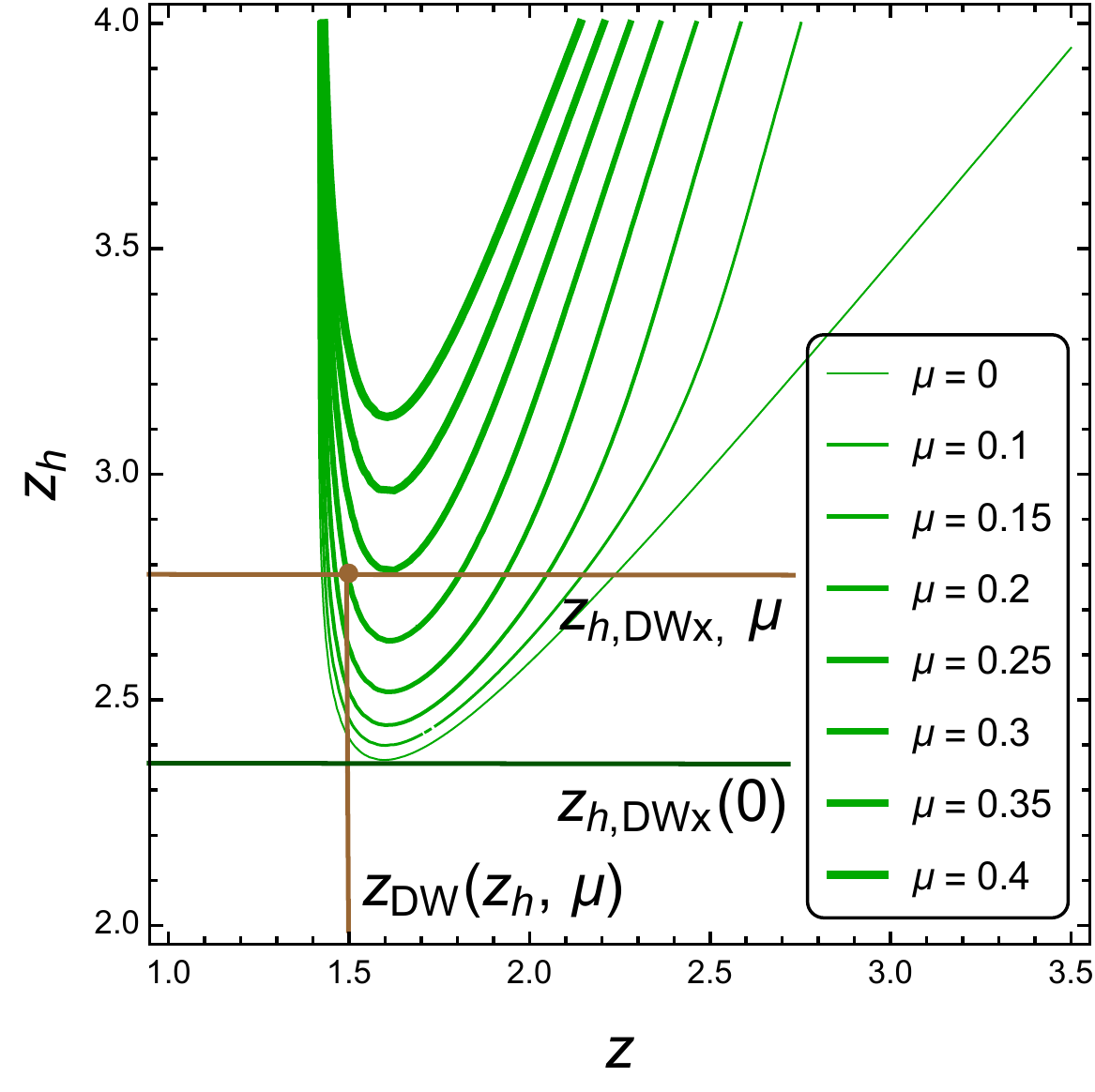} \qquad
  \includegraphics[scale=0.65]{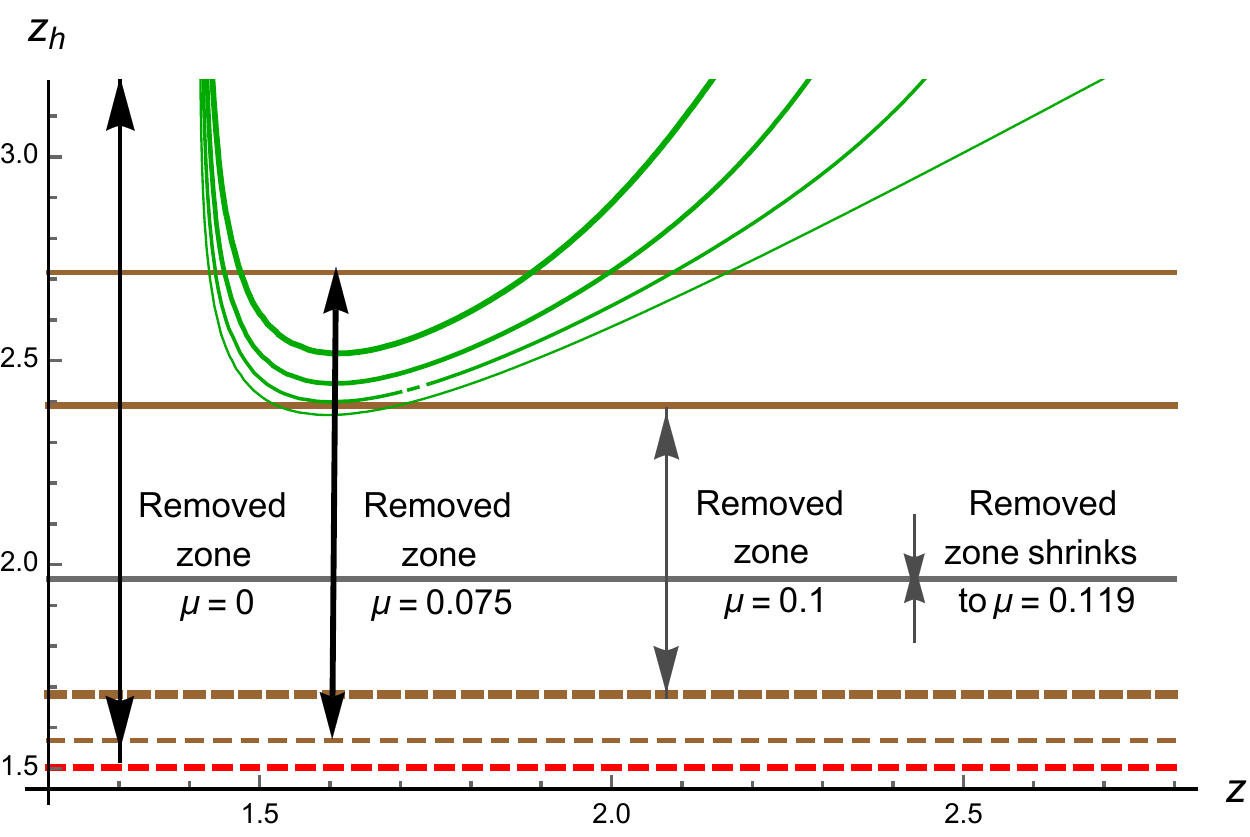}\\
  A \hspace{200pt} B
  \caption{A) The location of extreme points of the effective potential
    in the isotropic case $\nu = 1$, $c = - 1$ for different $z_h$ and
    $\mu$ are shown by green lines with different thickness, depending
    on the values of the chemical potential. Dynamical walls'
    positions are located on the left parts of these curves. B)
    Locations of instability zones corresponding to different $\mu$
    are shown by the black lines with arrows.}
  \label{Fig:DDWzh-mu}
\end{figure}

\begin{figure}[h!]
  \centering
  \includegraphics[scale=0.53]{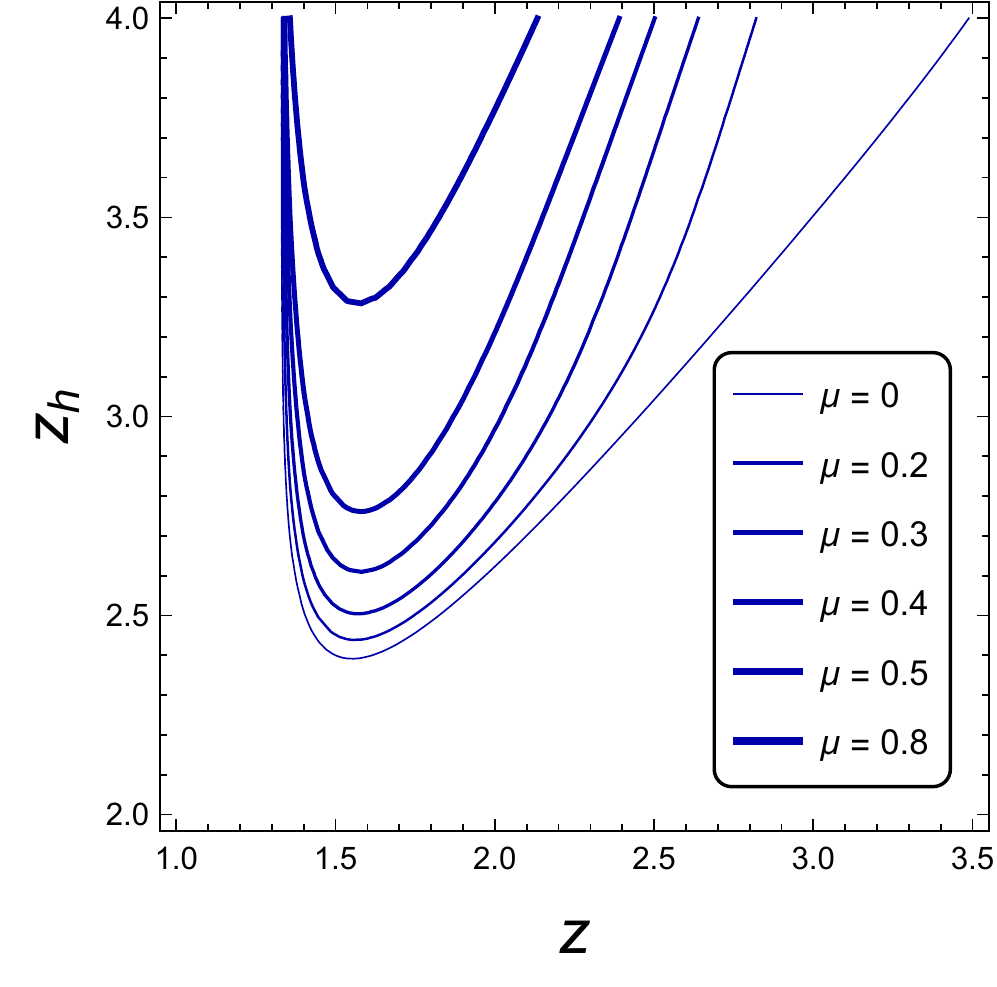} \qquad
  \includegraphics[scale=0.65]{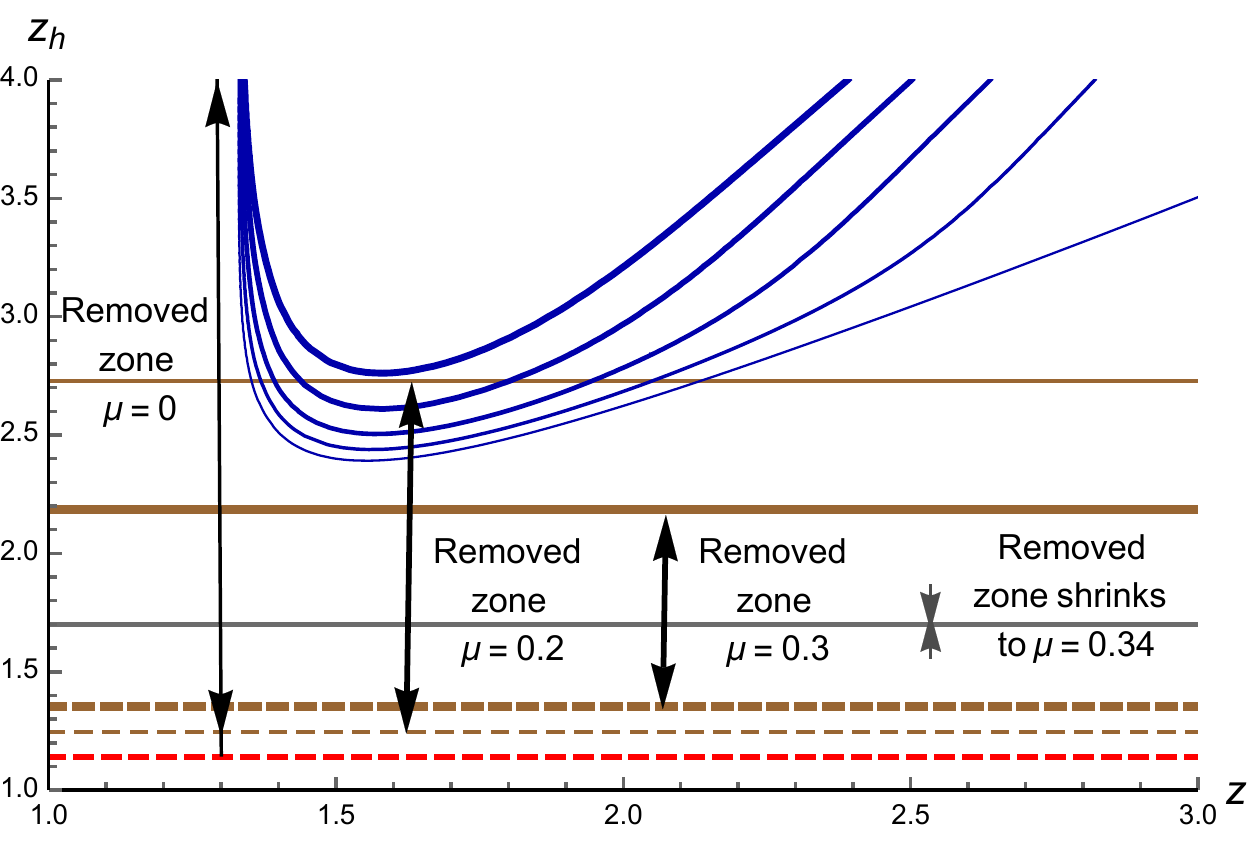}\\
  A \hspace{200pt} B \qquad \ \\ \ \\
  \includegraphics[scale=0.58]{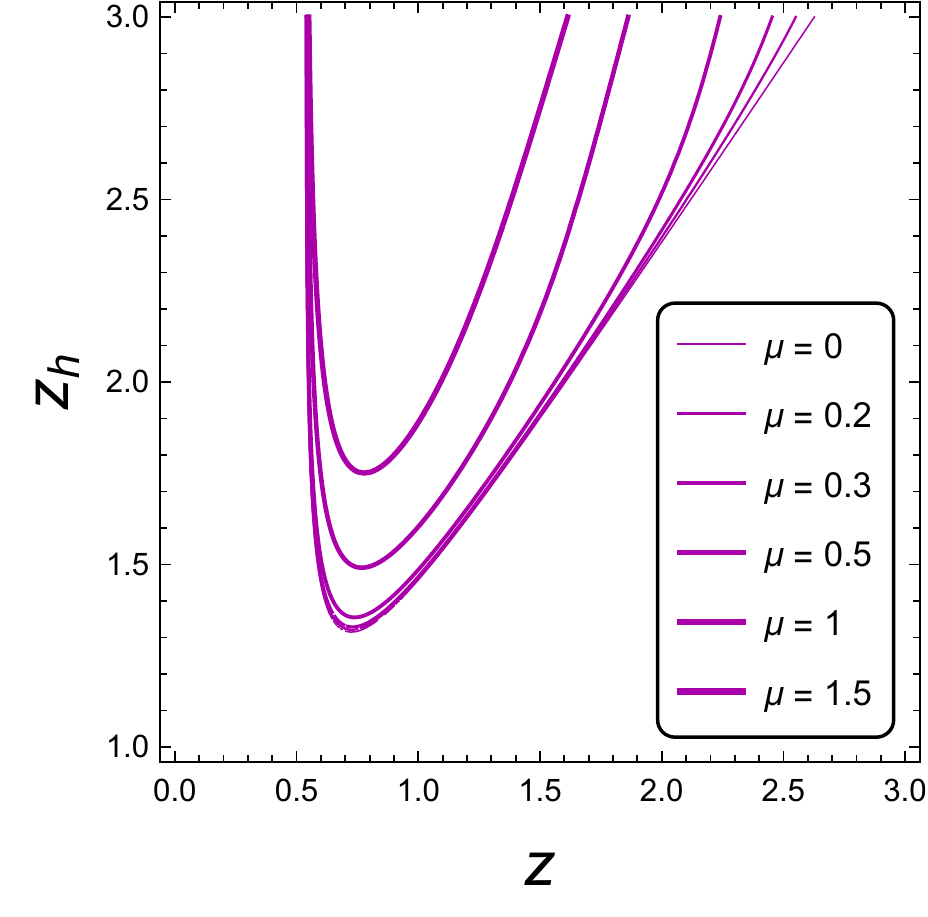} \qquad
  \includegraphics[scale=0.65]{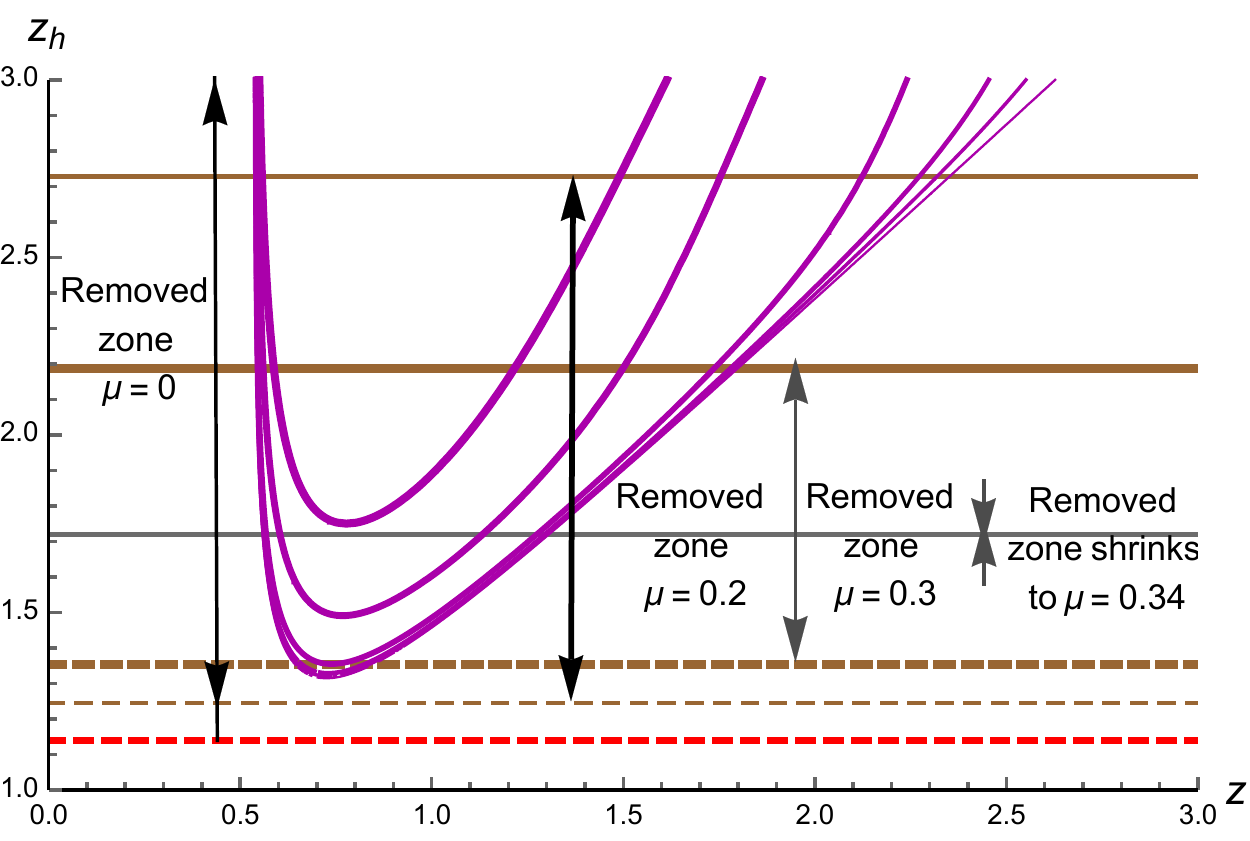}\\
  C \hspace{200pt} D \qquad \ 
  \caption{The location of extreme points of effective potential in
    the anisotropic case $\nu = 4.5 $, $ c = - 1$ for different $z_h$,
    $\mu$ and different orientations of quark pairs are shown by lines
    of different thickness depending on the values of the chemical
    potential, and colors, for the Wilson longitudinal line $W_x$ (A
    and B) with dark blue and for the Wilson transverse line $W_y$
    with dark purple lines (B and D). B) and D). The location of the
    instability zones corresponding to different $\mu$ is shown by
    black lines with arrows.}
     \label{Fig:DDW45zh-mu}
\end{figure}

All previous considerations can be summarized in contour
plots. Namely, we can draw the contours for the
locations of the effective
potentials' derivatives ${\cal V}'_{xy} = 0$ in the $(z,
z_h)$-plane, keeping $c = -1$ and considering two
cases $\nu = 1$ (Fig.\ref{Fig:DDWzh-mu}) and $\nu = 4.5$
(Fig.\ref{Fig:DDW45zh-mu}). Different contours correspond to different
values of $\mu$.
Dynamical walls' positions correspond to the minimal $z$
for fixed horizon $z_h$ and chemical potential $\mu$
(Fig.\ref{Fig:DDWzh-mu}.A).
The positions of the dynamical walls in the
anisotropic case differ for longitudinal $x$ and transversal $y_{1,2}$
quark pair orientation (Fig.\ref{Fig:DDW45zh-mu}.A and C,
correspondingly). Transversal case is characterized by smaller values
of $z_h$ and minimal $z$ for the same chemical potential
$\mu$.

\begin{figure}[h!]
  \centering
  \includegraphics[scale=0.6]{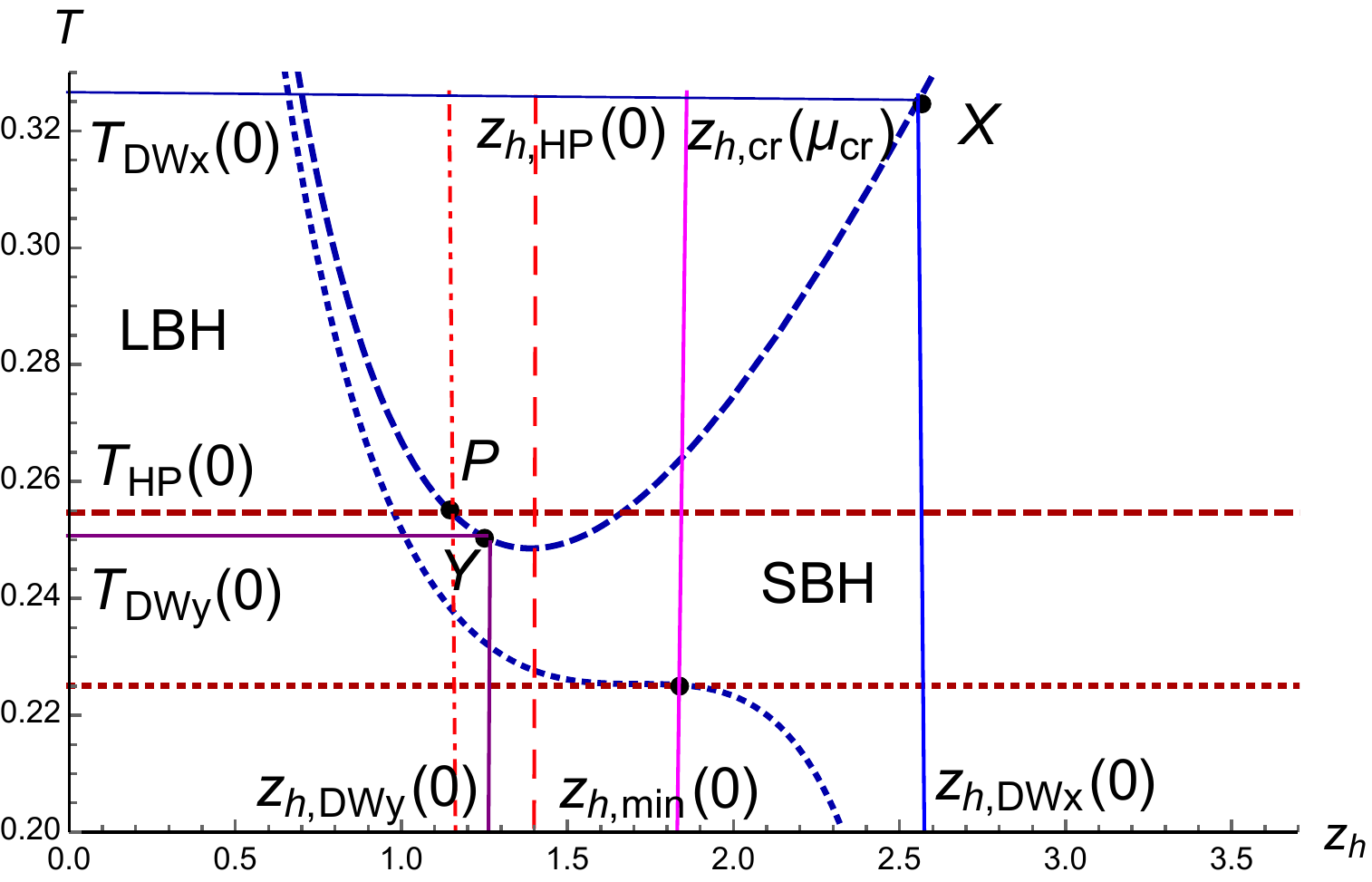}
  \caption{The plot explains the difference between phase diagrams for
    Wilson lines $W_{x}$  and $W_y$ for $\nu = 4.5$. The coordinates
    of points X,Y and P are $(z_{h,DWx}(0),T_{DWx}(0))$,
    $(z_{h,HP}(0),T_{HP}(0))$ and $(z_{h,DWx}(0),T_{DWx}(0))$. We see
    that P is in the SBH region, and X in the LBH region. The red
    dashed and point-dashed lines, and the magenta line are the same
    as at Fig.\ref{Fig:DDW2}.}
  \label{Fig:diff-WxWy}
\end{figure}

On Fig.\ref{Fig:DDWzh-mu}.B and Fig.\ref{Fig:DDW45zh-mu}.B and D the
instability regions of the background for $\mu = 0$ are indicated by
black arrows. For fixed $0< \mu < \mu_{cr}$ they are shown as domains
between corresponding brown solid and dashed lines of the same
thickness. The thick magenta line on Fig.\ref{Fig:DDW45zh-mu}.D shows 
the value of $z_h$, at which the transition between large and small
black holes disappears and the ``removed zone'' shrinks to $z_{h, cr}$
for the curve $\mu = \mu_{cr}$. For isotropic case   $\mu_{cr}^{(iso)}
= 0.119$ and for anisotropic one at $\mu_{cr}^{(anis)} = 0.34$ and it
happens at $z_h = z_{h, cr}^{(iso)} = 1.97$ and at $z_h = z_{h,
  cr}^{(aniso)} = 1.70$ in the isotropic and anisotropic cases,
respectively. Note, the line $z_h = z_{h, cr}^{(iso)}$ is below
$z_{h,DW}^{(iso)}$, the value of horizon for which the DW can appear
in the isotropic case. In the anisotropic case the line $z_h = z_{h,
  cr}^{(aniso)}$  is below values of $z_{h,DWx}$, for which the DW for
$W_x$ can appear, $z_{h, cr}^{(aniso)} < z_{h,DWx}$, but above the
values of horizons, for which  the DW can appear for $W_y$,
i.e. $z_{h, cr}^{(aniso)}>z_{h,DWy}$ (see Fig.\ref{Fig:diff-WxWy} and
discussion below).


\begin{figure}[h!]
  \centering
  \includegraphics[scale=0.5]{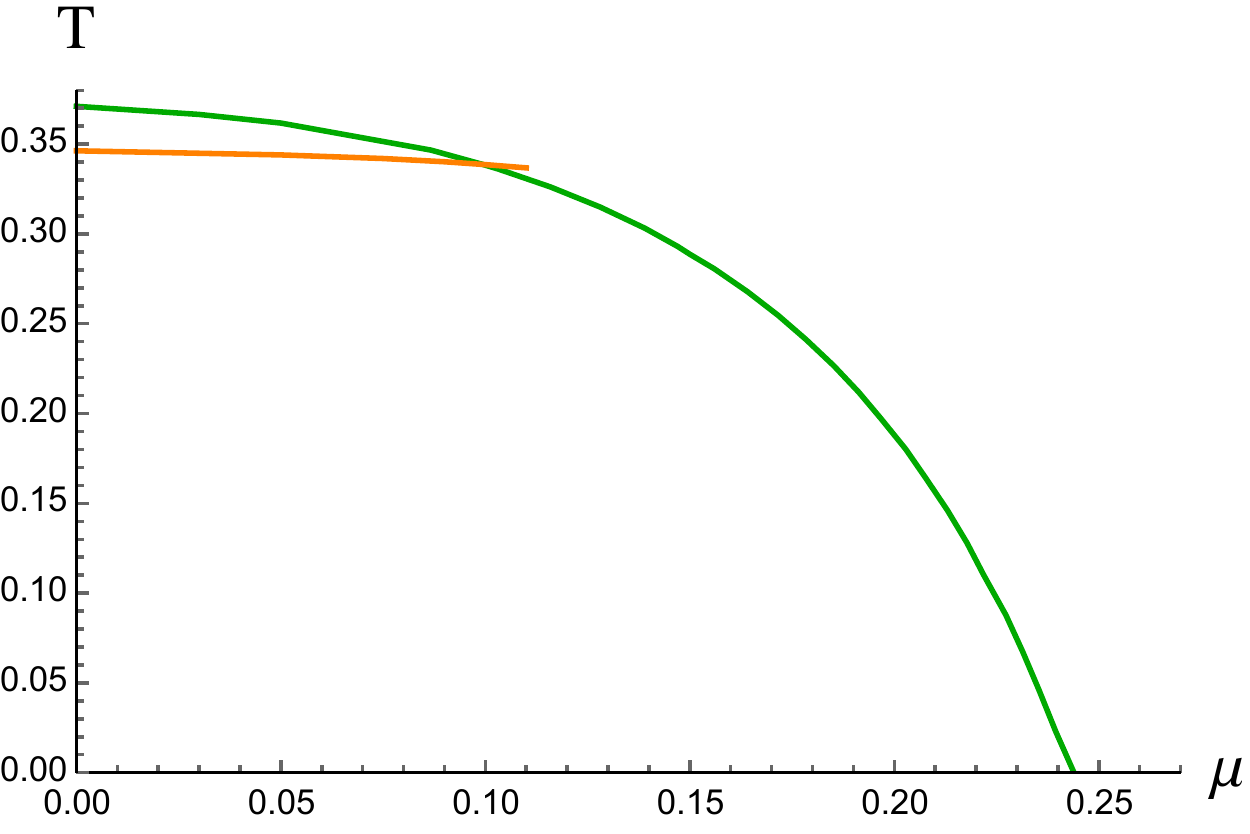} \qquad
  \includegraphics[scale=0.5]{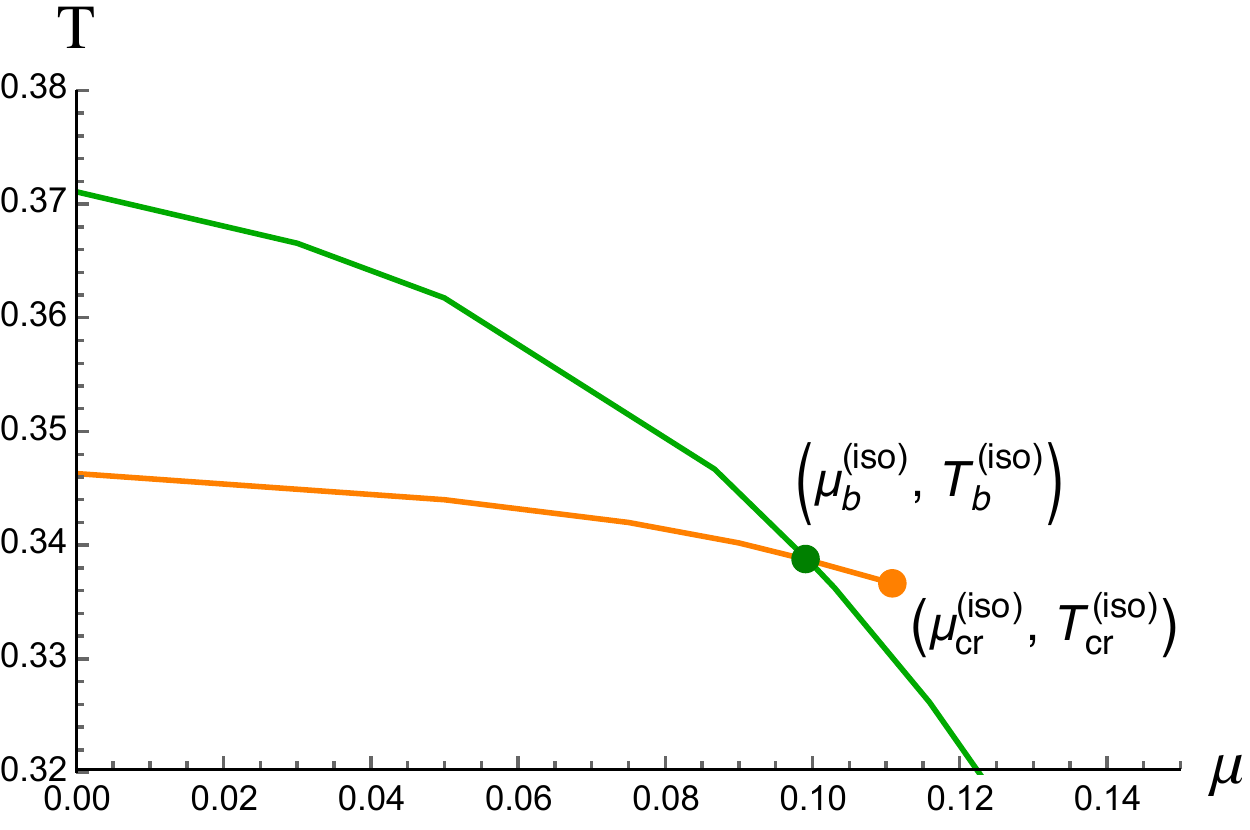} \\
  A \hspace{200pt} B
  \caption{Confinement-deconfinement phase transition in the isotropic
    case, $\nu = 1$, $c = -1$ for the Wilson line $W_{iso}$ (green)
    and the background (orange) (A). Plot (B) is a zoom of (A) near
    the intersection point $(\mu_b^{(iso)},T_b^{(iso)})$ and critical
    point $(\mu_{cr}^{(iso)},T_{cr}^{(iso)})$.}
  \label{Fig:Conf-Deconf-iso}
\end{figure} 
\begin{figure}[h!]
  \centering
  \includegraphics[scale=0.3]{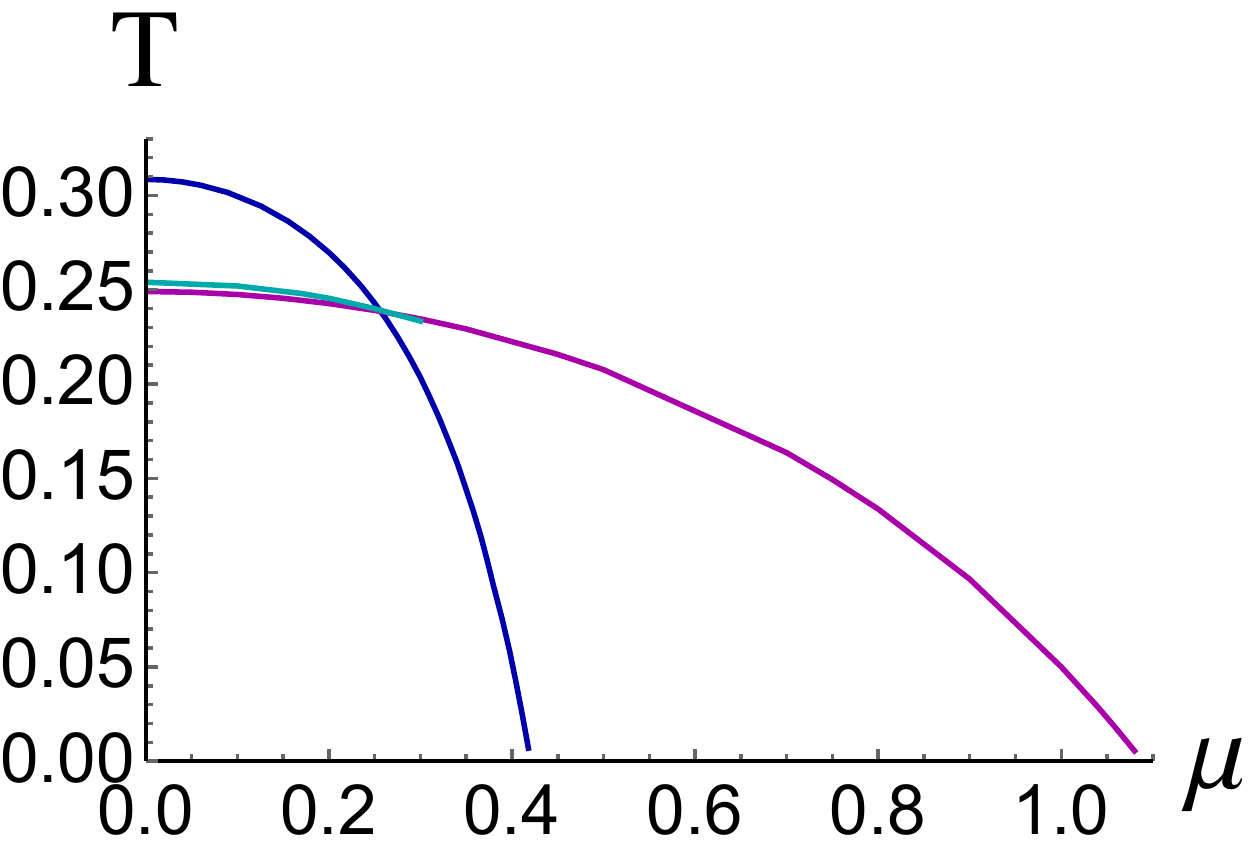} \qquad 
  \includegraphics[scale=0.8]{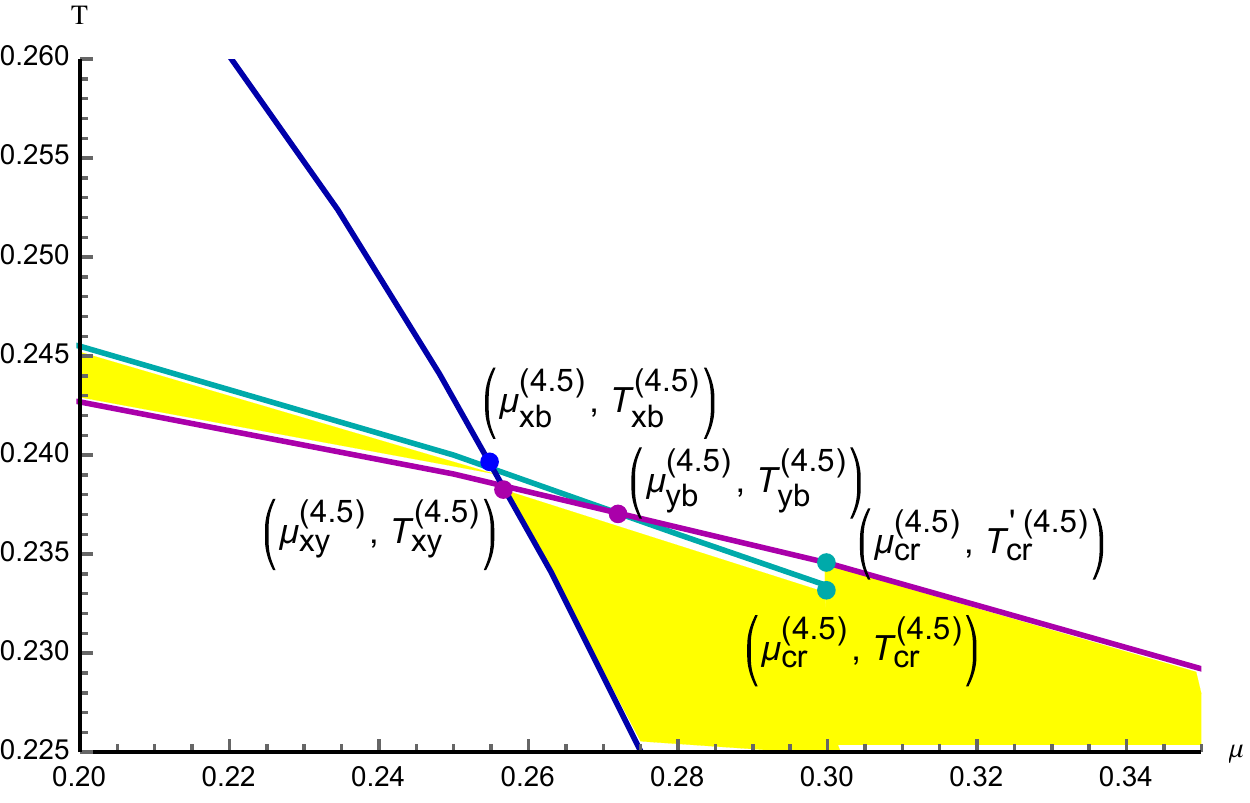}\\
  A \hspace{200pt} B\\$\,$\\
  \includegraphics[scale=0.3]{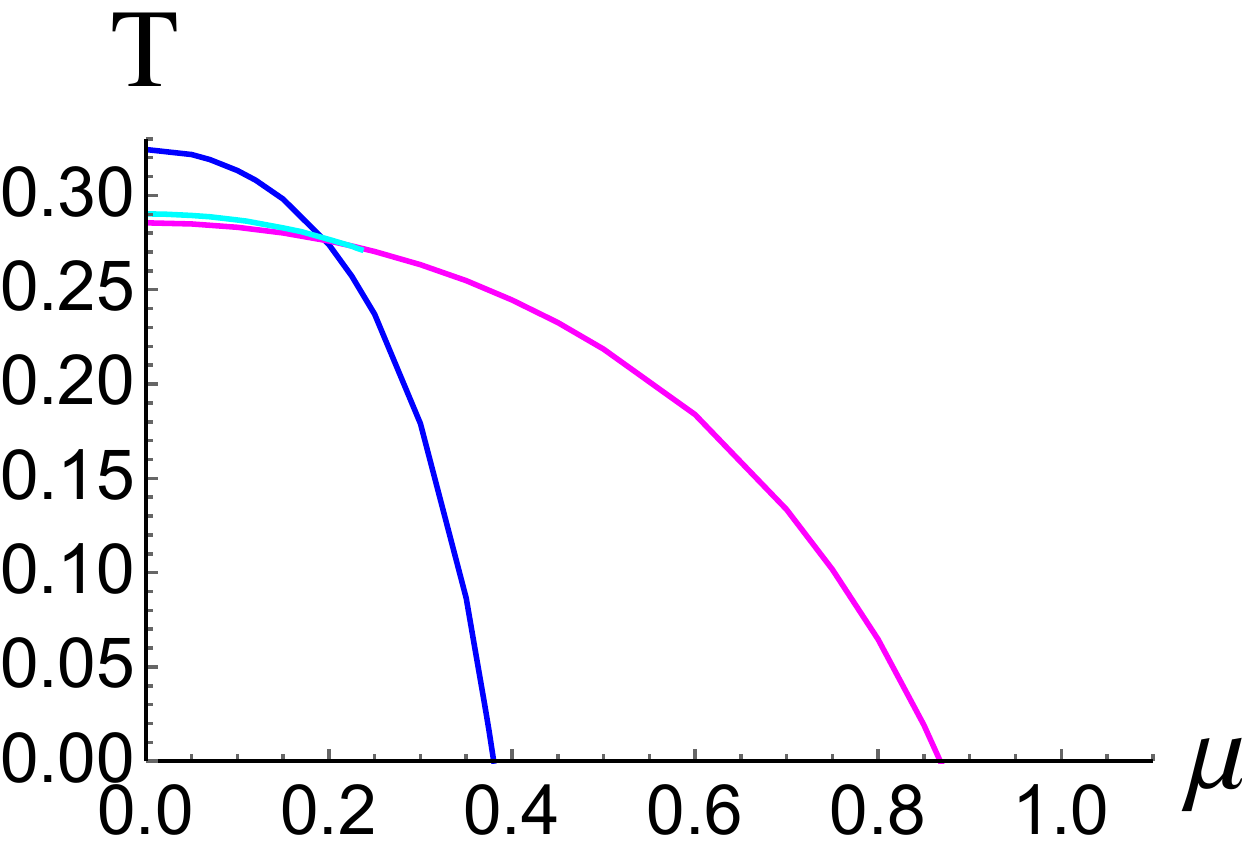}\qquad 
  \includegraphics[scale=0.8]{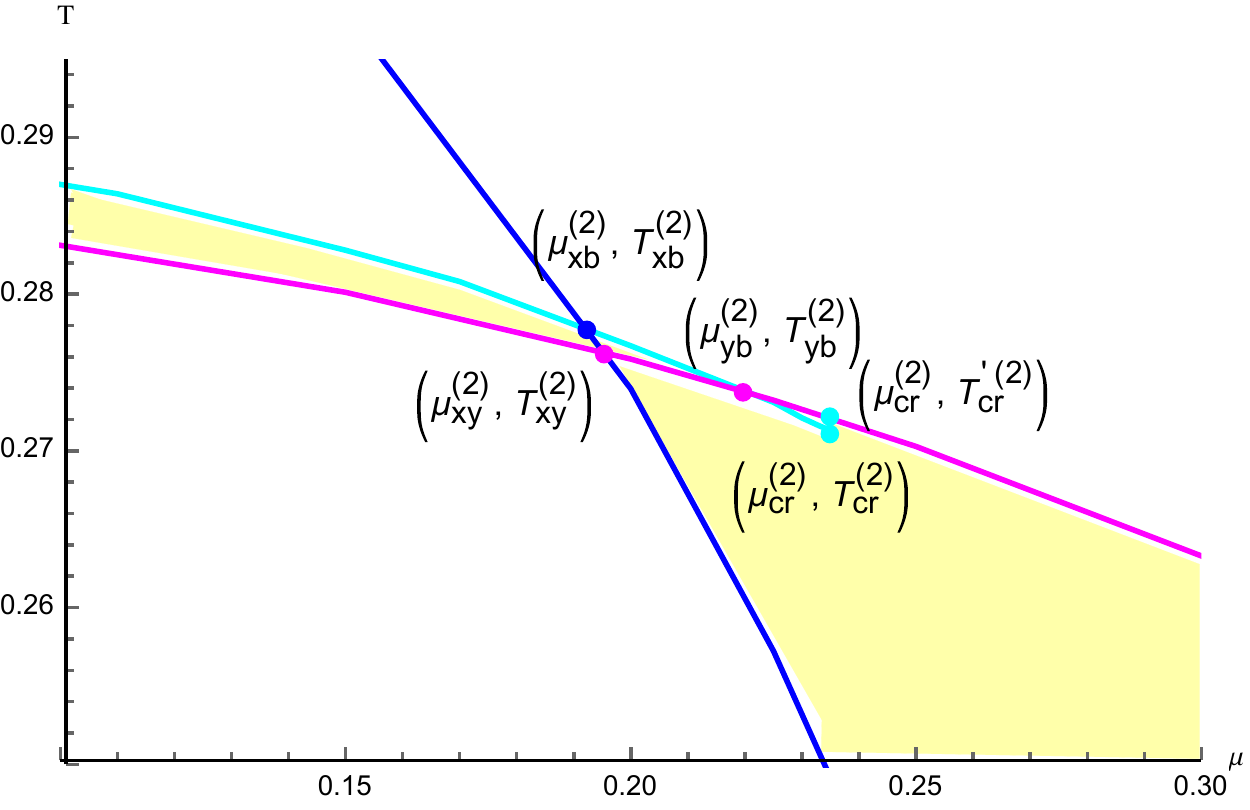}\\
  C \hspace{200pt} D
  \caption{Confinement-deconfinement phase transition in the
    anisotropic case, $\nu = 4.5$, $c = -1$, for the Wilson line $W_x$
    (blue), the Wilson line $W_y$ (magenta) and the background (cyan)
    (A). Plot (B) is a zoom of (A) near the critical points
    $(\mu_{xb},T_{xb})$, $(\mu_{yb},T_{yb})$ and FOPT
    $(\mu_{b},T_{b})\to (\mu_{b},T^{\prime}_{b})$. Plots (C) and (D)
    are the analogues of (A) and (B) for $\nu = 2$.}
  \label{Fig:Conf-Deconf-anis}
\end{figure}

\begin{figure}[h!]
  \centering
  \includegraphics[scale=1.1]{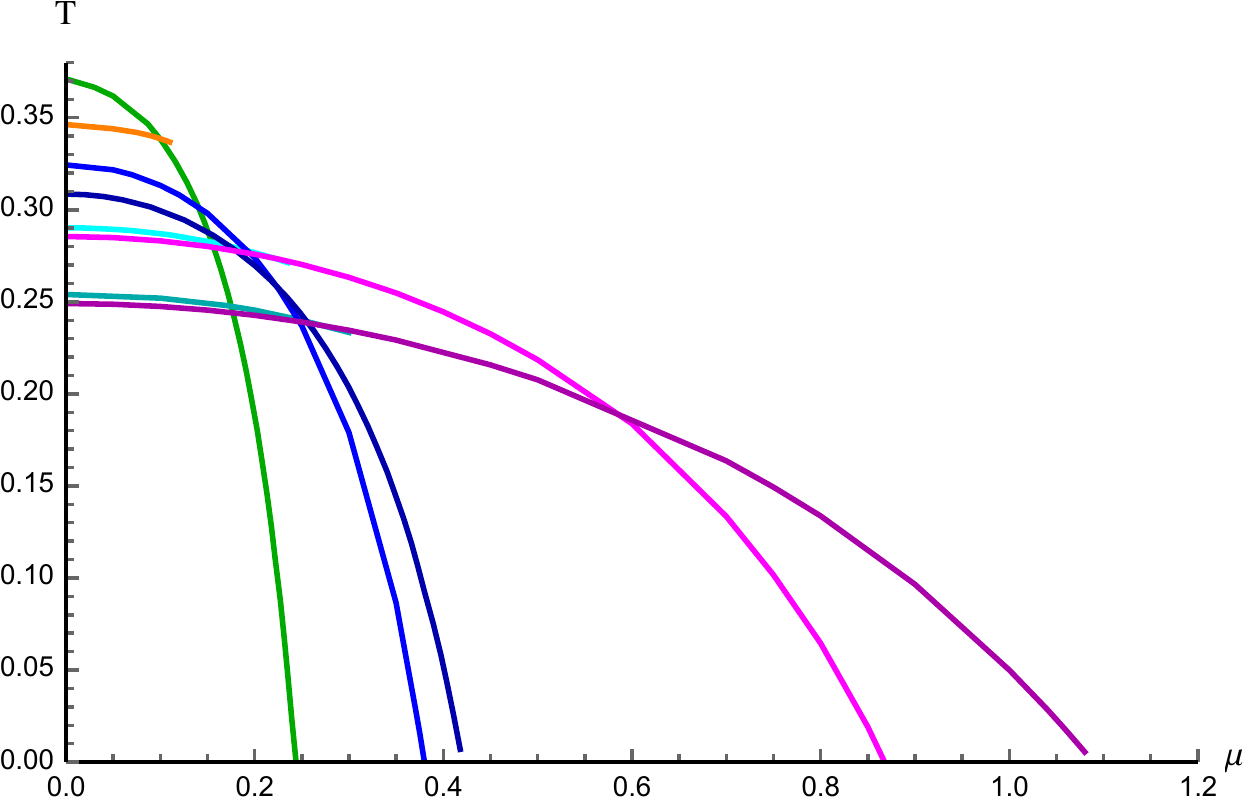}
  \caption{Confinement-deconfinement phase transitions of Wilson lines
    $W_{iso}$ for $\nu = 1$ (green), $W_x$ and $W_y$ for $\nu = 2$
    (blue and magenta) and $\nu = 4.5$ (dark blue and dark magenta);
    the background transition lines for $\nu = 1$ (orange), $\nu = 2$
    (cyan) and $\nu = 4.5$ (dark cyan).}
  \label{Fig:Conf-Deconf-isotropization}
\end{figure}

\subsubsection{Phase transition lines and critical points}\label{Sec:PTL}

Confinement/deconfinement phase transition for the isotropic case $\nu
= 1$ is shown on Fig.\ref{Fig:Conf-Deconf-iso}. Note that for zero
chemical potential the Hawking-Page temperature is less than the
temperature of the confinement/deconfinement transition temperatute,
$T_{HP}(0) < T_{CD}(0)$. The temperature of the black hole to black
hole transition $T_{BB}(\mu)$ is less than the temperature of the
confinement/deconfinement transition $T_{CD}(\mu)$ for $0 < \mu <
\mu_b^{(iso)}$, i.e. $T_{BB}(\mu) < T_{CD}(\mu)$, and $T_{BB}(\mu) >
T_{CD}(\mu)$ for $\mu_b^{(iso)} < \mu < \mu_{cr}^{(iso)}$ (let us
remind that the background transition line stops at
$(\mu_{cr}^{(iso)},T_{cr}^{(iso)})$). Therefore,  the phase transition
line for $0 < \mu < \mu_b^{(iso)}$ is determined by the background
transition line (the orange lines at Fig.\ref{Fig:Conf-Deconf-iso}) 
and for $\mu \ge \mu_b^{(iso)}$ by   the isotropic
confinement/deconfinement transition line (the green lines at 
Fig.\ref{Fig:Conf-Deconf-iso}). This is in agreement with results of
the previous studies \cite{yang2015} and refs. therein, where it has
been argued that the transition for $\mu < \mu_b^{(iso)}$ is the
first-order phase transition (FOPT) and for $\mu > \mu_b^{(iso)}$ is
a smooth one.

The plots in Fig.\ref{Fig:diff-WxWy}  explain the difference between
phase diagrams of Wilson lines $W_x$  and $W_y$ for $\nu = 4.5$.  In
this plot we see that for $\mu = 0$ the minimal value of the horizon
$z_{h,DWx}$, for which the DW can appears for $W_x$, corresponds to
small black hole, meanwhile the same horizon for $W_y$, $z_{h,DWy}$,
corresponds to large black hole. By this reason we have
\be
  T_{DWy}(0) < T_{HP}(0) < T_{DWx}(0).
\ee
Explicit numerical calculations show that 
\bea
  &T_{CDy}(\mu) < T_{BB}(\mu) < T_{CDx}(\mu) \quad {\mbox{for}}
  \quad 0 < \mu < \mu_{xb}.
\eea

The confinement/deconfinement  transition for the anisotropic case
$\nu = 4.5$ is shown on Fig.\ref{Fig:Conf-Deconf-anis}.A and B. The
phase diagram for the longitudinal Wilson line $W_x$ is depicted by
the blue lines, for transversal lines $W_y$ by the magenta lines and
for the anisotropic background by the cyan lines. Wilson lines can
also have arbitrary orientations, that corresponds to a modification
of blue and magenta lines to some intermediate configuration.

The case $\nu = 2$ can be considered as intermediate between the
isotropic one and our main case $\nu = 4.5$
(Fig.\ref{Fig:Conf-Deconf-anis}.C and
D). Fig.\ref{Fig:Conf-Deconf-isotropization} combines all the three
cases $\nu = 1, 2, 4.5$ thus showing isotropization.

\newpage
\section{Conclusion and discussion}

We have considered 5-dimensional Einstein-dilaton-two-Maxwell-scalar
system. We have found anisotropic solutions for this system by using
the potential reconstruction method, i.e. choosing the corresponding
dilaton and Maxwell potentials for the given background. This method
has been used for isotropic cases in \cite{1301.0385, yang2015,
  Li:2017tdz} and refs therein.

Our anisotropic background is the deformed AdS$_5$ that has the UV
boundary with two suppressed transversal coordinates and the IR
boundary with the suppressed time and longitudinal
coordinates. One can say that in this background two different
3-dimensional reductions, one in the UV domain and the other in the IR
domain, are realized. The UV reduction is realized by the suppression
of the original transversal coordinates, meanwhile the IR reduction is
obtained by the suppression of the longitudinal and temporal
coordinates. For the corresponding isotropic solution there is no
3-dim reduction neither in UV nor in IR regions.

In our calculations the warped factor is chosen in
  such a way that the explicit analytical calculations can be
performed. This solution can be generalized to provide a more
realistic model. In this case, similar to the isotropic case, the
solution can be given only in terms of quadratures. We solved the
equations of motion to obtain a family of the black hole solutions by
modifying the initial potential corresponding to zero temperature. In
this construction the special boundary conditions for the dilaton
field are chosen, namely we have required that the dilaton field is
zero at the horizon.

We have also studied the thermodynamical properties
of the constructed black hole background and found the large/small
black hole phase transitions at the temperature magenta
$T_{BB}(\mu)$. This result is presented in Fig.\ref{Fig:BCKFT}. At 
$\mu = 0$ and for $T < T_{HP}(0)$, the black hole dissolves to thermal
gas which is thermodynamically stable for $T < T_{HP}(0)$. When the
system cools down with the chemical potential less than the critical
value $\mu_{cr}$, the background undergoes the
phase transition from a large to a small black hole. This is a
generalization of the corresponding effect in the isotropic case
\cite{9902170,1012.1864,1108.2029,1301.0385, yang2015}. We have found
that $T_{BB}^{(aniso)}(\mu) < T_{BB}^{(iso)}(\mu)$ and the value of
the critical chemical potential, value up to which this phase
transition exists, is bigger in the anisotropic case as the compare to
the isotropic one, $\mu^{(\nu)}_{cr} > \mu_{cr}^{(iso)}$, see
Fig.\ref{Fig:Conf-Deconf-isotropization}. Also, we have found that the
point $(\mu^{(\nu)}_{cr},T^{(\nu)}_{cr})$ for $\nu\to 1$ goes smoothly
to $(\mu_{cr}^{(iso)}, T_{cr}^{(iso)})$.

We have studied the behavior of the temporal Wilson loops in the
constructed background. For this purpose we have considered open
strings in these backgrounds and identified the two ends of an open
string as a quark and antiquark pair in the dual holographic QCD. As
in the isotropic case, the equations of motion for this open string
support two configurations -- U-shape and straight-shape of the open
strings. The U-shape for large distances between
  quarks provides the quark confinement and is realized in the
presence of the dynamical wall (DW). We have found the domains in the
$(z_h,\mu)$ planes, where the DW can appear for the longitudinal and
transversal orientation of the temporal Wilson loops. In these regions
the open strings cannot exceed the DW, even the separation of the
quark and antiquark goes to infinite and the quark confinement takes
place. We have found that the phase diagram depends on the
orientation, cf.\cite{Arefeva:2016rob}. Taking
  into account the instability zones of the anisotropic background,
we have found more complicated confinement/deconfinement phase
diagrams for different oriented temporal Wilson loops and the  details
are the following:
\begin{itemize}
\item In the case of the longitudinal orientation, $W_{Tx}$, parts of
  regions near zero values of the chemical potential, $\mu <
  \mu_{xb}$, enter to the instability regions of our background,
  where the small black holes collapse to large ones. Here the horizon
  suddenly blows up to pass the critical value $z_{h,DWx}(\mu)$ (see
  Fig.\ref{Fig:Tvszh} and Fig.\ref{Fig:DDW45zh-mu}, at the last plots
  these jumps are indicated by the arrows), so that the confinement
  phase transforms to the deconfinement one by a phase
  transition. While the chemical potential is greater than the
  critical value $\mu > \mu_{xb}$, the black hole horizon grows
  gradually and continuously passes the critical horizon,
  corresponding to $(\mu_{cr},T_{cr})$, so that the confinement phase
  transforms to the deconfinement phase smoothly. In other words, the
  confinement-deconfinement line is determined by the probe string
  behavior itself. It is worth to notice that the similar situation
  takes place in the isotropic case.
\item In the case of the transversal orientation, $W_y$, situation
  is more interesting. It happens, that the
  background phase transition line for small $\mu$, $\mu < \mu_{yb}$,
  is located above  the phase transition line for the Wilson line, and
  for small $\mu$ we have a smooth confinement-deconfinement phase
  transition. For $\mu_{yb} < \mu < \mu_{cr}$ we fall in the zone of
  instability of the background and the first order phase transition
  takes place.
\end{itemize}

\ \\

As to the future investigations,   the following natural
questions to static and non-static properties of our
model are worth noting. As to static properties, it is natural to
\begin{itemize} 
\item investigate the opportunity to fix our holographic QCD model by
  a  suitable choice of the function $P(z)$ in \eqref{bP}, so that in
  the isotropic limit it would fit the Cornell potential known by
  lattice QCD;  it would be interesting to perform calculations on an
  anisotropic lattice and compare these future results with our model;
\item consider more general anisotropic backgrounds and derive the
  corresponding aniso\-tropic RG flows; 
\item study the Regge spectrum for mesons, adding the probe gauge
  fields to the backgrounds; we expect that similarly to the
  isotropic case, the gauge potential can be fixed requiring the
  linear Regge spectrum for mesons;
\item consider estimations for direct photons;
\item evaluate transport coefficients and their dependence on the
  anisotropy.
\end{itemize}

As to the thermalization processes, which are the
  main motivations of our considered of the
anisotropic background, we suppose to reexamine
\begin{itemize}
\item the shock wave collisions in the anisotropic background with the
  warped factor;
\item thermalization times for 2-point correlators; for the no-dilaton
  case this question has been addressed in \cite{1503.02185,
    Aref'eva:2016doe};
\item time dependence of the transport coefficients, for the
  no-dilaton model see \cite{Arefeva:2016rob}.
\end{itemize}

In this paper we have studied a particular anisotropic model specified
by the anisotropy parameter $\nu$ and in all plots we take $\nu =
4.5$, since just this case reproduces the total multiplicity
dependence on energy, ${\cal M} \sim s^{0.155}$. It would be
interesting to find isotropization of our solution and it is natural
to expect that in this case the both phase transition lines, the
large/small black hole transition and the string
confinement/deconfinement transition, will smoothly move to their
isotropic parters. We leave these matters to future works.

\section*{Acknowledgments}

This work was presented at the Helmholtz International Summer School
"Hadron Structure and Hadronic Matter, and Lattice QCD",
20.08.2017-2.09.2017, BLTP JINR, Dubna, Russia, the 9th mathematical
physics meeting: "Summer School and Conference on Modern Mathematical
Physics", 18.09.2017-23.09.2017, Belgrade, "IV Russian-Spanish
Congress Particle and Nuclear Physics at all Scales and Cosmology",
4.09.2017-8.09,2017, Dubna.

$$\,$$

\newpage
{\bf \Large {Appendix}}

\appendix
\section {Equations of motion}\label{Sec:Appendix}

\begin{gather}
  \begin{split}
    00: \quad &b'' - \cfrac{(b')^2}{2 b} + b' \left( \cfrac{g'}{2 g} -
      \cfrac{2}{\nu z} \right) + \cfrac{2 b}{z^2} \left( \cfrac{1}{3}
      + \cfrac{2}{3 \nu} + \cfrac{1}{\nu^2} \right) - \cfrac{b g'}{3
      z g} \left( 1 + \cfrac{2}{\nu} \right) + \cfrac{b}{6} \
    (\phi')^2 + \cfrac{b^2 V}{3 z^2 g} \ + \\
    &\quad + \cfrac{z^2 f_1 (A_t')^2}{6 g} + \cfrac{q^2 z^{- 2 +
        \frac{4}{\nu}} f_2}{6 g} = 0, \\
    11: \quad &g'' + \cfrac{3 g b''}{b} - \cfrac{3 g (b')^2}{2 b^2} -
    \cfrac{6 g b'}{b \nu z} - g' \left( \cfrac{2}{z} + \cfrac{4}{\nu
        z} -  \cfrac{3 b'}{b} \right) + \cfrac{2 g}{z^2} \left( 1 +
      \cfrac{2}{\nu} + \cfrac{3}{\nu^2} \right) + \cfrac{g}{2} \
    (\phi')^2 + \cfrac{b V}{z^2} \ + \\
    &\quad - \cfrac{z^2 f_1 (A_t')^2}{2b} + \cfrac{q^2 z^{- 2 +
        \frac{4}{\nu}} f_2}{2 b} = 0, \\
    22: \quad &g'' + \cfrac{3 g b''}{b} - \cfrac{3 g (b')^2}{2 b^2} -
    3 \ \cfrac{g b'}{z b} \left( 1 + \cfrac{1}{\nu} \right) - g'
    \left( \cfrac{4}{z} + \cfrac{2}{\nu z} - \cfrac{3 b'}{b} \right) 
    + \cfrac{2 g}{z^2} \left( 3 + \cfrac{2}{\nu} + \cfrac{1}{\nu^2}
    \right) + \cfrac{g}{2} \ (\phi')^2 \ + \\ 
    &\quad + \cfrac{b V}{z^2} - \cfrac{z^2 f_1 (A_t')^2}{2b} -
    \cfrac{q^2 z^{- 2 + \frac{4}{\nu}} f_2}{2 b} = 0, \\
    44: \quad &\quad - \cfrac{6 (b')^2}{b^2} + \cfrac{12 b'}{z b}
    \left( 1 + \cfrac{1}{\nu} \right) + \cfrac{g'}{g} \left(
      \cfrac{2}{z} + \cfrac{4}{\nu z} - \cfrac{3 b'}{b} \right) -
    \cfrac{4}{z^2} \left( 1 + \cfrac{4}{\nu} + \cfrac{1}{\nu^2}
    \right) + (\phi')^2 - \cfrac{2 b V}{z^2 g} \ - \\
    &\quad - \cfrac{z^2 f_1 (A_t')^2}{b g} - \cfrac{q^2 z^{- 2 +
        \frac{4}{\nu}} f_2}{b g} = 0,
  \end{split}\label{eq:2.23}
\end{gather}
where $'= \partial/\partial z$.

These equations can be transformed to the following equations
\bea
  &&g'' + g' \left( \cfrac{3 b'}{2 b} - \cfrac{1}{z} - \cfrac{2}{\nu
      z} \right) - \cfrac{z^2}{b} \ f_1 (A_t')^2 = 0, \label{1-st} \\
  && 2 g' \left( 1 - \cfrac{1}{\nu} \right) + g \left( 1 -
    \cfrac{1}{\nu} \right) \left( \cfrac{3 b'}{b} - \cfrac{4}{z} -
    \cfrac{4}{\nu z} \right) + \cfrac{q^2 z^{- 1 + \frac{4}{\nu}}
    f_2}{b} = 0, \label{2-nd} \\
  &&b'' - \cfrac{3 (b')^2}{2 b} + \cfrac{2 b'}{z} - \cfrac{4 b}{3 \nu
    z^2} \left( 1 - \cfrac{1}{\nu} \right) + \cfrac{b}{3} \ (\phi')^2
  = 0, \label{3th} \\
  && \cfrac{b''}{b} + \cfrac{(b')^2}{2b^2} + \cfrac{3 b'}{b} \left(
    \cfrac{g'}{2 g} - \cfrac{1}{z} - \cfrac{1}{\nu z} \right) -
  \cfrac{g'}{3 z g}  \frac{5\nu +4}{\nu} +
  \cfrac{4(2\nu^2+3\nu+1)}{3z^2 \nu^2}  + \cfrac{g''}{3 g} + \cfrac{2
    b V}{3 z^2 g} = 0. \nn \\ \label{4th}
\eea
Substituting expressions for $g^{\prime\prime}$ from \eqref{1-st} and
$b^{\prime\prime}$ from \eqref{3th} into \eqref{4th} we get the
expression for $V$ without $g^{\prime\prime}$ and $b^{\prime\prime}$
\bea
  V = - \, \frac{z^4 A_t^{\prime \,2} f_1}{2 b^2} - \frac{3 z^2 b'
    g'}{2 b^2} - \frac{3 z^2 g b^{\prime \, 2}}{b^3} + \frac{9 z g
    b'}{2 \nu b^2} + \frac{15 z g b'}{2 b^2} + \frac{z g'}{\nu b} +
  \frac{2 z g'}{b} + \frac{z^2 g \phi^{\prime \, 2}}{2b} - \frac{8
    g}{\nu b} - \frac{4 g}{b}, \nn \\ \label{Pot}
\eea
that is nothing but the constraint equation.
   
One can check explicitly, that from eqs. \eqref{1-st}--\eqref{3th} and
\eqref{Pot} follows the field equation for the scalar field
\eqref{eq:2.25}. Indeed, differentiating \eqref{Pot} we get
\bea
  &&\frac{b}{z^2 g} \frac{\partial V}{\partial \phi} = \frac{1}{2 \nu
    z^2 b^3 g   \phi'} \label{V-prime} \\
  &&\Big\{2 z b b' \left(\nu z^3 A_t^{\prime \, 2} f_1 - 3 g \left(2
      \nu z b'' + (7 \nu + 3) b' \right) \right) - \nn \\
  &&- \, b^2 \Big[z (\nu  z (z^2 A_t^{\prime \, 2} \phi' f_1' + 3 b''
  g') + b' (3 \nu z g'' + (- 5 \nu - 7) g')) + \nn \\
  &&+ \, 2 \nu z^3 A_t' f_1(\phi) \left(z A_t'' + 2 A_t' \right) + g
  \left( b' \left( - 23 \nu + \nu z^2 \phi^{\prime \, 2} - 25 \right)
    - 3 (5 \nu + 3) z b'' \right) \Big] + \nn \\
  &&+ \, 18 \nu z^2 g b^{\prime \, 3} + b^3 \left[2 z \left((2 \nu +
      1) g'' + \nu g \phi' \left(z \phi'' + \phi' \right) \right) + g'
    \left( - 4 \nu + \nu z^2 \phi^{\prime \, 2} - 14 \right) \right]
  \Big\}. \nn
\eea
   
Substituting in the R.H.S. of \eqref{V-prime} $g^{\prime\prime}$ and
$b^{\prime\prime}$ from \eqref{1-st} and \eqref{3th}, respectively, we
get the following expression for $\frac{\partial V}{\partial \phi}$
without $g^{\prime\prime}$ and $b^{\prime\prime}$:
\bea
  &&\frac{b}{z^2 g} \frac{\partial V}{\partial \phi} = \frac{1}{4
    \nu^3 z^3 b^2 g \phi'} \Big\{ 2 \nu z b \Big[ \nu z \left(\nu z^3
    A_t^{\prime \, 2} \phi' f_1' + 2 (\nu - 1) b' g'
  \right) + \label{parV} \\
  &&+ \, 2 (\nu - 1) \nu z^3 A_t^{\prime \, 2} f_1 + g b' \left( - 7
    \nu^2 - 9 \nu + 3 \nu^2 z^2 \phi{\prime \, 2} + 16 \right) \Big] +
  9 (\nu - 1) \nu^2 z^2 g b^{\prime \, 2} + \nn \\
  &+& 2 b^2 \Big[2 \nu z g' \left(- 4 \nu + \nu^2 z^2 \phi{\prime \,
      2} + 4 \right) + g \left(4 \left(5 \nu^2 - 2 \nu - 3 \right) + 2
    \nu^3 z^3 \phi' \phi'' - 3 (\nu + 1) \nu^2 z^2 \phi{\prime \, 2}
  \right) \Big] \Big\}. \nn
\eea
In a similar way we get the expression for $f_2'$ without
$g^{\prime\prime}$ and $b^{\prime\prime}$. Indeed, we differentiate
\eqref{2-nd} and obtain
\bea
  \frac{q^2 z^{\frac{4}{\nu} - 2}}{2 b g } \, \frac{\partial
    f_2}{\partial \phi} = \frac{q^2 z^{\frac{4}{\nu} - 2} f_2'}{2 b g
    \phi'} = &-& \frac{(\nu - 1) }{2 \nu ^3 z^3 b g \phi'} \Big\{\nu z
  \left(5 \nu z b' g' - g \left[(\nu + 16) b' - 3 \nu z b'' \right)
  \right] + \nn \\
  &+& 2 b \left[\nu z \left(\nu z g'' - (\nu + 6) g' \right) + 8 (\nu
    + 1) g \right] \Big\}. \label{f2-prime}
\eea
Then we substitute there \eqref{1-st} and \eqref{3th} and get
\bea
  &&\frac{q^2 z^{\frac{4}{\nu} - 2}}{2 b g} \, \frac{\partial
    f_2}{\partial \phi} = \frac{(\nu - 1)}{4 \nu^3 z^3 b^2 g \phi'}
  \Big\{ 2 \nu z b \left( b' \left( (7 \nu + 16) g - 2 \nu z g'
    \right) - 2 \nu z^3 A_t{\prime \, 2} f_1(\phi) \right) - \nn \\
  &&- \, 9 \nu^2 z^2 g b^{\prime \, 2} + 2 b^2 \left(8 \nu z g' + g
    \left( - 20 \nu + \nu^2 z^2 \phi^{\prime \, 2} - 12 \right)
  \right) \Big\}. \label{f2-prime-m}
\eea
From \eqref{parV} and \eqref{f2-prime-m} we get
\bea 
  \cfrac{q^2 z^{- 2 + \frac{4}{\nu}}}{8 b g} \ \cfrac{\partial
    f_2}{\partial \phi} + \cfrac{b}{z^2 g} \ \cfrac{\partial
    V}{\partial \phi} =\phi'' + \phi' \left( \cfrac{g'}{g} + \cfrac{3
      b'}{2 b} - \cfrac{\nu + 2}{\nu z} \right) + \cfrac{z^2
    (A_t')^2}{2 b g} \ \cfrac{\partial f_1}{\partial
    \phi}. \label{eq:2.25mm}
\eea
This is nothing but E.O.M. for the scalar field.  
  
\section{Simplest solutions to system
  \eqref{1-st-t}--\eqref{A0}}\label{Sec:simplest-sol}

\subsection{Solutions for anisotropic metric  $ \nu \neq 1$ and $c =
  0$, $A_t \neq 0$}

The system \eqref{1-st-t}--\eqref{A0} for anisotropic metric $\nu \neq
1$ and $b(z) = 1$, $A_t \neq 0$ has the form
\bea
  g'' + g' \left( - \cfrac{1}{z} - \cfrac{2}{\nu z} \right) - z^2 \
  f_1 A_t^{\prime \,2 } &=& 0, \label{1-st-s} \\
  - \ \cfrac{4}{ \nu z^2} \left( 1 - \cfrac{1}{\nu} \right) +
  \phi^{\prime \, 2} &=& 0, \label{2-nd-s} \\
  2 g' \left( 1 - \cfrac{1}{\nu} \right) + g \left( 1 - \cfrac{1}{\nu}
  \right) \left( - \cfrac{4}{z} - \cfrac{4}{\nu z} \right) + q^2 z^{-
    1 + \frac{4}{\nu}} \, f_2 &=& 0, \label{3-rd-s} \\
  - \ V - \frac{z^4}{2} \, A_t^{\prime \, 2} f_1 + \frac{z g'}{\nu} +
  2 z g' + \frac{z^2 g \ \phi^{\prime \, 2}}{2} - \frac{8 g}{\nu} - 4
  g &=& 0. \label{4-th-s}
\eea
Substituting $\phi^{\prime \,2}$ from \eqref{2-nd-s} into
\eqref{4-th-s} we get
\be
  - V - \frac{z^4}{2} \, A_t^{\prime \, 2} f_1 + \frac{z g'}{\nu} + 2
  z g' - \cfrac{2g}{\nu^2} -\frac{6 g}{\nu} - 4 g = 0. \label{4th-ss}
\ee
The variation of the action \eqref{action} over scalar field $\phi$
and  components $A^{(1)}_{\mu}$ of the first Maxwell fields' leads to
the following EOM:
\bea
  \phi'' + \phi' \left( \cfrac{g'}{g}  - \cfrac{\nu + 2}{\nu z}
  \right) + \cfrac{z^2 A_t^{\prime \, 2}}{2 g} \ \cfrac{\partial
    f_1}{\partial \phi} - \cfrac{q^2 z^{- 2 + \frac{4}{\nu}}}{2 g} \
  \cfrac{\partial f_2}{\partial \phi} - \cfrac{b}{z^2 g} \
  \cfrac{\partial V}{\partial \phi} &=& 0, \label{phi-s} \\
  A_t'' + A_t' \left( \cfrac{f_1'}{f_1} - \cfrac{2 - \nu}{\nu z}
  \right) &=& 0. \label{A0-s}
\eea
The last equation \eqref{A0-s} gives
\be
  \log A_t' = - \log f_1 + \cfrac{2 - \nu}{\nu} \log z + \log C_1
  \Rightarrow \, A_t'= C_1 \frac{z^{ \frac{2 - \nu}{\nu}}}{f_1}.
\ee
If we take $f_1$ as in \eqref{f1} we get
\bea
  A_t'= C_1\frac{z^{ \frac{2 - \nu}{\nu}}}{z^{- 2 + \frac{2}{\nu}}} =
  C_1 z, \\
  A_t(z) = \frac{C_1}{2} z^2 + C_2.
\eea
Taking into account 
\bea
  A_t(0) &=& C_2 = \mu, \\
  A_t(z_h) &=& \frac{C_1}{2} z_h^2 + \mu = 0,
\eea
we get
\be
  A_t (z) = \mu \ (1 - z^2/z_h^2).
\ee
Substituting this to the blacking function equation \eqref{1-st-s} we
get
\be
  g'' + g' \left( - \cfrac{1}{z} - \cfrac{2}{\nu z} \right) - {4}\mu^2
  \frac{z^{2 + \frac{2}{\nu}}}{z_h^4} = 0.
\ee
    
From \eqref{2-nd-s} we get for the scalar field
\be
  \phi^\prime = \pm \ \sqrt{\cfrac{4}{\nu z^2} \left( 1 -
      \cfrac{1}{\nu} \right)} = \frac{2}{\nu \, z} \, \sqrt{\nu -
    1} \label{3th-tt}
\ee
and 
\be
  \phi = \frac{2}{\nu} \, \sqrt{\nu - 1} \ \log z + C_5.
\ee   

From \eqref{3-rd-s} we get $f_2$
\bea
  f_2 &=& - \ q^2 z^{- 1 + \frac{4}{\nu}}\, \Big[2 g' \left( 1 -
    \cfrac{1}{\nu} \right) + g \left( 1 - \cfrac{1}{\nu} \right)
  \left( - \cfrac{4}{z} - \cfrac{4}{\nu z} \right)
  \Big] = \label{2-nd-sss}
\eea
and from \eqref{4-th-s} we get
\bea
  V &=& - \ \frac{z^4}{2} \, A_t^{\prime \, 2} f_1 + \frac{z g'}{\nu}
  + 2 z g' + \frac{z^2 g \phi^{\prime \, 2}}{2} - \frac{8 g}{\nu} - 4g
  = \label{4th-ss}
\eea

\subsection{Solutions for anisotropic metric $\nu \neq 1$ and $c = 0$,
  $A_t = 0$}

From the previous consideration it is easy to reproduce the solution
has been found in \cite{AGG}.

From \eqref{1-st-t} we find 
\be
  g(z) = 1 - \cfrac{z^{2 + \frac{2}{\nu}}}{z_h^{2 +
      \frac{2}{\nu}}}, \label{blacking}
\ee
Second, we substitute \eqref{blacking} into \eqref{2-nd-t} and get
\be
  f_2(z) = \cfrac{4 z^{-4/\nu}}{q^2} \ \frac{(\nu - 1)(1 + 3 \nu + 2
      \nu^2)}{\nu^2 \ (1 + 2 \nu)}, \label{f2} 
\ee
Then we solve \eqref{3th-t}:
\be
  \phi(z) = C_5 \pm 2 \ \cfrac{\sqrt{\nu - 1}}{\nu} \
  \log(z), \label{phi}
\ee
Finally, we have to find $V$ from \eqref{4-th-t} and check the scalar
equation \eqref{eq:2.25}. Substituting \eqref{blacking} into
\eqref{4-th-t} we obtain
\be
  V(z) = - \ 2 \ \cfrac{(1 + \nu)(1 + 2 \nu)}{\nu^2}. \label{V}
\ee
Hence we get the solution given by \eqref{blacking}, \eqref{f2},
\eqref{phi} and \eqref{V}, that coincides with \cite{AGG}.

\subsection{Vacuum solutions}\label{App:zerotemp}

Without black brane, i.e. for $g = 1$ the EOM
\eqref{1-st-t}--\eqref{A0} transform to:
\bea
  \cfrac{z^2}{b} \, f_1 A_t^{\prime \,2} &=& 0, \label{1-st-t-01} \\
  b'' - \frac{3 (b')^2}{2 b} + \frac{2 b'}{z} - \frac{4b}{3 \nu z^2}
  \left( 1 - \frac{1}{\nu} \right) + \frac{b}{3} \, (\phi')^2 &=&
  0, \label{2-nd-t-01} \\
  \left( 1 - \cfrac{1}{\nu} \right) \left( \cfrac{3 b'}{b} -
    \cfrac{4}{z} - \cfrac{4}{\nu z} \right) + \cfrac{q^2 z^{- 1 +
      \frac{4}{\nu}}}{b} \, f_2 &=& 0, \label{3-rd-t-01} \\
  - V - \frac{z^4}{2 b^2}\,A_t^{\prime \, 2} f_1 - \frac{3 z^2
    b^{\prime \, 2}}{b^3} + \frac{9 z b'}{2 \nu b^2} + \frac{15 z
    b'}{2b^2} + \frac{z^2 \phi^{\prime \,2}}{2b} - \frac{8}{\nu b} -
  \frac{4}{b} &=& 0, \label{4-th-t-01} \\
  \phi'' + \phi' \left( \cfrac{3 b'}{2 b} - \cfrac{\nu + 2}{\nu z}
  \right) + \cfrac{z^2 A_t^{\prime \, 2}}{2 b} \ \cfrac{\partial
    f_1}{\partial \phi} - \cfrac{q^2 z^{- 2 + \frac{4}{\nu}}}{2 b} \
  \cfrac{\partial f_2}{\partial \phi} - \cfrac{b}{z^2} \
  \cfrac{\partial V}{\partial \phi} &=& 0, \label{eq:2.25-01} \\
  A_t'' + A_t' \left( \cfrac{b'}{2 b} + \cfrac{f_1'}{f_1} - \cfrac{2 -
      \nu}{\nu z}\right) &=& 0. \label{A0-01} 
\eea
For $b = e^{\frac{c z^2}{2}}$:
\bea
  z^2 \, e^{-\frac{cz^2}{2}} f_1 A_t^{\prime \,2} &=&
  0, \label{1-st-t-02} \\
  \cfrac{c^2 z^2}{2} - 3c + \cfrac{4}{3} \, \cfrac{\nu - 1}{\nu^2 z^2}
  - \cfrac{(\phi')^2}{3} &=&
  0, \label{2-nd-t-02} \\
  \left( 1 - \cfrac{1}{\nu} \right) \left( 3cz - \cfrac{4(\nu +
      1)}{\nu z} \right) + z^{- 1 + \frac{4}{\nu}} \,
  e^{-\frac{c z^2}{2}} f_2 \, q^2 &=& 0, \label{3-rd-t-02} \\
  \cfrac{3 c^2 z^2}{2} - c \left( 2 + \cfrac{3}{\nu} \right) +
  \cfrac{4}{3 z^2} \left( 2 + \cfrac{3}{\nu} + \cfrac{1}{\nu^2} 
  \right) + \cfrac{2 e^{\frac{c z^2}{2}}}{3 z^2} \, V &=&
  0, \label{4-th-t-02} \\
  \phi'' + \phi' \left( \cfrac{3 c z}{2} - \cfrac{\nu + 2}{\nu z}
  \right) + \cfrac{z^2 \, e^{-\frac{c z^2}{2}}}{2} \, \cfrac{\partial
    f_1}{\partial \phi} \, A_t^{\prime \, 2} - \cfrac{z^{- 2 +
      \frac{4}{\nu}} \, e^{-\frac{c z^2}{2}}}{2} \, \cfrac{\partial
    f_2}{\partial \phi} \ q^2 &-& \nn \\
  - \, \cfrac{e^{\frac{c z^2}{2}}}{z^2} \, \cfrac{\partial V}{\partial
    \phi} &=& 0, \label{eq:2.25-02} \\
  A_t'' + A_t' \left( \cfrac{c z}{2} - \cfrac{2 - \nu}{\nu z} +
    \cfrac{f_1'}{f_1} \right) &=& 0 \label{A0-02} 
\eea
The system \eqref{1-st-t-02}--\eqref{A0-02} has the solution:
\bea
  A_t &=& \mu, \label{Ag1} \\
  f_2 &=& z^{-\frac{4}{\nu}} \ \cfrac{1 - \nu}{\nu^2 q^2} \left( 3 c
    z^2 - 4 (\nu + 1) \right) e^{\frac{c z^2}{2}}, \label{f2g1} \\
  \phi &=& \cfrac{1}{2 \sqrt{2} \, \nu} \left\{ \sqrt{3 c^2 \nu^2 z^4
      - 18 c \, \nu^2 z^2 + 8 \, (\nu - 1)} - \sqrt{3 c^2 \nu^2 z_*^4
      - 18 c \, \nu^2 z_*^2 + 8 \, (\nu - 1)} \right. + \label{phig1}
  \\
  &+& \left. 2 \sqrt{2 \, (\nu - 1)} \ln \left( \cfrac{z^2}{z_*^2}
    \right) - 3 \, \sqrt{3} \, \nu \ln\left( \cfrac{\sqrt{3 c^2 \nu^2
          z^4 - 18 c \, \nu^2 z^2 + 8 \, (\nu - 1)} - \sqrt{3} \, \nu
        \, (3 - c z^2)}{\sqrt{3 c^2 \nu^2 z_*^4 - 18 c \, \nu^2 z_*^2
          + 8 \, (\nu - 1)} - \sqrt{3} \, \nu \, (3 - c z_*^2)}
    \right) \right. - \nn \\
  &-& \left. 2 \sqrt{2 \, (\nu - 1)} \ln{\left( \cfrac{9 c \, \nu^2
          z^2 - 8 \, (\nu - 1) - \sqrt{2 \, (\nu - 1)} \, \sqrt{3 c^2
            \nu^2 z^4 - 18 c \, \nu^2 z^2 + 8 \, (\nu - 1)}}{9 c \,
          \nu^2 z_*^2 - 8 \, (\nu - 1) - \sqrt{2 \, (\nu - 1)} \,
          \sqrt{3 c^2 \nu^2 z_*^4 - 18 c \, \nu^2 z_*^2 + 8 \, (\nu -
            1)}} \right)} \right\}, \nn \\
  V &=& - \, \frac{1}{4} \left[ 9 c^2 z^4 - 6 c z^2 \left( 2 +
      \frac{3}{\nu} \right) + 8 \left( 2 + \frac{3}{\nu} +
      \frac{1}{\nu^2} \right) \right] e^{-\frac{c
      z^2}{2}} \label{Vg1}.
\eea
The particular form of factor $f_1$ doesn't matter as it is coupled
with the constant function $A_t$, so all the terms containing $f_1$
also include $A_t' \equiv 0$. The scalar field $\phi$ isn't influenced
by the assumption $g = 1$ and coincides with \eqref{phicneg}. The only
difference is that $z_h$ loses its sense as without the black brane
there is no horizon any more. Therefore we should replace $z_h$ by
some $z_*$, whose main property is $\phi(z_*) = 0$.

If we substitute \eqref{f2g1}--\eqref{Vg1} into the equation
\eqref{eq:2.25-02} its left-hand side disappears proving that the
system \eqref{1-st-t-02}--\eqref{A0-02} still remains a
self-consistent one.
\begin{figure}[h!]
  \centering
  \includegraphics[scale=0.7]{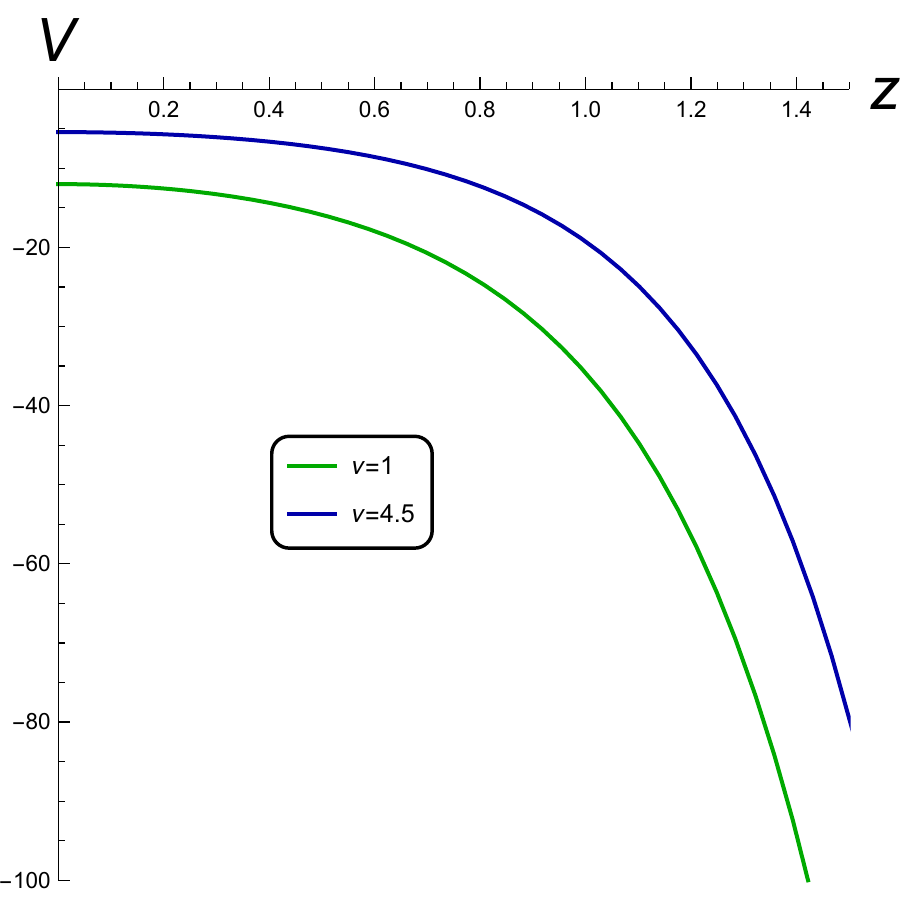} A\qquad
  \includegraphics[scale=0.7]{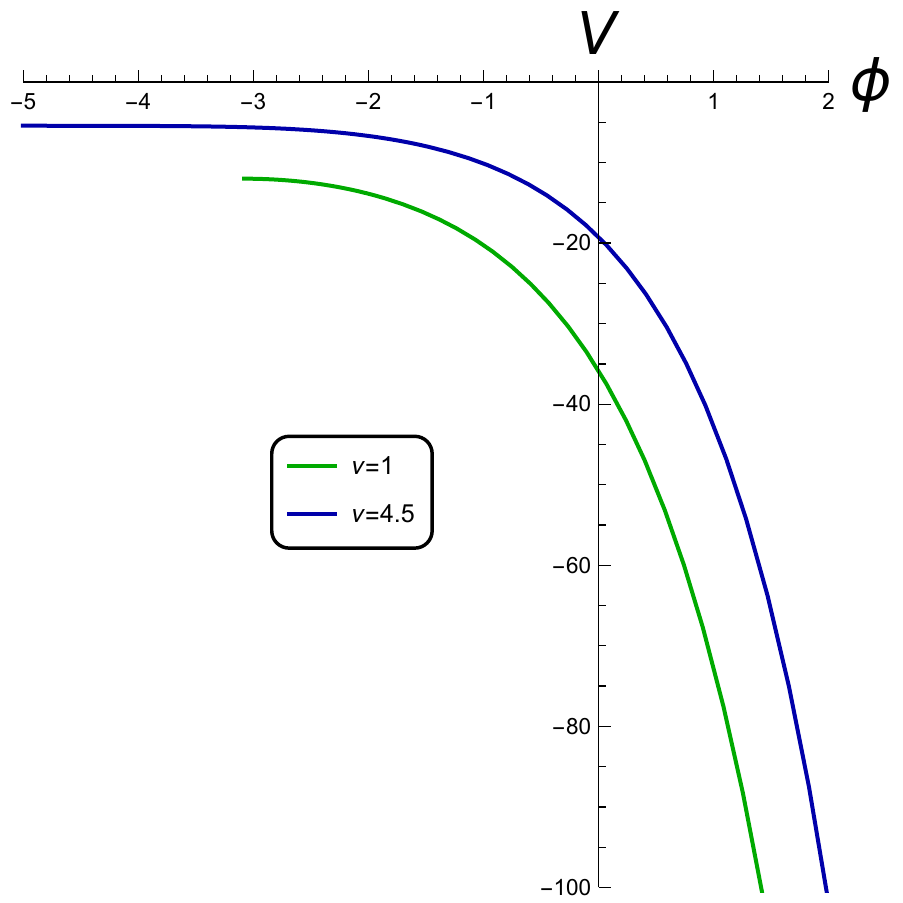}B
  \caption{Dilaton potentials that support vacuum solutions
    corresponding to isotropic and anisotropic cases $\nu = 4.5$ with
    $c = -1$ and initial conditions $\phi(1) = 0$ as function of $z$
    A) and $\phi$ B).}
  \label{Fig:Conf-Deconf}
\end{figure}

\end{document}